\documentclass[11pt]{article}
\usepackage{amsmath,amsfonts,amssymb,graphicx,theorem,multirow}
\usepackage[usenames]{color}
\usepackage{pstricks,cite} 
\usepackage{caption}
\usepackage{lscape}
\usepackage{graphicx}
\usepackage{float}

\usepackage[backref=false]{hyperref}
\numberwithin{equation}{section}
\numberwithin{figure}{section}
\usepackage[capitalise,noabbrev]{cleveref} 
\hypersetup{
colorlinks=true,
citecolor=red,
linkcolor=darkblue,
urlcolor=darkblue
}

\long\def\ignore#1{}

\definecolor{darkblue}{rgb}{0,0,.8}
\definecolor{red}{rgb}{1,0,0}
\definecolor{purple}{rgb}{1,0,1}
\definecolor{coloroflink}{rgb}{0.7,0,1}
\definecolor{coloroflink}{rgb}{0.180392, 0.545098, 0.341176}
\definecolor{darkpurple}{rgb}{1,.2,1}
\definecolor{newpurple}{rgb}{0.76,0.6,1}
\definecolor{pink}{rgb}{1,.7,.7}
\definecolor{lightblue}{rgb}{.61,.61,1}
\definecolor{midblue}{rgb}{.7,.7,1}
\definecolor{lightmidlightblue}{rgb}{.8,.8,1}
\definecolor{lightlightblue}{rgb}{.9,.9,1}
\definecolor{lightestblue}{rgb}{.96,.96,1}
\definecolor{lightpurple}{rgb}{1,.65,1}
\definecolor{darkgreen}{rgb}{0.180392, 0.545098, 0.341176}
\definecolor{mygray}{rgb}{.75,.75,.75}
\definecolor{lightlightgray}{rgb}{.85,.85,.85}
\definecolor{lightyellow}{rgb}{0.975,0.975,0.65}
\definecolor{lightgreen}{rgb}{1, 1., 1}

\theorembodyfont{\itshape} 
\theoremheaderfont{\scshape}
\theoremstyle{plain}

\newcommand{\nc}{\newcommand}
\nc{\bib}{\bibitem}

\nc{\proof}{{\scshape Proof.\ }} 				
\nc{\eproof}{{\hfill \rule{0.5em}{0.5em}\medskip}}		
\nc{\be}{\begin{equation}}
\nc{\ee}{\end{equation}}
\nc{\psun}{\psset{unit=0.4cm}}
\nc{\psdeux}{\psset{unit=0.3cm}}
\nc{\chit}{\protect\raisebox{0.25ex}{$\chi$}}
\def\facegrid#1#2{
\psframe[fillstyle=solid,fillcolor=lightlightblue,linewidth=0pt]#1#2
\psgrid[gridlabels=0pt,subgriddiv=1]#1#2}

\def\arxiv#1#2{\href{http://arxiv.org/abs/#1}{\textsf{arXiv:#1 #2}}}

\nc{\ir}{\mathrm{i}}
\nc{\eE}{\mathsf{e}} 
\nc{\dd}{\mathrm{d}}   
\nc{\qo}{\omega}
\nc{\Mod}{\textrm{ mod }}
\nc{\Tr}{\textrm{tr}}
\nc{\tr}{\textrm{tr}}
\nc{\tA}{\textrm{A}}
\nc{\tB}{\textrm{B}}
\nc{\tX}{\textrm{X}}
\nc{\tY}{\textrm{Y}}

\nc{\tl}{\mathsf{TL}}
\nc{\eptl}{\mathsf{\mathcal EPTL}}
\nc{\stan}{\mathsf{V}}
\nc{\irre}{\mathsf{I}}
\nc{\stanp}{\mathsf W}

\nc{\Db}{\mbox{\boldmath $D$}}
\nc{\Fb}{\mbox{\boldmath $D_\infty$}}
\nc{\Fbtp}{\mbox{$\boldsymbol{T}_{\!+\infty}$}}
\nc{\Fbtm}{\mbox{$\boldsymbol{T}_{\!-\infty}$}}
\nc{\Fbtpm}{\mbox{$\boldsymbol{T}_{\!\pm\infty}$}}
\nc{\Ib}{\mbox{\boldmath $I$}}
\nc{\Jb}{\mbox{\boldmath $J$}}
\nc{\Kb}{\mbox{\boldmath $K$}}
\nc{\Kbt}{\mbox{\boldmath $K$}}
\nc{\Lambdab}{\mbox{\boldmath $\Lambda$}}
\nc{\Tb}{\mbox{\boldmath $T$}}
\nc{\ab}{\mbox{\boldmath $\mathfrak a$}}
\nc{\bb}{\mbox{\boldmath $\mathfrak b$}}
\nc{\Ab}{\mbox{\boldmath $\mathfrak A$}}
\nc{\db}{\mbox{\boldmath $d$}}
\nc{\tb}{\mbox{\boldmath $t$}}
\nc{\ZdN}{Z_d^{\textrm{\tiny$(N)$}}(q)}
\nc{\ZdNP}{Z_d^{\textrm{\tiny$(N)$}}(q,\bar q)}
\nc{\XN}{\chit^{\textrm{\tiny$(N)$}}}
\nc{\ZcylMN}{Z_{\textrm{cyl}}^{\textrm{\tiny$(M,\!N)$}}}
\nc{\Zcyl}{Z_{\textrm{cyl}}}
\nc{\Zcylodd}{Z^{\text{\tiny($N$\,odd)}}_{\textrm{cyl}}}
\nc{\Zcyleven}{Z^{\text{\tiny($N$\,even)}}_{\textrm{cyl}}}
\nc{\ZtorMN}{Z_{\textrm{torus}}^{\textrm{\tiny$(M,\!N)$}}}
\nc{\Ztor}{Z_{\textrm{torus}}}
\nc{\Ztorodd}{Z^{\text{\tiny($N$\,odd)}}_{\textrm{torus}}}
\nc{\Ztoreven}{Z^{\text{\tiny($N$\,even)}}_{\textrm{torus}}}
\nc{\Ztoreveneven}{Z^{\text{\tiny($M$\,even,$N$\,even)}}_{\textrm{torus}}}
\nc{\Ztorevenodd}{Z^{\text{\tiny($M$\,even,$N$\,odd)}}_{\textrm{torus}}}
\nc{\Ztoroddeven}{Z^{\text{\tiny($M$\,odd,$N$\,even)}}_{\textrm{torus}}}
\nc{\Ztoroddodd}{Z^{\text{\tiny($M$\,odd,$N$\,odd)}}_{\textrm{torus}}}

\nc{\Zcdpeveneven}{Z^{\text{\tiny($M$\,odd, $N$\,even)}}_{\textrm{torus,\,CDP}}}
\nc{\Zcdpevenodd}{Z^{\text{\tiny($M$\,even, $N$\,odd)}}_{\textrm{torus,\,CDP}}}
\nc{\Zcdpoddeven}{Z^{\text{\tiny($M$\,odd, $N$\,even)}}_{\textrm{torus,\,CDP}}}
\nc{\Zcdpoddodd}{Z^{\text{\tiny($M$\,odd, $N$\,odd)}}_{\textrm{torus,\,CDP}}}
\nc{\Zcdpeven}{Z^{\text{\tiny($N$\,even)}}_{\textrm{torus,\,CDP}}}
\nc{\Zcdpodd}{Z^{\text{\tiny($N$\,odd)}}_{\textrm{torus,\,CDP}}}

\renewcommand{\ge}{\geqslant}
\renewcommand{\le}{\leqslant}

\def\facegrid#1#2{
\psframe[fillstyle=solid,fillcolor=lightlightblue,linewidth=0pt]#1#2
\psgrid[gridlabels=0pt,subgriddiv=1]#1#2}

\def\loopa{
\psframe[linewidth=.25pt](0,0)(1,1)
\psarc[linewidth=1.5pt,linecolor=blue](1,0){.5}{90}{180}
\psarc[linewidth=1.5pt,linecolor=blue](0,1){.5}{-90}{0}
}
\def\loopb{
\psframe[linewidth=.25pt](0,0)(1,1)
\psarc[linewidth=1.5pt,linecolor=blue](0,0){.5}{0}{90}
\psarc[linewidth=1.5pt,linecolor=blue](1,1){.5}{180}{270}
}

\def\sep{
\begin{pspicture}[shift=-0.2](-0.07,0)(0.07,0.6)
\psline[linestyle=dashed,dash=1pt 1pt]{-}(0,0)(0,0.6)
\end{pspicture}
}

\def\smallsep{
\begin{pspicture}[shift=-0.1](-0.07,0)(0.07,0.4)
\psline[linestyle=dashed,dash=1pt 1pt]{-}(0,0)(0,0.4)
\end{pspicture}
}

\def\whitecircle{\pscircle[linewidth=0.5pt,linecolor=black,fillstyle=solid,fillcolor=white](0,0){.15}}
\def\blackcircle{\pscircle[linewidth=0.5pt,linecolor=black,fillstyle=solid,fillcolor=black](0,0){.15}}

\def\Aone#1#2{
\mathcal A^{#1}_{#2}
}
\def\Adom#1#2#3{
\mathcal D^{#1}_{#2,#3}
}
\def\Abarone#1#2{
\bar {\mathcal A}^{#1}_{#2}
}
\def\AA#1#2#3#4#5{
\mathcal D^{#1,#2}_{#3,#4,#5}
}
\def\AAAB#1#2#3#4#5#6{
\mathcal D^{#1,#2\,(#6)}_{#3,#4,#5}
}

\def\BB#1#2#3#4#5{
\mathcal A^{#1,#2}_{#3,#4,#5}
}

\def\BBAB#1#2#3#4#5#6{
\mathcal A^{#1,#2\,(#6)}_{#3,#4,#5}
}
\def\Ang#1#2{
\Big\langle\,\begin{matrix}#1\\#2\end{matrix}\,\Big\rangle
}
\def\ang#1#2{
\big\langle\begin{smallmatrix}#1\\[0.05cm]#2\end{smallmatrix}\big\rangle
}

\def\qbinom#1#2{
\left[
\begin{matrix}
#1 \\ #2
\end{matrix}
\right]_{\!q}
}
\def\smallqbinom#1#2{
\left[
\begin{smallmatrix}
#1 \\[0.05cm] #2
\end{smallmatrix}
\right]_{q}
}
\def\specialqbinom#1#2{
\left[
\begin{matrix}
#1 \\[0.1cm] #2
\end{matrix}
\right]_{\!q}
}

\def\qqbinom#1#2{
\left[
\begin{matrix}
#1 \\ #2
\end{matrix}
\right]_{\!\bar q}
}

\def\specialqqbinom#1#2{
\left[
\begin{matrix}
#1 \\[0.1cm] #2
\end{matrix}
\right]_{\!\bar q}
}

\def\qnarayana#1#2#3{
\left\{
\begin{matrix}
#1 \\ #2,\,#3
\end{matrix}
\right\}_{\!q}
}
\def\smallqnarayana#1#2#3{
\big\{
\begin{smallmatrix}
#1 \\[0.05cm] #2,\,#3
\end{smallmatrix}
\big\}_{\!q}
}

\begin{document}

\topmargin -15mm
\oddsidemargin 05mm

%
%

\title{\mbox{}\vspace{-.2in}
\bf 
Conformal partition functions of critical percolation \\[0.5cm]
from $\boldsymbol{D_3}$ Thermodynamic Bethe Ansatz equations
}

\date{}
\maketitle

\begin{center}
{\vspace{-5mm}\Large Alexi Morin-Duchesne$^\ast$, Andreas Kl\"umper$^\dagger$, Paul A.~Pearce$^\ddagger$}
\\[.5cm]
{\em {}$^\ast$Institut de Recherche en Math\'ematique et Physique\\ Universit\'e catholique de Louvain, Louvain-la-Neuve, B-1348, Belgium}
\\[.2cm]
{\em {}$^\dagger$Fachbereich C Physik, Bergische Universit\"at Wuppertal, 42097 Wuppertal, Germany}
\\[.2cm]
{\em {}$^\dagger$School of Mathematics and Statistics, University of Melbourne\\
Parkville, Victoria 3010, Australia}
\\[.4cm] 
{\tt alexi.morin-duchesne\,@\,uclouvain.be}
\qquad
{\tt kluemper\,@\,uni-wuppertal.de}
\qquad
{\tt papearce\,@unimelb.edu.au}

\end{center}

%
%

\vspace{0.5cm}

\begin{abstract}
Using the planar Temperley-Lieb algebra, critical bond percolation on the square lattice can be reformulated as a loop model. 
In this form, it is incorporated as ${\cal LM}(2,3)$ in the Yang-Baxter integrable family of logarithmic minimal models ${\cal LM}(p,p')$. We consider this model of percolation in the presence of boundaries and with periodic boundary conditions. Inspired by Kuniba, Sakai and Suzuki, we rewrite the recently obtained infinite $Y$-system of functional equations. In this way, we obtain nonlinear integral equations in the form of a closed finite set of TBA equations described by a $D_3$ Dynkin diagram. 
Following the methods of Kl\"umper and Pearce, we solve the TBA equations for the conformal finite-size corrections. For the ground states of the standard modules on the strip, these agree with the known central charge $c=0$ and conformal weights $\Delta_{1,s}$ for $s\in {\mathbb Z_{\ge 1}}$ with $\Delta_{r,s}=\big((3r-2s)^2-1\big)/24$. For the periodic case, the finite-size corrections agree with the conformal weights $\Delta_{0,s}$, $\Delta_{1,s}$ with $s\in\frac{1}{2}{\mathbb Z_{\ge 0}}$.
These are obtained analytically using Rogers dilogarithm identities. We incorporate all finite excitations by formulating empirical selection rules for the patterns of zeros of all the eigenvalues of the standard modules. 
We thus obtain the conformal partition functions on the cylinder and the modular invariant partition function (MIPF) on the torus. By applying $q$-binomial and $q$-Narayana identities, it is shown that our refined finitized characters on the strip agree with those of Pearce, Rasmussen and Zuber. For percolation on the torus, the MIPF is a non-diagonal sesquilinear form in affine $u(1)$ characters given by the $u(1)$ partition function $Z_{2,3}(q)=Z_{2,3}^{\text{Circ}}(q)$. 
The $u(1)$ operator content is ${\cal N}_{\Delta,\bar\Delta}=1$ for $\Delta=\bar\Delta=-\frac{1}{24}, \frac{35}{24}$ and ${\cal N}_{\Delta,\bar\Delta}=2$ for $\Delta=\bar\Delta=\frac{1}{8}, \frac{1}{3}, \frac{5}{8}$ and $(\Delta,\bar\Delta)=(0,1), (1,0)$. 
This result is compatible with the general conjecture of Pearce and Rasmussen, namely $Z_{p,p'}(q)=Z^{\text{Proj}}_{p,p'}(q)+n_{p,p'} Z^{\text{Min}}_{p,p'}(q)$ with $n_{p,p'}\in {\mathbb Z}$, 
where the minimal partition function is $Z^{\text{Min}}_{2,3}(q)=1$ and the lattice derivation fixes $n_{2,3}=-1$.

\vspace{0.75cm}

\noindent Keywords: percolation, solvable lattice models, conformal field theory.
\end{abstract}

%
%

\newpage

\tableofcontents
\clearpage

%
\section{Introduction}
%

In 1957, Broadbent and Hammersley~\cite{BroadHamm57} introduced a lattice percolation model as a mathematical model of the physical process of a fluid flowing through a random medium. Most importantly, they showed that their model exhibits a phase transition characterized by a critical probability threshold $p=p_c$. Comprehensive reviews of percolation theory can be found in \cite{KestonPerc82,Stauffer92,Grimmet97,Saberi16}.

One of the challenging goals in percolation is to understand the precise thermodynamic behavior of the model in the vicinity of the critical point. This behavior is believed to be {\em conformally invariant} and {\em universal}. Invariance under conformal maps implies invariance under translation, rotation and local scaling transformations. Universality implies that the critical behavior, characterized by critical exponents, depends on the lattice dimensionality but is otherwise insensitive to the details of the lattice model (for example the lattice structure or the choice of site versus bond percolation). 
In this paper, we view critical bond percolation on the two-dimensional square lattice as a Yang-Baxter integrable~\cite{BaxBook} loop model and solve exactly the associated $D_3$ $Y$-system for the conformal spectra to establish how it fits into the framework of logarithmic conformal field theory~\cite{FMS,Gurarie,SpecialIssue}. 

In bond percolation on the square lattice, the bonds $j$ of the lattice are {\em open or occupied} ($\sigma_j=1$) with a probability $p$ and {\em closed or empty} ($\sigma_j=0$) with a probability $1-p$. A typical bond configuration is shown in \cref{fig:percoconf}. In this description, the configuration space representing the local degrees of freedom is $\Omega=\{0,1\}^{{\mathbb Z}^2}$. The ``spins" $\sigma_j$ are independent identically distributed random variables. As a consequence, the usual observables given by the correlations of these spins factorize and are trivial. Accordingly, the statistical weight $W(\sigma)$ of a configuration $\sigma$ is
\begin{equation}
W(\sigma)=p^\text{\#\! bonds}(1-p)^\text{\#\! empty bonds},\qquad Z=\sum_\sigma W(\sigma)=1
\end{equation}
and the partition function $Z$ is trivial. 

In fact, the interesting physical behavior resides in the properties of connected clusters. The probability $P(p)$ that the origin is part of an infinite connected cluster is called the {\em percolation} probability. For $p<p_c$,  the open bonds are sparsely distributed at random throughout the lattice with no large clusters and $P(p)=0$. For $p>p_c$, the percolation probability is strictly positive: $P(p)>0$.
The percolation probability is thus the order parameter for an order-disorder phase transition. 
For bond percolation on the square lattice, it has been proved that $P(p_c)=0$~\cite{Harris} and that the critical threshold is precisely $p_c=1/2$~\cite{Kesten80}. 
More generally, the interesting physical observables \cite{VJS12,DPSV13} include the probabilities $P(j_1,j_2,\ldots,j_n)$ that the bonds $j_1,j_2,\ldots,j_n$ all lie in the same connected cluster.

The behavior of connected clusters is captured by introducing degrees of freedom in the form of planar non-crossing loop segments representing non-local {\em connectivities}.
Mathematically, the local properties of loop segments are encoded in the planar Temperley-Lieb algebra~\cite{TL71,Jones}. 
In critical percolation, the loop segments can close to form loops with an assigned statistical weight or fugacity $\beta=1$. This description of percolation is sometimes referred to as hull percolation~\cite{HullPerc88,DuplantierHull}. On the square lattice, there is a one-to-one mapping between bond configurations and loop configurations. This is illustrated by an example in \cref{fig:percoconf}. On rectangular lattices, each connected cluster is surrounded by loop segments, and crucially for crossing probabilities~\cite{Cardy92,LPSA94}, a connected cluster spans the lattice if and only if the surrounding loop segments also span the lattice. 

The critical point $p=p_c$ of percolation marks a second order phase transition. From the viewpoint of statistical mechanics, the universality class of such a phase transition is characterized by critical exponents. The first few critical exponents considered for percolation are related to the number of clusters per site, the percolation probability, the truncated mean cluster size, the cluster volume and the correlation length respectively:
\begin{subequations}
\begin{alignat}{3}
\alpha&=\frac{2\Delta_t-1}{\Delta_t-1}=-\frac{2}{3},\qquad&&  
\beta=\frac{\Delta_h}{1-\Delta_t}=\frac{5}{36},\qquad  
&&\gamma=\frac{2\Delta_h-1}{\Delta_t-1}=\frac{43}{18},\\[0.2cm]
\delta&=\frac{1-\Delta_h}{\Delta_h}=\frac{91}{5},\qquad&&
\nu=\frac{1}{2(1-\Delta_t)}=\frac{4}{3}.
\end{alignat}
\end{subequations}
For example,  as $p\to p_c^+$, 
the percolation probability behaves as $P(p)\sim (p-p_c)^\beta$. 
Only two of these exponents are independent. The others are related by scaling relations~\cite{CardyRGScaling} to the thermal and magnetic conformal weights
\begin{equation}
\Delta_t=\Delta_{2,1}=\frac{5}{8},\qquad \Delta_h=\Delta_{\frac{1}{2},0}=\frac{5}{96}.
\end{equation}
The values of these critical exponents were originally conjectured by den Nijs~\cite{denNijs79} and Nienhuis, Riedel and Schick~\cite{NienhuisRS80} based on Coulomb gas arguments~\cite{CoulombGas1,CoulombGas2,FSZ87} by viewing percolation as the $Q\to 1$ limit of the critical $Q$-state Potts model.
The $Q$-state Potts model with $Q=1$ is indeed trivial. It has a unique frozen state and the partition function is trivially $Z=1$, so a $Q\to 1$ limit is needed~\cite{Delfino1,Delfino2} to recover the critical exponents. In general, the critical line of the Coulomb gas maps onto the critical line of the six-vertex model and is parameterised by the crossing parameter $\lambda\in (0,\pi)$. It is related to the loop fugacity by $\beta=\sqrt{Q}=2\cos\lambda$, with percolation corresponding to $\lambda=\frac \pi 3$.

The statistical behavior of percolation shares many commonalities with the model of critical dense polymers, which has a loop fugacity $\beta=0$. In polymers, the non-local degrees of freedom are extended segments of polymer chains which are not allowed to form closed loops. The study of polymers and percolation as Conformal Field Theories (CFTs) began with Saleur and Duplantier~\cite{Saleur86,Saleur87a,Saleur87b,Duplantier86,DupSaleur86,SaleurDup87,Saleur92} in the mid-eighties. A CFT is a continuum theory that describes directly the {\em universal}\/ properties of a critical statistical system (characterized by a linear system size $N$, a lattice spacing $a$ and a continuum coordinate $R$) in the continuum scaling limit ($N\to\infty$, $a\to 0$, $Na\to R$). 
The conformal symmetry of percolation is described by a Virasoro algebra with central charge $c=0$. 
As a CFT, the Coulomb critical line is an $su(2)$ affine Wess-Zumino-Witten CFT with effective central charge $c_\text{eff}=1$. The nature of a critical point on this line is very different if $\lambda/\pi$ is rational compared to generic points where $\lambda/\pi$ is irrational. If $\frac{\lambda}{\pi}=\frac{p'-p}{p'}$ is rational, the theory admits a higher symmetry algebra described by the $sl_2$ loop algebra~\cite{DeguchiFabMcCoy,Deguchi}. These points are characterized by two integers $p,p'$ satisfying 
\be
\label{eq:ppp}
1\le p<p',\qquad \gcd(p,p')=1
\ee 
and are dense along the critical line, with each point representing a different CFT. At each of these points, there are additional eigenvalue degeneracies and the theory is logarithmic. Percolation corresponds to the point with $(p,p')=(2,3)$. 

To set it in context, the loop model of critical bond percolation is the $(p,p')=(2,3)$ member of the family of logarithmic minimal models ${\cal LM}(p,p')$~\cite{PRZ06} with conformal data consisting of the central charge $c$ and Virasoro Kac conformal weights $\Delta_{r,s}$
\be
\label{logMin}
c= 1 - \frac{6(p'-p)^2}{pp'}, \qquad \Delta_{r,s}=\frac{(p' r-p s)^2-(p'-p)^2}{4pp'}.
\ee
Since these loop models are defined in terms of the diagrammatic action of local operators on a vector space of link states, the logarithmic minimal models are intrinsically quantum in nature. The choice of the vector space of link states is an integral part of the definition of the model. But this space of states is not a Hilbert space since the inner product is not positive-definite.
The infinitely extended Virasoro Kac table of conformal weights for percolation is shown in \cref{fig:VirKac} for $r, s \in \mathbb Z_{> 0}$. 
In this paper, we will encounter conformal weights $\Delta_{r,s}$ with $r=0,1$ and $s \in \frac12 \mathbb Z_{>0}$.
Additional physical conformal weights are given by allowing $r$ or $s$ or both in this Kac formula to be half-integers \cite{PRT14,BPT2016} or even possibly to take values in $\mathbb Q$ \cite{CJV17}.
The central charge and conformal weights, given by the Kac formula, vary continuously with the parameter $\lambda\in (0,\pi)$.

In analogy to the rational minimal models ${\cal M}(m,m')$, the logarithmic minimal models ${\cal LM}(p,p')$ are coset CFTs~\cite{PRcosetBranch13}. This analogy is the origin of the name but, in contradistinction to the unitary minimal models ${\cal M}(m'\!-\!1,m')$, the logarithmic minimal models are all nonunitary. 
In particular, ${\cal LM}(2,3)$ is a nonunitary coset CFT with
\begin{equation}
c=0,\qquad \Delta_\text{min}=-1/24,\qquad c_\text{eff}=c-24\Delta_\text{min}=1,\qquad \Delta_{r,s}=\frac{(3r-2s)^2-1}{24}.
\end{equation}
Tellingly, since $\frac{\lambda}{\pi}=\tfrac{1}{3}$, critical percolation is a {\em logarithmic} CFT~\cite{MathieuRidout} and not a {\em rational}\/ CFT~\cite{MooreSeiberg}.  The infinitely extended Virasoro Kac table of percolation in \cref{fig:VirKac}
displays the conformal weights of 
an infinite number of Virasoro scaling operators. If a theory is rational, there can only be a finite number of conformal weights associated with a finite number of scaling operators and the associated Virasoro 
(or extended) 
representations must be irreducible and close among themselves under fusion. In contrast, logarithmic CFTs are characterized~\cite{Gurarie,MDSA2011,MDSA13no2} by the existence of reducible yet indecomposable representations of the Virasoro algebra. On the strip, there is a single copy of the Virasoro algebra but, on a torus, there are two chiral copies of the Virasoro algebra and conformal invariance extends~\cite{CardyModInv} to include invariance under the modular group. For simple rational CFTs, such as the $c<1$ $A$-$D$-$E$ models~\cite{CIZ}, conformal and modular invariance together suffice to uniquely determine the conformal torus partition function. This is not the case for general logarithmic minimal models.

Strikingly, conformality was only rigorously established in 2001 by Smirnov~\cite{Smirnov01} for critical site percolation on the triangular lattice. This mathematical approach, which is built on random conformally invariant fractal curves, entails the identification of the models \eqref{logMin} with $\mbox{SLE}_\kappa$ (Schramm-Loewner Evolution) with $\kappa = \frac{4p'}{p}$. For percolation, with $(p,p')=(2,3)$ and $\kappa=6$, the fractal dimensions $d=2(1-\Delta)$ of various fractal geometric curves are known~\cite{Saleur87a,SaleurDup87,AAMRH,JS2005,JS2006,SAPR2009} including those of chordal SLE paths, hulls (H), cluster mass (C), external perimeter (EP) and red bonds (RB):
\begin{subequations}
\begin{align}
(\Delta_\text{path}^\text{SLE},\Delta_\text{H},\Delta_\text{C},\Delta_\text{EP},\Delta_\text{RB})
&=(\Delta_{p,p'\pm1},\Delta_{p,p'\pm 1},\Delta_{\frac12(p\pm1),\frac12p'},\Delta_{p\pm1,p'},\Delta_{p,p'\pm2})
=(\tfrac{1}{8},\tfrac{1}{8},\tfrac{5}{96},\tfrac{1}{3},\tfrac{5}{8}),\\[4pt]
&(d_\text{path}^\text{SLE},d_\text{H},d_\text{C},d_\text{EP},d_\text{RB})
=(\tfrac{7}{4},\tfrac{7}{4},\tfrac{91}{48},\tfrac{4}{3},\tfrac{3}{4}).
\end{align}
\end{subequations}
The value $d_\text{EP}=\frac{4}{3}$ was conjectured by Mandelbrot~\cite{Mandelbrot} and much later proved by Lawler, Schramm and Werner~\cite{LSW01}. The value $d_\text{path}^\text{SLE}=\frac{7}{4}$ was proved by Beffara~\cite{Beffara}.

The incorporation of critical dense polymers ${\cal LM}(1,2)$~\cite{PR07,PRVcyl2010,MD11,PRV1210,MDPR13,PRT14} and critical percolation ${\cal LM}(2,3)$ into the framework of the family of logarithmic minimal models ${\cal LM}(p,p')$~\cite{PRZ06,PRcosetBranch13,PRcosetGraphs11,MDPR14,PTC2015,MDRD2015,BPT2016} establishes that these models are Yang-Baxter integrable. The transfer matrices of the logarithmic minimal models are built from so called {\it transfer tangles} of the planar Temperley-Lieb algebra~\cite{TL71,Jones}, which we respectively denote by $\Db(u)$ and $\Tb(u)$ for the boundary and the periodic cases. The finite-size corrections to the eigenvalues $D(u)$ and $T(u)$ of the transfer matrices provide a direct way to access the central charge and conformal weights analytically. Indeed, for large horizontal system size $N$, the leading eigenvalues of the transfer matrices behave as
\begin{subequations}
\begin{alignat}{2}
\hspace{-0.2cm}-\ln D(u) &=  2N f_{\text{b}}(u) +  f_{\text{s}}(u) + \tfrac{2 \pi}{N} \sin(\tfrac{\pi u}{\lambda}) \big(\!-\tfrac c {24} + \Delta + k \big) + o(\tfrac1N)\label{eq:Dexpansion},\\
\hspace{-0.2cm}-\ln T(u) &=  N f_{\text{b}}(u) +  \tfrac{2 \pi }{N} \Big(\sin(\tfrac{\pi u}{\lambda})(\Delta + \bar \Delta + k + \bar k- \tfrac c {12}) + \ir \cos(\tfrac{\pi u}{\lambda}) (\Delta - \bar \Delta + k - \bar k) \Big) + o(\tfrac1N),
\label{eq:Texpansion}
\end{alignat} 
\end{subequations}
where $f_{\text{b}}(u)$ and $f_{\text{s}}(u)$ are the {\em non-universal\/} bulk and surface free energies, and $k,\bar k$ are integers. Yang-Baxter integrability on the lattice means that $f_{\text{b}}(u)$ and $f_{\text{s}}(u)$ can be calculated exactly. In addition, by solving $T$- and $Y$-systems~\cite{BR1989,Zam1991a,Zam1991b,PK91,KP92,KunibaNS9309,KunibaNS9310,KunibaNS1010} satisfied by the commuting transfer matrices, one can calculate analytically the $\frac1N$ term to obtain {\em universal}\/ quantities such as the central charge, conformal weights and conformal partition functions. 
The $T$-system takes~\cite{KLWZ1997} the form of a bilinear Hirota equation and is the master equation of integrability. 
Two key steps~\cite{PK91,KP92} in the process of solving the system are, first, to derive the $Y$-system from the $T$-system and, second, to use analyticity properties of the eigenvalues to convert the $Y$-system of functional equations into non-linear integral equations in the form of Thermodynamic Bethe Ansatz (TBA) equations~\cite{YangYang,T71,G71,Zam90,Zam91}. These latter works on TBA focused on the ground state. The approach of Kl\"umper and Pearce~\cite{PK91,KP92}, which we follow closely here, applies to all finite excitations and enables the analytic calculation of conformal partition functions. 
The $T$-system is non-universal but the $Y$-system, which relates to the conformal spectra, is universal~\cite{CMP02} in the sense that it holds for all boundary conditions and topologies. 
So these calculations can be carried out with periodic boundary conditions or in the presence of boundaries on the strip~\cite{Sklyanin,BPO96}. 
This program has been carried to completion~\cite{PK91,KP92, KP91,BPO96,OPW97,BDP15} for prototypical $c<1$ $A$-type rational minimal models. 

Within the lattice approach, our longer term goal is to extend these calculations, based on functional equations and TBA, to the general logarithmic minimal models ${\cal LM}(p,p')$. 
The $T$- and $Y$-systems for the general ${\cal LM}(p,p')$ models were obtained recently in \cite{MDPR14}. 
These hierarchies of functional equations are infinite but the $Y$-system can be truncated to a finite $D$-type $Y$-system following the methods of \cite{KSS98}. 
In this paper, we start with critical percolation ${\cal LM}(2,3)$ as a prototypical example with $p>1$. This model admits a set of TBA equations encoded by a $D_3\simeq A_3$ Dynkin diagram. 
Our specific goals are to calculate analytically, for critical percolation, the following quantities: 
\begin{enumerate}
\item[(i)] the central charge and conformal weights using dilogarithm identities;\\[-20pt]
\item[(ii)] the finitized characters on the strip for half-arc boundary conditions and an arbitrary number of defects;\\[-20pt]
\item[(iii)] the cylinder
conformal partition functions with half-arc boundary conditions;\\[-20pt]
\item[(iv)] the modular invariant partition function (MIPF) on the torus.
\end{enumerate}
This program has been completed~\cite{PR07,PRVcyl2010,MD11,PRV1210,MDPR13,PRT14} for critical dense polymers ${\cal LM}(1,2)$. In this case, 
the task was simplified because the transfer matrices satisfy a trivial $Y$-system in the form of an inversion identity similar to that of the (free-fermionic) Ising model~\cite{BaxBook,OPW1996}.
The analysis of critical dense polymers introduced combinatorial constructs to enumerate patterns of zeros, namely single- and double-column
diagrams and $q$-Narayana polynomials. Remarkably, these reappear in generalizing the calculations to critical percolation. 
Similarly, because of the occurrence of non-contractible loops and winding on the cylinder, a modified trace~\cite{MDPR13} (analogous to the Markov trace~\cite{Jones83} on the strip) is needed to obtain the MIPF of critical percolation as was the case for critical dense polymers.
We also stress that, as for critical dense polymers, the MIPF that we find for $\mathcal{LM}(2,3)$ is obtained from the scaling limit of the loop model on a torus of size $M\times N$, with $M$ and $N$ even, where each non-contractible loop is weighted by a fugacity $\alpha = 2$.

\def\loopa{
\psframe[linewidth=.25pt](0,0)(1,1)
\psarc[linewidth=1.5pt,linecolor=blue](1,0){.5}{90}{180}
\psarc[linewidth=1.5pt,linecolor=blue](0,1){.5}{-90}{0}
}
\def\loopb{
\psframe[linewidth=.25pt](0,0)(1,1)
\psarc[linewidth=1.5pt,linecolor=blue](0,0){.5}{0}{90}
\psarc[linewidth=1.5pt,linecolor=blue](1,1){.5}{180}{270}
}
\def\loopab{
\psframe[linewidth=.25pt](0,0)(1,1)
\psarc[linewidth=1.5pt,linecolor=blue](1,0){.5}{90}{180}
\psarc[linewidth=1.5pt,linecolor=blue](0,1){.5}{-90}{0}
\psline[linewidth=3pt,linecolor=blue](0,0)(1,1)
}
\def\loopar{
\psframe[linewidth=.25pt](0,0)(1,1)
\psarc[linewidth=1.5pt,linecolor=blue](1,0){.5}{90}{180}
\psarc[linewidth=1.5pt,linecolor=blue](0,1){.5}{-90}{0}
\psline[linewidth=3pt,linecolor=red](0,0)(1,1)
}
\def\loopbb{
\psframe[linewidth=.25pt](0,0)(1,1)
\psarc[linewidth=1.5pt,linecolor=blue](0,0){.5}{0}{90}
\psarc[linewidth=1.5pt,linecolor=blue](1,1){.5}{180}{270}
\psline[linewidth=3pt,linecolor=blue](1,0)(0,1)
}
\def\loopbr{
\psframe[linewidth=.25pt](0,0)(1,1)
\psarc[linewidth=1.5pt,linecolor=blue](0,0){.5}{0}{90}
\psarc[linewidth=1.5pt,linecolor=blue](1,1){.5}{180}{270}
\psline[linewidth=3pt,linecolor=red](1,0)(0,1)
}
\def\psqa#1{\raisebox{-1.5\unitlength}{
\begin{pspicture}[shift=0](0,0)(4,4)
\pspolygon[linewidth=.25pt,fillstyle=solid,fillcolor=lightlightblue](0,0)(4,0)(4,4)(0,4)
\psarc[linewidth=2pt,linecolor=blue](4,0){2}{90}{180}
\psarc[linewidth=2pt,linecolor=blue](0,4){2}{-90}{0}
\rput(2,2){\small $#1$}
\end{pspicture}}}
\def\psqb#1{\raisebox{-1.5\unitlength}{
\begin{pspicture}[shift=0](0,0)(4,4)
\pspolygon[linewidth=.25pt,fillstyle=solid,fillcolor=lightlightblue](0,0)(4,0)(4,4)(0,4)
\psarc[linewidth=2pt,linecolor=blue](0,0){2}{0}{90}
\psarc[linewidth=2pt,linecolor=blue](4,4){2}{180}{270}
\rput(2,2){\small $#1$}
\end{pspicture}}}

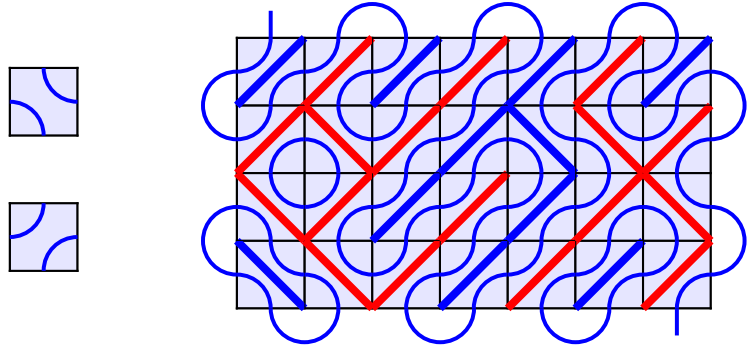
\begin{figure}[p]
\begin{center}
\psset{unit=.9}
\begin{pspicture}[shift=0.5cm](0,-.3)(1,3)
\facegrid{(0,0)}{(1,1)}
\facegrid{(0,2)}{(1,3)}
\rput(0,0){\loopa}
\rput(0,2){\loopb}
\end{pspicture}
\qquad \qquad
\begin{pspicture}(-0.5,-.3)(7.5,4)
\facegrid{(0,0)}{(7,4)}
\rput[bl](0,0){\loopbb}
\rput[bl](1,0){\loopbr}
\rput[bl](2,0){\loopar}
\rput[bl](3,0){\loopab}
\rput[bl](4,0){\loopar}
\rput[bl](5,0){\loopab}
\rput[bl](6,0){\loopar}
\rput[bl](0,1){\loopbr}
\rput[bl](1,1){\loopar}
\rput[bl](2,1){\loopab}
\rput[bl](3,1){\loopar}
\rput[bl](4,1){\loopab}
\rput[bl](5,1){\loopar}
\rput[bl](6,1){\loopbr}
\rput[bl](0,2){\loopar}
\rput[bl](1,2){\loopbr}
\rput[bl](2,2){\loopar}
\rput[bl](3,2){\loopab}
\rput[bl](4,2){\loopbb}
\rput[bl](5,2){\loopbr}
\rput[bl](6,2){\loopar}
\rput[bl](0,3){\loopab}
\rput[bl](1,3){\loopar}
\rput[bl](2,3){\loopab}
\rput[bl](3,3){\loopar}
\rput[bl](4,3){\loopab}
\rput[bl](5,3){\loopar}
\rput[bl](6,3){\loopab}
\psarc[linecolor=blue,linewidth=1.5pt](0,1){.5}{90}{270}
\psarc[linecolor=blue,linewidth=1.5pt](0,3){.5}{90}{270}
\psarc[linecolor=blue,linewidth=1.5pt](7,1){.5}{270}{90}
\psarc[linecolor=blue,linewidth=1.5pt](7,3){.5}{270}{90}
\multiput(0,0)(2,0){3}{\psarc[linecolor=blue,linewidth=1.5pt](1,0){.5}{180}{360}}
\multiput(0,0)(2,0){3}{\psarc[linecolor=blue,linewidth=1.5pt](2,4){.5}{0}{180}}
\psline[linecolor=blue,linewidth=1.5pt](.5,4)(.5,4.4)
\psline[linecolor=blue,linewidth=1.5pt](6.5,-.4)(6.5,0)
\end{pspicture}
\end{center}
\caption{The two Temperley-Lieb face tiles of hull percolation. Bond percolation configurations 
on the rotated blue square lattice are mapped one-to-one onto loop configurations. The unoccupied faces show the dual red bonds. The connected clusters of blue bonds are surrounded by loop segments. Closed loops have a fugacity $\beta=1$. In the scaling limit, the curve that ties the two defects behaves as a chordal SLE path with fractal dimension $d_\text{path}^\text{SLE}=2(1-\Delta_{2,2})=\frac{7}{4}$.}
\label{fig:percoconf}
\end{figure}

\def\vvdots{\mathinner{\mkern1mu\raise1pt\vbox{\kern7pt\hbox{.}}\mkern2mu
  \raise4pt\hbox{.}\mkern2mu\raise7pt\hbox{.}\mkern1mu}}

\begin{figure}
\psset{unit=0.9cm}
\begin{equation*}
\begin{pspicture}(0,-.3)(7,11)
\psframe[linewidth=0pt,fillstyle=solid,fillcolor=lightestblue](0,0)(7,11)
\psframe[linewidth=0pt,fillstyle=solid,fillcolor=lightlightblue](1,0)(2,11)
\psframe[linewidth=0pt,fillstyle=solid,fillcolor=lightlightblue](3,0)(4,11)
\psframe[linewidth=0pt,fillstyle=solid,fillcolor=lightlightblue](5,0)(6,11)
\psframe[linewidth=0pt,fillstyle=solid,fillcolor=lightlightblue](0,2)(7,3)
\psframe[linewidth=0pt,fillstyle=solid,fillcolor=lightlightblue](0,5)(7,6)
\psframe[linewidth=0pt,fillstyle=solid,fillcolor=lightlightblue](0,8)(7,9)
\multiput(0,0)(0,3){3}{\multiput(0,0)(2,0){3}{\psframe[linewidth=0pt,fillstyle=solid,fillcolor=midblue](1,2)(2,3)}}
\psgrid[gridlabels=0pt,subgriddiv=1]
\rput(.5,10.6){$\vdots$}\rput(1.5,10.6){$\vdots$}\rput(2.5,10.6){$\vdots$}
\rput(3.5,10.6){$\vdots$}\rput(4.5,10.6){$\vdots$}\rput(5.5,10.6){$\vdots$}
\rput(6.5,10.5){$\vvdots$}
\rput(.5,9.5){$12$}\rput(1.5,9.5){$\frac{65}8$}\rput(2.5,9.5){$5$}
\rput(3.5,9.5){$\frac{21}8$}\rput(4.5,9.5){$1$}\rput(5.5,9.5){$\frac{1}8$}\rput(6.525,9.5){$\cdots$}
\rput(.5,8.5){$\frac{28}3$}\rput(1.5,8.5){$\frac{143}{24}$}\rput(2.5,8.5){$\frac{10}3$}
\rput(3.5,8.5){$\frac{35}{24}$}\rput(4.5,8.5){$\frac 13$}\rput(5.5,8.5){$-\frac{1}{24}$}
\rput(6.525,8.5){$\cdots$}
\rput(.5,7.5){$7$}\rput(1.5,7.5){$\frac {33}8$}\rput(2.5,7.5){$2$}
\rput(3.5,7.5){$\frac{5}8$}\rput(4.5,7.5){$0$}\rput(5.5,7.5){$\frac{1}8$}
\rput(6.525,7.5){$\cdots$}
\rput(.5,6.5){$5$}\rput(1.5,6.5){$\frac {21}8$}\rput(2.5,6.5){$1$}
\rput(3.5,6.5){$\frac{1}8$}\rput(4.5,6.5){$0$}\rput(5.5,6.5){$\frac{5}8$}\rput(6.525,6.5){$\cdots$}
\rput(.5,5.5){$\frac{10}3$}\rput(1.5,5.5){$\frac {35}{24}$}\rput(2.5,5.5){$\frac 13$}
\rput(3.5,5.5){$-\frac{1}{24}$}\rput(4.5,5.5){$\frac 13$}\rput(5.5,5.5){$\frac{35}{24}$}
\rput(6.525,5.5){$\cdots$}
\rput(.5,4.5){$2$}\rput(1.5,4.5){$\frac 58$}\rput(2.5,4.5){$0$}\rput(3.5,4.5){$\frac{1}8$}
\rput(4.5,4.5){$1$}\rput(5.5,4.5){$\frac{21}8$}\rput(6.525,4.5){$\cdots$}
\rput(.5,3.5){$1$}\rput(1.5,3.5){$\frac 18$}\rput(2.5,3.5){$0$}\rput(3.5,3.5){$\frac{5}8$}
\rput(4.5,3.5){$2$}\rput(5.5,3.5){$\frac{33}8$}\rput(6.525,3.5){$\cdots$}
\rput(.5,2.5){$\frac 13$}\rput(1.5,2.5){$-\frac 1{24}$}\rput(2.5,2.5){$\frac 13$}
\rput(3.5,2.5){$\frac{35}{24}$}\rput(4.5,2.5){$\frac{10}3$}\rput(5.5,2.5){$\frac{143}{24}$}
\rput(6.525,2.5){$\cdots$}
\rput(.5,1.5){$0$}\rput(1.5,1.5){$\frac 18$}\rput(2.5,1.5){$1$}\rput(3.5,1.5){$\frac{21}8$}
\rput(4.5,1.5){$5$}\rput(5.5,1.5){$\frac{65}8$}\rput(6.525,1.5){$\cdots$}
\rput(.5,.5){$0$}\rput(1.5,.5){$\frac 58$}\rput(2.5,.5){$2$}\rput(3.5,.5){$\frac{33}8$}\rput(4.5,.5){$7$}
\rput(5.5,.5){$\frac{85}8$}\rput(6.525,.5){$\cdots$}
{\color{blue}
\rput(.5,-.5){$1$}
\rput(1.5,-.5){$2$}
\rput(2.5,-.5){$3$}
\rput(3.5,-.5){$4$}
\rput(4.5,-.5){$5$}
\rput(5.5,-.5){$6$}
\rput(6.5,-.5){$r$}
\rput(-.5,.5){$1$}
\rput(-.5,1.5){$2$}
\rput(-.5,2.5){$3$}
\rput(-.5,3.5){$4$}
\rput(-.5,4.5){$5$}
\rput(-.5,5.5){$6$}
\rput(-.5,6.5){$7$}
\rput(-.5,7.5){$8$}
\rput(-.5,8.5){$9$}
\rput(-.5,9.5){$10$}
\rput(-.5,10.5){$s$}}
\end{pspicture}
\hspace{3cm}
\begin{pspicture}(-0.5,-0.3)(5,15)
\psframe[fillstyle=solid,fillcolor=white,linewidth=0pt](0,0)(5,15)
\multiput(1,2)(0,2){7}{\psframe[fillstyle=solid,fillcolor=lightlightblue,linewidth=0pt](0,0)(1,1)}
\multiput(2,2)(0,2){7}{\psframe[fillstyle=solid,fillcolor=lightlightblue,linewidth=0pt](0,0)(1,1)}
\multiput(3,2)(0,2){7}{\psframe[fillstyle=solid,fillcolor=lightlightblue,linewidth=0pt](0,0)(1,1)}
\multiput(4,2)(0,2){7}{\psframe[fillstyle=solid,fillcolor=lightlightblue,linewidth=0pt](0,0)(1,1)}
\psgrid[gridlabels=0pt,subgriddiv=1](0,0)(5,15)
{\color{blue}
\rput(0.5,-0.5){$0$}
\rput(1.5,-0.5){$1$}
\rput(2.5,-0.5){$2$}
\rput(3.5,-0.5){$3$}
\rput(4.5,-0.5){$r$}
\rput(-0.5,0.5){$0$}
\rput(-0.5,1.5){$\frac12$}
\rput(-0.5,2.5){$1$}
\rput(-0.5,3.5){$\frac32$}
\rput(-0.5,4.5){$2$}
\rput(-0.5,5.5){$\frac52$}
\rput(-0.5,6.5){$3$}
\rput(-0.5,7.5){$\frac72$}
\rput(-0.5,8.5){$4$}
\rput(-0.5,9.5){$\frac92$}
\rput(-0.5,10.5){$5$}
\rput(-0.5,11.5){$\frac{11}2$}
\rput(-0.5,12.5){$6$}
\rput(-0.5,13.5){$\frac{13}2$}
\rput(-0.5,14.5){$s$}
}
\rput(0.5,0.5){$-\frac1{24}$}
\rput(0.5,1.5){$0$}
\rput(0.5,2.5){$\frac18$}
\rput(0.5,3.5){$\frac13$}
\rput(0.5,4.5){$\frac58$}
\rput(0.5,5.5){$1$}
\rput(0.5,6.5){$\frac{35}{24}$}
\rput(0.5,7.5){$2$}
\rput(0.5,8.5){$\frac{21}{8}$}
\rput(0.5,9.5){$\frac{10}{3}$}
\rput(0.5,10.5){$\frac{33}{8}$}
\rput(0.5,11.5){$5$}
\rput(0.5,12.5){$\frac{143}{24}$}
\rput(0.5,13.5){$7$}
\rput(0.5,14.60){$\vdots$}
\rput(1.5,0.5){$\frac13$}
\rput(1.5,1.5){$\frac18$}
\rput(1.5,2.5){$0$}
\rput(1.5,3.5){$-\frac1{24}$}
\rput(1.5,4.5){$0$}
\rput(1.5,5.5){$\frac18$}
\rput(1.5,6.5){$\frac13$}
\rput(1.5,7.5){$\frac58$}
\rput(1.5,8.5){$1$}
\rput(1.5,9.5){$\frac{35}{24}$}
\rput(1.5,10.5){$2$}
\rput(1.5,11.5){$\frac{21}{8}$}
\rput(1.5,12.5){$\frac{10}{3}$}
\rput(1.5,13.5){$\frac{33}{8}$}
\rput(1.5,14.60){$\vdots$}
\rput(2.5,0.5){$\frac{35}{24}$}
\rput(2.5,1.5){$1$}
\rput(2.5,2.5){$\frac{5}{8}$}
\rput(2.5,3.5){$\frac{1}{3}$}
\rput(2.5,4.5){$\frac{1}{8}$}
\rput(2.5,5.5){$0$}
\rput(2.5,6.5){$-\frac{1}{24}$}
\rput(2.5,7.5){$0$}
\rput(2.5,8.5){$\frac{1}{8}$}
\rput(2.5,9.5){$\frac{1}{3}$}
\rput(2.5,10.5){$\frac{5}{8}$}
\rput(2.5,11.5){$1$}
\rput(2.5,12.5){$\frac{35}{24}$}
\rput(2.5,13.5){$2$}
\rput(2.5,14.60){$\vdots$}
\rput(3.5,0.5){$\frac{10}{3}$}
\rput(3.5,1.5){$\frac{21}{8}$}
\rput(3.5,2.5){$2$}
\rput(3.5,3.5){$\frac{35}{24}$}
\rput(3.5,4.5){$1$}
\rput(3.5,5.5){$\frac{5}{8}$}
\rput(3.5,6.5){$\frac{1}{3}$}
\rput(3.5,7.5){$\frac{1}{8}$}
\rput(3.5,8.5){$0$}
\rput(3.5,9.5){$-\frac{1}{24}$}
\rput(3.5,10.5){$0$}
\rput(3.5,11.5){$\frac{1}{8}$}
\rput(3.5,12.5){$\frac{1}{3}$}
\rput(3.5,13.5){$\frac{5}{8}$}
\rput(3.5,14.60){$\vdots$}
\multiput(0,0)(0,1){14}{\rput(4.5,0.525){$\cdots$}}
\rput(4.5,14.50){$\vvdots$}
\end{pspicture}
\end{equation*}
\caption{{\it Left}: Infinitely extended Virasoro Kac table of conformal weights $\Delta_{r,s}$ for critical percolation ${\cal LM}(2,3)$ for $r,s \in \mathbb Z_{>0}$. 
{\it Right}: The conformal weights $\Delta_{r,s}$ for critical percolation with $r \in \mathbb Z_{\ge 0}$ and $s \in \frac12\mathbb Z_{\ge 0}$. The entries $(r,s)$ with $r,s \in \mathbb Z_{>0}$ of the original Kac table are indicated by shading.
}
\label{fig:VirKac}
\end{figure}
For logarithmic minimal models ${\cal LM}(p,p')$ with $p>1$, the MIPF is not uniquely determined by conformal and modular invariance. 
The conjectured form~\cite{PRcosetGraphs11} for these MIPFs is
\begin{subequations}
\begin{alignat}{2}
Z_{p,p'}(q)&=Z^{\text{Proj}}_{p,p'}(q)+n_{p,p'} Z^{\text{Min}}_{p,p'}(q),\label{Zppa}\\
&=\tfrac{1}{2}(1+n_{p,p'})Z_{1,pp'}^{\text{Circ}}(q)+\tfrac{1}{2}(1-n_{p,p'})Z_{p,p'}^{\text{Circ}}(q),\qquad n_{p,p'}\in {\mathbb Z},
\label{Zpp}
\end{alignat}
\end{subequations}
where the integer $n_{p,p'}$ is undetermined and the projective partition function $Z^{\text{Proj}}_{p,p'}(q)$ is defined in~\cite{PRcosetGraphs11}. 
The $u(1)$ modular invariant partition functions, corresponding to a compactified boson on $S^1$ with radius $R=\sqrt\frac{2p'}{p}$, are
\be
 Z^{\text{Circ}}_{p,p'}(q)=\sum_{j=0}^{2n-1} 
   \varkappa^n_j(q)\varkappa^n_{\omega_0 j}(\bar q)
\label{ZCirc}
\ee
where $n=pp'$ and the $u(1)$ characters $\varkappa_j^n(q)$ are given by \eqref{eq:u1chars}. 
The Bezout number $\omega_0$ is defined by
\be
 \omega_0=r_0p'+s_0p\quad (\mbox{mod $2n$})
\ee
in terms of the Bezout pair $(r_0,s_0)$ which is uniquely determined by the conditions
\be
 r_0p'-s_0p=1,\qquad\quad 1\le r_0\le p-1,\quad 1\le s_0\le p'-1,\quad p s_0<p'r_0.
\ee
For $p=1$, $Z_{1,p'}(q)=Z^{\text{Circ}}_{1,p'}(q)$ is the diagonal $u(1)$ partition function

\begin{equation}
 Z^{\text{Circ}}_{1,p'}(q) = \sum_{j=0}^{p'} d^{p'}_j |\varkappa_{j}^{p'}(q)|^2,\qquad 
\label{eq:dj}
d_j^{n}=
\left\{\begin{array}{ll}
1,&j=0, \,n,\\
2,&\mbox{otherwise}
\end{array}\right.
\end{equation}
and $Z^{\text{Min}}_{p,p'}(q)=\tfrac{1}{2}(Z^{\text{Circ}}_{1,pp'}(q)-Z^{\text{Circ}}_{p,p'}(q))$
 implies $Z^{\text{Min}}_{1,p'}(q)=0$. 
For $p>1$, $Z^{\text{Circ}}_{p,p'}(q)$ is a non-diagonal $u(1)$ partition function.
For critical percolation, $Z^{\text{Min}}_{2,3}(q)=1$ and our analytic derivation of the MIPF from the lattice model shows that $n_{2,3} = -1$. 
The MIPF of critical percolation is therefore given by the non-diagonal $u(1)$ partition function
\begin{equation}
Z_{2,3}(q)=Z^{\text{Circ}}_{2,3}(q)
\end{equation}
where the Bezout number giving the Bezout conjugation is $\omega_0=5$.

The layout of the paper is as follows. \cref{sec:CFTdata} recalls the conformal data for bond percolation which is referred to in the rest of the paper. \cref{sec:boundarycase} contains our computations and results for bond percolation on the strip with vacuum boundary conditions. We recall the definition of the Temperley-Lieb algebra $\tl_N$ and the transfer tangle $\Db(u)$ in \cref{sec:TLandD} and review the standard modules over this algebra in \cref{sec:stanmod}. In \cref{sec:fusion}, we give the definition of the fused transfer matrices and present the fusion hierarchy relations. We write down the corresponding $T$- and $Y$-systems in \cref{sec:functional}. In \cref{sec:props}, we analyse the analyticity properties of the eigenvalues of the transfer matrices in terms of their patterns of zeros. In \cref{sec:fsc}, we transform the $T$- and $Y$-systems into TBA equations and solve for the finite-size corrections of the finite excitations characterized by their patterns of zeros. The results are expressed in terms of sums of Rogers dilogarithms which are evaluated in \cref{sec:Rogers}. In \cref{sec:cfgs}, we specialise the result to the ground states of the standard modules and reproduce the conformal weights in the 
$(1,s)$ column of the Kac table. In \cref{sec:colconfs}, we review the construction of single- and double-column diagrams which were previously introduced in the analysis of critical dense polymers. In \cref{sec:excitations}, we formulate a set of empirical selection rules which describe, in terms of column diagrams, the patterns of zeros for the full set of eigenvalues of the standard modules. We use these to write down explicit expressions for the finite-size characters. These are simplified to the known finitized Kac characters in \cref{sec:charids} using identities derived in \cref{sec:qids}. In \cref{sec:cylPF}, we combine the partition functions of the standard modules using the Markov trace to obtain the conformal cylinder partition function.

In \cref{sec:periodiccase}, we present our results for periodic boundary conditions. \cref{sec:PTLandT,sec:Pstanmod,sec:Pfusion,sec:Pfunctional,sec:Pprops,sec:Pfsc,sec:Pcfgs} follow the same presentation as \cref{sec:TLandD,sec:stanmod,sec:fusion,sec:functional,sec:props,sec:fsc,sec:cfgs}, presenting the corresponding results for the periodic case. In \cref{sec:fsgf}, we write down empirical selection rules that describe the full set of finite excitations in the standard modules over the periodic Temperley-Lieb algebra $\eptl_N(\alpha, \beta)$. These allow us to write down explicit expressions for the spectrum generating functions, which are collected in \cref{sec:Pcpf}. Also in \cref{sec:Pcpf}, the behavior of these generating functions in the scaling limit is extracted using the identities derived in \cref{sec:qids}. In \cref{sec:mipf}, we combine the previous results using the equivalent of the Markov trace for the torus, compute the modular invariant/covariant partition functions and write the result in terms of $u(1)$ characters. \cref{sec:conclusion} presents a discussion of our results and an overview of future avenues to be explored. 
The torus partition functions for critical dense polymers and sample patterns of zeros for critical percolation are collected in Appendices C and D respectively.

%
\section{Conformal data of critical percolation}\label{sec:CFTdata}
%

For critical percolation, the central charge $c$ and the Virasoro Kac conformal weights $\Delta_{r,s}$ are given by
\be
c = 0, \qquad \Delta_{r,s} = \frac{(3r-2s)^2 -1}{24},
\ee
with $r,s \in \mathbb Z_{>0}$. These are organised in the infinitely extended Kac table in the left panel of \cref{fig:VirKac}. The conformal weights with $r=0,1$ and $s$ taking half-integer are given in the right panel.

In terms of the lattice data, the modular nome is given by
\be\label{eq:qbdy}
q = \exp\Big(\!-\! 2\pi \delta \sin(3u)\Big)
\ee
for the boundary case and by 
\be\label{eq:qper}
q = \exp\Big(\!-\! 2\pi \ir \delta\, \eE^{-3\ir u}\Big)
\ee
for the periodic case, where the aspect ratio is
\be
\delta = \lim_{M,N\rightarrow \infty} \frac MN.
\ee
The finitized Kac characters are given by
\be
\label{eq:finchar}
\XN_{r,s}(q) = q^{\Delta_{r,s}-c/24} \bigg(\qbinom{N}{\frac{N-s+r}2}-q^{rs} \qbinom{N}{\frac{N-s-r}2}\bigg)
\ee
and yield the conformal Kac characters in the scaling limit:
\be  
\chit_{r,s}(q) = \lim_{N\rightarrow \infty}\XN_{r,s}(q) = q^{\Delta_{r,s}-c/24}\,\frac{(1-q^{rs})}{(q)_\infty}
\ee
where 
\be\label{eq:Poch}
(q)_\infty = {\prod_{i=1}^\infty(1-q^i)}.
\ee
The $u(1)$ characters are given by
\begin{equation}
\varkappa_j^{n}(q) = \varkappa_j^{n}(q,1), \qquad \varkappa_j^{n}(q,z)=\frac{\Theta_{j,n}(q,z)}{q^{1/24}(q)_\infty}=\frac{q^{-1/24}}{(q)_\infty} \sum_{k\in{\mathbb Z}} z^kq^{(j+2kn)^2/4n},
\label{eq:u1chars}
\end{equation}
with $n = pp' = 6$ for percolation. 

%
\section{Critical percolation with strip boundary conditions}\label{sec:boundarycase}
%

\subsection{The transfer tangle and the Temperley-Lieb algebra}\label{sec:TLandD}

The dense loop model of critical percolation is a Temperley-Lieb model described in terms of the elementary face operator
\be
\psset{unit=.9cm}
\begin{pspicture}[shift=-.42](1,1)
\facegrid{(0,0)}{(1,1)}
\psarc[linewidth=0.025]{-}(0,0){0.16}{0}{90}
\rput(.5,.5){$u$}
\end{pspicture}
\ =\ 
s_1(-u)\ \begin{pspicture}[shift=-.45](1,1)
\facegrid{(0,0)}{(1,1)}
\rput[bl](0,0){\loopa}
\end{pspicture}
\;+\,s_0(u)\
\begin{pspicture}[shift=-.42](1,1)
\facegrid{(0,0)}{(1,1)}
\rput[bl](0,0){\loopb}
\end{pspicture}
\label{eq:faceop}
\ee
where 
\be
s_k(u)=\frac{\sin (u+k\lambda)}{\sin\lambda}.
\ee
Here $\lambda$ is the crossing parameter and is related to the loop fugacity $\beta$ by $\beta = 2 \cos \lambda$. For critical percolation, $\lambda = \frac \pi 3$ and therefore $\beta =1$.
The double-row transfer tangle is defined as
\be
\Db (u)= (-1)^N
\ 
\psset{unit=0.9}
\begin{pspicture}[shift=-1.7](-0.5,-0.8)(5.5,2.0)
\facegrid{(0,0)}{(5,2)}
\psarc[linewidth=0.025]{-}(0,0){0.16}{0}{90}
\psarc[linewidth=0.025]{-}(0,1){0.16}{0}{90}
\psarc[linewidth=0.025]{-}(1,0){0.16}{0}{90}
\psarc[linewidth=0.025]{-}(1,1){0.16}{0}{90}
\psarc[linewidth=0.025]{-}(4,0){0.16}{0}{90}
\psarc[linewidth=0.025]{-}(4,1){0.16}{0}{90}
\rput(2.5,0.5){$\ldots$}
\rput(2.5,1.5){$\ldots$}
\rput(3.5,0.5){$\ldots$}
\rput(3.5,1.5){$\ldots$}
\psarc[linewidth=1.5pt,linecolor=blue]{-}(0,1){0.5}{90}{-90}
\psarc[linewidth=1.5pt,linecolor=blue]{-}(5,1){0.5}{-90}{90}
\rput(0.5,.5){$u$}
\rput(0.5,1.5){\small$\lambda\!-\!u$}
\rput(1.5,.5){$u$}
\rput(1.5,1.5){\small$\lambda\!-\!u$}
\rput(4.5,.5){$u$}
\rput(4.5,1.5){\small$\lambda\!-\!u$}
\rput(2.5,-0.5){$\underbrace{\qquad \qquad \qquad \qquad \qquad  \qquad}_N$}
\end{pspicture}\ 
\label{eq:Du}
\ee
where $u$ is the spectral parameter. The tangle $\Db(u)$ is a linear combination of connectivity diagrams and therefore an element of the Temperley-Lieb algebra \cite{TL71} $\tl_N(\beta)$ at $\beta =1$:
\be\label{eq:TLdiag}
 \tl_N(\beta)=\big\langle I,\,e_j ;\,j=1,\ldots,N-1\big\rangle,\qquad
 I=\,
\begin{pspicture}[shift=-0.55](0.0,-0.65)(2.0,0.45)
\pspolygon[fillstyle=solid,fillcolor=lightlightblue,linewidth=0pt](0,-0.35)(2.0,-0.35)(2.0,0.35)(0,0.35)
\rput(1.4,0.0){\small$...$}
\psline[linecolor=blue,linewidth=1.5pt]{-}(0.2,0.35)(0.2,-0.35)\rput(0.2,-0.55){$_1$}
\psline[linecolor=blue,linewidth=1.5pt]{-}(0.6,0.35)(0.6,-0.35)\rput(0.6,-0.55){$_2$}
\psline[linecolor=blue,linewidth=1.5pt]{-}(1.0,0.35)(1.0,-0.35)\rput(1.0,-0.55){$_3$}
\psline[linecolor=blue,linewidth=1.5pt]{-}(1.8,0.35)(1.8,-0.35)\rput(1.8,-0.55){$_N$}
\end{pspicture} 
\ ,\qquad
 e_j=\,
 \begin{pspicture}[shift=-0.55](0.0,-0.65)(3.2,0.45)
\pspolygon[fillstyle=solid,fillcolor=lightlightblue,linewidth=0pt](0,-0.35)(3.2,-0.35)(3.2,0.35)(0,0.35)
\rput(0.6,0.0){\small$...$}
\rput(2.6,0.0){\small$...$}
\psline[linecolor=blue,linewidth=1.5pt]{-}(0.2,0.35)(0.2,-0.35)\rput(0.2,-0.55){$_1$}
\psline[linecolor=blue,linewidth=1.5pt]{-}(1.0,0.35)(1.0,-0.35)
\psline[linecolor=blue,linewidth=1.5pt]{-}(2.2,0.35)(2.2,-0.35)
\psline[linecolor=blue,linewidth=1.5pt]{-}(3.0,0.35)(3.0,-0.35)\rput(3.0,-0.55){$_{N}$}
\psarc[linecolor=blue,linewidth=1.5pt]{-}(1.6,0.35){0.2}{180}{0}\rput(1.35,-0.55){$_j$}
\psarc[linecolor=blue,linewidth=1.5pt]{-}(1.6,-0.35){0.2}{0}{180}\rput(1.85,-0.55){$_{j+1}$}
\end{pspicture} \ .
\ee
The algebra 
$\tl_N(\beta)$ is a unital associative algebra whose defining relations are
\be
e_j^2=\beta e_j, \qquad e_j e_{j\pm1} e_j = e_j, \qquad e_i e_j = e_j e_i \qquad (|i-j|>1).
\label{eq:TLdef}
\ee
The transfer tangle satisfies a number of relations, in particular the crossing symmetry $\Db(\lambda - u) = \Db(u)$, the periodicity symmetry $\Db(u+\pi) = \Db(u)$, the commutativity property $[\Db(u),\Db(v)] = 0$ and the initial condition $\Db(u=0) = (-1)^N \Ib$.
We sometimes denote the identity connectivity using the bold letter $\Ib$.

The braid transfer matrix is also an element of $\tl_N(\beta)$. It is defined by
\be
\label{eq:braidD}
\Db_\infty= \ 
\psset{unit=0.9}
\begin{pspicture}[shift=-1.7](-0.5,-0.8)(5.5,2.0)
\facegrid{(0,0)}{(5,2)}
\rput(2.5,0.5){$\ldots$}
\rput(2.5,1.5){$\ldots$}
\rput(3.5,0.5){$\ldots$}
\rput(3.5,1.5){$\ldots$}
\psarc[linewidth=1.5pt,linecolor=blue]{-}(0,1){0.5}{90}{-90}
\psarc[linewidth=1.5pt,linecolor=blue]{-}(5,1){0.5}{-90}{90}
\rput(0,0){\psline[linewidth=1.5pt,linecolor=blue]{-}(0.0,0.5)(1,0.5)
\psline[linewidth=1.5pt,linecolor=blue]{-}(0.5,0)(0.5,0.35)
\psline[linewidth=1.5pt,linecolor=blue]{-}(0.5,0.65)(0.5,1)}
\rput(1,0){\psline[linewidth=1.5pt,linecolor=blue]{-}(0.0,0.5)(1,0.5)
\psline[linewidth=1.5pt,linecolor=blue]{-}(0.5,0)(0.5,0.35)
\psline[linewidth=1.5pt,linecolor=blue]{-}(0.5,0.65)(0.5,1)}
\rput(4,0){\psline[linewidth=1.5pt,linecolor=blue]{-}(0.0,0.5)(1,0.5)
\psline[linewidth=1.5pt,linecolor=blue]{-}(0.5,0)(0.5,0.35)
\psline[linewidth=1.5pt,linecolor=blue]{-}(0.5,0.65)(0.5,1)}
\rput(0,1){\psline[linewidth=1.5pt,linecolor=blue]{-}(0.5,0.0)(0.5,1)
\psline[linewidth=1.5pt,linecolor=blue]{-}(0,0.5)(0.35,0.5)
\psline[linewidth=1.5pt,linecolor=blue]{-}(0.65,0.5)(1,0.5)}
\rput(1,1){\psline[linewidth=1.5pt,linecolor=blue]{-}(0.5,0.0)(0.5,1)
\psline[linewidth=1.5pt,linecolor=blue]{-}(0,0.5)(0.35,0.5)
\psline[linewidth=1.5pt,linecolor=blue]{-}(0.65,0.5)(1,0.5)}
\rput(4,1){\psline[linewidth=1.5pt,linecolor=blue]{-}(0.5,0.0)(0.5,1)
\psline[linewidth=1.5pt,linecolor=blue]{-}(0,0.5)(0.35,0.5)
\psline[linewidth=1.5pt,linecolor=blue]{-}(0.65,0.5)(1,0.5)}
\rput(2.5,-0.5){$\underbrace{\qquad \qquad \qquad \qquad \qquad  \qquad}_N$}
\end{pspicture}\
\ee
where the elementary braid operators are given by
\be
\label{eq:braidops}
\psset{unit=.9cm}
\begin{pspicture}[shift=-.4](1,1)
\facegrid{(0,0)}{(1,1)}
\psline[linewidth=1.5pt,linecolor=blue]{-}(0,0.5)(1,0.5)
\psline[linewidth=1.5pt,linecolor=blue]{-}(0.5,0)(0.5,0.35)
\psline[linewidth=1.5pt,linecolor=blue]{-}(0.5,0.65)(0.5,1)
\end{pspicture}
\ = \ e^{-\ir\tfrac{\pi-\lambda}2}\;\begin{pspicture}[shift=-.4](1,1)
\facegrid{(0,0)}{(1,1)}
\rput[bl](0,0){\loopa}
\end{pspicture}\;+\,e^{\ir\tfrac{\pi-\lambda}2}\;
\begin{pspicture}[shift=-.4](1,1)
\facegrid{(0,0)}{(1,1)}
\rput[bl](0,0){\loopb}
\end{pspicture}\ ,
\qquad \qquad 
\psset{unit=.9cm}
\begin{pspicture}[shift=-.4](1,1)
\facegrid{(0,0)}{(1,1)}
\psline[linewidth=1.5pt,linecolor=blue]{-}(0,0.5)(0.35,0.5)
\psline[linewidth=1.5pt,linecolor=blue]{-}(0.65,0.5)(1,0.5)
\psline[linewidth=1.5pt,linecolor=blue]{-}(0.5,0)(0.5,1)
\end{pspicture}
\ =\
e^{\ir\tfrac{\pi-\lambda}2}\;
\begin{pspicture}[shift=-.4](1,1)
\facegrid{(0,0)}{(1,1)}
\rput[bl](0,0){\loopa}
\end{pspicture}\;+\,e^{-\ir\tfrac{\pi-\lambda}2}\;
\begin{pspicture}[shift=-.4](1,1)
\facegrid{(0,0)}{(1,1)}
\rput[bl](0,0){\loopb}
\end{pspicture}\ .
\ee
The braid transfer matrix is obtained as the $u \rightarrow \ir \infty$ limit of $\Db(u)$:
\be
\Db_\infty = \lim_{u \rightarrow \ir \infty}  \bigg(\frac{\eE^{\ir(\pi-\lambda)}}{s_{0}(u)^2}\bigg)^N (-1)^N\Db(u).
\ee
We note that $\Db_\infty$ is also obtained by taking the limit $u\to -\ir \infty$ of $\Db(u)$.

\subsection{Standard modules}\label{sec:stanmod}

The representation theory of the Temperley-Lieb algebra was investigated by Jones \cite{Jones83}, Martin \cite{M91}, Goodman and Wenzl \cite{GW93} and Westbury \cite{W95} and was recently reviewed by Ridout and Saint-Aubin \cite{RSA14}. In the following, we study the action of $\Db(u)$ on a family of finite-dimensional modules over $\tl_N(\beta)$: the standard modules $\stan_N^d$. These modules are constructed on the vector spaces generated from link states with $d$ defects, with $0 \le d \le N$ and $d \equiv N \text{ mod }2$, and have dimension
\be
\dim \stan_N^d = \binom{N}{\frac{N-d}2}-\binom{N}{\frac{N-d-2}2}.
\ee
For example, for $N = 6$ and $d=2$, there are nine link states:
\be
\begin{array}{c}
\psset{unit=0.8cm}
\begin{pspicture}[shift=-0.0](-0.0,0)(2.4,0.5)
\psline{-}(0,0)(2.4,0)
\psarc[linecolor=darkgreen,linewidth=1.5pt]{-}(0.4,0){0.2}{0}{180}
\psarc[linecolor=darkgreen,linewidth=1.5pt]{-}(1.2,0){0.2}{0}{180}
\psline[linecolor=darkgreen,linewidth=1.5pt]{-}(1.8,0)(1.8,0.5)
\psline[linecolor=darkgreen,linewidth=1.5pt]{-}(2.2,0)(2.2,0.5)
\end{pspicture}\ , \ \ 
\begin{pspicture}[shift=-0.0](-0.0,0)(2.4,0.5)
\psline{-}(0,0)(2.4,0)
\psarc[linecolor=darkgreen,linewidth=1.5pt]{-}(0.4,0){0.2}{0}{180}
\psarc[linecolor=darkgreen,linewidth=1.5pt]{-}(1.6,0){0.2}{0}{180}
\psline[linecolor=darkgreen,linewidth=1.5pt]{-}(1.0,0)(1.0,0.5)
\psline[linecolor=darkgreen,linewidth=1.5pt]{-}(2.2,0)(2.2,0.5)
\end{pspicture}\ , \ \ 
\begin{pspicture}[shift=-0.0](-0.0,0)(2.4,0.5)
\psline{-}(0,0)(2.4,0)
\psarc[linecolor=darkgreen,linewidth=1.5pt]{-}(0.4,0){0.2}{0}{180}
\psarc[linecolor=darkgreen,linewidth=1.5pt]{-}(2.0,0){0.2}{0}{180}
\psline[linecolor=darkgreen,linewidth=1.5pt]{-}(1.0,0)(1.0,0.5)
\psline[linecolor=darkgreen,linewidth=1.5pt]{-}(1.4,0)(1.4,0.5)
\end{pspicture}\ , \ \ 
\begin{pspicture}[shift=-0.0](-0.0,0)(2.4,0.5)
\psline{-}(0,0)(2.4,0)
\psarc[linecolor=darkgreen,linewidth=1.5pt]{-}(0.8,0){0.2}{0}{180}
\psarc[linecolor=darkgreen,linewidth=1.5pt]{-}(1.6,0){0.2}{0}{180}
\psline[linecolor=darkgreen,linewidth=1.5pt]{-}(0.2,0)(0.2,0.5)
\psline[linecolor=darkgreen,linewidth=1.5pt]{-}(2.2,0)(2.2,0.5)
\end{pspicture}\ , \ \ 
\begin{pspicture}[shift=-0.0](-0.0,0)(2.4,0.5)
\psline{-}(0,0)(2.4,0)
\psarc[linecolor=darkgreen,linewidth=1.5pt]{-}(0.8,0){0.2}{0}{180}
\psarc[linecolor=darkgreen,linewidth=1.5pt]{-}(2.0,0){0.2}{0}{180}
\psline[linecolor=darkgreen,linewidth=1.5pt]{-}(0.2,0)(0.2,0.5)
\psline[linecolor=darkgreen,linewidth=1.5pt]{-}(1.4,0)(1.4,0.5)
\end{pspicture}\ , \\[0.3cm]
\psset{unit=0.8cm}
\begin{pspicture}[shift=-0.0](-0.0,0)(2.4,0.5)
\psline{-}(0,0)(2.4,0)
\psline[linecolor=darkgreen,linewidth=1.5pt]{-}(0.2,0)(0.2,0.5)
\psline[linecolor=darkgreen,linewidth=1.5pt]{-}(0.6,0)(0.6,0.5)
\psarc[linecolor=darkgreen,linewidth=1.5pt]{-}(1.2,0){0.2}{0}{180}
\psarc[linecolor=darkgreen,linewidth=1.5pt]{-}(2.0,0){0.2}{0}{180}
\end{pspicture}\ , \ \ 
\begin{pspicture}[shift=-0.0](-0.0,0)(2.4,0.5)
\psline{-}(0,0)(2.4,0)
\psarc[linecolor=darkgreen,linewidth=1.5pt]{-}(0.8,0){0.2}{0}{180}
\psbezier[linecolor=darkgreen,linewidth=1.5pt]{-}(0.2,0)(0.2,0.7)(1.4,0.7)(1.4,0)
\psline[linecolor=darkgreen,linewidth=1.5pt]{-}(1.8,0)(1.8,0.5)
\psline[linecolor=darkgreen,linewidth=1.5pt]{-}(2.2,0)(2.2,0.5)
\end{pspicture}\ , \ \ 
\begin{pspicture}[shift=-0.0](-0.0,0)(2.4,0.5)
\psline{-}(0,0)(2.4,0)
\psarc[linecolor=darkgreen,linewidth=1.5pt]{-}(1.2,0){0.2}{0}{180}
\psbezier[linecolor=darkgreen,linewidth=1.5pt]{-}(0.6,0)(0.6,0.7)(1.8,0.7)(1.8,0)
\psline[linecolor=darkgreen,linewidth=1.5pt]{-}(0.2,0)(0.2,0.5)
\psline[linecolor=darkgreen,linewidth=1.5pt]{-}(2.2,0)(2.2,0.5)
\end{pspicture}\ , \ \ 
\begin{pspicture}[shift=-0.0](-0.0,0)(2.4,0.5)
\psline{-}(0,0)(2.4,0)
\psarc[linecolor=darkgreen,linewidth=1.5pt]{-}(1.6,0){0.2}{0}{180}
\psbezier[linecolor=darkgreen,linewidth=1.5pt]{-}(1.0,0)(1.0,0.7)(2.2,0.7)(2.2,0)
\psline[linecolor=darkgreen,linewidth=1.5pt]{-}(0.2,0)(0.2,0.5)
\psline[linecolor=darkgreen,linewidth=1.5pt]{-}(0.6,0)(0.6,0.5)
\end{pspicture}\ .
\end{array}
\ee

The standard modules are defined by the defect-preserving action of the Temperley-Lieb connectivity diagrams on the link patterns. To compute this action, one draws the link state above the connectivity diagram and reads the new link state from the bottom nodes. A multiplicative factor of $\beta$ is then added for each closed loop. The result is set to zero if the number of defects of the new link state is smaller than that of the original link state.
Here are examples to illustrate:
\be
\begin{pspicture}[shift=-0.55](0,-0.65)(1.6,0.95)
\pspolygon[fillstyle=solid,fillcolor=lightlightblue,linewidth=0pt](0,-0.35)(1.6,-0.35)(1.6,0.35)(0,0.35)
\psline[linecolor=blue,linewidth=1.5pt]{-}(0.2,0.35)(0.2,-0.35)
\psline[linecolor=blue,linewidth=1.5pt]{-}(0.6,0.35)(0.6,-0.35)
\psarc[linecolor=blue,linewidth=1.5pt]{-}(1.2,0.35){0.2}{180}{0}
\psarc[linecolor=blue,linewidth=1.5pt]{-}(1.2,-0.35){0.2}{0}{180}
\psline{-}(0,0.35)(1.6,0.35)
\psarc[linecolor=darkgreen,linewidth=1.5pt]{-}(0.8,0.35){0.2}{0}{180}
\psbezier[linecolor=darkgreen,linewidth=1.5pt]{-}(0.2,0.35)(0.2,1.05)(1.4,1.05)(1.4,0.35)
\end{pspicture} \ = \ 
\begin{pspicture}[shift=0.0](0,0.35)(1.6,0.95)
\psline{-}(0,0.35)(1.6,0.35)
\psarc[linecolor=darkgreen,linewidth=1.5pt]{-}(0.4,0.35){0.2}{0}{180}
\psarc[linecolor=darkgreen,linewidth=1.5pt]{-}(1.2,0.35){0.2}{0}{180}
\end{pspicture} \, ,
\qquad
\begin{pspicture}[shift=-0.55](0,-0.65)(1.6,0.95)
\pspolygon[fillstyle=solid,fillcolor=lightlightblue,linewidth=0pt](0,-0.35)(1.6,-0.35)(1.6,0.35)(0,0.35)
\psline[linecolor=blue,linewidth=1.5pt]{-}(0.2,0.35)(0.2,-0.35)
\psline[linecolor=blue,linewidth=1.5pt]{-}(0.6,0.35)(0.6,-0.35)
\psarc[linecolor=blue,linewidth=1.5pt]{-}(1.2,0.35){0.2}{180}{0}
\psarc[linecolor=blue,linewidth=1.5pt]{-}(1.2,-0.35){0.2}{0}{180}
\psline{-}(0,0.35)(1.6,0.35)
\psarc[linecolor=darkgreen,linewidth=1.5pt]{-}(1.2,0.35){0.2}{0}{180}
\psline[linecolor=darkgreen,linewidth=1.5pt]{-}(0.2,0.35)(0.2,0.85)
\psline[linecolor=darkgreen,linewidth=1.5pt]{-}(0.6,0.35)(0.6,0.85)
\end{pspicture} \ = \beta \ 
\begin{pspicture}[shift=0.0](0,0.35)(1.6,0.95)
\psline{-}(0,0.35)(1.6,0.35)
\psarc[linecolor=darkgreen,linewidth=1.5pt]{-}(1.2,0.35){0.2}{0}{180}
\psline[linecolor=darkgreen,linewidth=1.5pt]{-}(0.2,0.35)(0.2,0.85)
\psline[linecolor=darkgreen,linewidth=1.5pt]{-}(0.6,0.35)(0.6,0.85)
\end{pspicture}\, ,
\qquad
\begin{pspicture}[shift=-0.55](0,-0.65)(1.6,0.95)
\pspolygon[fillstyle=solid,fillcolor=lightlightblue,linewidth=0pt](0,-0.35)(1.6,-0.35)(1.6,0.35)(0,0.35)
\psline[linecolor=blue,linewidth=1.5pt]{-}(0.2,0.35)(0.2,-0.35)
\psline[linecolor=blue,linewidth=1.5pt]{-}(0.6,0.35)(0.6,-0.35)
\psarc[linecolor=blue,linewidth=1.5pt]{-}(1.2,0.35){0.2}{180}{0}
\psarc[linecolor=blue,linewidth=1.5pt]{-}(1.2,-0.35){0.2}{0}{180}
\psline{-}(0,0.35)(1.6,0.35)
\psarc[linecolor=darkgreen,linewidth=1.5pt]{-}(0.4,0.35){0.2}{0}{180}
\psline[linecolor=darkgreen,linewidth=1.5pt]{-}(1.0,0.35)(1.0,0.85)
\psline[linecolor=darkgreen,linewidth=1.5pt]{-}(1.4,0.35)(1.4,0.85)
\end{pspicture} \ = 0.
\ee 
The standard modules play a key role in the representation theory of $\tl_N(\beta)$. In particular, they generate a complete set of irreducible modules for generic values of $\beta$. 

The case $\beta = 1$ is however not generic: The standard modules remain indecomposable, but depending on $d$ some of them are reducible. Let us refer to the integers $d \equiv 2 \text{ mod } 3$ in the set $\{0, \dots, N\}$ as {\it critical integers} and to the maximal such integer as $\hat d$. For $d$ critical, the standard module $\stan_N^d$ is irreducible: $\stan_N^d \simeq \irre_N^d$. For $d \equiv 0,1 \text{ mod } 3$, $\stan_N^d$ typically has two composition factors: an irreducible submodule $R_N^d$ and an irreducible quotient $\irre_N^d$. The submodule $R_N^d$ is isomorphic to $\irre_N^{d'}$ where 
$d' = d+4$ for $d \equiv 0 \text{ mod } 3$ and $d' = d+2$ for $d \equiv 1 \text{ mod } 3$. In other words, $d'$ is the integer obtained by reflecting $d$ with respect to the next critical integer. The structure of $\stan_N^d$ in this case is easily understood from its Loewy diagram, which we write as $\stan_N^d \simeq \irre_N^d \rightarrow \irre_N^{d'}$. The arrow indicates that the states in $\irre_N^{d'}$ can be obtained from those in $\irre_N^d$ by the action of $\tl_N(\beta = 1)$, but not the other way around. If $d'>\hat d$, then $R_N^d$ is trivial and the corresponding standard module is irreducible: $\stan_N^d\simeq \irre_N^d$.

The modules $\irre_N^d$, with $0 \le d \le N$ and $d \equiv N \text{ mod } 2$,  form a complete list of non-isomorphic irreducible modules of $\tl_N(\beta =1)$. Their dimensions are given by
\be
\dim \irre_N^d = 
\left\{
\begin{array}{ll}\displaystyle
\dim \stan_N^d & d \equiv 2 \text{ mod } 3,\\[0.2cm]
\displaystyle
\sum_{k\ge 0} \dim \stan_N^{d+6k} - \sum_{k\ge 0} \dim \stan_N^{d'+6k} & d \equiv 0,1 \text{ mod } 3,
\end{array}
\right.
\ee
where it is understood that $\dim \stan_N^d = 0$ for $d>N$.
These dimensions are displayed in \cref{tab:dim} for $1\le N \le 10$. We note that $\irre_N^0$ and $\irre_N^1$ are always one-dimensional.

\begin{figure}
\begin{equation*}
\begin{pspicture}(0,0)(0.1,0.1)
\psline[linecolor=gray,linestyle=dashed,dash=1.5pt 2pt]{-}(4.91,-3.2)(4.91,-2.95)
\psline[linecolor=gray,linestyle=dashed,dash=1.5pt 2pt]{-}(4.91,-2.55)(4.91,-1.8)
\psline[linecolor=gray,linestyle=dashed,dash=1.5pt 2pt]{-}(4.91,-1.4)(4.91,-0.65)
\psline[linecolor=gray,linestyle=dashed,dash=1.5pt 2pt]{-}(4.91,-0.25)(4.91,0.5)
\psline[linecolor=gray,linestyle=dashed,dash=1.5pt 2pt]{-}(4.91,0.9)(4.91,1.65)
\psline[linecolor=gray,linestyle=dashed,dash=1.5pt 2pt]{-}(4.91,2.1)(4.91,2.6)
\psline[linecolor=gray,linestyle=dashed,dash=1.5pt 2pt]{-}(9.505,-3.2)(9.505,-2.4)
\psline[linecolor=gray,linestyle=dashed,dash=1.5pt 2pt]{-}(9.505,-1.95)(9.505,-1.25)
\psline[linecolor=gray,linestyle=dashed,dash=1.5pt 2pt]{-}(9.505,-0.8)(9.505,-0.10)
\psline[linecolor=gray,linestyle=dashed,dash=1.5pt 2pt]{-}(9.505,0.35)(9.505,2.6)
\psline[linecolor=gray,linestyle=dashed,dash=1.5pt 2pt]{-}(13.685,-3.2)(13.685,-2.95)
\psline[linecolor=gray,linestyle=dashed,dash=1.5pt 2pt]{-}(13.685,-2.55)(13.685,-1.8)
\psline[linecolor=gray,linestyle=dashed,dash=1.5pt 2pt]{-}(13.685,-1.4)(13.685,2.6)
\end{pspicture}
\begin{array}{c | cccccccccccccc}
N\backslash d & 0 & 1 & 2 & 3 & 4 & 5 & 6 & 7 & 8 & 9 & 10
\\
\hline\\[-0.3cm]
1 & & 1 & \phantom{1 \rightarrow 1} &\phantom{1 \rightarrow 1} &\phantom{1 \rightarrow 1} & \phantom{1 \rightarrow 1} &\phantom{1 \rightarrow 1} &\phantom{1 \rightarrow 1} & \phantom{1 \rightarrow 1} & \phantom{1 \rightarrow 1}& \phantom{1 \rightarrow 1}\\[0.1cm]
2 & 1 && 1 & & & & & & & \\[0.1cm]
3 && 1 \rightarrow 1&& 1 && && && \\[0.1cm]
4 & 1\rightarrow 1 && 3 && 1 && && && \\[0.1cm]
5 && 1\rightarrow 4&& 4 && 1 && && \\[0.1cm]
6 & 1\rightarrow 4 && 9 && 4\rightarrow 1&& 1&& && \\[0.1cm]
7 && 1\rightarrow 13&& 13\rightarrow 1&& 6 && 1 && \\[0.1cm]
8 & 1\rightarrow 13&& 28 && 13\rightarrow 7&& 7 && 1 && \\[0.1cm]
9 && 1\rightarrow 41&& 41\rightarrow 7&& 27 && 7 \rightarrow 1&& 1\\[0.1cm]
10 & 1\rightarrow 41 && 90 && 41\rightarrow 34 && 34 \rightarrow 1&& 9 && 1 \\[0.1cm]
\end{array}
\end{equation*}
\captionof{table}{The Loewy diagrams for the modules $\stan_N^d$ for $1 \le N \le 10$, where each composition factor $\irre_N^d$ is indicated by its dimension. Vertical dashed lines are drawn below the critical integers.}
\label{tab:dim}
\end{figure}

\subsection{Fused transfer matrices and the fusion hierarchy}\label{sec:fusion}

Starting from the transfer tangle $\Db^1(u) = \Db(u)$, one can construct a family of fused transfer tangles $\Db^n(u)$ satisfying the fusion hierarchy relations
\begin{equation}
 \Db^n_0\Db^{1}_n=\frac{s_{n-3}(2u)s_{2n}(2u)}{s_{n-2}(2u)s_{2n-1}(2u)} f_n \Db^{n-1}_0 + \frac{s_{n-1}(2u)s_{2n-2}(2u)}{s_{n-2}(2u)s_{2n-1}(2u)}f_{n-1}\, \Db^{n+1}_0, \qquad n \ge 0,
 \label{eq:fushier}
\end{equation}
where
\be
\Db_k^n = \Db^n(u + k \lambda), \qquad \Db_0^0 = f_{-1} \Ib, \qquad \Db^{-1}_k = 0, \qquad f_k = (-1)^{N}s_{k}(u)^{2N}.
\ee
These transfer tangles were constructed in terms of fused face operators in \cite{MDPR14}. They commute as elements of $\tl_N(\beta)$: $[\Db^m(u),\Db^n(v)] = 0$. The fusion hierarchy relations were proven using the diagrammatic calculus of the Temperley-Lieb algebra. Of particular relevance for our investigation in later sections is the fused transfer tangle with fusion label $n=2$. For $\beta\ne 0$, it is constructed from the Wenzl-Jones projector on two sites,
\be
\begin{pspicture}[shift=-0.05](0,-0.15)(1.0,0.15)
\pspolygon[fillstyle=solid,fillcolor=pink](0,-0.15)(1.0,-0.15)(1.0,0.15)(0,0.15)(0,0.15)
\rput(0.5,0){$_{2}$}
\end{pspicture} 
\ =\,
\begin{pspicture}[shift=-0.25](-0.0,-0.35)(0.8,0.35)
\psline[linecolor=blue,linewidth=1.5pt]{-}(0.2,0.35)(0.2,-0.35)
\psline[linecolor=blue,linewidth=1.5pt]{-}(0.6,0.35)(0.6,-0.35)
\end{pspicture} - \frac{1}{\beta}
\begin{pspicture}[shift=-0.25](-0.,-0.35)(0.8,0.35)
\psarc[linecolor=blue,linewidth=1.5pt]{-}(0.4,0.35){0.2}{180}{0}
\psarc[linecolor=blue,linewidth=1.5pt]{-}(0.4,-0.35){0.2}{0}{180}
\end{pspicture} \ ,
\ee
and the $1 \times 2$ fused face operator
\be\label{eq:1x2face}
\psset{unit=.9cm}
\begin{pspicture}[shift=-0.4](0,0)(1,1)
\pspolygon[fillstyle=solid,fillcolor=lightlightblue](0,0)(1,0)(1,1)(0,1)(0,0)
\psarc[linewidth=0.025]{-}(0,0){0.16}{0}{90}
\rput(0.5,0.75){\tiny{$_{1\times 2}$}}
\rput(0.5,0.5){$u$}
\end{pspicture} \ = \frac1{s_0(u)} \ \,
\begin{pspicture}[shift=-0.9](-0.3,0)(1.3,2)
\facegrid{(0,0)}{(1,2)}
\rput(0.5,0.55){$u$}
\rput(0.5,1.55){$u\!+\!\lambda$}
\pspolygon[fillstyle=solid,fillcolor=pink](0,0.1)(0,1.9)(-0.3,1.9)(-0.3,0.1)(0,0.1)\rput(-0.15,1){$_2$}
\pspolygon[fillstyle=solid,fillcolor=pink](1,0.1)(1,1.9)(1.3,1.9)(1.3,0.1)(1,0.1)\rput(1.15,1){$_2$}
\psarc[linewidth=0.025]{-}(0,0){0.16}{0}{90}
\psarc[linewidth=0.025]{-}(0,1){0.16}{0}{90}
\end{pspicture} \  =  s_1(-u) \ \,
\begin{pspicture}[shift=-0.9](-0.3,0)(1.3,2)
\facegrid{(0,0)}{(1,2)}
\rput(0,0){\loopa}
\rput(0,1){\loopa}
\pspolygon[fillstyle=solid,fillcolor=pink](0,0.1)(0,1.9)(-0.3,1.9)(-0.3,0.1)(0,0.1)\rput(-0.15,1){$_2$}
\pspolygon[fillstyle=solid,fillcolor=pink](1,0.1)(1,1.9)(1.3,1.9)(1.3,0.1)(1,0.1)\rput(1.15,1){$_2$}
\end{pspicture} \  + s_1(u) \ \,
\begin{pspicture}[shift=-0.9](-0.3,0)(1.3,2)
\facegrid{(0,0)}{(1,2)}
\rput(0,0){\loopb}
\rput(0,1){\loopb}
\pspolygon[fillstyle=solid,fillcolor=pink](0,0.1)(0,1.9)(-0.3,1.9)(-0.3,0.1)(0,0.1)\rput(-0.15,1){$_2$}
\pspolygon[fillstyle=solid,fillcolor=pink](1,0.1)(1,1.9)(1.3,1.9)(1.3,0.1)(1,0.1)\rput(1.15,1){$_2$}
\end{pspicture} \ \, .
\ee
The normalisation is different from that appearing in \cite{MDPR14} and is instead chosen such that the fusion hierarchy relations \eqref{eq:fushier} are identical to those of the rational models, see for instance \cite{CMP02}.
The $1\times 2$ fused transfer tangle is defined as
\be 
\psset{unit=.9cm}
\Db^{2}(u)= 
\begin{pspicture}[shift=-1.6](-0.7,-0.7)(5.7,2)
\psarc[linewidth=6pt,linecolor=blue]{-}(0,1){0.5}{90}{-90}\psarc[linewidth=4pt,linecolor=white]{-}(0,1){0.5}{90}{-90}
\psarc[linewidth=6pt,linecolor=blue]{-}(5,1){0.5}{-90}{90}\psarc[linewidth=4pt,linecolor=white]{-}(5,1){0.5}{-90}{90}
\facegrid{(0,0)}{(5,2)}
\psarc[linewidth=0.025]{-}(0,0){0.16}{0}{90}
\psarc[linewidth=0.025]{-}(0,1){0.16}{0}{90}
\psarc[linewidth=0.025]{-}(1,0){0.16}{0}{90}
\psarc[linewidth=0.025]{-}(1,1){0.16}{0}{90}
\psarc[linewidth=0.025]{-}(4,0){0.16}{0}{90}
\psarc[linewidth=0.025]{-}(4,1){0.16}{0}{90}
\rput(0.5,0.75){\tiny{$_{1\times 2}$}}\rput(0.5,1.75){\tiny{$_{1\times 2}$}}
\rput(1.5,0.75){\tiny{$_{1\times 2}$}}\rput(1.5,1.75){\tiny{$_{1\times 2}$}}
\rput(4.5,0.75){\tiny{$_{1\times 2}$}}\rput(4.5,1.75){\tiny{$_{1\times 2}$}}
\rput(2.5,0.5){$\ldots$}
\rput(2.5,1.5){$\ldots$}
\rput(3.5,0.5){$\ldots$}
\rput(3.5,1.5){$\ldots$}
\rput(0.5,.5){$u$}
\rput(0.52,1.45){\small$-u$}
\rput(1.5,.5){$u$}
\rput(1.52,1.45){\small$-u$}
\rput(4.5,.5){$u$}
\rput(4.52,1.45){\small$-u$}
\rput(2.5,-0.5){$\underbrace{\qquad \qquad \qquad \qquad \qquad  \qquad}_N$}
\end{pspicture} \ .
\label{eq:D2u}
\ee  
It satisfies the relation $\Db^2(-u)=\Db^2(u)$, the periodicity property $\Db(u+\pi)=\Db(u)$, the commutativity property $[\Db^2(u),\Db^n(v)] = 0$, $n=1,2$, and most importantly the fusion hierarchy relation \eqref{eq:fushier} 
with $n=1$.
The construction of $\Db^n(u)$ for $n>2$ in terms of diagrams uses similar ideas.

For rational values of the crossing parameter, that is for $\lambda =\frac{(p'-p)\pi}{p'}$ with $(p,p')$ a pair of integers satisfying \eqref{eq:ppp},
it was shown in \cite{MDPR14} that the fused transfer tangles satisfy the closure relation
\be
\Db^{p'}_0 = \Db^{p'-2}_1 + 2(-1)^{p'-p} f_{-1} \Ib.
\label{eq:clo}
\ee
Critical percolation corresponds to $(p,p') = (2,3)$, in which case this is just
\be
\Db^{3}_0 = \Db^{1}_1 - 2 f_{-1} \Ib.
\label{eq:cloperco}
\ee
The proof given in \cite{MDPR14} relies on diagrammatic manipulations performed on the tangles $\Db^n(u)$ as elements of $\tl_N(\beta)$. If follows that \eqref{eq:cloperco} holds for all modules, and in particular for $\stan_N^d$.

\subsection[$T$-systems and $Y$-systems]{The $\boldsymbol T$-system and the $\boldsymbol Y$-system}
\label{sec:functional}
Using the fusion hierarchy relations and a recursive argument \cite{KP92}, one can show that the transfer tangles satisfy a set of functional relations known as the $T$-system:
\begin{equation}
 \Db^{n}_0\Db^{n}_1 = \frac{s_{-2}(2u)s_{2n}(2u)}{s_{n-2}(2u)s_{n}(2u)} f_{-1} f_n \Ib +  \frac{s_{n-1}(2u)^2}{s_{n-2}(2u)s_{n}(2u)} \Db^{n+1}_0\Db^{n-1}_1, \qquad n \ge 0.
 \label{eq:tsys}
\end{equation}
The $T$-system holds for generic values of $\beta$. By defining
\be
\db^n(u) = \frac{s_{n-1}(2u)^2}{s_{-2}(2u)s_{2n}(2u)} \frac{\Db_1^{n-1}\Db_0^{n+1}}{f_{-1}f_n}, \qquad n \ge 0,
\ee
and $\db^n_k = \db^n(u+k \lambda)$, one finds that these tangles satisfy a set of non-linear equations known as the (universal) $Y$-system:
\be
 \db^{n}_0 \db^{n}_1=\big(\Ib+ \db^{n-1}_1\big)\big(\Ib+ \db^{n+1}_0\big), \qquad n \ge 1,
 \label{eq:ysys}
\ee
where $\db^0_k = 0$.

\begin{figure}[h] 
\centering
\begin{tabular}{ccc}
$
A_\infty: \quad
\begin{pspicture}[shift=-0.9](0,-1)(6,1)
\multiput(0,0)(1,0){4}{\psline{-}(0,0)(1,0)}
\rput(4,0){\psline{-}(0,0)(0.5,0)}
\multiput(0,0)(1,0){5}{\pscircle[linewidth=1.5pt,linecolor=black,fillstyle=solid,fillcolor=white](0,0){.175}}
\rput(0,0.375){\scriptsize$1$}\rput(1,0.375){\scriptsize$2$}\rput(2,0.375){\scriptsize$3$}\rput(3,0.375){\scriptsize$4$}\rput(4,0.375){\scriptsize$\cdots$}
\rput(5,0){$\cdots$}
\end{pspicture}
$
&
\qquad
&
$
D_{p'}: \quad
\begin{pspicture}[shift=-0.9](0,-1)(5,1)
\multiput(0,0)(1,0){3}{\psline{-}(0,0)(1,0)}
\psline{-}(3,0)(3.7,0.7)\psline{-}(3,0)(3.7,-0.7)
\multiput(0,0)(1,0){4}{\pscircle[linewidth=1.5pt,linecolor=black,fillstyle=solid,fillcolor=white](0,0){.175}}
\rput(0,0.35){\scriptsize$1$}\rput(1,0.375){\scriptsize$2$}\rput(2,0.375){\scriptsize$\dots$}\rput(2.9,0.375){\scriptsize$p'\!-\!2$}
\pscircle[linewidth=1.5pt,linecolor=black,fillstyle=solid,fillcolor=white](3.7,0.7){.175}
\pscircle[linewidth=1.5pt,linecolor=black,fillstyle=solid,fillcolor=white](3.7,-0.7){.175}
\rput(3.7,1.06){\scriptsize$p'\!-\!1$}
\rput(3.7,-1.10){\scriptsize$p'$}
\rput(3.7,0.7){\tiny$+$}
\rput(3.7,-0.7){\tiny$-$}
\end{pspicture}
$
\end{tabular}
\caption{The Dynkin diagrams for $A_\infty$ and $D_{p'}$.}
\label{fig:Dynkin}
\end{figure}
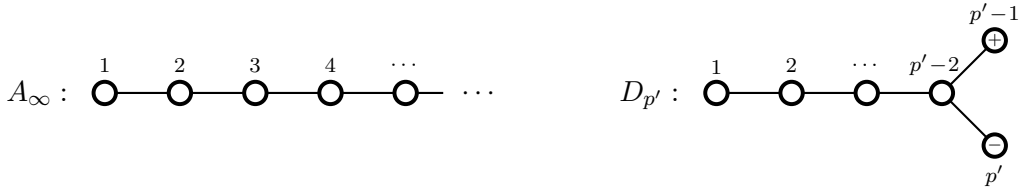

For rational values of $\lambda$, this set of non-linear relations closes finitely. Indeed, it was found in \cite{MDPR14} that the tangles $\db^n_k$ satisfy a linear, four-term closure relation. After a careful analysis, we find that this closure relation is not convenient for extracting eigenvalue solutions for $\Db(u)$. Inspired by ideas applied to vertex models \cite{KSS98}, one instead defines the tangle
\be
\Kb_0 = \frac{(-1)^{p'-p}}{f_{-1}}\Db_1^{p'-2}
\label{eq:Ktang}
\ee
and, using \eqref{eq:clo} and \eqref{eq:tsys}, finds the following alternative closure relations
\be
\Ib+ \db^{p'-1}_0 = (\Ib+\Kb_0)^2, \qquad \Kb_0\Kb_1 = \Ib+\db_1^{p'-2}.
\label{eq:yclo}
\ee
The closed $Y$-system thus consists of the relations \eqref{eq:ysys} for $n = 1, \dots, p'-2$ along with the relations \eqref{eq:yclo}. This truncates the initial infinite $Y$-system, which corresponds to a one-sided $A_\infty$ Dynkin diagram, to a finite $Y$-system described by a Dynkin diagram of type $D_{p'}$, see \cref{fig:Dynkin}.

For $(p,p')=(2,3)$, the Dynkin diagram is $D_3 \simeq A_3$ and the finite $Y$-system only
involves three tangles: 
the tangles $\Kb$ and $\db^1$, which are scalar multiples of $\Db^1(u)$ and $\Db^2(u)$ respectively, and the identity $\Ib$. It takes the form 
\begin{equation}
\db_0^1\db_1^1 = (\Ib+\Kb_0)^2, \qquad \Kb_0\Kb_1 = \Ib + \db_1^1.
\label{eq:ysysperco1}
\end{equation}
By defining
\begin{subequations}
\label{eq:aU}
\begin{alignat}{3}
&\ab^1(x) = \db_0^1(\tfrac {\ir x} 3), \qquad  &&\ab^2(x) = \Kb_0(\tfrac {\ir x} 3- \tfrac \pi 6), \\ &\Ab^1(x) = \Ib+\ab^1(x), \qquad  &&\Ab^2(x) = \big(\Ib+\ab^2(x)\big)^2,
\end{alignat}
\end{subequations}
the $Y$-system is written in a symmetric form:
\begin{equation}
\ab^1(x-\tfrac{\ir \pi}2)\ab^1(x+\tfrac{\ir \pi}2) = \Ab^2(x),\qquad
\ab^2(x-\tfrac{\ir \pi}2)\ab^2(x+\tfrac{\ir \pi}2) = \Ab^1(x).
\label{eq:finaly}
\end{equation}

\subsection{Properties of the eigenvalues}\label{sec:props}

The functional relations given in the previous section were derived quite generally in \cite{MDPR14} using the diagrammatic calculus of the Temperley-Lieb algebra. The eigenvalues of the corresponding tangles are solutions to these relations in any given representation. In \cref{sec:fsc}, we use these relations to extract the finite-size corrections for the eigenvalues of $\Db(u)$ in the modules $\stan_N^d$. The analysis is based on some properties of the eigenvalues of $\Db^n(u)$, $\db^1(u)$ and $\Kb(u)$ in $\stan_N^d$, which we respectively denote by $D^n(u)$, $d^1(u)$ and $K(u)$. This section details these properties.

\paragraph{Razumov-Stroganov eigenvalues.}
A simple solution to \eqref{eq:ysysperco1} is $K_0 = -1$ and $d_0^1 = 0$, which corresponds to 
\be 
D^1(u) = f_{-2} = (-1)^N \bigg(\frac{\sin (u+\frac\pi3)}{\sin \frac\pi3}\bigg)^{2N}, \qquad D^2(u) = 0.
\label{eq:RSeig}
\ee
This solution appears once, as the ground state, 
in the spectrum of the standard module $\stan_N^d$ for $d=0$ if $N$ is even,  and for $d=1$ if $N$ is odd. This is the celebrated Razumov-Stroganov eigenvalue for the loop model with strip boundary conditions \cite{BGN01,MBNM04}. It corresponds to the unique eigenvalue of the trivial rational model of percolation, or alternatively to the one-dimensional irreducible representations of $\tl_N(\beta = 1)$, $\irre_N^0$ or $\irre_N^1$, see \cref{sec:stanmod}. In this case, the expansion \eqref{eq:Dexpansion} can be computed exactly and the $\frac1N$ finite-size corrections are exactly zero, consistent with $c= \Delta = 0$. 

\begin{figure}[h] 
\centering
\begin{tabular}{cc}
$D^1(u):$ & $D^2(u):$\\[0.1cm]
\includegraphics[width=.45\textwidth]{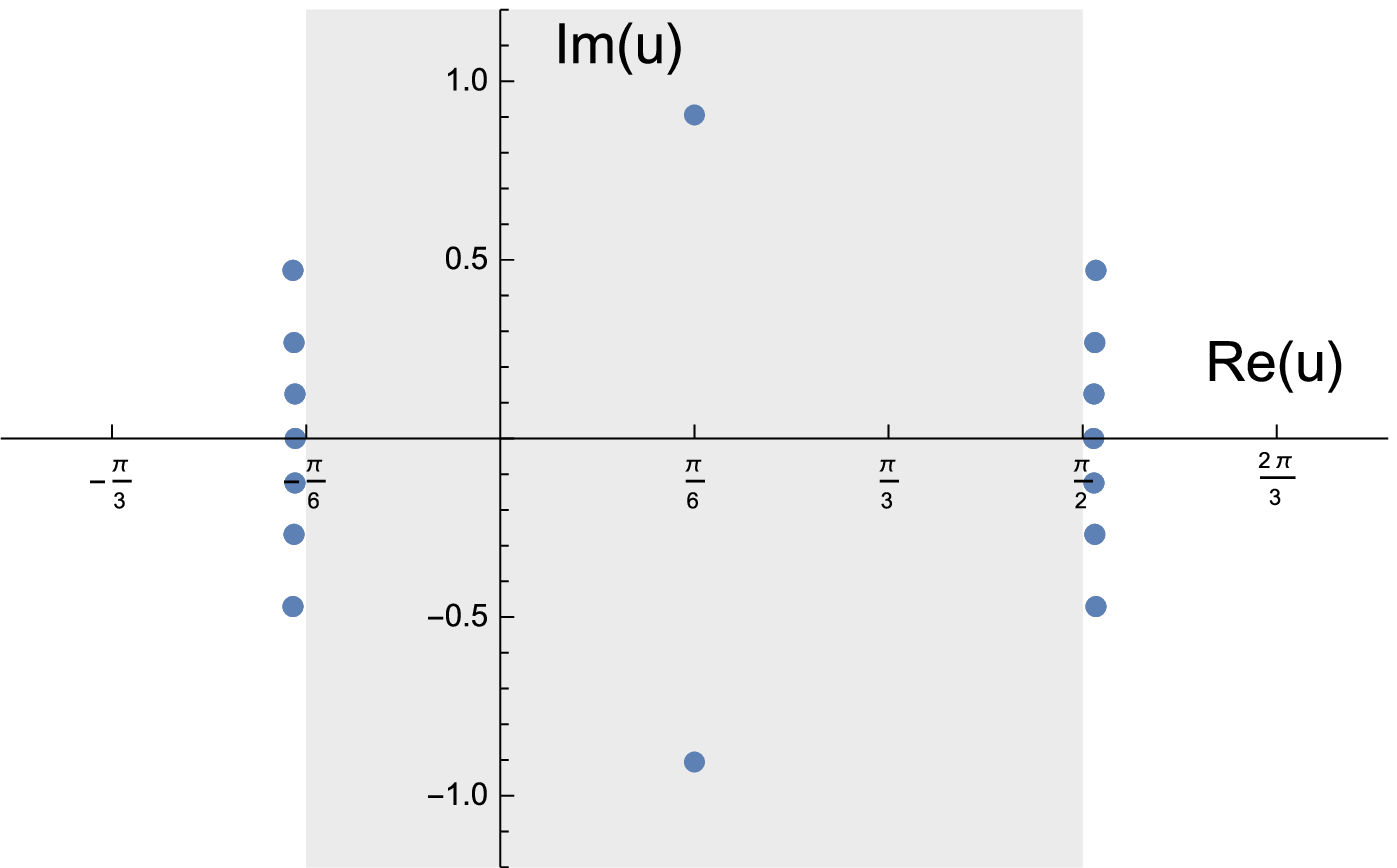}\qquad \qquad&
\includegraphics[width=.45\textwidth]{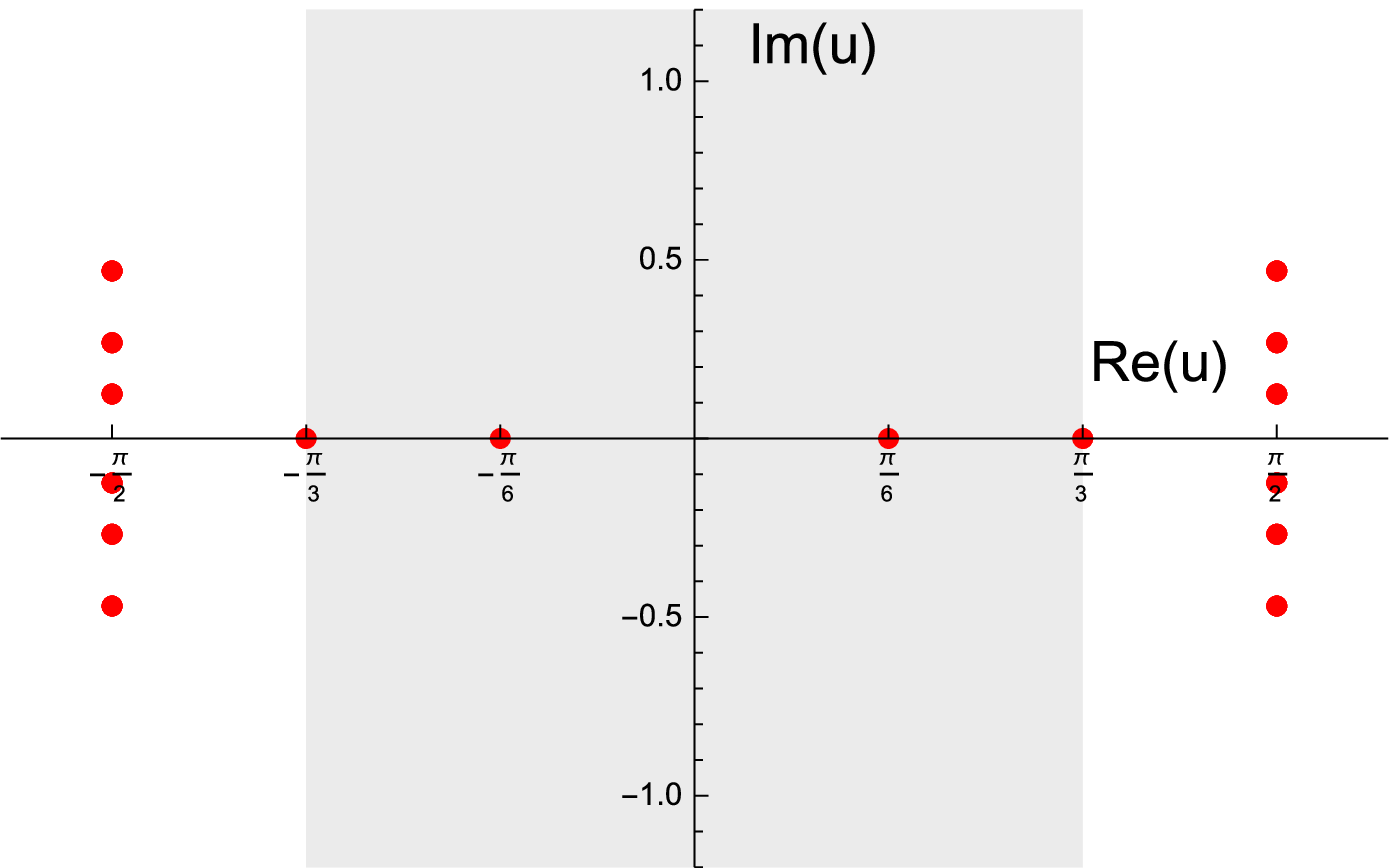} 
\end{tabular}
\caption{The location of the zeros for the ground state in $\stan_8^2$. The analyticity strips are shaded in gray. There is one pair of complex conjugate zeros in the analyticity strip for $D^1(u)$ and four real zeros on the real line for $D^2(u)$.}
\label{fig:eigpatterns}
\end{figure}

\begin{figure}[h] 
\centering
\begin{tabular}{cc}
$D^1(u):$ & $D^2(u):$\\[0.1cm]
\includegraphics[width=.45\textwidth]{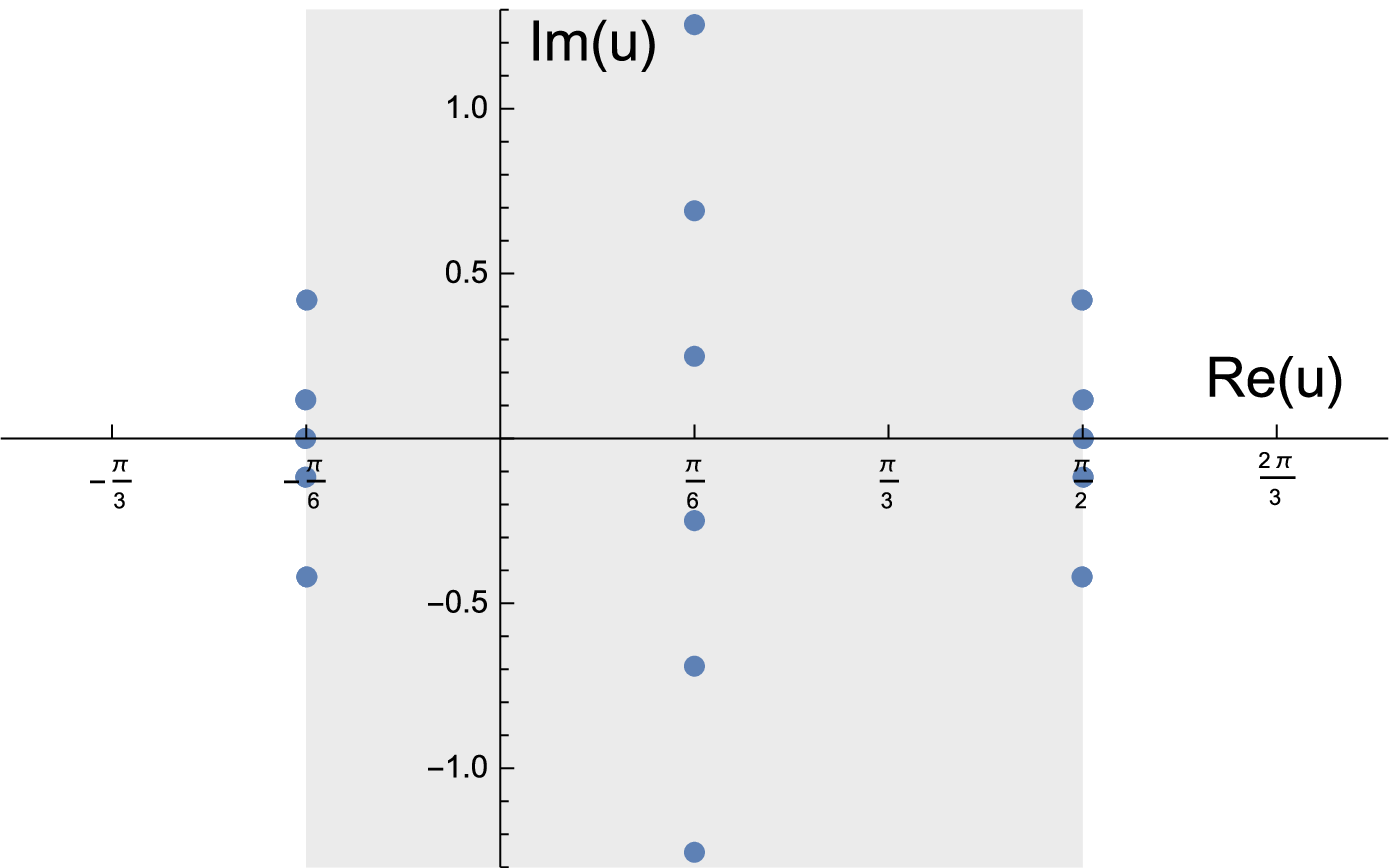}\qquad \qquad&
\includegraphics[width=.45\textwidth]{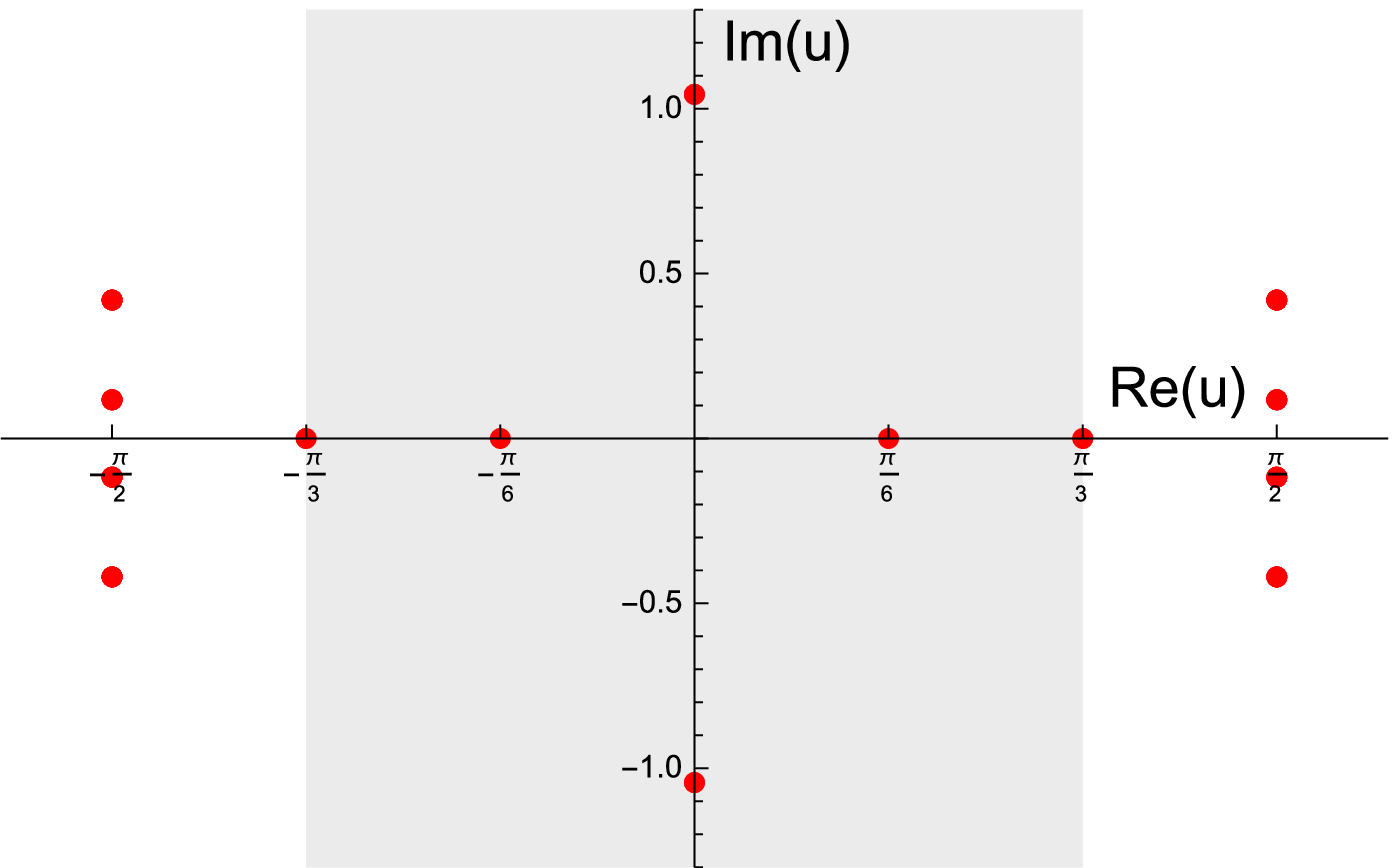} 
\end{tabular}
\caption{The location of the zeros for the fourth excited state in $\stan_8^2$. There are three pairs of complex zeros in the analyticity strip for $D^1(u)$. For $D^2(u)$, there are four real zeros, as well as one pair of complex conjugate zeros, each of order $2$, in the center of the analyticity strip.}
\label{fig:eigpatterns2}
\end{figure}

\paragraph{Patterns of zeros and analyticity strips.}
For the other eigenvalues, the exact solutions to the $Y$-system are unknown, but as shown in \cref{sec:fsc}, it is possible to compute the finite-size corrections. This requires knowledge about the analytic behavior of the eigenvalues, which we formulate empirically based on exact computations for small system sizes. Our computer implementation produces the zeros and analyticity data of the eigenvalues in the standard modules up to $N=12$. We find that the leading eigenvalues have the following analyticity strips:
\be 
D^1(u): \quad -\frac \pi 6< \text{Re}(u) < \frac {\pi} 2, \qquad D^2(u): \quad -\frac \pi 3< \text{Re}(u) < \frac {\pi} 3.
\ee
Let us be more precise as to what this means. From the definitions \eqref{eq:Du} and \eqref{eq:D2u}, $D^1(u)$ and $D^2(u)$ are Laurent polynomials in the variable $z = \eE^{\ir u}$, with minimal and maximal power $-2N$ and $2N$. Their eigenvalues share this property and thus as functions of $z$ have at most $4N$ zeros, and no poles except at $z = 0$. Due to the property $\Db^{n}(u+\pi)=\Db^{n}(u)$, in the complex $u$ plane, there are at most $2N$ zeros in any vertical strip of width $\pi$ and the pattern is repeated periodically. For $\Db^1(u)$, the leading eigenvalues in each $\stan_N^d$ have a finite number
of zeros inside the analyticity strip. The other zeros are located either on the edges of this strip, that is for $\text{Re}(u) = -\frac\pi6$ and $\frac\pi2$, or outside the analyticity strip at $\text{Re}(u) =-\frac\pi3,\frac{2\pi}3$. As $N$ grows, the number of zeros of these leading eigenvalues inside the analyticity strip of $\Db^1(u)$ remains unchanged, whereas the number of zeros on the edges increases. For $\Db^2(u)$, the analyticity strip also contains finitely many zeros, and as $N$ increases, the extra zeros accumulate outside the analyticity strip on the lines $\text{Re}(u) = \pm\frac\pi2$. Two examples of patterns of zeros are given in \cref{fig:eigpatterns,fig:eigpatterns2}. 

We observe that all the eigenvalues of $D^2(u)$ share single real zeros at $u = \pm \frac\pi 6,\pm \frac\pi 3$. These zeros can be understood from the fusion hierarchy equation \eqref{eq:fushier} for $n=1$. We also see, for instance in \cref{fig:eigpatterns,fig:eigpatterns2}, that $D^1(u)$ has a zero near $u = \frac \pi 2$. Its location is however not exactly at $u = \frac \pi 2$. Indeed, by specializing \eqref{eq:fushier} to $n=1$ and $u = \frac\pi2$, we find 
\be
\Db^1(\tfrac\pi2)^2 = (\tfrac{1}{3})^{2N} \Ib.
\ee
Using a diagrammatic argument, one can show that $\Db^1(\tfrac \pi 2) =(-\tfrac13)^N \Ib$. Thus, $D^1(u)$ has a zero near but not directly at $u = \frac \pi 2$. Its location in fact varies slightly for each eigenvalue.

The other zeros of $D^1(u)$ and $D^2(u)$ are not common to all the eigenvalues and come in complex conjugate pairs. From this observation, we infer that these eigenvalues are real for $\text{Im}(u) = 0$. Empirically, we also find that pairs of complex zeros inside the analyticity strips all lie on the central vertical line, that is respectively at $\text{Re}(u) = \frac \pi 6$ and $\text{Re}(u) = 0$ for $D^1(u)$ and $D^2(u)$. The degeneracy of these zeros is always one for $D^1(u)$, but can be one or two for $D^2(u)$. For instance, the eigenvalue whose zeros are shown in the right panel of \cref{fig:eigpatterns2} has one pair of double zeros with $\text{Re}(u) = 0$. 
In all cases, the patterns of zeros are symmetric with respect to a reflection about the central vertical line of the analyticity strip. It follows that the eigenvalues are real on the central line of the analyticity strips.

Because the zeros are symmetric under a reflection about the real axis, we depict an eigenvalue with a pattern diagram that includes only the zeros of the lower-half plane. In these diagrams, we omit the zeros on the real axis as they are common to all eigenvalues. For instance, the eigenvalues in \cref{fig:eigpatterns,fig:eigpatterns2} are represented by the patterns
\be
\begin{array}{rl}
\psun
\begin{pspicture}[shift=-1.2](-0.4,-2.4)(1.4,0.5)
\psline{-}(0,-1.9)(0,0)(1,0)(1,-1.9)
\psdots[dotsize=0.09cm](0,-0.3)(0,-0.6)(0,-0.9)
\psdots[dotsize=0.09cm](0.5,-1.2)
\psdots[dotsize=0.09cm](1,-0.3)(1,-0.6)(1,-0.9)
\end{pspicture}
\psun
\begin{pspicture}[shift=-1.2](-0.4,-2.4)(1.4,0.5)
\psline{-}(0,-1.9)(0,0)(1,0)(1,-1.9)
\end{pspicture}
\end{array} 
\quad \textrm{and} \quad
\begin{array}{rl}
\psun
\begin{pspicture}[shift=-1.2](-0.4,-2.4)(1.4,0.5)
\psline{-}(0,-1.9)(0,0)(1,0)(1,-1.9)
\psdots[dotsize=0.09cm](0,-0.3)(0,-0.9)
\psdots[dotsize=0.09cm](0.5,-0.6)(0.5,-1.2)(0.5,-1.5)
\psdots[dotsize=0.09cm](1,-0.3)(1,-0.9)
\end{pspicture}
\psun
\begin{pspicture}[shift=-1.2](-0.4,-2.4)(1.4,0.5)
\psline{-}(0,-1.9)(0,0)(1,0)(1,-1.9)
\psdots[linecolor=black,fillcolor=lightgray,dotstyle=o,dotsize=0.09cm](0.5,-0.95)
\end{pspicture}
\end{array}\ .
\ee
Black and gray circles respectively denote single and double zeros, and the analyticity strip is delimited by the vertical segments. More examples are given in \cref{sec:pats}. 

The analyticity properties of $K(u)$ and $d^1(u)$ follow readily from those of $D^1(u)$ and $D^2(u)$. From the definition \eqref{eq:Ktang}, we see that $K(u)$ has a pole of order $2N$ at $u = \frac \pi3$, whereas $d^1(u)$ has poles of order $2N$ at $u =-\frac\pi3, \frac\pi 3$, a zero of order $2N+2$ at $u=0$ and neither poles nor zeros at $u = \pm \frac \pi 6$. The resulting analyticity strips 
\be
K(u): \quad -\frac \pi 2< \text{Re}(u) < \frac {\pi} 6, \qquad
d^1(u): \quad -\frac \pi 3< \text{Re}(u) < \frac {\pi} 3
\ee
have width $2 \lambda$. A similar analysis on small system sizes reveals that the functions $1+K(u)$ and $1+d^1(u)$ are analytic and non-zero in the following strips of width $\lambda$:
\be
1+K(u): \quad -\frac \pi 3< \text{Re}(u) < 0, \qquad 1+d^1(u): \quad -\frac \pi 6< \text{Re}(u) < \frac {\pi} 6.
\ee
In these cases, the analyticity strips are entirely free of zeros and poles.

Finally, in terms of the variables defined in \eqref{eq:aU}, the analyticity strips take the following elegant forms:
\begin{subequations}
\begin{alignat}{4}
&\mathfrak a^1(x): &&\quad - \pi < \text{Im}(x) < \pi , \qquad
&&\mathfrak a^2(x): &&\quad -\pi < \text{Im}(x) < \pi, \\[0.15cm]
&\mathfrak A^1(x): &&\quad -\frac \pi 2< \text{Im}(x) < \frac \pi 2, \qquad
&&\mathfrak A^2(x): &&\quad -\frac \pi 2< \text{Im}(x) < \frac {\pi} 2.
\end{alignat}
\end{subequations}
In terms of $x$, the patterns of zeros are rotated by 90 degrees, the central line of the analyticity strip coincides with the real axis and the pairs of complex zeros inside the analyticity strips lie on this axis. 

\paragraph{Braid limit.} The braid limit $\Db^{n}_\infty$ is obtained by multiplying $\Db^n(u)$ by a suitable trigonometric function and taking the limit $u\rightarrow \ir \infty$. For $n=1$, $\Db^1_\infty\equiv \Db_\infty$ is defined in \eqref{eq:braidD}. For $n=2$, 
\be
\Db^2_\infty = \lim_{u \rightarrow \ir \infty}  \bigg(\frac{-\eE^{2\ir(\pi-\lambda)}}{s_{1}(u)^2}\bigg)^N \Db^2(u) = (\Db_\infty^1)^2 - \Ib.
\ee
The last equality is obtained by applying the braid limit to \eqref{eq:fushier} for $n=1$.
Likewise, the braid tangles $\db^1_\infty$, $\Kb_\infty$, $\ab^1_\infty$ and $\ab^2_\infty$ are defined as
\be
\db^1_\infty = \lim_{u \rightarrow\pm \ir \infty}  \db^1(u) = \Db_\infty^2, \qquad \Kb_\infty = \lim_{u \rightarrow\pm \ir \infty} \Kb(u) = - \Db_\infty, \qquad \ab^n_\infty = \lim_{x \rightarrow \pm \infty} \ab^n(x).
\ee
For generic $\beta$, the matrix representatives of the braid transfer tangles on $\stan_N^d$ are scalar multiples of the identity matrix \cite{MDPR14}, and the corresponding scalars depend only on $d$. On a given standard module $\stan_N^d$, each eigenvalue of  $\db^1(u)$ has the same braid limit, and likewise for $\Kb(u)$. For $\beta = 1$, the braid behavior is
\begin{subequations}
\begin{alignat}{3}
&d \equiv 0,1 \text{ mod } 3:\qquad && D^1_\infty = 1, \qquad && D^2_\infty = 0, \\
&d \equiv 2 \text{ mod } 3:\qquad && D^1_\infty = -2, \qquad && D^2_\infty = 3,
\end{alignat}
\end{subequations}
which implies that
\begin{subequations}
\begin{alignat}{3}
&d \equiv 0,1 \text{ mod } 3:\qquad && d^1_\infty = \mathfrak a^1_\infty=  0, \qquad && K_\infty = \mathfrak a^2_\infty= -1, \label{eq:braid01}\\
&d \equiv 2 \text{ mod } 3:\qquad && d^1_\infty =\mathfrak a^1_\infty= 3, \qquad && K_\infty = \mathfrak a^2_\infty= 2.\label{eq:braid02}
\end{alignat}
\end{subequations}
One can readily check that these satisfy \eqref{eq:finaly}. 

The analysis of the finite-size corrections in \cref{sec:fsc} in the case $d \equiv 0,1 \Mod 3$ requires a refinement of \eqref{eq:braid01} to two subcases, characterised by the rate of convergence of $1+\mathfrak a^2(x)$ to zero as $x\rightarrow \pm \infty$. From our numerical investigation, we find the following subcases:\footnote{We note that the Razumov-Stroganov eigenvalue does not fit in any of the two subcases. The analysis presented here holds for all the other eigenvalues.}
\be
\text{Subcase A}: \quad \lim_{x \rightarrow \pm \infty}  \eE^{\frac{2|x|}3} \big(1+ \mathfrak a^2(x)\big) = \kappa, \qquad  \text{Subcase B}: \quad \lim_{x \rightarrow \pm \infty}  \eE^{\frac{4|x|}3} \big(1+ \mathfrak a^2(x)\big) = \kappa',
\label{eq:subcases}
\ee
where $\kappa,\kappa'$ are non-zero real constants. We argue that $\kappa,\kappa'>0$. Indeed, as noted earlier, on the real $x$ axis, the function $1+ \mathfrak a^2(x)$ is real but never zero. Our numerics reveal that there are always one or more values $x=x^j$ where $\mathfrak a^2(x)$ vanishes, implying that $1+\mathfrak a^2(x^j) = 1$. From the previous observations, $1+\mathfrak a^2(x)$ is positive everywhere on the real $x$ axis, so $\kappa,\kappa' >0$.

Together, \eqref{eq:finaly} and \eqref{eq:subcases} determine the rate of convergence of $\mathfrak a^1(x)$ to zero as $x\rightarrow \pm \infty$:
\be
\text{Subcase A}: \quad \lim_{x \rightarrow \pm \infty}  \eE^{\frac{2|x|}3} \mathfrak a^1(x) = -\kappa, \qquad  \text{Subcase B}: \quad \lim_{x \rightarrow \pm \infty}  \eE^{\frac{4|x|}3} \mathfrak a^1(x) = \kappa'.
\label{eq:subcasesd1}
\ee
For $d\equiv 0,1 \Mod 3$, we include the letter A or B at the bottom of the patterns of zeros to indicate which subcase the corresponding eigenvalue belongs to, see for example \cref{fig:GSpatterns,fig:pats84}.

\paragraph{Bulk behavior.} The bulk limit of the eigenvalues is obtained by increasing the system size $N$ while keeping the spectral parameter $u$ finite and within the analyticity strip. Importantly, it comes into play in \cref{sec:fsc} as the asymptotic behavior at $x = -\infty$ of the scaling functions, defined in \eqref{eq:scalingfunctions}. In this limit, $d^1(u)$ and $K(u)$ converge to constants:
\be
d^1_{\text{bulk}}(u) = 0, \qquad K_{\text{bulk}}(u) = -1.
\label{eq:bulkb}
\ee
Indeed, for finite values of $u$ near the origin, the behavior of $d^1(u)$ is governed by the zero of order $2N+2$ at $u = 0$, so the function is approximately zero for large $N$. In the bulk limit, \eqref{eq:ysysperco1} becomes $K_{\text{bulk}}(u)^2 = 1$, so $K_{\text{bulk}}(u) \in \{+1, -1\}$. For a given pattern of zeros, one can deduce the values of $K_{\text{bulk}}(u)$ from the braid value $K_\infty$ and the number of zeros in the central line of the analyticity strip. For instance, let us consider the eigenvalue of \cref{fig:eigpatterns2}. The braid limit is $K_\infty = 2$. On the central line of the analyticity strip (namely $\text{Re}(u) = \frac\pi6$ for $D^1(u)$, corresponding to $\text{Re}(u) =-\frac \pi 6$ for $K_0(u)$), there are three zeros with $\text{Im}(u)>0$. We conclude that $K(-\frac \pi 6)$ is negative and that $K_{\text{bulk}}(u) = -1$. Applying the same logic to the pattern of zeros of \cref{fig:eigpatterns} would also yield $K_{\text{bulk}}(u) = -1$.

The same reasoning can be repeated for each eigenvalue. Empirically, 
we find that every pattern of zeros has the same asymptotic behavior: $K_{\text{bulk}}(u) =
-1$. 

\subsection{Finite-size corrections}\label{sec:fsc}

The eigenvalues $D(u)$ can be factored into a bulk, a surface and a finite-size correction as
\be
\label{eq:Dseparation}
D(u) = D_{\text{b}}(u) D_{\text{s}}(u) D_{\text{f}}(u).
\ee
From \eqref{eq:tsys}, the bulk and surface contributions satisfy the inversion relations
\be
D_{\text{b}}(u)D_{\text{b}}(u+\lambda) = f_{-1}f_1,
\qquad
D_{\text{s}}(u)D_{\text{s}}(u+\lambda) = \frac{s_{-2}(2u)s_2(2u)}{s_{-1}(2u)s_{1}(2u)},
\label{eq:inv.identities}
\ee
whereas the finite-size correction satisfies
\be
D_{\text{f}}(u)D_{\text{f}}(u+\lambda) = 1+d^1(u).
\label{eq:finitepart}
\ee
The solutions to \eqref{eq:inv.identities} give the bulk and surface free energies and were obtained in \cite{PRZ06} for generic $\beta$. For $\beta =1$, the right-hand side of the inversion identity for $D_{\text{s}}(u)$ equals $1$ and the solution is $D_{\text{s}}(u)=1$. In this section, we derive the finite-size corrections of $D_{\text{f}}(u)$ in the standard modules $\stan_N^d$. We do so by using the methods developed in \cite{PK91,KP92}. The solution holds for any finite excitation and works for all three cases, namely $d \equiv 0,1 \textrm{ mod } 3$\, (A), $d \equiv 0,1 \textrm{ mod } 3$\,(B) and $d \equiv 2 \textrm{ mod } 3$. For convenience, we work with the functions $\mathfrak a^1(x)$, $\mathfrak a^2(x)$, $\mathfrak A^1(x)$ and $\mathfrak A^2(x)$ in terms of which the $Y$-system is symmetric. We also define
\be
\mathfrak b(x) = D_{\textrm{f}}(\tfrac {\ir x} 3+ \tfrac \pi 6),
\ee
so that \eqref{eq:finitepart} becomes
\be
\mathfrak b(x - \tfrac{\ir \pi}2) \mathfrak b(x + \tfrac{\ir \pi}2) = \mathfrak A^1(x).
\label{eq:bbU}
\ee
We note that \eqref{eq:bbU} defines $\mathfrak b(x)$ up to a sign. We choose this sign to be such that 
\be
\mathfrak b_\infty = D^1_\infty.
\ee

\paragraph{TBA equations.} 
Let us denote by $w^1, w^2, \dots, w^{t^1}$ the values of $x$
where $\mathfrak a^1(x)=0$ is zero on the half-line $\text{Im}(x)=0$,
$\text{Re}(x)>0$. The same pattern is repeated symmetrically on the negative
part of the real axis: $\mathfrak a^1(\pm w^i) = 0$, $i = 1, \dots, t^1$. Likewise for
$\mathfrak a^2(x)$, we denote the positions of its zeros on the positive real $x$ axis by $x^1, x^2, \dots, x^{t^2}$.  The pattern is repeated symmetrically on the negative part of the
real axis: $\mathfrak a^2(\pm x^j) = 0$, $j = 1, \dots, t^2$. For convenience, we label the zeros such that $w^1\ge w^2\ge \dots
\ge w^{t^1}$ and $x^1> x^2 > \dots> x^{t^2}$, recalling that the zeros of
$\mathfrak a^2(x)$ are all distinct whereas those of $\mathfrak a^1(x)$ can be
twofold degenerate.

We introduce the finite-size correction functions $\ell^1(x)$ and $\ell^2(x)$ by writing
\begin{subequations}
\begin{alignat}{2}
\mathfrak a^1(x) &= \tanh^{2N+2}(\tfrac x 2)\cdot \bigg(\prod_{i = 1}^{t^1} \tanh(\tfrac{x-w^i}2)\tanh(\tfrac{x+w^i}2) \bigg)\cdot \ell^1(x),\\
\mathfrak a^2(x) &= \bigg(\prod_{j = 1}^{t^2} \tanh(\tfrac{x-x^j}2)\tanh(\tfrac{x+x^j}2) \bigg)\cdot \ell^2(x).
\end{alignat}
\end{subequations}
The functions $\ell^1(x)$, $\ell^2(x)$, $\mathfrak A^1(x)$ and $\mathfrak A^2(x)$ are analytic and non-zero in their corresponding analyticity strips, and satisfy
\be
\label{eq:Lsystem}
\ell^1(x-\tfrac{\ir \pi}2)\ell^1(x+\tfrac{\ir \pi}2) = \mathfrak A^2(x), \qquad \ell^2(x-\tfrac{\ir \pi}2)\ell^2(x+\tfrac{\ir \pi}2) = \mathfrak A^1(x).
\ee
From \eqref{eq:braid02}, $\ell^1(x)$, $\ell^2(x)$, $\mathfrak A^1(x)$ and $\mathfrak A^2(x)$ have constant asymptotics for $d \equiv 2 \textrm{ mod } 3$. One can therefore define the Fourier transforms of their logarithmic derivatives. From \eqref{eq:braid01}, for $d \equiv 0,1 \textrm{ mod }3$, $\ell^2(x)$, $\mathfrak A^1(x)$ and $\mathfrak A^2(x)$ also have non-zero asymptotics, but $\ell^1(x)$ does not. According to \eqref{eq:subcasesd1}, $\ell^1(x)$ behaves respectively as $\eE^{-2|x|/3}$ and $\eE^{-4|x|/3}$ as $x\rightarrow \pm \infty$ for the subcases A and B, so we instead define the Fourier transform of the second logarithmic derivative. To treat $d \equiv 0,1,2 \textrm{ mod } 3$ simultaneously, we consider the second logarithmic derivative of $\ell^1(x)$ in all cases:
\begin{subequations}
\begin{alignat}{2}
&\displaystyle L^1(k) = \frac{1}{2 \pi} \int_{-\infty}^\infty \dd x\, \big(\ln \ell^1(x)\big)''\eE^{-\ir k x}, \qquad &&\displaystyle \big(\ln \ell^1(x)\big)'' = \int_{-\infty}^\infty \dd k\, L^1(k)\, \eE^{\ir k x},\\[0.15cm]
&\displaystyle L^2(k) = \frac{1}{2 \pi} \int_{-\infty}^\infty \dd x\, \big(\ln \ell^2(x)\big)'\eE^{-\ir k x}, \qquad &&\displaystyle \big(\ln \ell^2(x)\big)' = \int_{-\infty}^\infty \dd k\, L^2(k)\, \eE^{\ir k x},\\[0.15cm]
&\displaystyle A^n(k) = \frac{1}{2 \pi} \int_{-\infty}^\infty \dd x\, \big(\ln \mathfrak A^n(x)\big)'\eE^{-\ir k x}, \qquad &&\displaystyle \big(\ln \mathfrak A^n(x)\big)' = \int_{-\infty}^\infty \dd k\, A^n(k)\, \eE^{\ir k x}, \qquad n=1,2.
\end{alignat}
\end{subequations}
Applying the Fourier transform to \eqref{eq:Lsystem}, the reverse transform and then integrating with respect to $x$, we obtain the TBA equations:
\begin{subequations}
\begin{alignat}{2}
\ln \mathfrak a^1(x) &= \ln \tanh^{2N+2}(\tfrac x 2) + \sum_{i = 1}^{t^1} \ln \Big(\!-\!\tanh(\tfrac{x-w^i}2)\tanh(\tfrac{x+w^i}2) \Big) + K \ast \ln \mathfrak A^2 + \phi\, x + \phi^1,\\
\ln \mathfrak a^2(x) &= \sum_{j = 1}^{t^2} \ln \Big(\!-\!\tanh(\tfrac{x-x^j}2)\tanh(\tfrac{x+x^j}2) \Big) + K \ast \ln \mathfrak A^1 + \phi^2,
\end{alignat}
\end{subequations}
where $\phi$, $\phi^1$ and $\phi^2$ are the integration constants. The kernel $K(x)$ is given by
\be
\label{eq:kernel}
K(x) = \frac{1}{2\pi \cosh x}
\ee
and the convolution of two functions is defined as 
\be
\label{eq:convolution}
(f \ast g) (x) = \int_{-\infty}^\infty \dd y \,f(x-y)\,g(y) = \int_{-\infty}^\infty \dd y \,f(y)\,g(x-y).
\ee
The constants $\phi$, $\phi^1$ and $\phi^2$ are evaluated below from the braid limits.

\paragraph{Scaling TBA equations.}
In the non-linear integral equations, the dependence on $N$ appears only algebraically, in the function $\ln \tanh^{2N+2}(\frac x 2)$. If $x$ is of order $\ln N$ and $N$ is large, this function has the following behavior:
\be
\lim_{N \rightarrow \infty} \tanh^{2N+2}\big(\pm \tfrac 12(x + \ln N)\big) = \exp(-4\, \eE^{-x}).
\ee
To compute the finite-size corrections, we assume that the following scaling limits also exist:
\be
\label{eq:scalingfunctions}
\mathsf a^n(x) = \lim_{N \rightarrow \infty} \mathfrak a^n\big(\!\pm\! (x + \ln N)\big), \qquad \mathsf A^n(x) = \lim_{N \rightarrow \infty} \mathfrak A^n\big(\!\pm\! (x + \ln N)\big), \qquad n = 1,2.
\ee
Because the patterns of zeros are symmetric in the upper and lower parts of the complex $u$ plane, the functions $\mathfrak a^n(x)$ and $\mathfrak A^n(x)$ are even in $x$. In \eqref{eq:scalingfunctions}, the limits are therefore independent of the choice of the signs $\pm$. Let us denote the zeros of $\mathsf a^1(x)$ and $\mathsf a^2(x)$ by $z^i$ and $y^j$, namely
\be
z^i = w^i - \ln N, \qquad y^j = x^j - \ln N.
\ee
In the scaling limit, the non-linear equations become
\begin{subequations}\label{eq:scalNLIE}
\begin{alignat}{2}
\ln \mathsf a^1(x) &= -4 \,\eE^{-x} + \sum_{i=1}^{t^1} \ln\big(\!-\!\tanh (\tfrac{x-z^i}2) \big) + K \ast \ln \mathsf A^2 + \phi\, x + \phi^1,\\
\ln \mathsf a^2(x) &= \sum_{j=1}^{t^2} \ln\big(\!-\!\tanh (\tfrac{x-y^j}2) \big)+ K \ast \ln \mathsf A^1 + \phi^2.
\end{alignat}
\end{subequations}

\paragraph{Evaluation of the constants.}
For $d \equiv 2 \textrm{ mod } 3$, we fix the branch cuts of the logarithms by fixing
\be
\ln \mathsf a^1(x)\xrightarrow{x \rightarrow \infty} \ln 3, \qquad \ln \mathsf a^2(x)\xrightarrow{x \rightarrow \infty} \ln 2, \qquad \ln \Big(-\tanh(\tfrac{x-y}2)\Big)\xrightarrow{x \rightarrow \infty} \ir \pi. \ee
With this choice, the constants $\phi^1$ and $\phi^2$ are evaluated using the braid limit. Indeed, using
\be
\frac1{2\pi}\int_{-\infty}^\infty  \frac{\dd x}{\cosh x} = \frac12,
\ee
we find $K \ast \ln \mathsf A^1 \xrightarrow{x \rightarrow \infty} \ln 2$, $K \ast \ln \mathsf A^2 \xrightarrow{x \rightarrow \infty} \ln 3$ and 
\be
d \equiv 2 \textrm{ mod } 3: \qquad \phi = 0, \qquad \phi^1 = -\ir \pi t^1, \qquad \phi^2 = -\ir \pi t^2.
\label{eq:const2}
\ee

For $d \equiv 0,1 \textrm{ mod } 3$\,(A), according to \eqref{eq:subcasesd1}, the large $x$ behavior of $\mathfrak a^1(x)$ and $\mathfrak a^2(x)$ is given by
\be
\mathfrak a^1(x) \xrightarrow{x\rightarrow \infty} -\kappa\, \eE^{-\frac{2x}3}, \qquad 1+\mathfrak a^2(x) \xrightarrow{x\rightarrow \infty} \kappa\, \eE^{-\frac{2x}3},
\label{eq:largexbehavior1}
\ee
with $\kappa>0$. The branches of the logarithms are fixed using the convention
\be
\ln \mathsf a^1(x) \xrightarrow{x \rightarrow \infty} -\frac{2x}3 + \ln \kappa + \ir \pi, \qquad \ln \mathsf a^2(x) \xrightarrow{x \rightarrow \infty} \ir \pi.
\ee
The braid behavior of $K \ast \ln \mathsf A^2$ can be evaluated explicitly:
\be
K \ast \ln \mathsf A^2 \xrightarrow{x \rightarrow \infty}\frac{1}{2\pi}\int_{-\infty}^\infty \dd y\, \frac{\ln \Big(\kappa^2 \eE^{-\frac{4(x-y)}3}\Big)}{\cosh y} = \ln \kappa - \frac{2x}3.
\ee
We find that the constants are
\be
d \equiv 0,1 \textrm{ mod } 3\ (\textrm{A}):\qquad \phi = 0, \qquad \phi^1 = -\ir \pi (t^1-1), \qquad \phi^2 = -\ir \pi(t^2-1).
\label{eq:const01A}
\ee

Finally, for $d \equiv 0,1 \textrm{ mod } 3$ subcase B, according to \eqref{eq:subcasesd1}, the large $x$ behavior of the eigenvalues is 
\be
\mathfrak a^1(x) \xrightarrow{x\rightarrow \infty} \kappa'\, \eE^{-\frac{4x}3}, \qquad 1+\mathfrak a^2(x) \xrightarrow{x\rightarrow \infty} \kappa'\, \eE^{-\frac{4x}3},
\ee
with $\kappa'>0$. The branches of the logarithms are fixed using the conventions
\be
\ln \mathfrak a^1(x) \xrightarrow{x \rightarrow \infty} -\frac{4x}3 + \ln \kappa', \qquad \ln \mathfrak a^2(x) \xrightarrow{x \rightarrow \infty} \ir \pi,
\ee
the braid behavior of $K \ast \ln \mathsf A^2$ is 
\be
K \ast \ln \mathsf A^2 \xrightarrow{x \rightarrow \infty}\frac{1}{2\pi}\int_{-\infty}^\infty \dd y\, \frac{\ln \Big((\kappa')^2 \eE^{-\frac{8(x-y)}3}\Big)}{\cosh y} = \ln \kappa' - \frac{4x}3
\ee
and the integration constants are evaluated to 
\be
d \equiv 0,1 \textrm{ mod } 3\ (\textrm{B}):\qquad \phi = 0, \qquad \phi^1 = -\ir \pi t^1, \qquad \phi^2 = -\ir \pi(t^2-1).
\label{eq:const01B}
\ee

\paragraph{Finite-size corrections.}
To apply the Fourier transform and its inverse to \eqref{eq:bbU}, one needs to remove the zeros of $\mathfrak b(x)$ on the real $x$ axis by dividing by $\prod_{j = 1}^{t^2} \tanh(\tfrac{x-x^j}2)\tanh(\tfrac{x+x^j}2)$. Because $\mathfrak b(x)$ has constant asymptotics, a single derivative of its logarithm is required. The result is
\be
\ln \mathfrak b(x) =  \sum_{j = 1}^{t^2} \ln \Big(\!-\!\tanh(\tfrac{x-x^j}2)\tanh(\tfrac{x+x^j}2) \Big) + K \ast \ln \mathfrak A^1 + \psi
\ee
where $\psi$ is the integration constant.
With the branch choices 
\be
\ln \mathfrak b(x) \xrightarrow{x \rightarrow \infty} 
\left\{\begin{array}{cl}
0 &\quad d \equiv 0,1 \textrm{ mod } 3,\\[0.1cm]
\ln 2 + \ir \pi &\quad d \equiv 2 \textrm{ mod } 3,
\end{array}\right.
\ee
the constant $\psi$ is evaluated using the braid limit and found to be $-\ir \pi t^2$ and $-\ir\pi (t^2-1)$ for $d \equiv 0,1 \textrm{ mod } 3$ and $d \equiv 2 \textrm{ mod } 3$ respectively. Because $\psi$ is independent of $N$, it does not contribute to the finite-size corrections. These are written in terms of the scaling functions as follows:
\begin{alignat}{2}
\ln \mathfrak b(x) - \psi &= \sum_{j = 1}^{t^2} \ln \Big(\!-\!\tanh(\tfrac{x-x^j}2)\tanh(\tfrac{x+x^j}2) \Big)
\nonumber\\&\hspace{0.8cm}
+ \frac{1}{2\pi} \int_{-\ln N}^\infty \dd y \Big( \frac{\ln\mathfrak A^1(y+\ln N)}{\cosh(x-y-\ln N)}+\frac{\ln\mathfrak A^1(-y-\ln N)}{\cosh(x+y+\ln N)}\Big) \nonumber\\
& \simeq -\frac{2 \cosh x}N \bigg(2 \sum_{j=1}^{t^2}\eE^{-y^j}- \frac1\pi \int_{-\infty}^\infty \dd y\, \eE^{-y} \ln \mathsf A^1(y)\bigg).\label{eq:FSC1}
\end{alignat}

\paragraph{Zeros of $\boldsymbol{\mathfrak a^1(x)}$ and $\boldsymbol{\mathfrak a^2(x)}$.} The next step is to rewrite $\eE^{-y^j}$  in terms of integrals involving the scaling functions. From \eqref{eq:finaly}, we find that
\be
\mathfrak a^2(x^j) = 0 \quad \Rightarrow \quad \mathfrak a^1(x^j-\tfrac {\ir \pi}2) = -1 \quad \Rightarrow \quad \mathsf a^1(y^j-\tfrac {\ir \pi}2) = -1.
\ee
Taking the logarithm and using \eqref{eq:scalNLIE}, we find
\begin{alignat}{2}
(2 k^j-1)\pi &= \ir \ln \mathsf a^1 (y^j - \tfrac{\ir \pi }2) \nonumber \\&
= 4\, \eE^{-y^j} + \ir \sum_{i=1}^{t^1} \ln\big(\!-\!\tanh \tfrac12(y^j-z^i-  \tfrac{\ir\pi}2 )\big) + \frac1{2\pi} \int_{-\infty}^\infty \dd y\, \frac{\ln \mathsf A^2(y)}{\sinh(y-y^j)} + \ir \phi^1 \label{eq:integerK}
\end{alignat}
where the $k^j$ are integers. Isolating $\eE^{-y^j}$ from this equation and replacing it in \eqref{eq:FSC1} produces an expression for the finite-size corrections wherein the zeros of $\mathsf a^1(x)$ and $\mathsf a^2(x)$ appear in terms of the expression $\ln\big(\!-\!\tanh \tfrac12(y^j-z^i-  \tfrac{\ir\pi}2 )\big)$. We can also write these in integral form:
\be
\mathfrak a^1(w^i) = 0 \quad \Rightarrow \quad \mathfrak a^2(w^i-\tfrac {\ir \pi}2) = -1 \quad \Rightarrow \quad \mathsf a^2( z^i-\tfrac {\ir \pi}2) = -1
\ee
and
\begin{alignat}{2}
(2 \ell^i-1)\pi &= \ir \ln \mathsf a^2 (z^i - \tfrac{\ir \pi}2) 
= \ir \sum_{j=1}^{t^2} \ln \big(\!-\!\tanh \tfrac12(z^i-y^j -\tfrac{\ir \pi}2)\big) + \frac{1}{2 \pi} \int_{-\infty}^\infty \dd y \frac{\ln \mathsf A^1(y)}{\sinh(y-z^i)}+ \ir\phi^2
\label{eq:integerL}
\end{alignat}
where the $\ell^i$ are integers.
From our choice of branches for the logarithms, we have
\be
\label{eq:tanh.id}
\ir \ln\big(\!-\!\tanh\tfrac12(z^i-y^j - \tfrac{\ir \pi}2) \big) = - \pi - \ir \ln\big(\!-\!\tanh\tfrac12(y^j-z^i - \tfrac{\ir \pi}2) \big)
\ee
which we use to isolate the expression $\ln\big(\!-\!\tanh \tfrac12(y^j-z^i-  \tfrac{\ir\pi}2 )\big)$ in \eqref{eq:integerL}. We obtain
\begin{alignat}{2}
\ln \mathfrak b(x) - \psi  \simeq -\frac{2 \pi \cosh x}{N} \bigg( &\sum_{j=1}^{t^2}(k^j-\tfrac12- \tfrac{\ir\phi^1}{2 \pi}) + \sum_{i=1}^{t^1}(\ell^i-\tfrac12 + \tfrac {t^2}2- \tfrac{\ir\phi^2}{2 \pi}) - \frac 1 {\pi^2} \int_{-\infty}^\infty \dd y \,\eE^{-y} \ln \mathsf A^1(y) \nonumber\\
& - \frac{1}{4\pi^2} \sum_{i=1}^{t^1}\int_{-\infty}^{\infty} \dd y\, \frac{\ln \mathsf A^1(y)}{\sinh(y-z^i)} - \frac{1}{4\pi^2} \sum_{j=1}^{t^2}\int_{-\infty}^{\infty} \dd y\, \frac{\ln \mathsf A^2(y)}{\sinh(y-y^j)}\bigg).\label{eq:FSC2pap}
\end{alignat}

\paragraph{Dilogarithm technique.} To evaluate \eqref{eq:FSC2pap}, we consider the integral
\be 
\mathcal J = \int_{-\infty}^\infty \dd y \Big((\ln \mathsf a^1)' \ln \mathsf A^1  -\ln |\mathsf a^1| (\ln \mathsf A^1)' \Big) + \int_{-\infty}^\infty \dd y \Big((\ln \mathsf a^2)' \ln \mathsf A^2  -\ln |\mathsf a^2| (\ln \mathsf A^2)' \Big)
\ee
where $\ln |\mathsf a^1|$ and $\ln |\mathsf a^2|$ are real for all $x$ and are thus given by
\be
\label{eq:ln|a|}
\ln |\mathsf a^n|(x) = \ln \mathsf a^n(x) + \theta^n(x), \qquad n = 1,2.
\ee
Here, the $\theta^n(x)$ are step functions defined for $x \in \mathbb R$. Starting from $x=+\infty$ and moving to the left on the $x$ axis, the $\theta^n(x)$
decrease by $\ir \pi$ each time a zero of the corresponding type is crossed ($z^i$ for $\theta^1(x)$ 
and $y^j$ for $\theta^2(x)$). The values at the right endpoints are 
$\theta^1(x) = 0$ for $x > z_1$ and $\theta^2(x) = 0$ for $x > y_1$, consistent with our choice of branches for the logarithms. 

The integral $\mathcal J$ can be evaluated in two ways. For the first, one uses the non-linear integral equations \eqref{eq:scalNLIE} and the symmetries of $K(x)$ to obtain
\begin{alignat}{2}
\mathcal J &= 4\int_{-\infty}^\infty \dd y \,  \eE^{-y} \big(\ln \mathsf A^1 + (\ln \mathsf A^1)'\big) +  \int_{-\infty}^\infty \dd y \sum_{i=1}^{t^1} \Big[\ln\big(\!-\! \tanh (\tfrac{y-z^i}2)\big) \Big]' \ln \mathsf A^1 \nonumber\\
& - \int_{-\infty}^\infty \dd y \Big(\sum_{i=1}^{t^1} \ln \big(\!-\!\tanh (\tfrac{y-z^i}2)\big) + \phi^1 - \theta^1(y) \Big) \big(\ln \mathsf A^1\big)' + \int_{-\infty}^\infty \dd y \sum_{j=1}^{t^2}  \Big[\ln \big(\!-\!\tanh (\tfrac{y-y^j}2)\big) \Big]' \ln \mathsf A^2 \nonumber\\ 
& - \int_{-\infty}^\infty \dd y \Big(\sum_{j=1}^{t^2} \ln \big(\!-\!\tanh (\tfrac{y-y^j}2)\big) + \phi^2- \theta^2(y) \Big) \big(\ln \mathsf A^2\big)'.
\end{alignat}
The integrals involving derivatives of $\ln \mathsf A^1$ and $\ln \mathsf A^2$ are transformed using integration by parts. For each one, it can be argued using the non-linear integral equations that the surface terms are zero. This yields
\be
\mathcal J = 8 \int_{-\infty}^\infty \dd y\, \eE^{-y} \ln \mathsf A^1(y) + 2 \sum_{i=1}^{t^1} \int_{-\infty}^\infty\dd y\, \frac{\ln \mathsf A^1(y)}{\sinh(y-z^i)} + 2\sum_{j=1}^{t^2} \int_{-\infty}^\infty \dd y\, \frac{\ln \mathsf A^2(y)}{\sinh(y-y^j)},
\ee
which is precisely the combination of integrals needed to compute the finite-size corrections:
\be
\ln \mathfrak b(x) - \psi  \simeq -\frac{2 \pi \cosh x}{N} \bigg( \sum_{j=1}^{t^2}(k^j-\tfrac12- \tfrac{\ir\phi^1}{2 \pi}) + \sum_{i=1}^{t^1}(\ell^i-\tfrac12 + \tfrac {t^2}2- \tfrac{\ir\phi^2}{2 \pi}) - \frac{\mathcal J}{8 \pi^2}\bigg).
\label{eq:lnbalmostfinal}
\ee
The second way of performing the integrals consists in changing the integration variable from $y$ to $\mathsf a$. For the integral involving $\mathsf a^1(x)$, we obtain
\begin{alignat}{2}
\int_{-\infty}^\infty \dd y \Big((&\ln \mathsf a^1)' \ln \mathsf A^1  -\ln |\mathsf a^1| (\ln \mathsf A^1)' \Big) = \int_{-\infty}^\infty \dd y\, \frac{\dd \mathsf a^1}{\dd y} \,\bigg( \frac{\ln(1+ \mathsf a^1)}{\mathsf a^1} - \frac{\ln |\mathsf a^1|}{1+ \mathsf a^1} \bigg)\nonumber\\
& = \bigg[\int_{-\infty}^{z_{t^1}}+\int_{z_{t^1}}^{z_{t^1-1}}+ \dots + \int_{z_2}^{z_1}+ \int_{z_1}^{\infty}\bigg] \dd y\, \frac{\dd \mathsf a^1}{\dd y}\, \bigg( \frac{\ln(1+ \mathsf a^1)}{\mathsf a^1} - \frac{\ln |\mathsf a^1|}{1+ \mathsf a^1} \bigg) \nonumber\\
& =  \bigg[\int_{0}^{0}+\int_{0}^{0}+ \dots + \int_{0}^{0}+ \int_{0}^{\mathsf a^1(\infty)}\bigg] \dd \mathsf a \,\Big( \frac{\ln(1+ \mathsf a)}{\mathsf a} - \frac{\ln |\mathsf a|}{1+ \mathsf a} \Big) \nonumber\\
& = 2L_+(\mathsf a^1(\infty)) = 2 L\big(\tfrac {\mathsf a^1(\infty)}{1+\mathsf a^1(\infty)}\big)
\end{alignat}
where the Rogers dilogarithm functions are given by
\begin{subequations}
\begin{alignat}{2}
L(x)&=-\frac12 \int_{0}^x \dd y \, \Big(\frac{\ln (1-y)}y + \frac{\ln y}{1-y}  \Big),\\ 
L_+(x)&= \frac12 \int_0^x \dd y \, \Big(\frac{\ln (1+y)}y - \frac{\ln y}{1+y}\Big)=L\Big(\frac{x}{1+x}\Big).
\end{alignat}
\end{subequations}
Recalling that $\mathsf A^2(x) = (1 + \mathsf a^2(x))^2$, the integral involving $\mathsf a^2(x)$ is computed with the same arguments:
\begin{alignat}{2}
\int_{-\infty}^\infty \dd y \Big((&\ln \mathsf a^2)' \ln \mathsf A^2  -\ln |\mathsf a^2| (\ln \mathsf A^2)' \Big) = 2\int_{-\infty}^\infty \dd y\, \frac{\dd \mathsf a^2}{\dd y} \,\bigg( \frac{\ln(1+ \mathsf a^2)}{\mathsf a^2} - \frac{\ln |\mathsf a^2|}{1+ \mathsf a^2} \bigg)\nonumber\\
& = 2\bigg[\int_{-\infty}^{y_{t^2}}+\int_{y_{t^2}}^{y_{t^2-1}}+ \dots + \int_{y_2}^{y_1}+ \int_{y_1}^{\infty}\bigg] \dd y\, \frac{\dd \mathsf a^2}{\dd y}\, \bigg( \frac{\ln(1+ \mathsf a^2)}{\mathsf a^2} - \frac{\ln |\mathsf a^2|}{1+ \mathsf a^2} \bigg) \nonumber\\
& =  2\bigg[\int_{-1}^{0}+\int_{0}^{0}+ \dots + \int_{0}^{0}+ \int_{0}^{\mathsf a^2(\infty)}\bigg] \dd \mathsf a \,\Big( \frac{\ln(1+ \mathsf a)}{\mathsf a} - \frac{\ln |\mathsf a|}{1+ \mathsf a} \Big) \nonumber\\
& = 4L_+(\mathsf a^2(\infty)) +4L(1) = 4 L\big(\tfrac {\mathsf a^2(\infty)}{1+\mathsf a^2(\infty)}\big) + 4 L(1).
\end{alignat}
Putting these results together, we find
\be
\mathcal J = 4 L(1) + 2 L\big(\tfrac {\mathsf a^1(\infty)}{1+\mathsf a^1(\infty)}\big) + 4 L\big(\tfrac {\mathsf a^2(\infty)}{1+\mathsf a^2(\infty)}\big) =
\left\{\begin{array}{ll}
\mathcal K_{-1}(\tfrac{2\pi}3) = 0 &\quad d \equiv 0,1 \textrm{ mod } 3,\\[0.1cm]
\mathcal K_{-1}(0) =\frac{4 \pi^2}3 &\quad d \equiv 2 \textrm{ mod } 3,
\end{array}
\right.
\ee
where the integral $\mathcal K_{-1}(\gamma)$ is computed in \cref{sec:Rogers}.
From \eqref{eq:lnbalmostfinal}, the final result for the finite-size corrections is 
\be
\label{eq:finalfsc}
\ln \mathfrak b(x) - \psi \simeq
-\frac{2 \pi\cosh x}N 
\Big(
\sum_{j=1}^{t^2} k^j+\sum_{i=1}^{t^1} \ell^i + \tau
\Big),
\quad 
\tau = 
\left\{\begin{array}{ll}
- \tfrac12t^1t^2 &\,  d \equiv 0,1\text{ mod }3\,\textrm{(A)}, \\[0.15cm]
- \tfrac12(t^2+t^1t^2) &\,  d \equiv 0,1\text{ mod }3\,\textrm{(B)}, \\[0.15cm]
- \tfrac12(t^1+t^2+t^1t^2)-\frac16 &\,  d \equiv 2\text{ mod }3,
\end{array}\right.
\ee
where we used the values of $\phi^1$ and $\phi^2$ given in \eqref{eq:const2}, \eqref{eq:const01A} and \eqref{eq:const01B}.
Comparing with \eqref{eq:Dexpansion} specialised to $c=0$, we find that the conformal dimension of the corresponding conformal state is the content of the parenthesis, namely:
\be
\label{eq:DelKL}
\Delta =  \sum_{j=1}^{t^2}k^j+\sum_{i=1}^{t^1}\ell^i + \tau.
\ee

\subsection{Solution for the ground states}\label{sec:cfgs}

For the ground state eigenvalue in $\stan_N^d$, our experimentations on small system sizes reveal that the corresponding pattern of zeros is characterised by
\be
t^1 = \left\{
\begin{array}{ll}
\frac{d-3}3 & d \equiv 0\text{ mod }3,\\[0.15cm]
\frac{d-4}3 & d \equiv 1\text{ mod }3,\\[0.15cm]
\frac{d-2}3 & d \equiv 2\text{ mod }3, 
\end{array}\right.\qquad
t^2 = \left\{
\begin{array}{cl}
\frac{2d}3 & d \equiv 0\text{ mod }3,\\[0.15cm]
\frac{2d-2}3 & d \equiv 1\text{ mod }3,\\[0.15cm]
\frac{2d-1}3 & d \equiv 2\text{ mod }3, 
\end{array}\right. 
\ee
and by $\frac{N-d}2$ (pairs of) zeros lying on its boundary lines for $D^1(u)$, see \cref{fig:GSpatterns}. Each single zero of $D^2(u)$ is joined by a pair of single zeros, sitting at the same height on the edges of the analyticity strip. 

For $d \equiv 0 \text{ mod } 3$ and $d \equiv 1 \text{ mod } 3$, we find empirically that the ground state eigenvalue respectively belongs to the subcase A and B. The integers $k^j$ and $\ell^i$ are not fixed by the technique used in \cref{sec:fsc}. They can instead be estimated from \eqref{eq:integerK} and \eqref{eq:integerL} using exact diagonalisation on small system sizes. For the ground state of $\stan_N^d$, we find, again empirically, that the $k^j$ and $\ell^i$ are given by
\be
k^j = \left\{
\begin{array}{cl}
j-1 & d \equiv 0\text{ mod }3,\\[0.1cm]
j & d \equiv 1\text{ mod }3,\\[0.1cm]
j & d \equiv 2\text{ mod }3, 
\end{array}\right. \qquad \ell^i = i.
\ee
\cref{fig:GSpatterns} features the patterns of zeros for the ground states for $N=14,15$.\footnote{Note that for $N=14,15$, our program can only produce the pattern diagrams for $d \ge 8$. For $d<8$, the corresponding pattern diagrams given in \cref{fig:GSpatterns} were inferred from our understanding of the patterns for smaller system sizes.}
For $d=0$ and $d=1$, the second analyticity strip is colored in gray, indicating that $D^2(u)=0$ for all $u$ for the Razumov-Stroganov eigenvalue. The integers $k^j$ and $\ell^i$ given there are those 
obtained 
using the method described above. To correctly compute the conformal weight from \eqref{eq:DelKL}, we recall that $t^1$ counts the zeros in the second analyticity strip, and $t^2$ counts the zeros in the first. This is because, up to prefactors and shifts in the arguments,  $\mathfrak a^1(x)$ corresponds to $D^2(u)$ whereas $\mathfrak a^2(x)$ corresponds to $D^1(u)$.

Carrying out the sum in \eqref{eq:DelKL} with the data given above, we find in all three cases ($d \equiv 0,1,2 \Mod 3$) that the conformal weight of the ground state of $\stan_N^d$ is
\be
\Delta = \frac{d(d-1)}6 = \Delta_{1,d+1}.
\ee

%
\subsection{Single- and double-column diagrams}
\label{sec:colconfs}
%

Our classification of the patterns of zeros for the excited states in \cref{sec:excitations} uses the column diagrams introduced in \cite{PR07}. Here we recall the definitions and some basic results in a self-consistent manner.

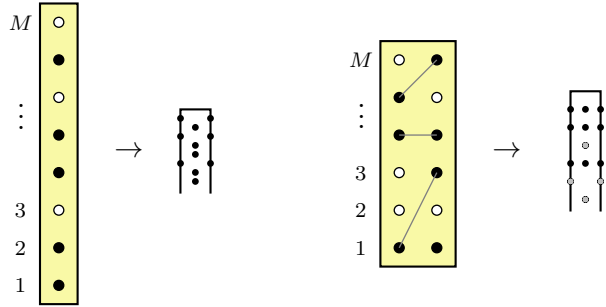
\begin{figure}[h!]
\begin{equation*}
\psset{unit=0.5}
\begin{pspicture}[shift=-3.9](0,0)(1,8)
\pspolygon[fillstyle=solid,fillcolor=lightyellow](0,0)(1,0)(1,8)(0,8)
\rput(0.5,0.5){\blackcircle}\rput(-0.5,0.5){$_1$}
\rput(0.5,1.5){\blackcircle}\rput(-0.5,1.5){$_2$}
\rput(0.5,2.5){\whitecircle}\rput(-0.5,2.5){$_3$}
\rput(0.5,3.5){\blackcircle}
\rput(0.5,4.5){\blackcircle}\rput(-0.5,5.2){$\vdots$}
\rput(0.5,5.5){\whitecircle}
\rput(0.5,6.5){\blackcircle}
\rput(0.5,7.5){\whitecircle}\rput(-0.5,7.5){$_M$}
\end{pspicture} 
\quad \rightarrow \quad 
{\psun
\begin{pspicture}[shift=-1.2](-0.0,-2.8)(1.0,0.5)
\psline{-}(0,-2.8)(0,0)(1,0)(1,-2.8)
\psdots[dotsize=0.09cm](0,-0.3)(0,-0.9)(0,-1.8)
\psdots[dotsize=0.09cm](0.5,-0.6)(0.5,-1.2)(0.5,-1.5)(0.5,-2.1)(0.5,-2.4)
\psdots[dotsize=0.09cm](1,-0.3)(1,-0.9)(1,-1.8)
\end{pspicture}
}
\ \hspace{2cm}\ 
\begin{pspicture}[shift=-2.9](0,0)(2,6)
\pspolygon[fillstyle=solid,fillcolor=lightyellow](0,0)(2,0)(2,6)(0,6)
\rput(0.5,0.5){\blackcircle}\rput(-0.5,0.5){$_1$}
\rput(0.5,1.5){\whitecircle}\rput(-0.5,1.5){$_2$}
\rput(0.5,2.5){\whitecircle}\rput(-0.5,2.5){$_3$}
\rput(0.5,3.5){\blackcircle}\rput(-0.5,4.2){$\vdots$}
\rput(0.5,4.5){\blackcircle}
\rput(0.5,5.5){\whitecircle}\rput(-0.5,5.5){$_M$}
\rput(1.5,0.5){\blackcircle}
\rput(1.5,1.5){\whitecircle}
\rput(1.5,2.5){\blackcircle}
\rput(1.5,3.5){\blackcircle}
\rput(1.5,4.5){\whitecircle}
\rput(1.5,5.5){\blackcircle}
\psline[linewidth=0.5pt,linecolor=gray]{-}(0.5,4.5)(1.5,5.5)
\psline[linewidth=0.5pt,linecolor=gray]{-}(0.5,3.5)(1.5,3.5)
\psline[linewidth=0.5pt,linecolor=gray]{-}(0.5,0.5)(1.5,2.5)
\end{pspicture}
\quad \rightarrow \quad 
{\psun
\begin{pspicture}[shift=-1.8](-0.4,-4.0)(1.4,0.5)
\psline{-}(0,-4.0)(0,0)(1,0)(1,-4.0)
\psdots[linecolor=black,fillcolor=lightgray,dotstyle=o,dotsize=0.09cm](0.5,-1.8)(0.5,-3.6)
\psdots[linecolor=black,fillcolor=lightgray,dotstyle=o,dotsize=0.09cm](0,-3.0)(1,-3.0)
\psdots[dotsize=0.09cm](0,-0.6)(0,-1.2)(0,-2.4)
\psdots[dotsize=0.09cm](0.5,-0.6)(0.5,-1.2)(0.5,-2.4)
\psdots[dotsize=0.09cm](1,-0.6)(1,-1.2)(1,-2.4)
\end{pspicture}
}
\end{equation*}
\caption{{\it Left}: 
A single-column diagram in $\Aone{8}{5}$ and its corresponding pattern of zeros. Its signature is $\{7,5,4,2,1\}$ and its weight is $q^{19}$. {\it Right}: A double-column diagram in $\Aone{6}{3,4}$ and its corresponding pattern of zeros. The three line segments drawn on the configuration have non-negative slopes, so the condition of dominance is satisfied. The signatures are $L=\{5,4,1\}$ and $R=\{6,4,3,1\}$ and the weight is $q^{24}$.}
\label{fig:columnconf}
\end{figure}

\paragraph{Single-column diagrams.}
A single-column diagram in the set $\Aone{M}{m}$ is a vertical array of $M$ sites of which $m$ are occupied and the other $M-m$ are unoccupied. We draw occupied and unoccupied sites in black and white respectively, as in the example of \cref{fig:columnconf}. The sites are assigned the height labels $1, \dots, M$ starting from the bottom. The {\it signature} $S = \{S_1, \dots, S_m\}$ of a single-column diagram in $\Aone{M}{m}$ is the list of the heights of its occupied sites in decreasing order. The {\it energy} of a single-column diagram is $E = \sum_{j=1}^m S_j$ and its {\it weight} is $q^{E}$. An example is given in the left panel of \cref{fig:columnconf}.

The generating function $\ang{M}{m}$ is defined as the sum of the weights over the single-column diagrams in $\Aone{M}{m}$. It is a polynomial in $q$ and can be computed using the following recursive argument. By removing the lowest site of a configuration in $\Aone{M}{m}$, we obtain a new configuration containing $M-1$ sites, with either $m$ or $m-1$ occupied sites depending on whether the lowest site of the original configuration was occupied. Moreover, the labels of the new configuration range from $2$ to $M$. The generating function then satisfies the recursion relation
\be
\Ang{M}{m} = q^m \bigg(\Ang{M-1}{m}+\Ang{M-1}{m-1}\bigg).
\ee
Here, the factor $q^m$ corrects for the energy difference arising due to the relabelling of the sites. For the second term, it also includes a factor of $q^1$ for the energy contribution of the lowest site of the original single-column diagram. Along with the boundary conditions $\ang{M}{0}=1$ and $\ang{M}{M}=q^{\frac12M(M+1)}$, the recursion relation fixes $\ang{M}{m}$ completely to
\be
\label{eq:AMm}
\Ang{M}{m} = q^{\frac12m(m+1)}\qbinom{M}{m}
\ee
where the {\it Gaussian polynomial} (or {\it $q$-binomial\/}) is defined as
\be
\label{eq:Gausspoly}
\qbinom{M}{m} = \frac{[M][M-1] \cdots [M-m+1]}{[m][m-1]\cdots[1]}, \qquad [m] = \frac{1-q^m}{1-q}.
\ee
The factor $q^{\frac12m(m+1)}$ in \eqref{eq:AMm} is identified as the energy of the single-column diagram in $\Aone{M}{m}$ with all occupied sites at the bottom. We note that if the height labels of the single-column diagram are $a, \dots, M+a-1$ instead of $1, \dots, M$, then the energies of all the eigenvalues are shifted by $m(a-1)$ and the generating function is $q^{m(a-1)} \ang{M}{m} = q^{\frac12m(m+2a-1)}\smallqbinom{M}{m}$.

A single-column diagram is mapped to a pattern of zeros of $D^1(u)$ using the following rule: (i)~an occupied site at height $j$ produces a zero of order one in the center of the analyticity strip; (ii)~an unoccupied site produces two zeros, each of order one, lying on the edges of the analyticity strip. This map is illustrated in the left panel of \cref{fig:columnconf}.

\paragraph{Double-column diagrams.}
We define $\Adom{M}{m}{n}$ to be the set of double-column diagrams that satisfy the condition of {\it dominance}. Such diagrams are made of two single-column configurations of $M$ sites drawn side by side, with respectively $m$ and $n$ occupied sites in the left and right columns. An example is given in the right panel of \cref{fig:columnconf}. We respectively denote by $L$ and $R$ the signatures of the left and right column. The energy of a double-column diagram is $E=\sum_{i=1}^m L_i + \sum_{j=1}^n R_j$ and its weight is $q^E$.

A double-column configuration in $\Adom{M}{m}{n}$ satisfies the condition of dominance if
\be\label{eq:dominance}
L_i \le R_i, \qquad j = 1, \dots, m.
\ee
This of course presupposes that $0\le m\le n\le M$. The criterion \eqref{eq:dominance} can be translated in terms of a diagrammatic rule for the double-column diagram. One draws $m$ non intersecting lines pairing the top $m$ occupied 
sites of each column starting from the top. The double-column diagram satisfies dominance if the slope of each line is non-negative.

The generating function for $\Adom{M}{m}{n}$ is denoted $\ang{M}{m,n}$. It is the sum of the weights of the diagrams in $\Adom{M}{m}{n}$. By removing the lowest row of a given configuration in $\Adom{M}{m}{n}$, we obtain a new configuration of height $M-1$ in which the occupation numbers are $(m,n)$, $(m-1,n)$, $(m,n-1)$ or $(m-1,n-1)$. Crucially, the resulting column configuration satisfies dominance in all cases. This is easy to see either from the definition \eqref{eq:dominance} or from the diagrammatic rule. As a result, the generating function satisfies the recursion relation
\be
\Ang{M}{m,n} = q^{m+n}\bigg(\Ang{M-1}{m,n}+\Ang{M-1}{m-1,n}+\Ang{M-1}{m,n-1}+\Ang{M-1}{m-1,n-1}\bigg),
\ee
where the prefactor $q^{m+n}$ compensates for the relabelling of the height labels and the energy contribution of the lowest sites. With the conditions 
\be
\Ang{M}{0,0}= 1, \quad \Ang{M}{0,M}= q^{\frac12M(M+1)},\quad \Ang{M}{M,0}= 0,\quad \Ang{M}{M,M}= q^{\frac12M(M+1)+\frac12N(N+1)},
\ee
the recursion relation fixes the generating functions entirely. The result is
\be
\label{eq:Amn}
\Ang{M}{m,n} = q^{\frac12m(m+1)+\frac12n(n+1)}\qnarayana{M}{m}{n}
\ee
where $\smallqnarayana{M}{m}{n}$ are the generalised $q$-Narayana numbers
\be
\qnarayana{M}{m}{n}
=q^{-M+n}\bigg(\qbinom{M}{m}\qbinom{M+1}{n+1}-\qbinom{M+1}{m}\qbinom{M}{n+1}\bigg).
\label{eq:qnar}
\ee
The factor $q^{\frac12m(m+1)+\frac12n(n+1)}$ in \eqref{eq:Amn} is the weight of the double-column diagram in $\Adom{M}{m}{n}$ with minimal energy. 

A double-column diagram is mapped to a pattern of zeros of $D^2(u)$ using the following rule: (i) if both sites at height $j$ are occupied, a zero of order two sits in the center of the analyticity strip; (ii) if both sites at height $j$ are unoccupied, a pair of double zeros is inserted on the edges of the analyticity strip; (iii) if one site is occupied and the other is not, this yields three zeros of order one, one of which is inserted in the center of the analyticity strip whereas the two others are inserted on the edges. An example is given in the right panel of \cref{fig:columnconf}.

%
\subsection{Solution for all the eigenvalues}
\label{sec:excitations}
%

In this section, we find closed expressions for the finitized spectrum generating functions $\ZdN$, defined by
\be
\ZdN = \sum_{\substack{\textrm{eigenstates}\\ \textrm{of }D^1(u)\textrm{ in } \stan_N^d}} q^{\Delta},
\ee 
where $q$ is the modular nome. We note that the sum is over the finite set of eigenvalues in $\stan_N^d$, characterized by their patterns of zeros, whereas $\Delta$ is the conformal weight of the corresponding pattern of zeros in the scaling limit.

\paragraph{Selection rules for the patterns of zeros.}
Our computation of $\ZdN$ is based on a conjecture for the selection rules for the eigenvalues in $\stan_N^d$ which we now formulate. These empirical rules give the patterns of zeros and the values taken by the integers $k^j$ and $\ell^i$ for each eigenvalue. Similar selection rules for the model of critical dense polymers on the strip were conjectured in \cite{PR07} and later proven in \cite{MD11}. The conjectured selection rules given below are supported by data produced with our computer implementation of the transfer matrices for $N \le 12$. For a given eigenvalue, our program outputs the corresponding patterns of zeros of $D^1(u)$ and $D^2(u)$. To illustrate, the data corresponding to all the eigenstates in $\stan_8^2$ and $\stan_8^4$ is given in \cref{fig:pats82,fig:pats84}. 

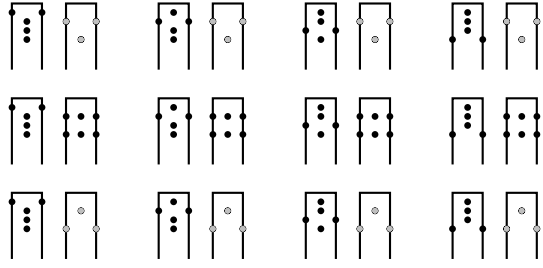
\begin{figure}[h!]
\begin{alignat*}{2}
&
\psun
\begin{pspicture}[shift=-0.8](-0.4,-2.2)(1.4,0.1)
\psline{-}(0,-2.2)(0,0)(1,0)(1,-2.2)
\psdots[dotsize=0.09cm](1,-0.3)
\psdots[dotsize=0.09cm](0.5,-0.6)(0.5,-0.9)(0.5,-1.2)
\psdots[dotsize=0.09cm](0,-0.3)
\end{pspicture}
\begin{pspicture}[shift=-0.8](-0.4,-2.2)(1.4,0.1)
\psline{-}(0,-2.2)(0,0)(1,0)(1,-2.2)
\psdots[linecolor=black,fillcolor=lightgray,dotstyle=o,dotsize=0.09cm](0,-0.6)
\psdots[linecolor=black,fillcolor=lightgray,dotstyle=o,dotsize=0.09cm](1,-0.6)
\psdots[linecolor=black,fillcolor=lightgray,dotstyle=o,dotsize=0.09cm](0.5,-1.2)
\end{pspicture} \quad\
\begin{pspicture}[shift=-0.8](-0.4,-2.2)(1.4,0.1)
\psline{-}(0,-2.2)(0,0)(1,0)(1,-2.2)
\psdots[dotsize=0.09cm](1,-0.6)
\psdots[dotsize=0.09cm](0.5,-0.3)(0.5,-0.9)(0.5,-1.2)
\psdots[dotsize=0.09cm](0,-0.6)
\end{pspicture}
\begin{pspicture}[shift=-0.8](-0.4,-2.2)(1.4,0.1)
\psline{-}(0,-2.2)(0,0)(1,0)(1,-2.2)
\psdots[linecolor=black,fillcolor=lightgray,dotstyle=o,dotsize=0.09cm](0,-0.6)
\psdots[linecolor=black,fillcolor=lightgray,dotstyle=o,dotsize=0.09cm](1,-0.6)
\psdots[linecolor=black,fillcolor=lightgray,dotstyle=o,dotsize=0.09cm](0.5,-1.2)
\end{pspicture} \quad\
\begin{pspicture}[shift=-0.8](-0.4,-2.2)(1.4,0.1)
\psline{-}(0,-2.2)(0,0)(1,0)(1,-2.2)
\psdots[dotsize=0.09cm](1,-0.9)
\psdots[dotsize=0.09cm](0.5,-0.3)(0.5,-0.6)(0.5,-1.2)
\psdots[dotsize=0.09cm](0,-0.9)
\end{pspicture}
\begin{pspicture}[shift=-0.8](-0.4,-2.2)(1.4,0.1)
\psline{-}(0,-2.2)(0,0)(1,0)(1,-2.2)
\psdots[linecolor=black,fillcolor=lightgray,dotstyle=o,dotsize=0.09cm](0,-0.6)
\psdots[linecolor=black,fillcolor=lightgray,dotstyle=o,dotsize=0.09cm](1,-0.6)
\psdots[linecolor=black,fillcolor=lightgray,dotstyle=o,dotsize=0.09cm](0.5,-1.2)
\end{pspicture} \quad\
\begin{pspicture}[shift=-0.8](-0.4,-2.2)(1.4,0.1)
\psline{-}(0,-2.2)(0,0)(1,0)(1,-2.2)
\psdots[dotsize=0.09cm](1,-1.2)
\psdots[dotsize=0.09cm](0.5,-0.3)(0.5,-0.6)(0.5,-0.9)
\psdots[dotsize=0.09cm](0,-1.2)
\end{pspicture}
\begin{pspicture}[shift=-0.8](-0.4,-2.2)(1.4,0.1)
\psline{-}(0,-2.2)(0,0)(1,0)(1,-2.2)
\psdots[linecolor=black,fillcolor=lightgray,dotstyle=o,dotsize=0.09cm](0,-0.6)
\psdots[linecolor=black,fillcolor=lightgray,dotstyle=o,dotsize=0.09cm](1,-0.6)
\psdots[linecolor=black,fillcolor=lightgray,dotstyle=o,dotsize=0.09cm](0.5,-1.2)
\end{pspicture} 
\\[0.2cm] &
\psun
\begin{pspicture}[shift=-0.8](-0.4,-2.2)(1.4,0.1)
\psline{-}(0,-2.2)(0,0)(1,0)(1,-2.2)
\psdots[dotsize=0.09cm](1,-0.3)
\psdots[dotsize=0.09cm](0.5,-0.6)(0.5,-0.9)(0.5,-1.2)
\psdots[dotsize=0.09cm](0,-0.3)
\end{pspicture}
\begin{pspicture}[shift=-0.8](-0.4,-2.2)(1.4,0.1)
\psline{-}(0,-2.2)(0,0)(1,0)(1,-2.2)
\psdots[dotsize=0.09cm](0,-0.6)(0,-1.2)
\psdots[dotsize=0.09cm](0.5,-0.6)(0.5,-1.2)
\psdots[dotsize=0.09cm](1,-0.6)(1,-1.2)
\end{pspicture} \quad\
\begin{pspicture}[shift=-0.8](-0.4,-2.2)(1.4,0.1)
\psline{-}(0,-2.2)(0,0)(1,0)(1,-2.2)
\psdots[dotsize=0.09cm](1,-0.6)
\psdots[dotsize=0.09cm](0.5,-0.3)(0.5,-0.9)(0.5,-1.2)
\psdots[dotsize=0.09cm](0,-0.6)
\end{pspicture}
\begin{pspicture}[shift=-0.8](-0.4,-2.2)(1.4,0.1)
\psline{-}(0,-2.2)(0,0)(1,0)(1,-2.2)
\psdots[dotsize=0.09cm](0,-0.6)(0,-1.2)
\psdots[dotsize=0.09cm](0.5,-0.6)(0.5,-1.2)
\psdots[dotsize=0.09cm](1,-0.6)(1,-1.2)
\end{pspicture} \quad \
\begin{pspicture}[shift=-0.8](-0.4,-2.2)(1.4,0.1)
\psline{-}(0,-2.2)(0,0)(1,0)(1,-2.2)
\psdots[dotsize=0.09cm](1,-0.9)
\psdots[dotsize=0.09cm](0.5,-0.3)(0.5,-0.6)(0.5,-1.2)
\psdots[dotsize=0.09cm](0,-0.9)
\end{pspicture}
\begin{pspicture}[shift=-0.8](-0.4,-2.2)(1.4,0.1)
\psline{-}(0,-2.2)(0,0)(1,0)(1,-2.2)
\psdots[dotsize=0.09cm](0,-0.6)(0,-1.2)
\psdots[dotsize=0.09cm](0.5,-0.6)(0.5,-1.2)
\psdots[dotsize=0.09cm](1,-0.6)(1,-1.2)
\end{pspicture}\quad \
\begin{pspicture}[shift=-0.8](-0.4,-2.2)(1.4,0.1)
\psline{-}(0,-2.2)(0,0)(1,0)(1,-2.2)
\psdots[dotsize=0.09cm](1,-1.2)
\psdots[dotsize=0.09cm](0.5,-0.3)(0.5,-0.6)(0.5,-0.9)
\psdots[dotsize=0.09cm](0,-1.2)
\end{pspicture}
\begin{pspicture}[shift=-0.8](-0.4,-2.2)(1.4,0.1)
\psline{-}(0,-2.2)(0,0)(1,0)(1,-2.2)
\psdots[dotsize=0.09cm](0,-0.6)(0,-1.2)
\psdots[dotsize=0.09cm](0.5,-0.6)(0.5,-1.2)
\psdots[dotsize=0.09cm](1,-0.6)(1,-1.2)
\end{pspicture}
\\[0.2cm] &
\psun
\begin{pspicture}[shift=-0.8](-0.4,-1.6)(1.4,0.1)
\psline{-}(0,-2.2)(0,0)(1,0)(1,-2.2)
\psdots[dotsize=0.09cm](1,-0.3)
\psdots[dotsize=0.09cm](0.5,-0.6)(0.5,-0.9)(0.5,-1.2)
\psdots[dotsize=0.09cm](0,-0.3)
\end{pspicture}
\begin{pspicture}[shift=-0.8](-0.4,-1.6)(1.4,0.1)
\psline{-}(0,-2.2)(0,0)(1,0)(1,-2.2)
\psdots[linecolor=black,fillcolor=lightgray,dotstyle=o,dotsize=0.09cm](0,-1.2)
\psdots[linecolor=black,fillcolor=lightgray,dotstyle=o,dotsize=0.09cm](1,-1.2)
\psdots[linecolor=black,fillcolor=lightgray,dotstyle=o,dotsize=0.09cm](0.5,-0.6)
\end{pspicture} \quad\
\begin{pspicture}[shift=-0.8](-0.4,-1.6)(1.4,0.1)
\psline{-}(0,-2.2)(0,0)(1,0)(1,-2.2)
\psdots[dotsize=0.09cm](1,-0.6)
\psdots[dotsize=0.09cm](0.5,-0.3)(0.5,-0.9)(0.5,-1.2)
\psdots[dotsize=0.09cm](0,-0.6)
\end{pspicture}
\begin{pspicture}[shift=-0.8](-0.4,-1.6)(1.4,0.1)
\psline{-}(0,-2.2)(0,0)(1,0)(1,-2.2)
\psdots[linecolor=black,fillcolor=lightgray,dotstyle=o,dotsize=0.09cm](0,-1.2)
\psdots[linecolor=black,fillcolor=lightgray,dotstyle=o,dotsize=0.09cm](1,-1.2)
\psdots[linecolor=black,fillcolor=lightgray,dotstyle=o,dotsize=0.09cm](0.5,-0.6)
\end{pspicture} \quad\
\begin{pspicture}[shift=-0.8](-0.4,-1.6)(1.4,0.1)
\psline{-}(0,-2.2)(0,0)(1,0)(1,-2.2)
\psdots[dotsize=0.09cm](1,-0.9)
\psdots[dotsize=0.09cm](0.5,-0.3)(0.5,-0.6)(0.5,-1.2)
\psdots[dotsize=0.09cm](0,-0.9)
\end{pspicture}
\begin{pspicture}[shift=-0.8](-0.4,-1.6)(1.4,0.1)
\psline{-}(0,-2.2)(0,0)(1,0)(1,-2.2)
\psdots[linecolor=black,fillcolor=lightgray,dotstyle=o,dotsize=0.09cm](0,-1.2)
\psdots[linecolor=black,fillcolor=lightgray,dotstyle=o,dotsize=0.09cm](1,-1.2)
\psdots[linecolor=black,fillcolor=lightgray,dotstyle=o,dotsize=0.09cm](0.5,-0.6)
\end{pspicture} \quad\
\begin{pspicture}[shift=-0.8](-0.4,-1.6)(1.4,0.1)
\psline{-}(0,-2.2)(0,0)(1,0)(1,-2.2)
\psdots[dotsize=0.09cm](1,-1.2)
\psdots[dotsize=0.09cm](0.5,-0.3)(0.5,-0.6)(0.5,-0.9)
\psdots[dotsize=0.09cm](0,-1.2)
\end{pspicture}
\begin{pspicture}[shift=-0.8](-0.4,-1.6)(1.4,0.1)
\psline{-}(0,-2.2)(0,0)(1,0)(1,-2.2)
\psdots[linecolor=black,fillcolor=lightgray,dotstyle=o,dotsize=0.09cm](0,-1.2)
\psdots[linecolor=black,fillcolor=lightgray,dotstyle=o,dotsize=0.09cm](1,-1.2)
\psdots[linecolor=black,fillcolor=lightgray,dotstyle=o,dotsize=0.09cm](0.5,-0.6)
\end{pspicture}
\end{alignat*}
\caption{The twelve patterns of zeros corresponding to $\AA{4}{2}{3}{1}{1}$.}
\label{fig:A42311}
\end{figure}

The selection rules for the patterns of zeros are described in terms of the single- and double-column diagrams discussed in \cref{sec:colconfs}. Let $(\sigma,\sigma')$ be a pair of column diagrams\footnote{We use the letter $M$ for both the number of sites of a column diagram and the vertical width of the lattice. It should be clear from the context which one is referred to, and likewise in \cref{sec:fsgf}.} with $\sigma \in \Aone{M}{m}$ and $\sigma' \in \Adom{L}{n}{\ell}$. We denote the set of such pairs by $\AA{M}{L}{m}{n}{\ell}$. An example is given in \cref{fig:A42311}. For $d \equiv 0,1 \Mod 3$, we indicate whether the patterns of zeros of the corresponding set belong to subcases A or B by writing $\AAAB{M}{L}{m}{n}{\ell}{\tA}$ or $\AAAB{M}{L}{m}{n}{\ell}{\tB}$. 

For $\stan_8^2$ and $\stan_8^4$, \cref{fig:pats82,fig:pats84} reveal that the patterns of zeros of $D^1(u)$ and $D^2(u)$ are encoded by the following sets:
\begin{subequations}
\begin{alignat}{2}
\stan_8^2&: \quad \AA{4}{0}{1}{0}{0}\,\cup\,\AA{4}{1}{3}{0}{0}\,\cup\,\AA{5}{1}{3}{1}{1}\,\cup\,\AA{5}{2}{5}{1}{1}  \,\cup\,\AA{6}{2}{5}{2}{2}\,\cup\,\AA{7}{3}{7}{3}{3},\\[0.2cm]
\stan_8^4&:\quad\AAAB{4}{0}{2}{0}{0}{\tB}\,\cup\,\AAAB{4}{1}{4}{0}{0}{\tB}\,\cup\,\AAAB{5}{1}{4}{1}{1}{\tB}\,\cup\,\AAAB{6}{2}{6}{2}{2}{\tB}\,\cup\,\AAAB{5}{1}{4}{0}{1}{\tA}  \,\cup\,\AAAB{6}{2}{6}{1}{2}{\tA}.
\end{alignat}
\end{subequations}
In general, for $d>1$, we conjecture that the full set of patterns of zeros in $\stan_N^d$ is given by the following sets:
\begin{subequations}\label{eq:SR}
\begin{alignat}{2}
d=3t&: \hspace{0.1cm}\bigcup_{i=0}^{\frac{N-d}2} \bigcup_{j=0}^{\big\lfloor\frac12\big(\frac{N-d}2-i\big)\big\rfloor} \AAAB{\frac{N+t}2 + i}{i+j+t-1}{2(i+j+t)}{i}{i+t-1}{\tA} \hspace{0.1cm} \cup \hspace{0.0cm} \bigcup_{i=0}^{\frac{N-d-4}2} \bigcup_{j=0}^{\big\lfloor\frac12\big(\frac{N-d-4}2-i\big)\big\rfloor} \AAAB{\frac{N+t}2 + i}{i+j+t}{2(i+j+t)+2}{i}{i+t}{\tB}\ ,\label{eq:SR0t}
\\
d=3t+1&: \hspace{0.1cm}\bigcup_{i=0}^{\frac{N-d}2} \bigcup_{j=0}^{\big\lfloor\frac12\big(\frac{N-d}2-i\big)\big\rfloor} \AAAB{\frac{N+t-1}2 + i}{i+j+t-1}{2(i+j+t)}{i}{i+t-1}{\tB}\hspace{0.1cm} \cup \hspace{0.0cm} \bigcup_{i=0}^{\frac{N-d-2}2} \bigcup_{j=0}^{\big\lfloor\frac12\big(\frac{N-d-2}2-i\big)\big\rfloor} \AAAB{\frac{N+t+1}2 + i}{i+j+t}{2(i+j+t)+2}{i}{i+t}{\tA}\ ,\label{eq:SR1t}
\\
d=3t+2&: \hspace{0.1cm}\bigcup_{i=0}^{\frac{N-d}2} \bigcup_{j=0}^{\big\lfloor\frac12\big(\frac{N-d}2-i\big)\big\rfloor} \AA{\frac{N+t}2 + i}{i+j+t}{2(i+j+t)+1}{i}{i+t}\ .\label{eq:SR2t}
\end{alignat}
For $d = 0$ and $d=1$, \eqref{eq:SR0t} and \eqref{eq:SR1t} are ill-defined because some indices are negative. In these cases, we instead have the following selection rules:
\begin{alignat}{2}
d=0&: \quad \AA{0}{0}{0}{0}{0} \hspace{0.1cm} \cup \hspace{0.1cm} \bigcup_{i=0}^{\frac{N-4}2} \bigcup_{j=0}^{\big\lfloor\frac12\big(\frac{N-4}2-i\big)\big\rfloor} \AAAB{\frac{N}2 + i}{i+j}{2(i+j)+2}{i}{i}{\tB}\ ,
\\
d=1&: \quad\AA{0}{0}{0}{0}{0}\hspace{0.1cm} \cup \hspace{0.1cm} \bigcup_{i=0}^{\frac{N-3}2} \bigcup_{j=0}^{\big\lfloor\frac12\big(\frac{N-3}2-i\big)\big\rfloor} \AAAB{\frac{N+1}2 + i}{i+j}{2(i+j)+2}{i}{i}{\tA}\ .
\end{alignat}
\end{subequations}
where $\AA{0}{0}{0}{0}{0}$ is the set that contains a unique element: the pattern of zeros corresponding to the Razumov-Stroganov eigenvalue.

\paragraph{Selection rules for the integers.}
As part of the conjectured selection rules, we also provide the prescription for the values taken by the numbers $k^j$ and $\ell^i$ for each eigenvalue. For $N\le12$, we have obtained these integers for each pattern of zeros by evaluating \eqref{eq:integerK} and \eqref{eq:integerL} with the finite-size spectra produced by our computer implementation. The precision of the approximate values of the $k^j$ and $\ell^i$ obtained in this way is remarkably good even for small system sizes: The error is less than $0.1$ in almost all the cases.  In \cref{fig:pats82,fig:pats84}, the 
values of these integers are given alongside the corresponding patterns of zeros. The prescription is as follows: for an element in $\AA{M}{L}{m}{n}{\ell}$,
\begin{itemize}
\item[(i)] the heights of the single-column diagram (corresponding to the $k^j$) are labelled from $0$ to $M-1$ for $d\equiv 0,1 \Mod 3\,$ (A), and from $1$ to $M$ for $d\equiv 0,1 \Mod 3\,$ (B) and $d\equiv 2 \Mod 3$; 
\item[(ii)] the heights of the double-column diagram (corresponding to the $\ell^i$) are labelled from $1$ to $L$ in all cases.
\end{itemize}

\paragraph{Finitized spectrum generating functions.}
The conformal weight $\Delta$ corresponding to a given pattern of zeros is given in \eqref{eq:DelKL}. The selection rules and the prescription for the integers allow us to write explicit expressions for the finitized spectrum generating functions. These are obtained as sums of the generating functions of the sets of column diagrams given in \eqref{eq:SR}. To do so, we compute the minimal conformal weights $\Delta_{\textrm{min}}$ for $\AA{M}{L}{m}{n}{\ell}$, $\AAAB{M}{L}{m}{n}{\ell}{\tA}$ and $\AAAB{M}{L}{m}{n}{\ell}{\tB}$ in terms of the energies $E$, $E^{(\tA)}$ and $E^{(\tB)}$ of the corresponding minimal configurations. Using \eqref{eq:DelKL} and the above prescription (i) and (ii), we find:
\begin{subequations}
\label{eq:Es}
\begin{alignat}{4}
&d \equiv 0,1 \Mod 3\,(\tA):\qquad&& \Delta_{\textrm{min}} = E^{(\tA)}= \tfrac12(m^2+n^2+\ell^2 - m + n+\ell - mn-m\ell),
\label{eq:EA}\\[0.15cm]
&d \equiv 0,1 \Mod 3\,(\tB):\qquad&& \Delta_{\textrm{min}} = E^{(\tB)}= \tfrac12(m^2+n^2+\ell^2 + n+\ell - mn-m\ell), \label{eq:EB}\\[0.15cm]
&d \equiv 2 \Mod 3:\qquad&& \Delta_{\textrm{min}} = E-\tfrac16 = \tfrac12(m^2+n^2+\ell^2- mn-m\ell)-\tfrac16.
\label{eq:justE}
\end{alignat}
\end{subequations}
The spectra generating functions $\ZdN$ are then obtained by summing $q^{\Delta_{\textrm{min}}} \smallqbinom{M}{m} \smallqnarayana{L}{n}{\ell}$ over the sets $\AA{M}{L}{m}{n}{\ell}$ given by \eqref{eq:SR}. In doing so, we note that the indices $i$ and $j$ run over all possible values for which the sets are well-defined. Equivalently, they run over all values such that the $q$-binomials in the generating functions have positive arguments, with the top one larger or equal to the bottom one. We therefore omit the indices of the sums over $i$ and $j$, understanding these sums as running over $\mathbb Z$, with only finitely many contributions. We obtain:
\begin{subequations}
\label{eq:char.identities}
\begin{alignat}{3}
&d = 3t: \quad
&&\ZdN 
= q^{\frac{d(d-1)}6} \sum_{i,j} q^{i^2+2j(i+j)+t(2i+3j)}
\qbinom{\frac{N+t}2+i}{2(i+j+t)}\qnarayana{i+j+t-1}{i}{i+t-1}\label{eq:char0}
\\
& &&\hspace{1.25cm} +
q^{\frac{(d+3)(d+4)}6} \sum_{i,j} q^{i(i+3)+2j(i+j+2)+t(2i+3j)}
\qbinom{\frac{N+t}2+i}{2(i+j+t)+2}\qnarayana{i+j+t}{i}{i+t},\nonumber
\\[0.3cm]
&d = 3t+1:\quad
&&\ZdN = q^{\frac{d(d-1)}6} \sum_{i,j} q^{i(i+1)+2j(i+j+\frac12)+t(2i+3j)}
\qbinom{\frac{N+t-1}2+i}{2(i+j+t)}\qnarayana{i+j+t-1}{i}{i+t-1}\label{eq:char1}
\\
& &&\hspace{1.25cm} +
q^{\frac{(d+1)(d+2)}6}\sum_{i,j} q^{i(i+2)+2j(i+j+\frac32)+t(2i+3j)}
\qbinom{\frac{N+t+1}2+i}{2(i+j+t)+2}\qnarayana{i+j+t}{i}{i+t},\nonumber
\\[0.3cm]
&d = 3t+2 : \quad
&&\ZdN = q^{\frac{d(d-1)}6} \sum_{i,j} q^{i(i+1)+2j(i+j+1)+t(2i+3j)}
\qbinom{\frac{N+t}2+i}{2(i+j+t)+1}\qnarayana{i+j+t}{i}{i+t}.\label{eq:char2}
\end{alignat}
\end{subequations}
For $d=0$ and $d=1$, \eqref{eq:char0} and \eqref{eq:char1} must be modified so that the first line is replaced by $1$. 

In \cref{sec:charids}, we show using $q$-binomial identities that these complicated expressions for $\ZdN$ can be simplified to yield the finitized Kac characters, namely:
\be
\ZdN =q^{d(d-1)/6}\bigg(\qbinom{N}{\frac{N-d}2}-q^{d+1}\qbinom{N}{\frac{N-d-2}2}\bigg) = \XN_{1,d+1}(q).
\ee
This holds for $d\equiv 0,1,2 \textrm{ mod } 3$.

\paragraph{Scaling limit.}
The behavior of the finitized character in the scaling limit is easily extracted using
\be
\qbinom{N}{\frac{N-d}2} \xrightarrow{\substack{N \rightarrow \infty}} \frac 1{(q)_\infty}
\ee
which holds for fixed $d$. We recall that $(q)_\infty$ is defined in \eqref{eq:Poch}. This yields
\be
\ZdN \xrightarrow{\substack{N \rightarrow \infty}}\chit_{1,d+1}(q).
\ee

As a final remark, we note that \eqref{eq:char.identities} provides expressions for finitized characters of the irreducible Virasoro representations $\mathsf I_{1,s}$. This is trivially true for $d \equiv 2 \textrm{ mod } 3$ because $\chit_{1,d+1} \simeq \mathsf I_{1,d+1}$ is already irreducible. For $d \equiv 0 \textrm{ mod } 3$, the first and second lines of \eqref{eq:char0} are respectively finitized characters for $\mathsf I_{1,d+1}$ and $\mathsf I_{1,d+5}$. For $d \equiv 1 \textrm{ mod } 3$, the first and second lines of \eqref{eq:char1} are respectively finitized characters for $\mathsf I_{1,d+1}$ and $\mathsf I_{1,d+3}$.

\subsection{Cylinder partition functions}\label{sec:cylPF}

The partition function $\ZcylMN$ of the loop model on the cylinder of size $2M\times N$ is obtained from the traces of $\Db(u)^M$ over $\stan_N^d$:
\be
\ZcylMN = \sum_{\substack{0 \le d \le N\\[0.05cm]d \equiv N\textrm{ mod }2}} U_{d}(\tfrac\alpha2)\,\tr \Db(u)^M.
\ee
This is the so-called Markov trace \cite{Jones83}, used to embed the Temperley-Lieb algebra on a cylinder. This was also used by Jacobsen and Richard \cite{RJ06} in the context of the $Q$-state Potts model. 
Here, $\alpha$ is the weight of the non-contractible loops and $U_k(x)$ is the $k$-th Chebyshev polynomial of the second type. The conformal cylinder partition function is obtained by removing the non-universal energy contributions and by taking the scaling limit:
\be
\eE^{2MN f_{\text{b}}(u)+M f_{\text{s}}(u)} \ZcylMN \xrightarrow{M,N \rightarrow \infty} \Zcyl(q)
\ee
where the ratio $\frac MN$ is taken to converge to a real number $\delta$, and the modular nome $q$ is given in terms of the lattice data in \eqref{eq:qbdy}. The scaling limit is taken separately for the two parities of $N$. We denote by $\Zcylodd(q)$ and $\Zcyleven (q)$ the corresponding partition functions. We then have
\be\label{eq:Zcylgen}
\Zcyl (q) = \lim_{N\rightarrow \infty}\sum_{\substack{0 \le d \le N\\[0.05cm] d \equiv N\textrm{\,mod\,}2}} \hspace{-0.2cm}U_{d}(\tfrac\alpha2) \ZdN =  \sum_{\substack{d \ge 0 \\[0.05cm] d \equiv N\textrm{\,mod\,}2}} \hspace{-0.2cm}U_{d}(\tfrac\alpha2)\chit_{1,d+1}(q).
\ee

For $\alpha = 1$, \eqref{eq:Zcylgen} simplifies to
\be
\Zcylodd(q)\big|_{\alpha = 1}=\Zcyleven (q)\big|_{\alpha = 1}=1
\ee
which is the trivial cylinder partition function of the rational model of percolation. For $\alpha = 2$, $U_{d}(1) = d+1$ and the partition functions can be written in terms of the $u(1)$ characters and their derivatives. Writing $\varkappa_j(q,z) = \varkappa^6_j(q,z)$ and $\varkappa_j(q) = \varkappa^6_j(q)$, we find
\begin{alignat}{2}
\Zcyleven(q) &+ \Zcylodd(q)\Big|_{\alpha = 2} = \frac{1}{(q)_\infty}\sum_{d\ge 0}(d+1) (q^{\Delta_{1,d+1}}-q^{\Delta_{1,d+4}})
= \frac{3}{(q)_\infty} \sum_{d \ge 1}q^{\Delta_{1,d+1}} 
\nonumber\\& = \frac{3}{2(q)_\infty} \sum_{d \in \mathbb Z}q^{\Delta_{1,d+1}}
=  \frac{3}{2(q)_\infty} \sum_{r=0}^5 \sum_{k\in \mathbb Z} q^{\Delta_{1,6k+r+1}}
= 3\big(\varkappa_1(q)+ \varkappa_3(q)+ \varkappa_5(q)\big)
\end{alignat}
and
\begin{alignat}{2}
\Zcyleven(q) &- \Zcylodd(q)\Big|_{\alpha = 2} = \frac{1}{(q)_\infty}\sum_{d\ge 0}(-1)^d(d+1) (q^{\Delta_{1,d+1}}-q^{\Delta_{1,d+4}})
= \frac{1}{(q)_\infty} \sum_{d \ge 1}(-1)^d(2d-1)q^{\Delta_{1,d+1}} 
\nonumber\\& 
= \frac{1}{(q)_\infty} \sum_{d \in \mathbb Z}(-1)^d d\, q^{\Delta_{1,d+1}}
=  \frac{1}{(q)_\infty} \sum_{r=0}^5 (-1)^r\sum_{k\in \mathbb Z}(6k+r) q^{\Delta_{1,6k+r+1}}
\nonumber\\& 
= -\varkappa_1(q)+ 3\varkappa_3(q)- 5\varkappa_5(q) -12\frac{\dd}{\dd z} \Big(\varkappa_1(q,z) - 3\varkappa_3(q,z) + \varkappa_5(q,z) \Big)\Big|_{z=1}\ .
\end{alignat}
The final result is
\begin{subequations}
\begin{alignat}{2}
&\Zcylodd (q)\big|_{\alpha = 2} &&= 2\varkappa_1(q)  + 4 \varkappa_5(q) +6\frac{\dd}{\dd z} \Big(\varkappa_1(q,z) - 3\varkappa_3(q,z) + \varkappa_5(q,z) \Big)\Big|_{z=1}\ ,\\[0.15cm]
&\Zcyleven (q)\big|_{\alpha = 2} &&= \varkappa_1(q) + 3\varkappa_3(q) - \varkappa_5(q) -6\frac{\dd}{\dd z} \Big(\varkappa_1(q,z) - 3\varkappa_3(q,z) + \varkappa_5(q,z) \Big)\Big|_{z=1}\ .
\end{alignat}
\end{subequations}

%
\section{Critical percolation with periodic boundary conditions}\label{sec:periodiccase}
%

\subsection{The transfer tangle and the enlarged periodic Temperley-Lieb algebra}\label{sec:PTLandT}

The transfer tangle with periodic boundary conditions is defined as
\be
\Tb (u)= \ir^{N}\ 
\psset{unit=0.9}
\begin{pspicture}[shift=-1.1](-0.2,-0.7)(5.2,1.0)
\facegrid{(0,0)}{(5,1)}
\psarc[linewidth=0.025]{-}(0,0){0.16}{0}{90}
\psarc[linewidth=0.025]{-}(1,0){0.16}{0}{90}
\psarc[linewidth=0.025]{-}(4,0){0.16}{0}{90}
\psline[linewidth=1.5pt,linecolor=blue]{-}(0,0.5)(-0.2,0.5)
\psline[linewidth=1.5pt,linecolor=blue]{-}(5,0.5)(5.2,0.5)
\rput(2.5,0.5){$\ldots$}
\rput(3.5,0.5){$\ldots$}
\rput(0.5,.5){$u$}
\rput(1.5,.5){$u$}
\rput(4.5,.5){$u$}
\rput(2.5,-0.5){$\underbrace{\qquad \qquad \qquad \qquad  \qquad \qquad}_N$}
\end{pspicture} \ ,
\label{eq:Tu}
\ee  
with the elementary face operator given in \eqref{eq:faceop}. 
Loop models on periodic geometries are usually described by the so-called {\it periodic Temperley-Lieb algebra}. This algebra 
was studied by Levy \cite{L91} and Martin and Saleur \cite{MS93} in the context of the Potts model. Its representation theory was also studied by Graham and Lehrer \cite{GL98}, Green \cite{G98} and Erdmann and Green \cite{EG99}. The terminology and precise definitions of this algebra tend to vary from author to author, and here we will work with the {\it enlarged} 
periodic Temperley-Lieb algebra $\eptl_N(\alpha, \beta)$, defined~\cite{PRVcyl2010} as 
\be
\eptl_N(\alpha, \beta) = \big\langle \Omega, \Omega^{-1},\,e_j;\,j=1,\ldots,N\big\rangle.\qquad  
\ee
It is a unital algebra, with the identity element $I$ obtained from the product of the generators $\Omega$ and $\Omega^{-1}$. 
The connectivities $I$ and $e_j$ for $1\le j\le N-1$ are given by the diagrams in \eqref{eq:TLdiag}, but drawn on a rectangle with periodic boundary conditions in the horizontal direction. The other generators are depicted as
\be
e_N= \
\begin{pspicture}[shift=-0.45](0,-0.55)(2.4,0.35)
\pspolygon[fillstyle=solid,fillcolor=lightlightblue,linewidth=0pt](0,-0.35)(2.4,-0.35)(2.4,0.35)(0,0.35)
\rput(0.2,-0.55){$_1$}\rput(0.6,-0.55){$_2$}\rput(1.0,-0.55){$_3$}\rput(2.2,-0.55){$_N$}
\rput(1.4,0.0){\small$...$}
\psarc[linecolor=blue,linewidth=1.5pt]{-}(0.0,0.35){0.2}{-90}{0}
\psarc[linecolor=blue,linewidth=1.5pt]{-}(0.0,-0.35){0.2}{0}{90}
\psline[linecolor=blue,linewidth=1.5pt]{-}(0.6,0.35)(0.6,-0.35)
\psline[linecolor=blue,linewidth=1.5pt]{-}(1.0,0.35)(1.0,-0.35)
\psline[linecolor=blue,linewidth=1.5pt]{-}(1.8,0.35)(1.8,-0.35)
\psarc[linecolor=blue,linewidth=1.5pt]{-}(2.4,-0.35){0.2}{90}{180}
\psarc[linecolor=blue,linewidth=1.5pt]{-}(2.4,0.35){0.2}{180}{-90}
\psframe[fillstyle=solid,linecolor=white,linewidth=0pt](-0.1,-0.4)(0,0.4)
\psframe[fillstyle=solid,linecolor=white,linewidth=0pt](2.4,-0.4)(2.5,0.4)
\end{pspicture} \ ,
\qquad
\begin{pspicture}[shift=-0.45](-0.7,-0.55)(2.0,0.35)
\rput(0.2,-0.55){$_1$}\rput(0.6,-0.55){$_2$}\rput(1.0,-0.55){$_3$}\rput(1.4,-0.55){\small$...$}\rput(1.8,-0.55){$_N$}
\pspolygon[fillstyle=solid,fillcolor=lightlightblue,linewidth=0pt](0,-0.35)(2.0,-0.35)(2.0,0.35)(0,0.35)
\multiput(0,0)(0.4,0){6}{\psbezier[linecolor=blue,linewidth=1.5pt]{-}(-0.2,-0.35)(-0.2,-0.0)(0.2,0.0)(0.2,0.35)}
\psframe[fillstyle=solid,linecolor=white,linewidth=0pt](-0.3,-0.4)(0,0.4)
\psframe[fillstyle=solid,linecolor=white,linewidth=0pt](2.0,-0.4)(2.4,0.4)
\rput(-0.55,0.042){$\Omega=$}
\end{pspicture} \ ,
\qquad
\begin{pspicture}[shift=-0.45](-1.1,-0.55)(2.0,0.35)
\rput(0.2,-0.55){$_1$}\rput(0.6,-0.55){$_2$}\rput(1.0,-0.55){$_3$}\rput(1.4,-0.55){\small$...$}\rput(1.8,-0.55){$_N$}
\pspolygon[fillstyle=solid,fillcolor=lightlightblue,linewidth=0pt](0,-0.35)(2.0,-0.35)(2.0,0.35)(0,0.35)
\multiput(0,0)(0.4,0){6}{\psbezier[linecolor=blue,linewidth=1.5pt]{-}(-0.2,0.35)(-0.2,-0.0)(0.2,0.0)(0.2,-0.35)}
\psframe[fillstyle=solid,linecolor=white,linewidth=0pt](-0.3,-0.4)(0,0.4)
\psframe[fillstyle=solid,linecolor=white,linewidth=0pt](2.0,-0.4)(2.4,0.4)
\rput(-0.75,0.07){$\Omega^{-1}=$}
\end{pspicture} \ .
\ee
The defining relations of $\eptl_N(\alpha, \beta)$ include \eqref{eq:TLdef}, wherein the indices are in the set $\{1, \dots, N\}$ and are taken modulo $N$, as well as
\be
\Omega \Omega^{-1}= \Omega^{-1} \Omega = I, \qquad
\Omega e_i \Omega^{-1} = e_{i-1}, \qquad
\Omega^{N} e_N = e_N \Omega^{N}, \qquad
(\Omega^{\pm 1} e_N)^{N-1} = \Omega^{\pm N} (\Omega ^{\pm 1} e_N).
\ee
For $N$ even, there are extra relations which remove the non-contractible loops in favor of weights $\alpha$:
\be
E \Omega^{\pm 1} E \,=\, \alpha E \qquad \textrm {where} \qquad E=\, e_2e_4\ldots e_{N-2}e_N.
\ee
We refer to this algebra as $\eptl_N(\alpha, \beta)$ for both parities of $N$, even if the parameter $\alpha$ does not come into play for $N$ odd. The transfer tangle $\Tb(u)$ is an element of $\eptl_N(\alpha, \beta)$. We are interested in the case $\lambda=\frac\pi 3$ and therefore $\beta = 1$ corresponding to the model of critical percolation. The parameter $\alpha$ remains free until \cref{sec:Pcfgs}, whereafter the case $\alpha=2$ is our main focus.

As an element of $\eptl_N(\alpha,\beta)$, $\Tb(u)$ satisfies a number of properties, namely: (i) the periodicity property $\Tb(u+\pi) = (-1)^N\Tb(u)$, (ii) the commutativity property $[\Tb(u),\Tb(v)] = 0$, and (iii) the specialisations $\Tb(u=0) = \ir^N\Omega$ and $\Tb(u=\lambda) = \ir^N\Omega^{-1}$.

The braid transfer matrices are also elements of $\eptl_N(\alpha,\beta)$. They are defined as 
\be
\label{eq:braidT}
\Fbtp=\ 
\psset{unit=.9cm}
\begin{pspicture}[shift=-1.1](-0.2,-0.7)(5.2,1.0)
\facegrid{(-0,0)}{(5,1)}
\psline[linewidth=1.5pt,linecolor=blue]{-}(-0.2,0.5)(2,0.5)
\psline[linewidth=1.5pt,linecolor=blue]{-}(0.5,0)(0.5,0.35)
\psline[linewidth=1.5pt,linecolor=blue]{-}(0.5,0.65)(0.5,1)
\psline[linewidth=1.5pt,linecolor=blue]{-}(1.5,0)(1.5,0.35)
\psline[linewidth=1.5pt,linecolor=blue]{-}(1.5,0.65)(1.5,1)
\rput(2.53,.51){$\dots$}
\rput(3.53,.51){$\dots$}
\psline[linewidth=1.5pt,linecolor=blue]{-}(4,0.5)(5.2,0.5)
\psline[linewidth=1.5pt,linecolor=blue]{-}(4.5,0)(4.5,0.35)
\psline[linewidth=1.5pt,linecolor=blue]{-}(4.5,0.65)(4.5,1)
\rput(2.5,-0.5){$\underbrace{\qquad \qquad \qquad \qquad  \qquad \qquad}_N$}
\end{pspicture}\ , \qquad
\Fbtm=\ 
\begin{pspicture}[shift=-1.1](-0.2,-0.7)(5.2,1.0)
\facegrid{(-0,0)}{(5,1)}
\rput(2.53,.51){$\dots$}
\rput(3.53,.51){$\dots$}
\psline[linewidth=1.5pt,linecolor=blue]{-}(0,0.5)(-0.2,0.5)
\psline[linewidth=1.5pt,linecolor=blue]{-}(5,0.5)(5.2,0.5)
\psline[linewidth=1.5pt,linecolor=blue]{-}(0.5,0)(0.5,1)
\psline[linewidth=1.5pt,linecolor=blue]{-}(1.5,0)(1.5,1)
\psline[linewidth=1.5pt,linecolor=blue]{-}(4.5,0)(4.5,1)
\psline[linewidth=1.5pt,linecolor=blue]{-}(0,0.5)(0.35,0.5)
\psline[linewidth=1.5pt,linecolor=blue]{-}(0.65,0.5)(1.35,0.5)
\psline[linewidth=1.5pt,linecolor=blue]{-}(1.65,0.5)(2,0.5)
\psline[linewidth=1.5pt,linecolor=blue]{-}(4,0.5)(4.35,0.5)
\psline[linewidth=1.5pt,linecolor=blue]{-}(4.65,0.5)(5,0.5)
\rput(2.5,-0.5){$\underbrace{\qquad \qquad \qquad \qquad  \qquad \qquad}_N$}
\end{pspicture}\ ,
\ee
with the braid face operators defined in \eqref{eq:braidops}. The two braid transfer matrices are not equal in general. They are obtained as the $u \rightarrow \pm \ir \infty$ limits of $\Tb(u)$ as follows:
\be
\Fbtp = \lim_{u \rightarrow \ir \infty}  \bigg(\frac{\eE^{\ir(\pi-\lambda)/2}}{s_{0}(u)}\bigg)^N (-\ir)^N\Tb(u), \qquad \Fbtm = \lim_{u \rightarrow -\ir \infty}  \bigg(\frac{\eE^{\ir(\pi-\lambda)/2}}{s_{1}(-u)}\bigg)^N (-\ir)^N\Tb(u).
\ee

\subsection{Standard modules}\label{sec:Pstanmod}

The standard modules $\stanp_N^d$ over $\eptl_N(\alpha, \beta)$ are defined on the vector space generated by the span of periodic link states on $N$ nodes with $d$ defects, where $N \equiv d \textrm{ mod }2$.  The construction  is similar to the one for $\tl_N(\beta)$ in \cref{sec:stanmod}. One difference is that the link states are drawn on a line segment with the left and right ends identified, and the loop segments can connect via the back (virtual cut) of the cylinder. The dimensions of these vector spaces are 
\be
\dim \stanp_N^d = \binom{N}{\frac{N-d}{2}}.
\ee
For example, the link states that generate $\stanp_5^1$ are
\be
\begin{array}{c}
\psset{unit=0.8cm}
\begin{pspicture}[shift=-0.0](-0.0,0)(2.0,0.5)
\psline{-}(0,0)(2.0,0)
\psarc[linecolor=darkgreen,linewidth=1.5pt]{-}(0.4,0){0.2}{0}{180}
\psarc[linecolor=darkgreen,linewidth=1.5pt]{-}(1.2,0){0.2}{0}{180}
\psline[linecolor=darkgreen,linewidth=1.5pt]{-}(1.8,0)(1.8,0.5)
\end{pspicture}\ , \ \ 
\begin{pspicture}[shift=-0.0](-0.0,0)(2.0,0.5)
\psline{-}(0,0)(2.0,0)
\psarc[linecolor=darkgreen,linewidth=1.5pt]{-}(0.4,0){0.2}{0}{180}
\psarc[linecolor=darkgreen,linewidth=1.5pt]{-}(1.6,0){0.2}{0}{180}
\psline[linecolor=darkgreen,linewidth=1.5pt]{-}(1.0,0)(1.0,0.5)
\end{pspicture}\ , \ \ 
\begin{pspicture}[shift=-0.0](-0.0,0)(2.0,0.5)
\psline{-}(0,0)(2.0,0)
\psarc[linecolor=darkgreen,linewidth=1.5pt]{-}(0.8,0){0.2}{0}{180}
\psarc[linecolor=darkgreen,linewidth=1.5pt]{-}(1.6,0){0.2}{0}{180}
\psline[linecolor=darkgreen,linewidth=1.5pt]{-}(0.2,0)(0.2,0.5)
\end{pspicture}\ , \ \ 
\begin{pspicture}[shift=-0.0](-0.0,0)(2.0,0.5)
\psline{-}(0,0)(2.0,0)
\psarc[linecolor=darkgreen,linewidth=1.5pt]{-}(0.0,0){0.2}{0}{90}
\psarc[linecolor=darkgreen,linewidth=1.5pt]{-}(2.0,0){0.2}{90}{180}
\psarc[linecolor=darkgreen,linewidth=1.5pt]{-}(1.2,0){0.2}{0}{180}
\psline[linecolor=darkgreen,linewidth=1.5pt]{-}(0.6,0)(0.6,0.5)
\end{pspicture}\ , \ \ 
\begin{pspicture}[shift=-0.0](-0.0,0)(2.0,0.5)
\psline{-}(0,0)(2.0,0)
\psarc[linecolor=darkgreen,linewidth=1.5pt]{-}(0.0,0){0.2}{0}{90}
\psarc[linecolor=darkgreen,linewidth=1.5pt]{-}(2.0,0){0.2}{90}{180}
\psarc[linecolor=darkgreen,linewidth=1.5pt]{-}(0.8,0){0.2}{0}{180}
\psline[linecolor=darkgreen,linewidth=1.5pt]{-}(1.4,0)(1.4,0.5)
\end{pspicture}\ , \\[0.3cm]
\psset{unit=0.8cm}
\begin{pspicture}[shift=-0.0](-0.0,0)(2.0,0.5)
\psline{-}(0,0)(2.0,0)
\psbezier[linecolor=darkgreen,linewidth=1.5pt]{-}(0.2,0)(0.2,0.7)(1.4,0.7)(1.4,0)
\psarc[linecolor=darkgreen,linewidth=1.5pt]{-}(0.8,0){0.2}{0}{180}
\psline[linecolor=darkgreen,linewidth=1.5pt]{-}(1.8,0)(1.8,0.5)
\end{pspicture}\ , \ \ 
\begin{pspicture}[shift=-0.0](-0.0,0)(2.0,0.5)
\psline{-}(0,0)(2.0,0)
\psbezier[linecolor=darkgreen,linewidth=1.5pt]{-}(0.6,0)(0.6,0.7)(1.8,0.7)(1.8,0)
\psarc[linecolor=darkgreen,linewidth=1.5pt]{-}(1.2,0){0.2}{0}{180}
\psline[linecolor=darkgreen,linewidth=1.5pt]{-}(0.2,0)(0.2,0.5)
\end{pspicture}\ , \ \ 
\begin{pspicture}[shift=-0.0](-0.0,0)(2.0,0.5)
\psline{-}(0,0)(2.0,0)
\psbezier[linecolor=darkgreen,linewidth=1.5pt]{-}(1.0,0)(1.0,0.7)(2.2,0.7)(2.2,0)
\psbezier[linecolor=darkgreen,linewidth=1.5pt]{-}(-0.2,0.54)(-0.12,0.43)(0.2,0.4)(0.2,0)
\psarc[linecolor=darkgreen,linewidth=1.5pt]{-}(1.6,0){0.2}{0}{180}
\psline[linecolor=darkgreen,linewidth=1.5pt]{-}(0.6,0)(0.6,0.5)
\psframe[fillstyle=solid,linecolor=white,linewidth=0pt](2.0,-0.1)(2.4,0.9)
\psframe[fillstyle=solid,linecolor=white,linewidth=0pt](0.0,-0.1)(-0.4,0.9)
\end{pspicture}\ , \ \ 
\begin{pspicture}[shift=-0.0](-0.0,0)(2.0,0.5)
\psline{-}(0,0)(2.0,0)
\psbezier[linecolor=darkgreen,linewidth=1.5pt]{-}(-0.12,0.53)(-0.07,0.545)(0.6,0.545)(0.6,0)
\psbezier[linecolor=darkgreen,linewidth=1.5pt]{-}(2.12,0.53)(2.07,0.545)(1.4,0.545)(1.4,0)
\psarc[linecolor=darkgreen,linewidth=1.5pt]{-}(0.0,0){0.2}{0}{90}
\psarc[linecolor=darkgreen,linewidth=1.5pt]{-}(2.0,0){0.2}{90}{180}
\psline[linecolor=darkgreen,linewidth=1.5pt]{-}(1.0,0)(1.0,0.5)
\psframe[fillstyle=solid,linecolor=white,linewidth=0pt](2.0,-0.1)(2.4,0.9)
\psframe[fillstyle=solid,linecolor=white,linewidth=0pt](0.0,-0.1)(-0.4,0.9)
\end{pspicture}\ , \ \ 
\begin{pspicture}[shift=-0.0](-0.0,0)(2.0,0.5)
\psline{-}(0,0)(2.0,0)
\psbezier[linecolor=darkgreen,linewidth=1.5pt]{-}(2.2,0.54)(2.12,0.43)(1.8,0.4)(1.8,0)
\psbezier[linecolor=darkgreen,linewidth=1.5pt]{-}(-0.2,0)(-0.2,0.7)(1.0,0.7)(1.0,0)
\psarc[linecolor=darkgreen,linewidth=1.5pt]{-}(0.4,0){0.2}{0}{180}
\psline[linecolor=darkgreen,linewidth=1.5pt]{-}(1.4,0)(1.4,0.5)
\psframe[fillstyle=solid,linecolor=white,linewidth=0pt](2.0,-0.1)(2.4,0.9)
\psframe[fillstyle=solid,linecolor=white,linewidth=0pt](0.0,-0.1)(-0.4,0.9)
\end{pspicture}\ .
\end{array}
\ee

The standard action of $\eptl_N(\alpha, \beta)$ on $\stanp_N^d$ is defect-preserving. For $d=0$, the resulting representations depend on $\beta$ and $\alpha$. For $d>0$, they depend on $\beta$ and a {\it twist parameter} $\omega$. To compute the action of a connectivity on a link state, one draws the link state above the connectivity and reads the new link state from the bottom nodes. If some defects have annihilated pairwise, the result is set to zero. Otherwise, a multiplicative factor of $\beta$ is inserted for each contractible loop. For $d=0$, a multiplicative factor of $\alpha$ is also inserted for each non-contractible loop. For $d>0$, a multiplicative factor of $\omega$ or $\omega^{-1}$ is inserted each time a defect crosses the virtual cut of the cylinder: $\omega$ if the defect travels to the left and $\omega^{-1}$ if it travels to the right. Here are examples of the standard action of $\eptl_N(\alpha, \beta)$:
\begin{subequations}
\begin{alignat}{2}
&
\begin{pspicture}[shift=-0.55](0,-0.65)(1.6,1.05)
\pspolygon[fillstyle=solid,fillcolor=lightlightblue,linewidth=0pt](0,-0.5)(1.6,-0.5)(1.6,0.5)(0,0.5)
\psarc[linecolor=blue,linewidth=1.5pt]{-}(1.2,-0.5){0.2}{0}{180}
\psarc[linecolor=blue,linewidth=1.5pt]{-}(0.4,0.5){0.2}{180}{0}
\psbezier[linecolor=blue,linewidth=1.5pt]{-}(0.2,-0.5)(0.2,0)(1.0,0)(1.0,0.5)
\psbezier[linecolor=blue,linewidth=1.5pt]{-}(0.6,-0.5)(0.6,0)(1.4,0)(1.4,0.5)
\psline{-}(0,0.5)(1.6,0.5)
\psarc[linecolor=darkgreen,linewidth=1.5pt]{-}(0,0.5){0.2}{0}{90}
\psbezier[linecolor=darkgreen,linewidth=1.5pt]{-}(0.6,0.5)(0.6,1.2)(1.8,1.2)(1.8,0.5)
\psarc[linecolor=darkgreen,linewidth=1.5pt]{-}(1.2,0.5){0.2}{0}{180}
\psframe[fillstyle=solid,linecolor=white,linewidth=0pt](-0.4,-0.5)(0,1.2)
\psframe[fillstyle=solid,linecolor=white,linewidth=0pt](1.6,-0.5)(2.0,1.2)
\end{pspicture} \ = \alpha \ 
\begin{pspicture}[shift=0.0](0,0.35)(1.6,0.95)
\psline{-}(0,0.35)(1.6,0.35)
\psarc[linecolor=darkgreen,linewidth=1.5pt]{-}(0.4,0.35){0.2}{0}{180}
\psarc[linecolor=darkgreen,linewidth=1.5pt]{-}(1.2,0.35){0.2}{0}{180}
\end{pspicture}\, ,
\qquad
&&
\begin{pspicture}[shift=-0.55](0,-0.65)(1.6,1.05)
\pspolygon[fillstyle=solid,fillcolor=lightlightblue,linewidth=0pt](0,-0.5)(1.6,-0.5)(1.6,0.5)(0,0.5)
\psarc[linecolor=blue,linewidth=1.5pt]{-}(0.8,0.5){0.2}{180}{0}
\psarc[linecolor=blue,linewidth=1.5pt]{-}(1.6,0.5){0.2}{180}{270}
\psarc[linecolor=blue,linewidth=1.5pt]{-}(0,0.5){0.2}{270}{0}
\psarc[linecolor=blue,linewidth=1.5pt]{-}(0.8,-0.5){0.2}{0}{180}
\psbezier[linecolor=blue,linewidth=1.5pt]{-}(0.2,-0.5)(0.2,0.2)(1.4,0.2)(1.4,-0.5)
\psframe[fillstyle=solid,linecolor=white,linewidth=0pt](-0.4,-0.5)(0,0.5)
\psframe[fillstyle=solid,linecolor=white,linewidth=0pt](1.6,-0.5)(2.0,0.5)
\psline{-}(0,0.5)(1.6,0.5)
\psline[linecolor=darkgreen,linewidth=1.5pt]{-}(0.2,0.5)(0.2,1.0)
\psline[linecolor=darkgreen,linewidth=1.5pt]{-}(0.6,0.5)(0.6,1.0)
\psarc[linecolor=darkgreen,linewidth=1.5pt]{-}(1.2,0.5){0.2}{0}{180}
\end{pspicture} \ = \ 0 \, ,
\\
&\begin{pspicture}[shift=-0.55](0,-0.65)(1.6,1.05)
\pspolygon[fillstyle=solid,fillcolor=lightlightblue,linewidth=0pt](0,-0.5)(1.6,-0.5)(1.6,0.5)(0,0.5)
\psarc[linecolor=blue,linewidth=1.5pt]{-}(0.8,-0.5){0.2}{0}{180}
\psarc[linecolor=blue,linewidth=1.5pt]{-}(0,0.5){0.2}{270}{0}
\psarc[linecolor=blue,linewidth=1.5pt]{-}(1.6,0.5){0.2}{180}{270}
\psbezier[linecolor=blue,linewidth=1.5pt]{-}(0.2,-0.5)(0.2,0)(1.0,0)(1.0,0.5)
\psbezier[linecolor=blue,linewidth=1.5pt]{-}(-0.2,-0.5)(-0.2,0)(0.6,0)(0.6,0.5)
\psbezier[linecolor=blue,linewidth=1.5pt]{-}(1.4,-0.5)(1.4,0)(2.2,0)(2.2,0.5)
\psline{-}(0,0.5)(1.6,0.5)
\psline[linecolor=darkgreen,linewidth=1.5pt]{-}(0.6,0.5)(0.6,1)
\psline[linecolor=darkgreen,linewidth=1.5pt]{-}(1.0,0.5)(1.0,1)
\psarc[linecolor=darkgreen,linewidth=1.5pt]{-}(0,0.5){0.2}{0}{90}
\psarc[linecolor=darkgreen,linewidth=1.5pt]{-}(1.6,0.5){0.2}{90}{180}
\psframe[fillstyle=solid,linecolor=white,linewidth=0pt](-0.4,-0.5)(0,1.2)
\psframe[fillstyle=solid,linecolor=white,linewidth=0pt](1.6,-0.5)(2.25,1.2)
\end{pspicture}  \ = \omega \beta \ 
\begin{pspicture}[shift=0.0](0,0.35)(1.6,0.95)
\psline{-}(0,0.35)(1.6,0.35)
\psarc[linecolor=darkgreen,linewidth=1.5pt]{-}(0.8,0.35){0.2}{0}{180}
\psline[linecolor=darkgreen,linewidth=1.5pt]{-}(0.2,0.35)(0.2,0.85)
\psline[linecolor=darkgreen,linewidth=1.5pt]{-}(1.4,0.35)(1.4,0.85)
\end{pspicture}\, , \qquad
&&\begin{pspicture}[shift=-0.55](0,-0.65)(1.6,1.05)
\pspolygon[fillstyle=solid,fillcolor=lightlightblue,linewidth=0pt](0,-0.5)(1.6,-0.5)(1.6,0.5)(0,0.5)
\psbezier[linecolor=blue,linewidth=1.5pt]{-}(0.6,-0.5)(0.6,0)(-0.2,0)(-0.2,0.5)
\psbezier[linecolor=blue,linewidth=1.5pt]{-}(1.0,-0.5)(1.0,0)(0.2,0)(0.2,0.5)
\psbezier[linecolor=blue,linewidth=1.5pt]{-}(1.4,-0.5)(1.4,0)(0.6,0)(0.6,0.5)
\psbezier[linecolor=blue,linewidth=1.5pt]{-}(1.8,-0.5)(1.8,0)(1,0)(1,0.5)
\psbezier[linecolor=blue,linewidth=1.5pt]{-}(2.2,-0.5)(2.2,0)(1.4,0)(1.4,0.5)
\psbezier[linecolor=blue,linewidth=1.5pt]{-}(0.2,-0.5)(0.2,0)(-0.6,0)(-0.6,0.5)
\psline{-}(0,0.5)(1.6,0.5)
\psline[linecolor=darkgreen,linewidth=1.5pt]{-}(1.4,0.5)(1.4,1)
\psline[linecolor=darkgreen,linewidth=1.5pt]{-}(1.0,0.5)(1.0,1)
\psarc[linecolor=darkgreen,linewidth=1.5pt]{-}(0.4,0.5){0.2}{0}{180}
\psframe[fillstyle=solid,linecolor=white,linewidth=0pt](-0.65,-0.5)(0,1.2)
\psframe[fillstyle=solid,linecolor=white,linewidth=0pt](1.6,-0.5)(2.25,1.2)
\end{pspicture}  \ = \omega^{-2} \ 
\begin{pspicture}[shift=0.0](0,0.35)(1.6,0.95)
\psline{-}(0,0.35)(1.6,0.35)
\psarc[linecolor=darkgreen,linewidth=1.5pt]{-}(1.2,0.35){0.2}{0}{180}
\psline[linecolor=darkgreen,linewidth=1.5pt]{-}(0.6,0.35)(0.6,0.85)
\psline[linecolor=darkgreen,linewidth=1.5pt]{-}(0.2,0.35)(0.2,0.85)
\end{pspicture}\, .
\end{alignat}
\end{subequations}
For generic values of $\beta$, $\alpha$ and $\omega$, the modules $\stanp_N^d$ are irreducible modules over $\eptl_N(\alpha,\beta)$.
For percolation, the indecomposable structures of $\stanp_N^0$ with $\alpha=1$ and $\stanp_N^{d>0}$ with $\omega^N=1$ are discussed in \cite{GRSV15}.

\subsection{Fused transfer matrices and the fusion hierarchy}\label{sec:Pfusion}
Using the Wenzl-Jones projectors for generic values of $\beta$, one can define a sequence of commuting fused transfer tangles $\Tb^n(u)$, starting with $\Tb^1(u) = \Tb(u)$, which satisfy the fusion hierarchy relations
\be 
\Tb^{n}_0\Tb^{1}_n =  h_n \Tb^{n-1}_0 +  h_{n-1} \Tb^{n+1}_0, \qquad n \ge 0,
\label{eq:fushierT}
\ee
where 
\be
\Tb_k^n = \Tb^n(u + k \lambda), \qquad \Tb_0^0 = h_{-1} \Ib, \qquad \Tb^{-1}_k = 0, \qquad h_k = s_{k}(u)^{N}.
\ee
These tangles indeed commute as elements of $\eptl_N(\alpha,\beta)$: $[\Tb^m(u),\Tb^n(v)] = 0$.

In particular, for $\beta\ne 0$, the fused transfer tangle $\Tb^2(u)$ is constructed from the $1\times2$ fused face operator, see \eqref{eq:1x2face}, as follows:
\be
\Tb^2 (u)= (-1)^N\ 
\psset{unit=0.9}
\begin{pspicture}[shift=-1.1](-0.2,-0.7)(5.2,1.0)
\facegrid{(0,0)}{(5,1)}
\psarc[linewidth=0.025]{-}(0,0){0.16}{0}{90}
\psarc[linewidth=0.025]{-}(1,0){0.16}{0}{90}
\psarc[linewidth=0.025]{-}(4,0){0.16}{0}{90}
\psline[linewidth=1.5pt,linecolor=blue]{-}(0,0.5)(-0.2,0.5)
\psline[linewidth=1.5pt,linecolor=blue]{-}(5,0.5)(5.2,0.5)
\rput(2.5,0.5){$\ldots$}
\rput(3.5,0.5){$\ldots$}
\rput(0.5,.5){$u$}\rput(0.5,0.75){\tiny{$_{1\times 2}$}}
\rput(1.5,.5){$u$}\rput(1.5,0.75){\tiny{$_{1\times 2}$}}
\rput(4.5,.5){$u$}\rput(4.5,0.75){\tiny{$_{1\times 2}$}}
\rput(2.5,-0.5){$\underbrace{\qquad \qquad \qquad \qquad  \qquad \qquad}_N$}
\end{pspicture} \ .
\ee  

For the rational values $\lambda = \frac{\pi(p'-p)}{p'}$, see \eqref{eq:ppp}, the hierarchy of fused transfer matrices closes \cite{MDPR14} at $n=p'$:
\be
\Tb^{p'}_0 = \Tb^{p'-2}_1 + 2(-\ir)^{N(p'-p)} h_{-1} \Jb, \qquad  \Jb = T_{p'}(\tfrac12 \Fbtp),
\ee
where $T_k(x)$ in the definition of $\Jb$ is the $k$-th Chebyshev polynomial of the first kind. For percolation, this becomes
\be
\Tb^{3}_0 = \Tb^{1}_1 + 2(-\ir)^{N} h_{-1} \Jb, \qquad  \Jb = T_3(\tfrac12 \Fbtp) = \tfrac12 (\Fbtp)^3-\tfrac32\Fbtp.
\ee
We note that the closure relation could equivalently be written in terms of the other braid transfer matrix $\Tb_{\!-\infty}$. 

\subsection[$T$-systems and $Y$-systems]{The $\boldsymbol T$-system and the $\boldsymbol Y$-system}\label{sec:Pfunctional}

The $T$-system for the periodic transfer matrices is found using the fusion hierarchy relations \eqref{eq:fushierT} and a recursive argument. It takes the form
\be
\Tb^{n}_0 \Tb^{n}_1 = h_{-1} h_n  \Ib +\Tb^{n+1}_0 \Tb^{n-1}_1, \qquad n\ge 0.
\ee
By defining 
\be
\tb^n(u) =\frac{\Tb_1^{n-1}\Tb_0^{n+1}}{h_{-1}h_n}, \qquad n \ge 0,
\ee
and $\tb^n_k = \tb^n(u+k \lambda)$, we find that the tangles $\tb^n(u)$ satisfy precisely the same $Y$-system as on the strip, namely:
\be
\label{eq:ysysP}
\tb^{n}_0 \tb^{n}_1=\big(\Ib+ \tb^{n-1}_1\big)\big(\Ib+ \tb^{n+1}_0\big), \qquad n \ge 1.
\ee
For rational values of $\lambda$, this $Y$-system can be written in terms of finitely many objects. Indeed, by defining $\Kbt(u)$ and $\Lambdab$ as follows:
\be
\Kbt(u) = \frac{\ir^{N(p'-p)}}{h_{-1}}\Tb_1^{p'-2}, \qquad \Jb =  \cos \Lambdab,
\label{eq:KtangT}
\ee
we find the closure relations:
\be
\Ib+ \tb^{p'-1}_0 = (\Ib+ \eE^{\ir \boldsymbol{\Lambda}}\Kbt_0)(\Ib+ \eE^{-\ir \boldsymbol{\Lambda}}\Kbt_0), \qquad \Kbt_0\Kbt_1 = 1+\tb_1^{p'-2}.
\label{eq:Pyclo}
\ee
The finite $Y$-system consists of the relations \eqref{eq:ysysP} for $n = 1, \dots, p'-2$ and \eqref{eq:Pyclo}. It is described by the Dynkin diagram $D_{p'-2}$ in \cref{fig:Dynkin}, with the new feature that the endpoint nodes to the right are distinguished by the factors $ \eE^{\pm\ir \boldsymbol{\Lambda}}$.

For percolation, the $Y$-system involves only the tangles $\tb^1(u)$, $\Kb(u)$, $\Ib$ and $\Lambdab$ and is written as
\be
\label{eq:ysysT}
\tb_0^1\tb_1^1 =  (\Ib+ \eE^{\ir \boldsymbol{\Lambda}}\Kbt_0)(\Ib+ \eE^{-\ir \boldsymbol{\Lambda}}\Kbt_0), \qquad \Kb_0\Kb_1 = \Ib + \tb_1^1.
\ee
By defining
\begin{subequations}
\label{eq:aUT}
\begin{alignat}{3}
&\ab^1(x) = \tb_0^1(\tfrac {\ir x} 3), \qquad  &&\ab^2(x) = (-1)^N\Kb_0(\tfrac {\ir x} 3- \tfrac \pi 6), \label{eq:aUT1}\\ 
&\Ab^1(x) = \Ib+\ab^1(x), \qquad  &&\Ab^2(x) = \big(\Ib+(-1)^N\eE^{\ir \boldsymbol{\Lambda}}\ab^2(x)\big)\big(\Ib+(-1)^N\eE^{-\ir \boldsymbol{\Lambda}}\ab^2(x)\big),
\end{alignat}
\end{subequations}
the $Y$-system takes a symmetric form:
\begin{equation}
\ab^1(x-\tfrac{\ir \pi}2)\ab^1(x+\tfrac{\ir \pi}2) = \Ab^2(x),\qquad
\ab^2(x-\tfrac{\ir \pi}2)\ab^2(x+\tfrac{\ir \pi}2) = \Ab^1(x).
\label{eq:finalyT}
\end{equation}

\subsection{Properties of the eigenvalues}\label{sec:Pprops}

\paragraph{$\boldsymbol{Y}$-system for the eigenvalues.} The tangles $\Tb_{\!\pm\infty}$ are not
proportional to the identity tangle, but nevertheless act as multiples of
the identity on the standard modules. For $\alpha=2$ and $\omega =1$, the (unique) eigenvalues of $\Tb_{\!+\infty}$ and $\Tb_{\!-\infty}$ on $\stanp_N^d$ are identical and were computed in \cite{MDSA13no2} for generic values of $\lambda$.
The generalisation to arbitrary $\alpha$ and $\omega$ is straightforward. For $\lambda=\frac \pi 3$, we obtain
\be
\label{eq:Jeigs}
T_{\pm\infty} = 2 \cos(\pm\gamma - \tfrac{\pi d}{3}),\qquad J = (-1)^d \cos (3 \gamma), \qquad \eE^{\ir \Lambda} = (-1)^d \eE^{3 \ir \gamma},
\ee
where we parameterise $\alpha$ and $\omega$ in terms of a single parameter $\gamma$ as follows:
\be
\alpha = 2 \cos \gamma, \qquad \omega = \eE^{\ir \gamma}.
\ee 
The $Y$-system satisfied by the eigenvalues in $\stanp_N^d$ is then given by \eqref{eq:finalyT} with 
\be
\mathfrak A^2(x) = \big(1+\eE^{3 \ir \gamma} \mathfrak a^2(x)\big)\big(1+\eE^{-3 \ir \gamma} \mathfrak a^2(x)\big).
\ee
The cancellation of the factors of $(-1)^d$ and $(-1)^N$ here explains the choice made in \eqref{eq:aUT1} to include an extra $(-1)^N$ in the definition of $\ab^2(x)$.

\paragraph{Razumov-Stroganov eigenvalues.}
For periodic boundary conditions, the Razumov-Stroganov eigenstate appears in two situations: (i) for $d=0$ with $\gamma = \pm \frac\pi3,\pm \frac{2\pi}3$, and (ii) for $d=1$ with $\gamma = 0$, $\pi$. This is consistent with the spin-chain results of \cite{RS01,RS01no2,S01}. Indeed, a simple solution to the $Y$-system \eqref{eq:ysysT} in these cases is $K_0 = -1$ and $t_0^1 = 0$, which corresponds to 
\be 
T^1(u) = -\frac{h_{-2}}{\ir^N} = -\ir^{N} \bigg(\frac{\sin (u+\frac\pi3)}{\sin \frac\pi3}\bigg)^{N}, \qquad T^2(u) = 0.
\label{eq:RSeigT}
\ee
In both cases (i) and (ii), the Razumov-Stroganov eigenvalue is non-degenerate and acts as the ground state of the corresponding standard module.

\begin{figure}[h] 
\centering
\begin{tabular}{cc}
$T^1(u):$ & $T^2(u):$\\[0.1cm]
\includegraphics[width=.45\textwidth]{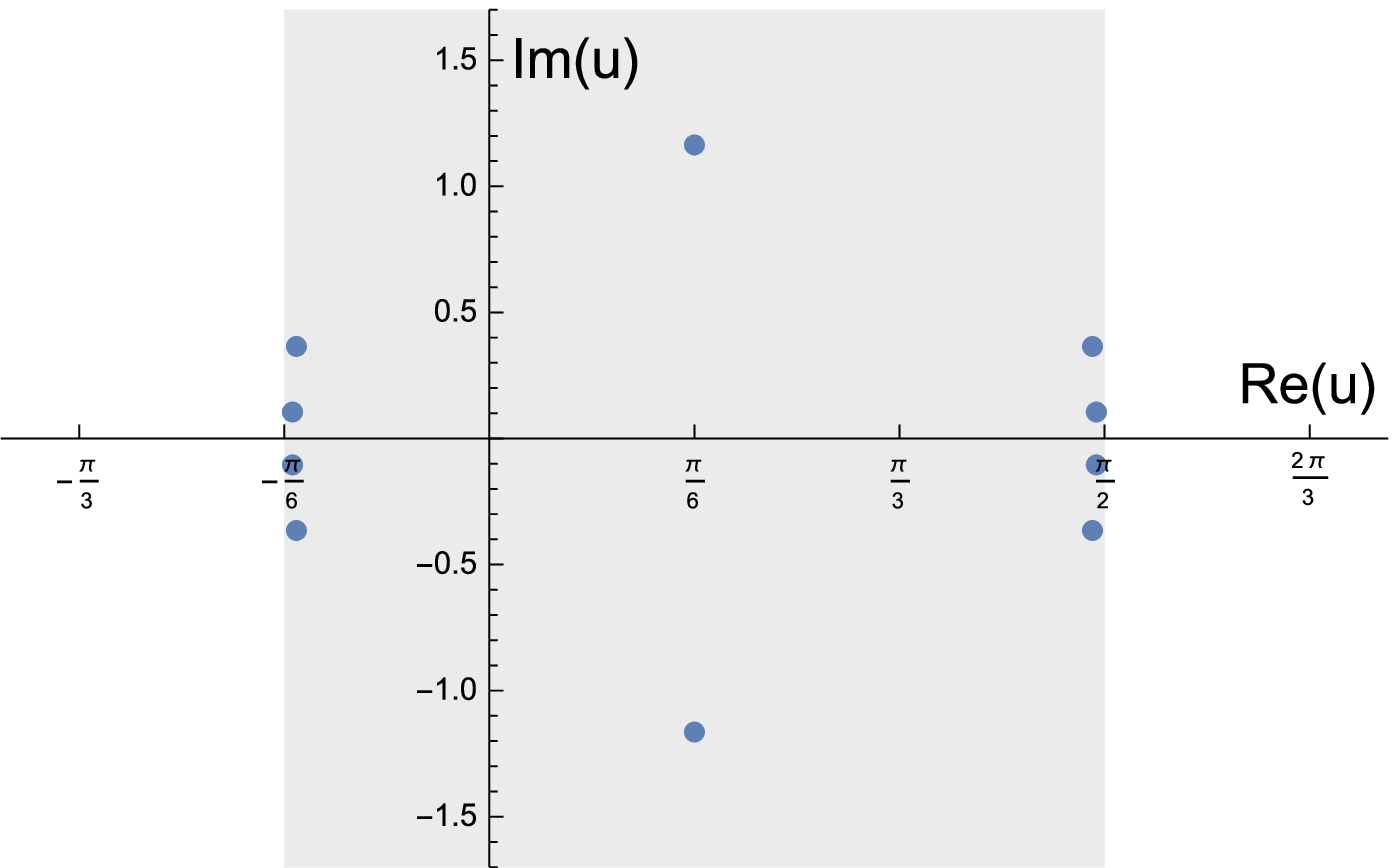}\qquad \qquad&
\includegraphics[width=.45\textwidth]{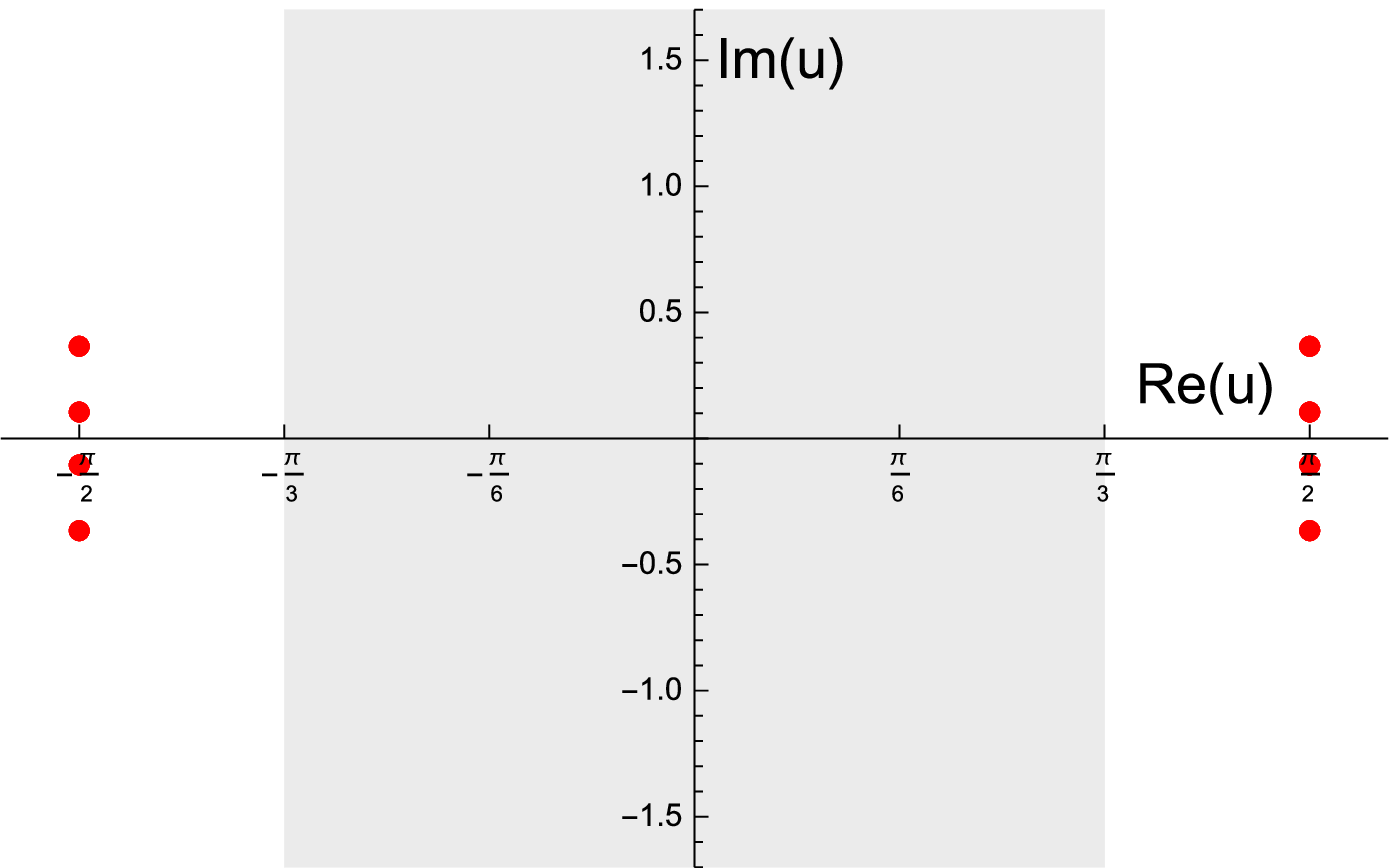} 
\end{tabular}
\caption{The location of the zeros for the ground state in $\stanp_{10}^2$ for $\omega = 1$. The analyticity strips are shaded in gray. There is one pair of complex zeros in the analyticity strip for $T^1(u)$ and none for $T^2(u)$.}
\label{fig:eigpatternsp}
\end{figure}

\begin{figure}[h] 
\centering
\begin{tabular}{cc}
$T^1(u):$ & $T^2(u):$\\[0.1cm]
\includegraphics[width=.45\textwidth]{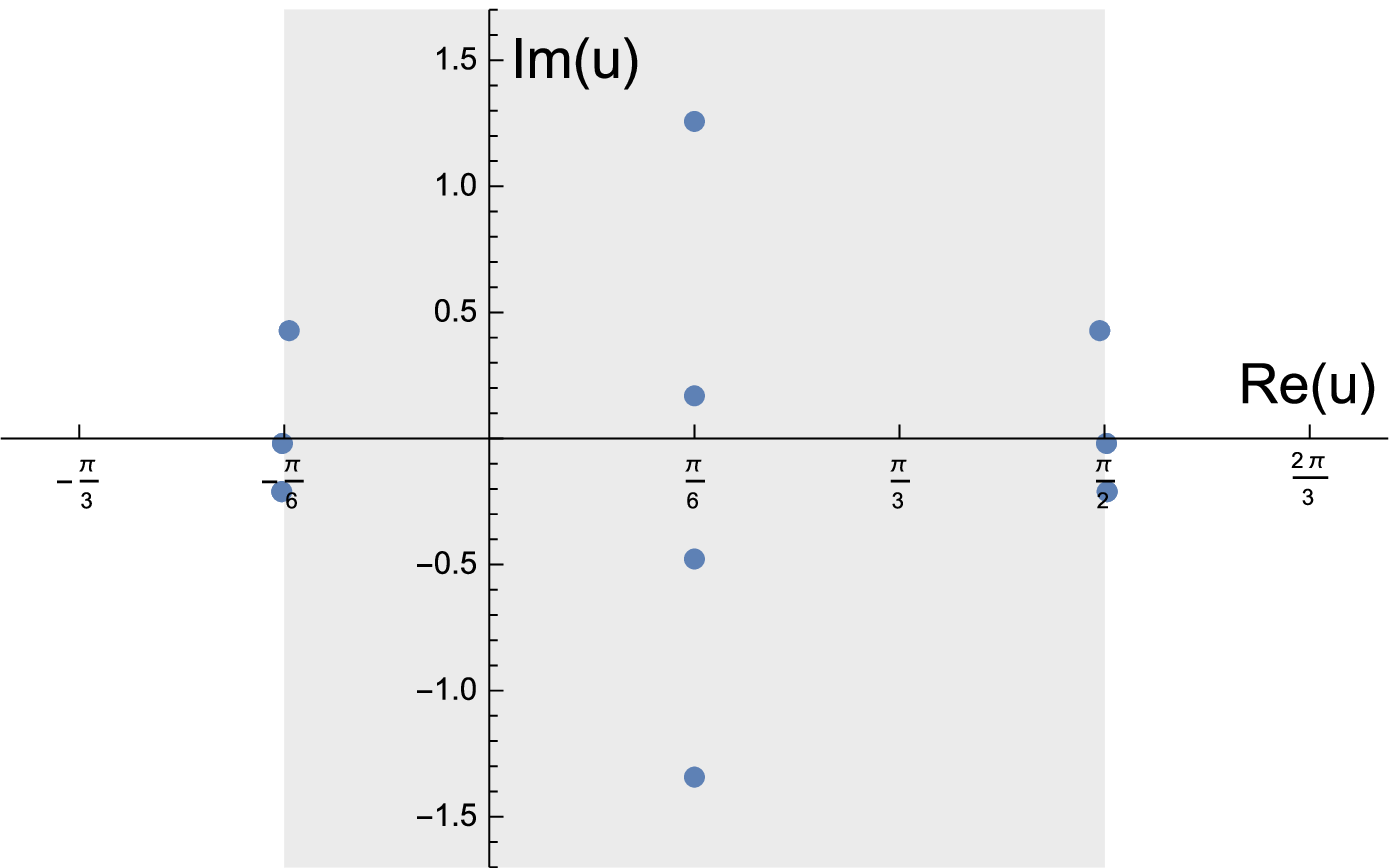}\qquad \qquad&
\includegraphics[width=.45\textwidth]{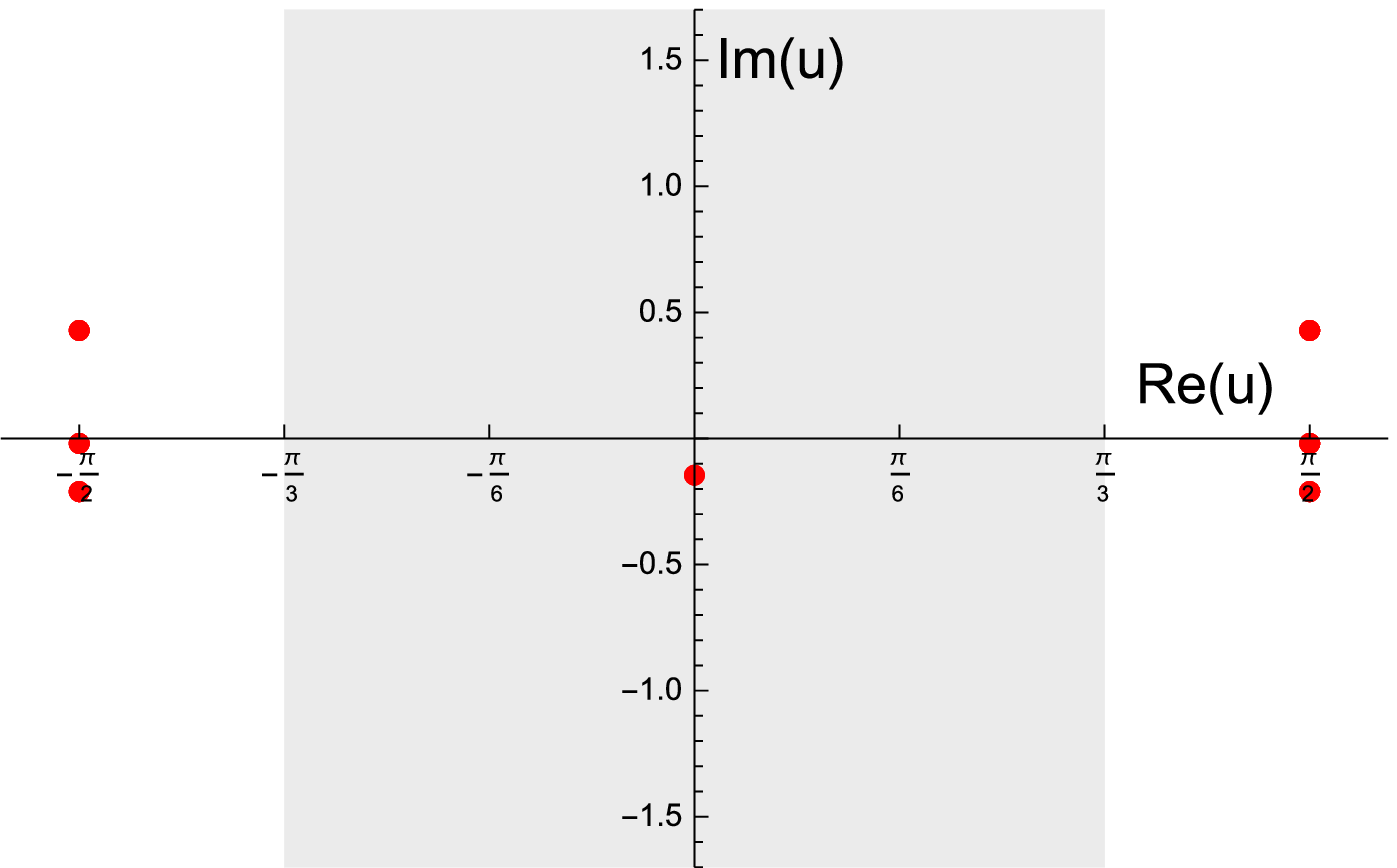} 
\end{tabular}
\caption{The location of the zeros of an excited state in $\stanp_{10}^2$ for $\omega = 1$. There are four complex zeros in the analyticity strip for $T^1(u)$ and they are not related by complex conjugation. For $T^2(u)$, there is a single double-zero on the central line of the analyticity strip.}
\label{fig:eigpatternsp2}
\end{figure}

\paragraph{Patterns of zeros and analyticity strips.}
The eigenvalues $T^1(u)$ and $T^2(u)$ are centered Laurent polynomials in the variable $z=\eE^{\ir u}$ with minimal and maximal powers $-N$ and $N$. Due to the periodicity properties of $\Tb^n(u)$, the patterns of zeros are periodic with period $\pi$ in the complex $u$ plane and each vertical strip of width $\pi$ has at most $N$ zeros. Examples of patterns of zeros are given in \cref{fig:eigpatternsp,fig:eigpatternsp2}. Empirically, we find that for the leading eigenvalues, only a finite number of these zeros are in the following analyticity strips:
\be 
T^1(u): \quad -\frac \pi 6< \text{Re}(u) < \frac {\pi} 2, \qquad T^2(u): \quad -\frac \pi 3< \text{Re}(u) < \frac {\pi} 3.
\ee

As opposed to the strip boundary conditions, none of the zeros are common to all the eigenvalues. This is due to the absence of surface terms in the fusion hierarchy relations \eqref{eq:fushierT}. In general, the zeros have three possible locations: (i) in the center of the analyticity strip, (ii) on one of the two edges of the analyticity strip, or (iii) outside the analyticity strip. We will be interested in zeros of type (i) and (ii). The zeros that lie in the center of the analyticity strip of $T^1(u)$ are always of order one. For $\gamma = 0$, the zeros in the center of the analyticity strip of $T^2(u)$ can be of order one or two. This degeneracy is lifted away from $\gamma = 0$, but may occur again for special values of $\gamma$. For generic values of $\gamma$, all zeros of $T^2(u)$ are of order one.

We observe that all the patterns of zeros are symmetric under a reflection with respect to the central line of the analyticity strips. This implies that the corresponding eigenvalues are real on this central line. In contrast, some patterns of zeros, for example the one in \cref{fig:eigpatternsp2}, are not symmetric with respect to a reflection about the real $u$ axis. In general, these eigenvalues are therefore complex for Im$(u)=0$. 

We shall see in \cref{sec:Pfsc} that, for the finite-size corrections, the zeros in the upper and lower parts of the $u$ plane are tied to contributions to $\Delta$ and $\bar\Delta$ respectively. For small values of $N$, we find that some zeros are very close to the real axis, and even coincide with this axis in certain instances. For the periodic boundary conditions, the line dividing the plane into two chiral halves need not necessarily coincide with the real axis. In fact, it can be chosen arbitrarily, as long as the convention does not impact the finite-size corrections for the leading eigenvalues. This is discussed further in \cref{sec:fsgf}. 

We encode the content of zeros in both the upper- and lower-half planes together in a pattern diagram. 
For the eigenvalues corresponding to \cref{fig:eigpatternsp,fig:eigpatternsp2}, the pattern diagrams are
\be
\psun
\begin{pspicture}[shift=-1.2](-0.4,-1.4)(1.4,1.4)
\psline{-}(0,-1.4)(0,1.4)\psline{-}(1,-1.4)(1,1.4)\psline[linestyle=dashed,dash=1pt 1pt]{-}(0,0)(1,0)
\psdots[dotsize=0.09cm](0,-0.3)(0,-0.6)(0,0.3)(0,0.6)
\psdots[dotsize=0.09cm](0.5,-0.9)(0.5,0.9)
\psdots[dotsize=0.09cm](1,-0.3)(1,-0.6)(1,0.3)(1,0.6)
\end{pspicture}
\begin{pspicture}[shift=-1.2](-0.4,-1.4)(1.4,1.4)
\psline{-}(0,-1.4)(0,1.4)\psline{-}(1,-1.4)(1,1.4)\psline[linestyle=dashed,dash=1pt 1pt]{-}(0,0)(1,0)
\end{pspicture}
\qquad \textrm{and} \qquad
\begin{pspicture}[shift=-1.2](-0.4,-1.4)(1.4,1.4)
\psline{-}(0,-1.4)(0,1.4)\psline{-}(1,-1.4)(1,1.4)\psline[linestyle=dashed,dash=1pt 1pt]{-}(0,0)(1,0)
\psdots[dotsize=0.09cm](0,0)(0,-0.3)(0,0.6)
\psdots[dotsize=0.09cm](0.5,-0.6)(0.5,-0.9)(0.5,0.3)(0.5,0.9)
\psdots[dotsize=0.09cm](1,0)(1,-0.3)(1,0.6)
\end{pspicture}
\begin{pspicture}[shift=-1.2](-0.4,-1.4)(1.4,1.4)
\psline{-}(0,-1.4)(0,1.4)\psline{-}(1,-1.4)(1,1.4)\psline[linestyle=dashed,dash=1pt 1pt]{-}(0,0)(1,0)
\psdots[linecolor=black,fillcolor=lightgray,dotstyle=o,dotsize=0.09cm](0.5,0)
\end{pspicture}\ .
\ee
We work with the following convention. For a given eigenvalue, if the first analyticity strip has a site, occupied or unoccupied, lying on the dashed separation line, it is considered to belong to the upper half-plane. The same convention was used in \cite{PRVcyl2010} for critical dense polymers. The pattern of zeros in the second strip is described below as a pair of single-column configurations. If the second analyticity strip has a site lying on the dashed separation line, then the corresponding site in the first column is considered to be in the upper-half of the plane and the site in the second column is considered to be in the lower-half. We will come back to this in \cref{sec:fsgf} and \cref{sec:Pcpf}.

The analyticity properties of $K(u)$ and $t^1(u)$ are deduced from those of $T^1(u)$ and $T^2(u)$. From \eqref{eq:KtangT}, we see that $K(u)$ has a pole of order $N$ at $u = \frac \pi3$. Likewise, $t^1(u)$ has poles of order $N$ at $u = -\frac\pi3, \frac\pi 3$ and a zero of order $N$ at $u=0$. The analyticity strips have width $2\lambda$:
\be
K(u): \quad -\frac \pi 2< \text{Re}(u) < \frac {\pi} 6, \qquad
t^1(u): \quad -\frac \pi 3< \text{Re}(u) < \frac {\pi} 3.
\ee

A similar analysis on small system sizes reveals that the functions $1+\eE^{\pm \ir \Lambda}K(u)$ and $1+t^1(u)$ are analytic and non-zero in the following strips:
\be
1+\eE^{\pm \ir \Lambda}K(u): \quad -\frac \pi 3< \text{Re}(u) < 0, \qquad 1+t^1(u): \quad -\frac \pi 6< \text{Re}(u) < \frac {\pi} 6.
\ee
We observe that these analyticity strips have width $\lambda$ and are entirely free of zeros and poles.

Finally, in terms of the variables defined in \eqref{eq:aU}, the analyticity strips take the following elegant forms:
\begin{subequations}
\begin{alignat}{4}
&\mathfrak a^1(x): &&\quad - \pi < \text{Im}(x) < \pi , \qquad
&&\mathfrak a^2(x): &&\quad -\pi < \text{Im}(x) < \pi, \\[0.15cm]
&\mathfrak A^1(x): &&\quad -\frac \pi 2< \text{Im}(x) < \frac \pi 2, \qquad
&&\mathfrak A^2(x): &&\quad -\frac \pi 2< \text{Im}(x) < \frac {\pi} 2.
\end{alignat}
\end{subequations}

\paragraph{Braid limit.} The braid transfer tangles $\Tb^n_{\!\pm\infty}$ are obtained as the $u\rightarrow \pm\ir \infty$ limit of the transfer tangle $\Tb^n(u)$. For $n=1$, $\Tb^1_{\!\pm\infty} \equiv \Tb_{\!\pm\infty}$ are defined in \eqref{eq:braidT}. For $n=2$, we have
\begin{subequations}
\begin{alignat}{2}
\Tb^2_{\!+\infty} &= \lim_{u \rightarrow \ir \infty}  \bigg(\frac{-\eE^{\ir(\pi-\lambda)}}{s_{1}(u)}\bigg)^N \Tb^2(u) = (\Fbtp)^2 - \Ib, \\ 
\Tb^2_{\!-\infty} &= \lim_{u \rightarrow - \ir \infty}  \bigg(\frac{\eE^{\ir(\pi-\lambda)}}{s_{1}(-u)}\bigg)^N \Tb^2(u) = (\Fbtm)^2 - \Ib. 
\end{alignat}
\end{subequations}
The last equalities are obtained by taking the proper $u \rightarrow \pm\ir \infty$ limits of \eqref{eq:fushierT} for $n=1$.
The braid limits of $\Kb_{\!\pm\infty}$ and $\tb^1_{\pm\infty}$ are likewise given by
\be
\Kb_{\!\pm\infty} = (-1)^N\ab^2_{\pm \infty}= \lim_{u \rightarrow \pm\ir \infty} \Kb(u) = \Tb^1_{\!\pm\infty}, \qquad \tb^1_{\pm\infty} = \ab^1_{\pm \infty}= \lim_{u \rightarrow \pm 
\ir \infty}  \tb^1(u) = \Tb^2_{\!\pm\infty}= (\Tb^1_{\!\pm\infty})^2-\Ib.
\ee

The matrix realisations on $\stanp_N^d$ of all these braid tangles are scalar multiples of the identity matrix. For $\beta = 1$, the result for $T^1_{\!\pm\infty}$ is given in \eqref{eq:Jeigs} and yields 
\be
\label{eq:braida1a2}
\mathfrak a^1_{\pm\infty} = 4 \cos^2(\pm\gamma+\tfrac{2\pi d}3)-1, \qquad \mathfrak a^2_{\pm\infty} = 2 \cos(\pm\gamma+\tfrac{2\pi d}3).
\ee
The eigenvalues thus have different braid behaviors in the lower- and upper-half planes for generic $\gamma$.

For $\gamma = 0$ and $d \equiv 0,1 \Mod 3$, we have $\mathfrak a^1_{\pm\infty} = 0$ and $\mathfrak a^2_{\pm \infty} = -1$. As for the strip case, the convergence of $1+\mathfrak a^2(x) $ to zero can follow different subcases. In the periodic case however, the convergence can be different for $x \rightarrow \infty$ and $x \rightarrow -\infty$. From our numerical data, we find the following four subcases:
\begin{alignat}{3}
\label{eq:subcasesAABB}
&\text{Subcase } \bar \tA:  \quad\lim_{x \rightarrow -\infty} \eE^{-\frac{2x}3} \big(1+ \mathfrak a^2(x)\big) = \kappa_-, \qquad  &&\text{Subcase } \bar \tB: \quad \lim_{x \rightarrow -\infty}  \eE^{-\frac{4x}3} \big(1+ \mathfrak a^2(x)\big) = \kappa'_-,\\[0.2cm]
&\text{Subcase A}: \quad\lim_{x \rightarrow +\infty} \eE^{\frac{2x}3} \big(1+ \mathfrak a^2(x)\big) = \kappa_+, \qquad  &&\text{Subcase B}: \quad \lim_{x \rightarrow +\infty}  \eE^{\frac{4x}3} \big(1+ \mathfrak a^2(x)\big) = \kappa'_+.
\end{alignat}
The constants $\kappa_\pm$ and $\kappa'_\pm$ are real and strictly 
positive, for all eigenvalues except the Razumov-Stroganov eigenvalues. The argument leading to this conclusion is identical to the one given below \eqref{eq:subcases}. From \eqref{eq:finalyT} and \eqref{eq:subcasesAABB}, the rate of convergence of  $\mathfrak a^1(x)$ to zero as $x \rightarrow \pm \infty$ is
\begin{alignat}{2}
&\text{Subcase } \bar \tA: \quad \lim_{x \rightarrow -\infty}  \eE^{-\frac{2x}3} \mathfrak a^1(x) = -\kappa_-, \qquad  &&\text{Subcase } \bar \tB: \quad \lim_{x \rightarrow -\infty}  \eE^{\frac{-4x}3} \mathfrak a^1(x) = \kappa'_-,\\
&\text{Subcase } \tA: \quad \lim_{x \rightarrow +\infty}  \eE^{\frac{2x}3} \mathfrak a^1(x) = -\kappa_+, \qquad  &&\text{Subcase } \tB: \quad \lim_{x \rightarrow +\infty}  \eE^{\frac{4x}3} \mathfrak a^1(x) = \kappa'_+.
\end{alignat}
An eigenstate is thus described by one of these four pairs: $(\bar\tA,\tA)$, $(\bar\tA,\tB)$, $(\bar\tB,\tA)$ or $(\bar\tB,\tB)$. In the pattern diagrams, this is indicated by small letters at the bottom and top of the analyticity strips, as in \cref{fig:GSpatternsp,fig:patsP62}.

\paragraph{Bulk behavior.} The eigenvalues of $\tb^1(u)$ and $\Kb(u)$ have the following bulk behaviors:
\be
t^1_{\text{bulk}}(u) = 0, \qquad K_{\text{bulk}}(u) = \sigma, \qquad \sigma \in \{-1,1\}.
\label{eq:bulkbP}
\ee
Indeed, the behavior of $t^1(u)$ near the origin is governed by the zero of order $N$ at $u = 0$, so the function is approximately zero in this neighborhood. From this remark, it follows that \eqref{eq:ysysperco1} is simply $K_{\text{bulk}}(u)^2 = 1$ in this neighborhood. 

For a given pattern of zeros, 
one can deduce the values of $\sigma$ from the braid value $K_{+\infty}$, and the number of zeros on the central line of the analyticity strip. For instance, let us consider the ground state of $\stanp_{10}^2$ for $\omega =1$, whose pattern is displayed in \cref{fig:eigpatternsp}. The braid behavior is $K_{+\infty} = -1$. In the upper half-plane, there is a single zero on the central line of the analyticity strip, so $K(-\frac\pi6)$ is positive. We deduce that $\sigma=1$. In contrast, applying the same reasoning to the eigenvalue corresponding to \cref{fig:eigpatternsp}, we find $\sigma=-1$. For periodic boundary conditions, both values of $\sigma$ are possible.

\subsection{Finite-size corrections}\label{sec:Pfsc}

The eigenvalues $T(u)$ can be factored as
\be
\label{eq:Tseparation}
T(u) = T_{\text{b}}(u) T_{\text{f}}(u).
\ee
Compared to \eqref{eq:Dseparation}, the surface term is absent because of the periodic boundary conditions. The bulk and finite contributions satisfy
\be
T_{\text{b}}(u)T_{\text{b}}(u+\lambda) = h_{-1}h_1,\qquad T_{\text{f}}(u)T_{\text{f}}(u+\lambda) = 1+t^1(u). \label{eq:inv.identities.P}
\ee
The solution to the first relation in \eqref{eq:inv.identities.P} gives a bulk free energy identical to the one for the strip. Below, we compute the finite-size correction to $T_{\text{f}}(u)$ for any eigenvalue in $\stanp_N^d$. 
We use the functions $\mathfrak a^1(x)$, $\mathfrak a^2(x)$, $\mathfrak A^1(x)$ and $\mathfrak A^2(x)$ for the computation, as well as
\be
\mathfrak b(x) = T_{\textrm{f}}(\tfrac {\ir x} 3+ \tfrac \pi 6).
\ee
The relation for $T_{\text{f}}(u)$ in \eqref{eq:inv.identities.P} then becomes
\be
\mathfrak b(x - \tfrac{\ir \pi}2) \mathfrak b(x + \tfrac{\ir \pi}2) = \mathfrak A^1(x).
\label{eq:bbUT}
\ee
Since \eqref{eq:bbUT} defines $\mathfrak b(x)$, only up to a sign, we choose the convention that
\be
\mathfrak b_{\pm\infty} = T^1_{\pm\infty}.
\ee
The rest of this section computes the finite-size corrections to $\mathfrak b(x)$ for any eigenvalue. The results could be obtained for the full interval $\gamma \in [-\pi,\pi]$, but for simplicity we consider $\gamma$ in a neighborhood of $\gamma = 0$, namely $\gamma \in (-\frac\pi3,\frac\pi3)$ for $d \equiv 0 \Mod 3$ and $\gamma \in (-\frac\pi6,\frac\pi6)$ for $d \equiv 1,2 \Mod 3$.

\paragraph{TBA equations.} 
For a given eigenstate of $\mathfrak a^1(x)$ and $\mathfrak a^2(x)$, we respectively denote by $t^1_+$ and $t^1_-$ the numbers of zeros $w_\pm^i$ of $\mathfrak a^1(x)$ in the upper and lower part of the plane. These zeros are ordered such that $w_\pm^i\ge w_\pm^{i+1}$. Likewise $t^2_+$ and $t^2_-$ count the zeros $x_\pm^j$ of $\mathfrak a^2(x)$ in the upper and lower half-planes, which are ordered as $x_\pm^j\ge x_\pm^{j+1}$. Depending on $\gamma$, some of the zeros may be degenerate.

The finite-size correction functions $\ell^1(x)$ and $\ell^2(x)$ are defined as
\begin{subequations}
\begin{alignat}{2}
\mathfrak a^1(x) &= \tanh^{N}(\tfrac x 2)\cdot \bigg(\prod_{i = 1}^{t_+^1}\tanh\big(\tfrac{x-w^i_+}2\big) \bigg)\bigg(\prod_{i = 1}^{t_-^1} \tanh\big(\tfrac{x-w^i_-}2\big)\bigg)\cdot \ell^1(x),\label{eq:nlieper1}\\
\mathfrak a^2(x) &= \bigg(\prod_{j = 1}^{t^2_+}\tanh\big(\tfrac{x-x^j_+}2\big) \bigg)\bigg(\prod_{j = 1}^{t^2_-} \tanh\big(\tfrac{x-x^j_-}2\big)\bigg)\cdot \ell^2(x).
\end{alignat}
\end{subequations}
The functions $\ell^1(x)$, $\ell^2(x)$, $\mathfrak A^1(x)$ and $\mathfrak A^2(x)$ are analytic and free of zeros in their corresponding analyticity strips, and satisfy
\be
\ell^1(x-\tfrac{\ir \pi}2)\ell^1(x+\tfrac{\ir \pi}2) = \mathfrak A^2(x), \qquad \ell^2(x-\tfrac{\ir \pi}2)\ell^2(x+\tfrac{\ir \pi}2) = \mathfrak A^1(x).
\ee
For $\gamma$ generic, the four functions have constant asymptotics, so we can define the Fourier transforms of their logarithmic derivatives:
\begin{subequations}
\begin{alignat}{2}
&\displaystyle L^n(k) = \frac{1}{2 \pi} \int_{-\infty}^\infty \dd x\, \big(\ln \ell^n(x)\big)'\eE^{-\ir k x}, \qquad 
&& \displaystyle A^n(k) = \frac{1}{2 \pi} \int_{-\infty}^\infty \dd x\, \big(\ln \mathfrak A^n(x)\big)'\eE^{-\ir k x},\\[0.25cm]
&\displaystyle \big(\ln \ell^n(x)\big)' = \int_{-\infty}^\infty \dd k\, L^n(k)\, \eE^{\ir k x}, \qquad 
&& \displaystyle \big(\ln \mathfrak A^n(x)\big)' = \int_{-\infty}^\infty \dd k\, A^n(k)\, \eE^{\ir k x}.
\end{alignat}
\end{subequations}
Applying the Fourier transform and its inverse to \eqref{eq:finalyT}, and subsequently integrating with respect to $x$, we find
\begin{subequations}
\label{eq:PTBA}
\begin{alignat}{2}
\ln \mathfrak a^1(x) &= \ln \tanh^{N}(\tfrac x 2) + \sum_{i = 1}^{t^1_+} \ln \Big(\tanh\big(\tfrac{x-w^i_+}2\big) \Big) + \sum_{i = 1}^{t^1_-} \ln \Big(\tanh\big(\tfrac{x-w^i_-}2\big)\Big) + K \ast \ln \mathfrak A^2 + \phi^1,\\
\ln \mathfrak a^2(x) &=  \sum_{j = 1}^{t^2_+} \ln \Big(\tanh\big(\tfrac{x-x^j_+}2\big) \Big) +\sum_{j = 1}^{t^2_-} \ln \Big(\tanh\big(\tfrac{x-x^j_-}2\big) \Big) +  K \ast \ln \mathfrak A^1 + \phi^2,
\end{alignat}
\end{subequations}
where $\phi^1$ and $\phi^2$ are the integration constants. The kernel $K(x)$ and the convolution are defined as for the strip, see \eqref{eq:kernel} and \eqref{eq:convolution}. 

We note that for special values of $\gamma$, the asymptotics of the functions $\ell^1(x)$, $\ell^2(x)$, $\mathfrak A^1(x)$ and $\mathfrak A^2(x)$ can be zero. In this case, one should instead define the Fourier transform of the second logarithmic derivative, which results in extra linear terms of the form $\phi\, x$ in \eqref{eq:PTBA}. However, this care turns out to be superfluous because the constants $\phi$ are evaluated to zero using the braid limit.
\paragraph{Scaling TBA equations.}
In the scaling regime, the function $\tanh^{N}(\frac x 2)$ has the following behavior:
\be
\lim_{N \rightarrow \infty} \tanh^{N}\big(\pm \tfrac 12(x + \ln N)\big) = (-1)^N\exp(-2\, \eE^{-x}).
\ee
We assume that the following scaling limits also exist:
\be
\mathsf a^n_\pm(x) = \lim_{N \rightarrow \infty} \mathfrak a^n\big(\!\pm\! (x + \ln N)\big), \qquad \mathsf A^n_\pm(x) = \lim_{N \rightarrow \infty} \mathfrak A^n\big(\!\pm\! (x + \ln N)\big), \qquad n = 1,2.
\ee
Because the functions $\mathfrak a^n(x)$ are not symmetric under the transformation $x \rightarrow -x$, the functions $\mathsf a^n_+(x)$ and $\mathsf a^n_-(x)$ are not equal. We denote the zeros of $\mathsf a^1_\pm(x)$ and $\mathsf a^2_\pm(x)$ by $z_\pm^i$ and $y_\pm^j$, namely
\be
z_\pm^i = w_\pm^i - \ln N, \qquad y_\pm^j = x_\pm^j - \ln N.
\ee
In the scaling regime, the non linear integral equations become
\begin{subequations}\label{eq:scalNLIET}
\begin{alignat}{2}
\ln \mathsf a_\pm^1(x) &= -2 \,\eE^{-x} + \sum_{i=1}^{t^1_\pm} \ln\big(\!-\!\tanh (\tfrac{x-z_\pm^i}2) \big) + K \ast \ln \mathsf A_\pm^2 +\phi^1_\pm,\\
\ln \mathsf a_\pm^2(x) &= \sum_{j=1}^{t^2_\pm} \ln\big(\!-\!\tanh (\tfrac{x-y^j_\pm}2) \big)+ K \ast \ln \mathsf A_\pm^1 + \phi^2_\pm.
\end{alignat}
\end{subequations}
Here, we consider the non linear integral equations to be independent in the upper and lower part of the planes, thus allowing two different sets of
constants, $(\phi^1_-,\phi^2_-)$ and $(\phi^1_+,\phi^2_+)$. 

\paragraph{Evaluation of the constants.}
For $d \equiv 0 \Mod 3$, we consider $\gamma \in (-\frac \pi3,\frac \pi3)$. On this interval, the braid limits \eqref{eq:braida1a2} are both positive. We fix the
branches of the logarithms by imposing
\be
\ln \big(\!-\tanh(\tfrac{x-y}2)\big)\xrightarrow{x \rightarrow \infty} \ir \pi
\ee
and 
\be
\label{eq:logbranches}
\ln \mathsf a_\pm^1(x)\xrightarrow{x \rightarrow \infty} \ln (4 \cos^2 \gamma-1), \qquad \ln \mathsf a_\pm^2(x)\xrightarrow{x \rightarrow \infty} \ln (2 \cos \gamma).
\ee
Applying the braid limit to \eqref{eq:scalNLIET}, we find that
\be
d \equiv 0 \Mod 3: \qquad\phi_\pm^1 = -\ir \pi t_\pm^1, \qquad \phi_\pm^2 = -\ir \pi t_\pm^2.
\ee

For $d \equiv 1,2 \Mod 3$, the interval $\gamma \in (-\frac \pi 6, \frac \pi 6)$ splits into three cases: $\gamma \in (-\frac {\pi}6,0)$, $\gamma \in (0,\frac \pi6)$ and $\gamma = 0$. Here, $\mathsf a_\pm^2(\infty)$ is negative in all three cases. Fixing the branches of the logarithms fixes the constants $\phi^2_\pm$:
\be
\ln \mathsf a_\pm^2(x)\xrightarrow{x \rightarrow \infty} \ln \big(-2 \cos (\pm\gamma+\tfrac{2\pi d}3)\big) + \ir \pi, \qquad \phi^2_\pm = -\ir \pi(t^1_\pm-1).
\ee
In stark contrast, $\mathsf a_\pm^1(\infty)$ takes opposite signs in the first and second interval and is zero for $\gamma = 0$. For $d \equiv 1 \Mod 3$, $\gamma \in (-\tfrac\pi6,0)$ as well as for $d \equiv 2 \Mod 3$, $\gamma \in (0,\tfrac\pi6)$, we have
\be
\begin{array}{ll} \ln \mathsf a^1_-(x) \xrightarrow{x \rightarrow \infty} \ln (4 \cos^2 (-\gamma+\tfrac{2\pi d}3)-1), \quad & \phi^1_- = -\ir \pi t^1_-, \\[0.15cm]
\ln \mathsf a^1_+(x) \xrightarrow{x \rightarrow \infty} \ln \big(1-4 \cos^2 (\gamma+\tfrac{2\pi d}3)\big) + \ir \pi, \quad & \phi^1_+ = -\ir \pi (t^1_+-1).
\end{array}
\ee
Likewise, for $d \equiv 1 \Mod 3$, $\gamma \in (0,\tfrac\pi6)$ and $d \equiv 2 \Mod 3$, $\gamma \in (-\tfrac\pi6,0)$, we have
\be
\begin{array}{ll} \ln \mathsf a^1_-(x) \xrightarrow{x \rightarrow \infty} \ln \big(1-4 \cos^2 (-\gamma+\tfrac{2\pi d}3)\big) + \ir \pi, \quad &\phi^1_- = -\ir \pi(t^1_--1),\\[0.15cm]
\ln \mathsf a^1_+(x) \xrightarrow{x \rightarrow \infty}  \ln \big(4 \cos^2 (\gamma+\tfrac{2\pi d}3)-1\big), \quad  &\phi^1_+ = -\ir \pi t^1_+.
\end{array}
\ee
For $\gamma = 0$, the result depends on the subcases $\bar \tA$ or $\bar \tB$ and $\tA$ or $\tB$: 
\begin{subequations}
\begin{alignat}{3}
&\textrm{Subcase } \bar \tA: \quad &&\ln \mathsf a^1_-(x) \xrightarrow{x \rightarrow -\infty} -\tfrac{2x}3 + \ln \kappa_- + \ir \pi && \qquad \phi^1_- = -\ir \pi (t^1_--1), \\[0.15cm]
&\textrm{Subcase } \bar \tB: \quad &&\ln \mathsf a^1_-(x) \xrightarrow{x \rightarrow -\infty} -\tfrac{4x}3 + \ln \kappa'_-&& \qquad \phi^1_- = -\ir \pi t^1_-, \\[0.15cm]
&\textrm{Subcase } \tA: \quad &&\ln \mathsf a^1_+(x) \xrightarrow{x \rightarrow \infty} -\tfrac{2x}3 + \ln \kappa_+ + \ir \pi && \qquad \phi^1_+ = -\ir \pi (t^1_+-1), \\[0.15cm]
&\textrm{Subcase } \tB: \quad &&\ln \mathsf a^1_-(x) \xrightarrow{x \rightarrow \infty} -\tfrac{4x}3 + \ln \kappa'_+&& \qquad \phi^1_+ = -\ir \pi t^1_+.
\end{alignat}
\end{subequations}
\paragraph{Finite-size corrections.}
We apply the Fourier transform to \eqref{eq:bbUT} by first removing the zeros of $\mathfrak b(x)$ on the real axis. This yields
\be
\ln \mathfrak b(x) =    \sum_{j = 1}^{t_+^2} \ln \Big(-\tanh\big(\tfrac{x-x^j_+}2\big) \Big) + \sum_{j = 1}^{t_-^2} \ln \Big(\tanh\big(\tfrac{x-x_-^j}2\big)\Big) + K \ast \ln \mathfrak A^1 + \psi.
\ee
The constant $\psi$ can be evaluated from the braid limits to be a multiple of $\pi$. It does not contribute to the finite-size corrections. These are written in terms of 
the scaling functions and their zeros as follows: 
\begin{alignat}{2}
\ln \mathfrak b(x) - \psi &= \sum_{j = 1}^{t^2_-} \ln \Big(\!\tanh\big(\tfrac{x-x_-^j}2\big) \Big)+\sum_{j = 1}^{t^2_+} \ln \Big(\!-\!\tanh\big(\tfrac{x-x^j_+}2\big) \Big)\nonumber\\&\hspace{0.8cm}+ \frac{1}{2\pi} \int_{-\ln N}^\infty \dd y \Big( \frac{\ln\mathfrak A^1(y+\ln N)}{\cosh(x-y-\ln N)}+\frac{\ln\mathfrak A^1(-y-\ln N)}{\cosh(x+y+\ln N)}\Big) \label{eq:FSC1T}\\
&\hspace{-1cm} \simeq - \frac1N \bigg(2 \sum_{j=1}^{t_-^2}\eE^{-x-y_-^j} + 2 \sum_{j=1}^{t_+^2}\eE^{x-y_+^j}-\frac{\eE^{-x}}\pi \int_{-\infty}^\infty \dd y\, \eE^{-y} \ln \mathsf A_-^1(y)- \frac{\eE^{x}}\pi \int_{-\infty}^\infty \dd y\, \eE^{-y} \ln \mathsf A_+^1(y)\bigg).\nonumber
\end{alignat}

\paragraph{Zeros of $\boldsymbol{\mathfrak a^1(x)}$ and $\boldsymbol{\mathfrak a^2(x)}$.} 
To rewrite $\eE^{-y_\pm^j}$ in terms of integrals involving the scaling functions, we use \eqref{eq:finalyT} and find
\be
\mathfrak a^2(x^j_\pm) = 0 \quad \Rightarrow \quad \mathfrak a^1(x^j_\pm-\tfrac {\ir \pi}2) = -1 \quad \Rightarrow \quad \mathsf a^1_\pm(y_\pm^j-\tfrac {\ir \pi}2) = -1.
\ee
Using \eqref{eq:scalNLIET}, we find
\begin{alignat}{2}
(2 k_\pm^j-1)\pi &= \ir \ln \mathsf a^1_\pm (y^j_\pm - \tfrac{\ir \pi }2) \nonumber \\&= 2\, \eE^{-y_\pm^j} + \ir \sum_{i=1}^{t_\pm^1} \ln\big(\!-\!\tanh \tfrac12(y_\pm^j-z_\pm^i-  \tfrac{\ir\pi}2 )\big) + \frac1{2\pi} \int_{-\infty}^\infty \dd y\, \frac{\ln \mathsf A_\pm^2(y)}{\sinh(y-y_\pm^j)} + \ir \phi^1_\pm
\end{alignat}
where the $k_\pm^j$ are integers. The terms $\ln\big(\!-\!\tanh \tfrac12(y_\pm^j-z_\pm^i-  \tfrac{\ir\pi}2 )\big)$ in this last expression are rewritten using \eqref{eq:finalyT}:
\be
\label{eq:integerKP}
\mathfrak a^1(w^i_\pm) = 0 \quad \Rightarrow \quad \mathfrak a^2(w^i_\pm-\tfrac {\ir \pi}2) = -1 \quad \Rightarrow \quad \mathsf a^2_\pm( z^j_\pm-\tfrac {\ir \pi}2) = -1.
\ee
Similarly,
\begin{alignat}{2}
(2 \ell_\pm^i-1)\pi &= \ir \ln \mathsf a^2_\pm (z_\pm^i - \tfrac{\ir \pi}2) \nonumber\\
& = \ir \sum_{j=1}^{t_\pm^2} \ln \big(\!-\!\tanh \tfrac12(z_\pm^i-y_\pm^j -\tfrac{\ir \pi}2)\big) + \frac{1}{2 \pi} \int_{-\infty}^\infty \dd y \frac{\ln \mathsf A^1_\pm(y)}{\sinh(y-z_\pm^i)}+ \ir \phi_\pm^2
\label{eq:integerLP}
\end{alignat}
where the $\ell_\pm^i$ are integers. We can then apply \eqref{eq:tanh.id} to each of the logarithms in \eqref{eq:integerLP}. By combining \eqref{eq:FSC1T}, \eqref{eq:integerKP} and \eqref{eq:integerLP}, we obtain
\begingroup
\allowbreak
\begin{alignat}{2}
\ln \mathfrak b(x) - \psi & \simeq -\frac{2 \pi}N \bigg(\eE^x \bigg[  \sum_{j=1}^{t_+^2}\big( k_+^j -\tfrac12 - \tfrac{\ir \phi^1_+}{2\pi}\big) + \sum_{i=1}^{t^1_+} \big(\ell_+^i-\tfrac12+\tfrac{t^2_+}2 -  \tfrac{\ir \phi^2_+}{2\pi}\big) - \frac 1 {2\pi^2} \int_{-\infty}^\infty \dd y \,\eE^{-y} \ln \mathsf A_+^1(y)  \nonumber\\ 
& \hspace{1.5cm}-  \frac{1}{4\pi^2} \sum_{i=1}^{t_+^1}\int_{-\infty}^{\infty} \dd y\, \frac{\ln \mathsf A_+^1(y)}{\sinh(y-z_+^i)}- \frac{1}{4\pi^2} \sum_{j=1}^{t_+^2}\int_{-\infty}^{\infty} \dd y\, \frac{\ln \mathsf A_+^2(y)}{\sinh(y-y_+^j)}  \bigg] \nonumber\\
&\hspace{1cm}+\eE^{-x} \bigg[ \sum_{j=1}^{t_-^2}\big( k_-^j -\tfrac12 - \tfrac{\ir \phi^1_-}{2\pi}\big) + \sum_{i=1}^{t^1_-} \big(\ell_-^i-\tfrac12+\tfrac{t^2_-}2 -  \tfrac{\ir \phi^2_-}{2\pi}\big)- \frac 1 {2\pi^2} \int_{-\infty}^\infty \dd y \,\eE^{-y} \ln \mathsf A_-^1(y)  \nonumber\\ 
&\hspace{1.5cm}-  \frac{1}{4\pi^2} \sum_{i=1}^{t_-^1}\int_{-\infty}^{\infty} \dd y\, \frac{\ln \mathsf A_-^1(y)}{\sinh(y-z_-^i)} - \frac{1}{4\pi^2} \sum_{j=1}^{t_-^2}\int_{-\infty}^{\infty} \dd y\, \frac{\ln \mathsf A_-^2(y)}{\sinh(y-y_-^j)}  \bigg] 
\bigg). \label{eq:FSC2}
\end{alignat}
\endgroup

\paragraph{Dilogarithm technique.}
To evaluate \eqref{eq:FSC2}, we consider the integrals
\be 
\label{eq:Jints}
\mathcal J_\pm = \int_{-\infty}^\infty \dd y \Big((\ln \mathsf a_\pm^1)' \ln \mathsf A_\pm^1  -\ln |\mathsf a_\pm^1| (\ln \mathsf A_\pm^1)' \Big) + \int_{-\infty}^\infty \dd y \Big((\ln \mathsf a_\pm^2)' \ln \mathsf A_\pm^2  -\ln |\mathsf a_\pm^2| (\ln \mathsf A_\pm^2)' \Big)
\ee
where $\ln |\mathsf a_\pm^1|$ and $\ln |\mathsf a_\pm^2|$ are real for all $x$ and given by
\be
\ln |\mathsf a_\pm^n|(x) = \ln \mathsf a^n_\pm(x) + \theta_\pm^n(x), \qquad n = 1,2.
\ee
The functions $\theta_+^n(x)$ and $\theta_-^n(x)$ are step functions defined for $x \in \mathbb R$ with the following behavior: starting at $x = \infty$ and progressing to the left on the real $x$ axis, they decrease by $\ir\pi$ each time a zero of the corresponding type is crossed ($z_\pm^i$ for $\theta^1_\pm(x)$ and $y_\pm^j$ for $\theta^2_\pm(x)$). The values at the endpoints are 
\be
\theta_\pm^1(x) = 0 \hspace{0.2cm} \quad\text{for } x > z_\pm^1, \qquad 
\theta_\pm^2(x) = 0 \hspace{0.2cm} \quad\text{for } x > y_\pm^1.
\ee

The integrals $\mathcal J_\pm$ can be evaluated in two ways. For the first, one uses the non-linear integral equations \eqref{eq:scalNLIE} and the symmetries of $K(x)$ and obtains
\begin{alignat}{2}
\mathcal J_\pm &= 2\int_{-\infty}^\infty \dd y \,  \eE^{-y} \big(\ln \mathsf A_\pm^1 + (\ln \mathsf A_\pm^1)'\big) +  \int_{-\infty}^\infty \dd y \sum_{i=1}^{t_\pm^1} \Big[\ln\big(\!-\! \tanh (\tfrac{y-z_\pm^i}2)\big) \Big]' \ln \mathsf A_\pm^1 \nonumber\\
& - \int_{-\infty}^\infty \dd y \Big(\sum_{i=1}^{t_\pm^1} \ln \big(\!-\!\tanh (\tfrac{y-z_\pm^i}2)\big) - \theta_\pm^1(y) - \ir \pi t_\pm^1\Big) \big(\ln \mathsf A_\pm^1\big)' + \int_{-\infty}^\infty \dd y \sum_{j=1}^{t_\pm^2}  \Big[\ln \big(\!-\!\tanh (\tfrac{y-y_\pm^j}2)\big) \Big]' \ln \mathsf A_\pm^2 \nonumber\\ 
& - \int_{-\infty}^\infty \dd y \Big(\sum_{j=1}^{t_\pm^2} \ln \big(\!-\!\tanh (\tfrac{y-y_\pm^j}2)\big) - \theta_\pm^2(y) - \ir \pi t_\pm^2\Big) \big(\ln \mathsf A_\pm^2\big)'.
\end{alignat}
The integrals involving derivatives of $\ln \mathsf A_\pm^1$ and $\ln \mathsf A_\pm^2$ are transformed using integration by parts. For each one, it can be argued using the non-linear integral equations that the surface terms are zero. This yields
\be
\mathcal J_\pm = 4 \int_{-\infty}^\infty \dd y\, \eE^{-y} \ln \mathsf A_\pm^1(y) + 2 \sum_{i=1}^{t_\pm^1} \int_{-\infty}^\infty\dd y\, \frac{\ln \mathsf A_\pm^1(y)}{\sinh(y-z_\pm^i)} + 2\sum_{j=1}^{t_\pm^2} \int_{-\infty}^\infty \dd y\, \frac{\ln \mathsf A_\pm^2(y)}{\sinh(y-y_\pm^j)},
\ee
which is precisely the combination of integrals needed to compute the finite-size corrections: 
\begin{alignat}{2}\label{eq:finalfscP}
\ln \mathfrak b(x) -\psi \simeq -\frac{2\pi}N \bigg(&\eE^x \Big[\sum_{j=1}^{t_+^2}\big( k_+^j -\tfrac12 - \tfrac{\ir \phi^1_+}{2\pi}\big) + \sum_{i=1}^{t^1_+} \big(\ell_+^i-\tfrac12+\tfrac{t^2_+}2 -  \tfrac{\ir \phi^2_+}{2\pi}\big) - \frac{\mathcal J_+}{8 \pi^2}\Big] \nonumber\\
&+ \eE^{-x} \Big[  \sum_{j=1}^{t_-^2}\big( k_-^j -\tfrac12 - \tfrac{\ir \phi^1_-}{2\pi}\big) + \sum_{i=1}^{t^1_-} \big(\ell_-^i-\tfrac12+\tfrac{t^2_-}2 -  \tfrac{\ir \phi^2_-}{2\pi}\big) - \frac{\mathcal J_-}{8 \pi^2}\Big]\bigg).
\end{alignat}
The second way of performing the integrals consists in changing the integration variable from $y$ to $\mathsf a_\pm$. The asymptotic behavior of the scaling functions is
\be
\mathsf a^1_\pm(-\infty) = 0, \quad \mathsf a^2_\pm(-\infty) = \sigma, \quad \mathsf a^1_\pm(\infty) = 4 \cos^2(\pm\gamma+\tfrac{2\pi d}3)-1, \quad \mathsf a^2_\pm(\infty) = 2 \cos (\pm\gamma+\tfrac{2\pi d}3)
\ee
with $\sigma \in \{-1,+1\}$. For the integrals involving $\mathsf a^1_\pm$, we find
\begin{alignat}{2}
\int_{-\infty}^\infty \dd y \Big((&\ln \mathsf a_\pm^1)' \ln \mathsf A_\pm^1  -\ln |\mathsf a_\pm^1| (\ln \mathsf A_\pm^1)' \Big) = \int_{-\infty}^\infty \dd y\, \frac{\dd \mathsf a_\pm^1}{\dd y} \,\bigg( \frac{\ln(1+ \mathsf a_\pm^1)}{\mathsf a_\pm^1} - \frac{\ln |\mathsf a_\pm^1|}{1+ \mathsf a_\pm^1} \bigg)\nonumber\\
& = \bigg[\int_{-\infty}^{z^\pm_{t_\pm^1}}+\int_{z^\pm_{t_\pm^1}}^{z^\pm_{t_\pm^1-1}}+ \dots + \int_{z^\pm_2}^{z^\pm_1}+ \int_{z^\pm_1}^{\infty}\bigg] \dd y\, \frac{\dd \mathsf a_\pm^1}{\dd y}\, \bigg( \frac{\ln(1+ \mathsf a_\pm^1)}{\mathsf a_\pm^1} - \frac{\ln |\mathsf a_\pm^1|}{1+ \mathsf a_\pm^1} \bigg) \nonumber\\
& =  \bigg[\int_{0}^{0}+\int_{0}^{0}+ \dots + \int_{0}^{0}+ \int_{0}^{\mathsf a^1_\pm(\infty)}\bigg] \dd \mathsf a \,\Big( \frac{\ln(1+ \mathsf a)}{\mathsf a} - \frac{\ln |\mathsf a|}{1+ \mathsf a} \Big) \nonumber\\
& =\int_{0}^{\mathsf a^1_\pm(\infty)} \dd \mathsf a \,\Big( \frac{\ln(1+ \mathsf a)}{\mathsf a} - \frac{\ln |\mathsf a|}{1+ \mathsf a} \Big). \label{eq:J1finalper}
\end{alignat}
Recalling that $\mathsf A^2_\pm = (1 + \eE^{3 \ir \gamma}\mathsf a^2_\pm)(1 + \eE^{-3 \ir \gamma}\mathsf a^2_\pm)$, the integrals involving $\mathsf a^2_\pm$ are obtained with the same arguments. We find
\begin{alignat}{2}
\int_{-\infty}^\infty \dd y \Big((\ln \mathsf a_\pm^2)' \ln \mathsf A_\pm^2  -\ln |\mathsf a_\pm^2| (\ln \mathsf A_\pm^2)' \Big)  &= \int_{\sigma}^{\mathsf a^2_\pm(\infty)} \dd \mathsf a \,\Big( \frac{\ln(1+ \eE^{3 \ir \gamma}\mathsf a)}{\mathsf a} - \frac{\eE^{3 \ir \gamma}\ln |\mathsf a|}{1+ \eE^{3 \ir \gamma} \mathsf a} \Big) \nonumber\\ &+  \int_{\sigma}^{\mathsf a^2_\pm(\infty)} \dd \mathsf a \,\Big( \frac{\ln(1+ \eE^{-3 \ir \gamma}\mathsf a)}{\mathsf a} - \frac{\eE^{-3 \ir \gamma}\ln |\mathsf a|}{1+ \eE^{-3 \ir \gamma} \mathsf a} \Big). \label{eq:J2finalper}
\end{alignat}
The sum of the integrals \eqref{eq:J1finalper} and \eqref{eq:J2finalper} is expressed as
\be
\mathcal J_\pm = 
\left\{\begin{array}{ll}
\mathcal K_\sigma (\gamma) & \quad d \equiv 0 \Mod 3, \\[0.2cm]
\mathcal K_\sigma (\frac{2\pi}3\pm\gamma) & \quad d \equiv 1 \Mod 3, \\[0.2cm]
\mathcal K_\sigma (\frac{2\pi}3\mp\gamma) & \quad d \equiv 2 \Mod 3, \\[0.2cm]
\end{array}
\right.
\label{eq:JpmT}
\ee
where $\mathcal K_\sigma (\gamma)$ is defined and evaluated in \cref{sec:Rogers}.
Comparing with \eqref{eq:Texpansion} specialised to $c=0$, we find that the first and second bracket in \eqref{eq:finalfscP} are identified with $\Delta$ and $\bar\Delta$:
\begin{subequations}
\begin{alignat}{2}
\Delta &= \sum_{j=1}^{t_+^2}\big( k_+^j -\tfrac12 - \tfrac{\ir \phi^1_+}{2\pi}\big) + \sum_{i=1}^{t^1_+} \big(\ell_+^i-\tfrac12+\tfrac{t^2_+}2 -  \tfrac{\ir \phi^2_+}{2\pi}\big) - \frac{\mathcal J_+}{8 \pi^2}, \\
\bar \Delta &= \sum_{j=1}^{t_-^2}\big( k_-^j -\tfrac12 - \tfrac{\ir \phi^1_-}{2\pi}\big) + \sum_{i=1}^{t^1_-} \big(\ell_-^i-\tfrac12+\tfrac{t^2_-}2 -  \tfrac{\ir \phi^2_-}{2\pi}\big) - \frac{\mathcal J_-}{8 \pi^2},
\end{alignat}
\end{subequations}
where the integers $k_\pm^j$ and $\ell_\pm^i$ are given in terms of the eigenvalue and its zeros by \eqref{eq:integerKP} and \eqref{eq:integerLP}.

\paragraph{Specialisation to $\boldsymbol{\gamma = 0}$.}
For $\gamma=0$, we have $\mathcal J_+ = \mathcal J_-$ and
\begin{subequations}
\label{eq:Deltagamma0}
\be
\Delta =  \sum_{j=1}^{t^2_+}k_+^j+\sum_{i=1}^{t^1_+}\ell_+^i + \tau_+ + \iota(\sigma), \qquad \bar\Delta =  \sum_{j=1}^{t^2_-}k_-^j+\sum_{i=1}^{t^1_-}\ell_-^i + \tau_- + \iota(\sigma)
\ee
with
\begin{alignat}{2}
\tau_\pm &= 
\left\{\begin{array}{cl}
-\frac12(t^1_\pm + t^2_\pm + t^1_\pm t^2_\pm) & d\equiv 0 \Mod 3,\\[0.1cm]
-\frac12t^1_\pm t^2_\pm & d \equiv 1,2 \Mod 3\, (\bar \tA \textrm{ or } \tA),\\[0.1cm]
-\frac12(t^2_\pm +  t^1_\pm t^2_\pm) & d \equiv 1,2 \Mod 3\, (\bar \tB \textrm{ or } \tB)
\end{array}\right. \\[0.2cm]
\iota(\sigma) &= -\frac{\mathcal J_\pm}{8\pi^2}  = 
\left\{\begin{array}{cll}
-\frac1{24} &d\equiv 0 \Mod 3,& \sigma = 1,\\[0.15cm]
-\frac1{6} &d\equiv 0 \Mod 3,& \sigma = -1,\\[0.15cm]
\frac1{8} & d\equiv 1,2 \Mod 3,& \sigma = 1,\\[0.15cm]
0 & d\equiv 1,2 \Mod 3,& \sigma = -1.\\[0.15cm]
\end{array}\right.
\end{alignat}
\end{subequations}

\subsection{Solution for the ground states}\label{sec:Pcfgs}

In this section, we study the conformal weights of the ground states of each $\stanp_N^d$ for $\gamma = 0$. 
For the ground state of $\stanp_N^d$, we observe that the pattern of zeros is
symmetric with respect to the horizontal axis, implying that $\Delta = \bar
\Delta$. The analysis of the bulk behavior of these eigenvalues (using the method discussed below \eqref{eq:bulkbP}) reveals that $\sigma = 1$ for $N$ even and $\sigma = -1$ for $N$ odd. Moreover, for $d \equiv 1,2 \text{ mod } 3$, we observe that the ground state eigenvalue respectively belongs to the subcases $(\bar\tB,\tB)$ and $(\bar\tA,\tA)$. For $N$ even, the zero patterns are characterised by
\begin{subequations}\label{eq:datat1t2}
\be\label{eq:datat1t2even}
t^1_\pm = \left\{
\begin{array}{cl}
\frac{d}6 & d \equiv 0\text{ mod }6,
\\[0.15cm]
\frac{d-2}6 & d \equiv 2\text{ mod }6,
\\[0.15cm]
\frac{d-4}6 & d \equiv 4\text{ mod }6, 
\end{array}\right.\quad\quad
t^2_\pm = \left\{
\begin{array}{cl}
\frac{d}3 & d \equiv 0\text{ mod }6,\\[0.15cm]
\frac{d+1}3 & d \equiv 2\text{ mod }6,\\[0.15cm]
\frac{d-1}3 & d \equiv 4\text{ mod }6.
\end{array}\right. 
\ee
For $N$ odd, we work with the convention described in \cref{sec:Pprops} that the single zeros on the axis are in the upper-half plane. We stress however that the opposite convention produces the same conformal weights. With this convention, we have
\be
\label{eq:datat1t2odd}
t^1_+ = \left\{
\begin{array}{cl}
\frac{d-1}6 & d \equiv 1\text{ mod }6,
\\[0.15cm]
\frac{d+3}6 & d \equiv 3\text{ mod }6,
\\[0.15cm]
\frac{d+1}6 & d \equiv 5\text{ mod }6, 
\end{array}\right.\quad
t^1_- = \left\{
\begin{array}{cl}
\frac{d-7}6 & d \equiv 1\text{ mod }6,
\\[0.15cm]
\frac{d-3}6 & d \equiv 3\text{ mod }6,
\\[0.15cm]
\frac{d-5}6 & d \equiv 5\text{ mod }6, 
\end{array}\right.\quad
t^2_\pm = \left\{
\begin{array}{cl}
\frac{d-1}3 & d \equiv 1\text{ mod }6,\\[0.15cm]
\frac{d}3 & d \equiv 3\text{ mod }6,\\[0.15cm]
\frac{d+1}3 & d \equiv 5\text{ mod }6.
\end{array}\right. 
\ee
\end{subequations}
In both cases, there are also $\frac{N-d}2$ pairs of zeros on the boundary edges of the analyticity strip of $T^1(u)$, split between the upper and lower halves, see the examples in \cref{fig:GSpatternsp}. Each zero in the center of the analyticity strip of $T^2(u)$ is of order one and is joined, at the same height, by a pair of single zeros sitting on the edges of the analyticity strip. The integers $k^j_\pm$ and $\ell^i_\pm$ are not fixed by the technique of \cref{sec:Pfsc}. We estimate them using our computer program by evaluating \eqref{eq:integerKP} and \eqref{eq:integerLP} on small system sizes. For the ground state of $\stanp_N^d$, we find:
\be\label{eq:kdata}
k^j_\pm = \left\{
\begin{array}{cl}
j & d \equiv 0\text{ mod }3,\\[0.1cm]
j & d \equiv 1\text{ mod }3,\\[0.1cm]
j-1 & d \equiv 2\text{ mod }3,
\end{array}\right. \qquad \ell^i_\pm = i.
\ee

The patterns of zeros for the ground states for $N=12,13$ are given in \cref{fig:GSpatternsp}. In this figure, for $d=1$, the analyticity strip of $T^2(u)$ for the Razumov-Stroganov eigenstate is colored in gray, indicating that $T^2(u)=0$. The zeros on the separation lines are divided between the upper and lower half-planes using the prescription discussed in \cref{sec:Pprops}.

Inserting the data \eqref{eq:datat1t2} and \eqref{eq:kdata} into \eqref{eq:Deltagamma0}, we find that in all cases, the ground state conformal weights are given by
\be
\Delta = \bar \Delta = \frac{d^2-1}{24} = \Delta_{0,\frac d 2}.
\ee

\subsection{Solution for all the eigenvalues}\label{sec:fsgf}

In this section, we describe the full spectrum in the standard modules $\stanp_N^d$ for $\gamma = 0$. We compute the finitized spectrum generating functions
\be
\label{eq:ZdNP}
\ZdNP = \sum_{\substack{\textrm{eigenstates}\\ \textrm{of }T^1(u)\textrm{ in } \stanp_N^d}} \varepsilon\, q^{\Delta}\bar q^{\bar \Delta}, 
\ee
where $\Delta$ and $\bar \Delta$ are the conformal weights of the limiting conformal states, $\varepsilon$ is the overall sign of the eigenvalue and the sum is over eigenstates characterised by the patterns of zeros. Indeed, for periodic boundary conditions, the eigenvalues can have either a positive or a negative overall sign. This sign is crucial in \cref{sec:mipf} for the partition function on the $M\times N$ torus with $M$ odd. Concretely, $\varepsilon$ is obtained from the bulk limit $K_{\textrm{bulk}} = \sigma$ and the vertical lattice width $M$ as $\varepsilon = \sigma^M$.

As for the strip, our derivation is built on a set of conjectured selection rules for the patterns of zeros in $\stanp_N^d$. This conjecture is supported by data produced with our computer implementation of the transfer matrices and the computation of the eigenvalues and patterns of zeros, for $1 \le N \le 12$. We present data for $\stanp_6^0$ and $\stanp_6^2$ in \cref{fig:patsP60,fig:patsP62}.

\paragraph{Selection rules for the patterns of zeros.}
The selection rules are expressed in terms of triples $(\sigma, \sigma',\sigma'')$ of single-column diagrams, with $\sigma \in \Aone{M}{m}$, $\sigma' \in \Aone{L}{n}$ and $\sigma'' \in \Aone{L}{\ell}$. We denote by $\BB{M}{L}{m}{n}{\ell}$ the set of such triples. In the selection rule, the configuration $\sigma$ describes the content in zeros of the first strip, whereas the pair $(\sigma', \sigma'')$ describes the content of the second strip. Crucially, in contrast with the boundary case, the constraint of dominance is {\it not} imposed on the pair $(\sigma', \sigma'')$. The cardinality of $\BB{M}{L}{m}{n}{\ell}$ is therefore simply given by the product of the cardinalities of $\Aone{M}{m}$, $\Aone{L}{n}$ and $\Aone{L}{\ell}$:
\be\label{eq:cardB}
\big|\BB{M}{L}{m}{n}{\ell}\big| = \binom{M}{m}\binom{L}{n}\binom{L}{\ell}.
\ee
For instance, $\BB{4}{2}{3}{1}{1}$ contains 16 configurations. These are precisely the configurations appearing in $\AA{4}{2}{3}{1}{1}$ given in \cref{fig:A42311}, except that those on the second row are twice degenerate.
For $d \equiv 1,2 \Mod 3$, we include $(\bar \tA, \tA)$, $(\bar \tA, \tB)$, $(\bar \tB, \tA)$ or $(\bar \tB, \tB)$ 
in the upper labels of $\BB{M}{L}{m}{n}{\ell}$ to specify which subcase the corresponding eigenvalues belong to.

We give a conjecture for the full content of zeros in the analyticity strips, with the separation between the upper and lower planes discussed below. For example, the patterns for $\stanp_6^0$ and $\stanp_6^2$ given in \cref{fig:patsP60,fig:patsP62} are encoded by the following sets of column configurations:
\begin{subequations}
\begin{alignat}{2}
\stanp_6^0&: \quad \BB{3}{0}{0}{0}{0}\,\cup\,\BB{3}{1}{2}{0}{0}\,\cup\,\BB{4}{1}{2}{1}{1}\,\cup\,\BB{4}{2}{4}{1}{1}  \,\cup\,\BB{5}{2}{4}{2}{2}\,\cup\,\BB{6}{3}{6}{3}{3},\\[0.2cm]
\stanp_6^2&:\quad\BBAB{4}{0}{2}{0}{0}{\bar \tA,\tA}\,\cup\,\BBAB{4}{1}{4}{0}{0}{\bar \tA,\tA}\,\cup\,\BBAB{5}{1}{4}{1}{1}{\bar\tA,\tA}\,\cup\,\BBAB{6}{2}{6}{2}{2}{\bar\tA,\tA}\,\cup\,\BBAB{4}{1}{4}{0}{1}{\bar\tA,\tB}  \,\cup\,\BBAB{4}{1}{4}{0}{1}{\bar\tB,\tA}.
\end{alignat}
\end{subequations}

More generally, for $d\neq 1$, we conjecture that the full set of patterns of zeros in $\stanp_N^d$ is described by the following sets:
\begin{subequations}\label{eq:SRp}
\begin{alignat}{2}
d=3t&: \hspace{0.2cm}\bigcup_{i=0}^{\frac{N-d}2}\,\, \bigcup_{j=0}^{\big\lfloor\frac12\big(\frac{N-d}2-i\big)\big\rfloor} \BB{\frac{N+t}2 + i}{i+j+t}{2(i+j+t)}{i}{i+t} \ ,\label{eq:SRp0t}
\\[0.2cm]
d=3t+1&: \hspace{0.2cm}\bigcup_{i=0}^{\frac{N-d}2}\,\, \bigcup_{j=0}^{\big\lfloor\frac12\big(\frac{N-d}2-i\big)\big\rfloor} \BBAB{\frac{N+t-1}2 + i}{i+j+t-1}{2(i+j+t)}{i}{i+t-1}{\bar\tB,\tB}
\cup\bigcup_{i=0}^{\frac{N-d-2}2}\,\, \bigcup_{j=0}^{\big\lfloor\frac12\big(\frac{N-d-2}2-i\big)\big\rfloor} \BBAB{\frac{N+t+1}2 + i}{i+j+t}{2(i+j+t+1)}{i}{i+t}{\bar\tA,\tB} \nonumber\\
&\cup\bigcup_{i=0}^{\frac{N-d-2}2}\,\, \bigcup_{j=0}^{\big\lfloor\frac12\big(\frac{N-d-2}2-i\big)\big\rfloor} \BBAB{\frac{N+t+1}2 + i}{i+j+t}{2(i+j+t+1)}{i}{i+t}{\bar\tB,\tA} \cup
 \bigcup_{i=0}^{\frac{N-d-4}2}\,\, \bigcup_{j=0}^{\big\lfloor\frac12\big(\frac{N-d-4}2-i\big)\big\rfloor} \BBAB{\frac{N+t+3}2 + i}{i+j+t+1}{2(i+j+t+2)}{i}{i+t+1}{\bar\tA,\tA}\ ,
\label{eq:SRp1t}
\\[0.2cm]
d=3t+2&: \hspace{0.2cm}\bigcup_{i=0}^{\frac{N-d}2}\,\, \bigcup_{j=0}^{\big\lfloor\frac12\big(\frac{N-d}2-i\big)\big\rfloor} \BBAB{\frac{N+t+2}2 + i}{i+j+t}{2(i+j+t+1)}{i}{i+t}{\bar\tA,\tA}
\cup  \bigcup_{i=0}^{\frac{N-d-4}2}\,\, \bigcup_{j=0}^{\big\lfloor\frac12\big(\frac{N-d-4}2-i\big)\big\rfloor} \BBAB{\frac{N+t+2}2 + i}{i+j+t+1}{2(i+j+t+2)}{i}{i+t+1}{\bar\tA,\tB} \nonumber\\
&\cup  \bigcup_{i=0}^{\frac{N-d-4}2}\,\, \bigcup_{j=0}^{\big\lfloor\frac12\big(\frac{N-d-4}2-i\big)\big\rfloor} \BBAB{\frac{N+t+2}2 + i}{i+j+t+1}{2(i+j+t+2)}{i}{i+t+1}{\bar\tB,\tA}
\cup \bigcup_{i=0}^{\frac{N-d-8}2}\,\, \bigcup_{j=0}^{\big\lfloor\frac12\big(\frac{N-d-8}2-i\big)\big\rfloor} \BBAB{\frac{N+t+2}2 + i}{i+j+t+2}{2(i+j+t+3)}{i}{i+t+2}{\bar\tB,\tB}\ .
\label{eq:SRp2t}\end{alignat}
For $d=1$, we instead have
\begin{alignat}{2}
d=1: \quad&\BB{0}{0}{0}{0}{0} \hspace{0.1cm}\cup\hspace{0.1cm} \bigcup_{i=1}^{\frac{N-1}2}\,\, \bigcup_{j=1}^{\big\lfloor\frac12\big(\frac{N-1}2-i\big)\big\rfloor} \BBAB{\frac{N-1}2 + i}{i+j-1}{2(i+j)}{i}{i-1}{\bar\tB,\tB}
\hspace{0.1cm}\cup\hspace{0.1cm} \bigcup_{i=0}^{\frac{N-3}2}\,\, \bigcup_{j=0}^{\big\lfloor\frac12\big(\frac{N-3}2-i\big)\big\rfloor} \BBAB{\frac{N+1}2 + i}{i+j}{2(i+j+1)}{i}{i}{\bar\tA,\tB} \nonumber\\
&
\cup\hspace{0.1cm} \bigcup_{i=0}^{\frac{N-3}2}\,\, \bigcup_{j=0}^{\big\lfloor\frac12\big(\frac{N-3}2-i\big)\big\rfloor} \BBAB{\frac{N+1}2 + i}{i+j}{2(i+j+1)}{i}{i}{\bar\tB,\tA}
\hspace{0.1cm}\cup\hspace{0.1cm} \bigcup_{i=0}^{\frac{N-5}2}\,\, \bigcup_{j=0}^{\big\lfloor\frac12\big(\frac{N-5}2-i\big)\big\rfloor} \BBAB{\frac{N+3}2 + i}{i+j+1}{2(i+j+2)}{i}{i+1}{\bar\tA,\tA}\ .
\label{eq:SRp1}
\end{alignat}
\end{subequations}
Using \eqref{eq:cardB}, one can use binomial identities to show that the sum of the cardinalities of the above sets equals $\dim \stanp_N^d$ in all cases.

\paragraph{Separation between upper and lower half-planes.}
Writing down explicit expressions for the spectrum generating functions requires understanding how the zeros are split between the upper and lower halves of the plane.  Let us denote by $(\Abarone{M_1}{m_1} \smallsep\Aone{M_2}{m_2})$ the set of single-column diagrams in $\Aone{M_1+M_2}{m_1+m_2}$ for which the subconfigurations of the lower and upper half planes respectively belong to $\Aone{M_1}{m_1}$ and $\Aone{M_2}{m_2}$. The bar in $\Abarone{M_1}{m_1}$ is a reminder that this factor contributes powers of $\bar q$ in $\ZdNP$.

We first describe the splitting of the heights in the first strip. A configuration of zeros of the first strip is described by a single-column diagram of $\Aone{M}{m}$. From the selection rules, we see that for the leading eigenvalues, $M$ is a number that grows linearly with the system size $N$, whereas $m$ remains small.\footnote{Here $M$ is the number of sites of the column diagram in $\Aone{M}{m}$, not the vertical width of the $M\times N$ lattice.} The zeros of these leading eigenvalues lie at a distance $\sim \ln N$ from the real axis and the separating line. There is thus some arbitrariness in the choice of the position of the separation. One can choose
\be
\Aone{M}{m} \quad \rightarrow \quad \bigcup_{k_1 = 0}^m\hspace{0.2cm} \Big(\Abarone{\lfloor \frac {M}2\rfloor - \epsilon}{k_1} \sep \Aone{\lfloor \frac {M+1}2 \rfloor + \epsilon}{m-k_1}\Big)
\label{eq:k1splitting}
\ee
where $\epsilon$ is an arbitrary fixed number that is much smaller than $M$. The union over $k_1$ allows for the zeros to be split between the upper and lower half-planes in any possible way. Depending on $\epsilon$, some eigenvalues are treated differently in terms of the contribution of their zeros to $\Delta$ and $\bar \Delta$. Indeed for finite $N$, the numbers $t^2_-$ and $t^2_+$ of zeros in each half-plane depends on $\epsilon$, and likewise for the resulting conformal weights \eqref{eq:Deltagamma0}. However, this does not occur for the leading eigenstates, namely those that correspond to the conformal states in the scaling limit. The finitized partition functions given below depend on $\epsilon$, but by varying $\epsilon$ (while keeping $\epsilon \ll M$), the resulting expressions only change by powers of $q$ and/or $\bar q$ that are linear in $N$. For $\epsilon \ll M$, the scaling limits of these partition functions are therefore independent of $\epsilon$. We note that the data presented in \cref{fig:patsP60,fig:patsP62} corresponds to $\epsilon = 0$.

Crucially, the bulk behavior $K_{\text{bulk}}=\sigma$ is a function of the variable $k_1$, which controls the separation between upper and lower halves in \eqref{eq:k1splitting}:
\be
\label{eq:sigmak1}
\sigma = 
\left\{\begin{array}{ll}
(-1)^{k_1}& d \equiv 0 \Mod 3,\\[0.1cm]
(-1)^{k_1+1} & d \equiv 1,2 \Mod 3.\\
\end{array}\right.
\ee
The value of $k_1$ therefore dictates both the selection of $\iota(\sigma)$ in \eqref{eq:Deltagamma0} and the sign $\varepsilon$ in \eqref{eq:ZdNP}.

The second strip does not allow as much arbitrariness in the choice of the separation line. This is because for the leading eigenvalues, the number of sites in the second strip does not scale with the system size $N$. Let us denote by $[(\Abarone{M_1}{m_1} \smallsep\Aone{M_2}{m_2})(\Abarone{L_1}{n_1} \smallsep\Aone{L_2}{m_2})(\Abarone{P_1}{\ell_1} \smallsep\Aone{P_2}{\ell_2})]$, with $L_1+L_2 = P_1 + P_2$, the set of configurations $(\sigma, \sigma', \sigma '')$ in $\BB{M_1+M_2}{L_1+L_2}{m_1+m_2}{n_1+n_2}{\ell_1+\ell_2}$ for which $\sigma \in (\Abarone{L_1}{n_1} \smallsep\Aone{L_2}{m_2})$, $\sigma' \in (\Abarone{L_1}{n_1} \smallsep\Aone{L_2}{m_2})$ and $\sigma'' \in (\Abarone{P_1}{\ell_1} \smallsep\Aone{P_2}{\ell_2})$. We discuss the cases $d \equiv 0 \Mod 3$ and $d \equiv 1,2 \Mod 3$ separately. 

For $d \equiv 0 \Mod 3$, we note that all sets in the selection rule \eqref{eq:SRp0t} are of the form $\BB{M}{L}{2L}{n}{\ell}$. The  number of occupied sites in the first strip is twice the number of sites in the second strip. From our numerical data, we find that the splitting of the zeros in the second strip follows the same rule: If $\sigma$ has $k_1$ zeros in the lower half of the first strip, then the total number of sites in the lower half-plane of the second strip (namely those of $\sigma'$ plus those of $\sigma''$) is also equal to $k_1$. The splitting is therefore as follows:
\be\label{eq:sep0}
\BB{M}{L}{2L}{n}{\ell} \quad \rightarrow \quad \bigcup_{k_1 = 0}^{2L}\bigcup_{k_2 = 0}^{n}\bigcup_{k_3 = 0}^{\ell}\hspace{0.2cm} 
\bigg[\Big(\Abarone{\lfloor \frac {M}2\rfloor - \epsilon}{k_1} \sep \Aone{\lfloor \frac {M+1}2 \rfloor + \epsilon}{2L-k_1}\Big)
\Big(\Abarone{\lfloor \frac {k_1}2\rfloor}{k_2} \sep \Aone{L-\lfloor \frac {k_1}2 \rfloor}{n-k_2}\Big)
\Big(\Abarone{\lfloor \frac {k_1+1}2\rfloor}{k_3} \sep \Aone{L-\lfloor \frac {k_1+1}2 \rfloor}{\ell-k_3}\Big)\bigg].
\ee 
Thus, if $k_1$ is even, the sites are split evenly between $\sigma'$ and $\sigma''$. If $k_1$ is odd, we choose the lower half-plane of $\sigma''$ to have an extra site compared to the lower half-plane of $\sigma'$. Had we chosen the opposite convention, the finitized characters given below would be slightly different, but their scaling limits would be identical.

For $d \equiv 1,2 \Mod 3$, the sets in the selection rules \eqref{eq:SRp1t} and \eqref{eq:SRp2t} are of the form $\BBAB{M}{L}{2(L+1)}{n}{\ell}{\tX,\tY}$ with $\tX, \tY \in \{\tA, \tB\}$. In this case, the splitting is such that the total number of sites in the second strip equals the number of occupied sites in the lower half of the first strip, minus one:
\be
\label{eq:sep12}
\BBAB{M}{L}{2(L+1)}{n}{\ell}{\tX,\tY} \quad \rightarrow \quad \bigcup_{k_1 = 0}^{2(L+1)}\bigcup_{k_2 = 0}^{n}\bigcup_{k_3 = 0}^{\ell}\hspace{0.2cm} 
\bigg[\Big(\Abarone{\lfloor \frac {M}2\rfloor - \epsilon}{k_1} \sep \Aone{\lfloor \frac {M+1}2 \rfloor + \epsilon}{2(L+1)-k_1}\Big)
\Big(\Abarone{\lfloor \frac {k_1-1}2\rfloor}{k_2} \sep \Aone{L-\lfloor \frac {k_1-1}2 \rfloor}{n-k_2}\Big)
\Big(\Abarone{\lfloor \frac {k_1}2\rfloor}{k_3} \sep \Aone{L-\lfloor \frac {k_1}2 \rfloor}{\ell-k_3}\Big)\bigg].
\ee 

In computing $\ZdNP$, each $[(\Abarone{M_1}{m_1} \smallsep\Aone{M_2}{m_2})(\Abarone{L_1}{n_1} \smallsep\Aone{L_2}{m_2})(\Abarone{P_1}{\ell_1} \smallsep\Aone{P_2}{\ell_2})]$ will contribute
\be\label{eq:Bgeneratingfunction}
\bar q^{\bar \Delta_{\textrm{min}}}q^{\Delta_{\textrm{min}}} \qqbinom{M_1}{m_1} \qqbinom{L_1}{n_1} \qqbinom{P_1}{\ell_1} \qbinom{M_2}{m_2}\qbinom{L_2}{n_2} \qbinom{P_2}{\ell_2}
\ee
where $\Delta_{\textrm{min}}$ and $\bar\Delta_{\textrm{min}}$ are the conformal weights of the minimal configurations.
\paragraph{Selection rules for the integers and spectrum generating functions.}
Remarkably, the prescription for the integers $k^j_\pm$ and $\ell^i_\pm$ is very similar to the one found for the strip. We formulate it based on data collected from our computer implementation, namely approximations to $k^j_\pm$ and $\ell^i_\pm$ obtained using \eqref{eq:integerKP} and \eqref{eq:integerLP}, for all the eigenvalues in $\stanp_N^d$ for $N\le12$. 
In \cref{fig:patsP60,fig:patsP62}, the 
values of these integers are displayed with each pattern of zeros for $\stanp_6^0$ and $\stanp_6^2$. The prescription is as follows. For any configuration in $\BB{M}{L}{m}{n}{\ell}$,
\begin{itemize}
\item[(i)] the heights of the single-column diagram (corresponding to the $k^j_\pm$) are labelled from $0$ to $M-1$ for $d\equiv 1,2 \Mod 3\, (\tA)$ or $(\bar \tA)$.  They are labeled from $1$ to $M$ for $d\equiv 0 \Mod 3$, and likewise for $d\equiv 1,2 \Mod 3\, (\tB)$ or $(\bar \tB)$; 
\item[(ii)] the heights of the double-column diagram (corresponding to the $\ell^i_\pm$) are labelled from $1$ to $L$ in all cases.
\end{itemize}

Using the selection rules and the formula \eqref{eq:Deltagamma0} for the conformal weights, we can formulate expressions for $\ZdNP$ by writing down the generating function for each set in \eqref{eq:SRp} and taking their sum over $i$ and $j$. Because of the similarities between the current prescription and the one in \cref{sec:excitations} for the strip, the energies of the minimal configuration of each set in \eqref{eq:SRp} is obtained from \eqref{eq:Es} under the suitable specifications of $m, n$ and $\ell$. The resulting expressions are complicated power-law series in $q$ and $\bar q$, involving products of three $q$-binomials and three $\bar q$-binomials, with sums over five integers:  $i$, $j$, $k_1$, $k_2$ and $k_3$. Moreover for $d \equiv 1,2 \Mod 3$, there are four contributions to the generating functions depending on the subcases $\bar \tA$ or $\bar \tB$ and $\tA$ or $\tB$. The resulting expressions are collected in \cref{sec:Pcpf}. 

\paragraph{Scaling behavior.}
Unlike in the boundary case, for the periodic case we are unable to directly simplify the expressions for the finitized conformal partition functions. We are nevertheless able to perform such simplifications in the scaling limit:
\be
\ZdNP\xrightarrow{M,N \rightarrow \infty} Z_d(q,\bar q).
\ee 
Indeed, in \cref{sec:Pcpf}, we use the $q$-binomial identities of \cref{sec:qids} to extract the following expressions for $Z_d(q,\bar q)$:
\begin{subequations}
\label{eq:Zdresults}
\begin{alignat}{2}
&d \equiv 0 \Mod 3 : \quad
&&Z_d(q,\bar q) = 
\frac1{(q)_\infty(\bar q)_\infty} \sum_{\ell\in \mathbb Z} \Big(q^{\Delta_{0,3\ell -d }} \bar q^{\Delta_{0,3 \ell}} + (-1)^M q^{\Delta_{1,3\ell - d}} \bar q^{\Delta_{1,3 \ell}} \Big),\label{eq:Z0results}\\[0.2cm]
&d \equiv 1 \Mod 3 : \quad
&&Z_d(q,\bar q) = 
\frac1{(q)_\infty(\bar q)_\infty} \sum_{\ell\in \mathbb Z} \Big( q^{\Delta_{0,3\ell -d +2}} \bar q^{\Delta_{0,3 \ell+2}} + (-1)^Mq^{\Delta_{1,3\ell - d+2}} \bar q^{\Delta_{1,3 \ell+2}} \Big),\label{eq:Z1results}\\[0.2cm]
&d \equiv 2 \Mod 3  : \quad
&&Z_d(q,\bar q) = 
\frac1{(q)_\infty(\bar q)_\infty} \sum_{\ell\in \mathbb Z} \Big(q^{\Delta_{0,3\ell -d +1}} \bar q^{\Delta_{0,3 \ell+1}} + (-1)^Mq^{\Delta_{1,3\ell - d+1}} \bar q^{\Delta_{1,3 \ell+1}} \Big).\label{eq:Z2results}
\end{alignat}
\end{subequations}

\subsection{Torus partition functions}\label{sec:mipf}

Computing the partition function of a model described by the periodic Temperley-Lieb algebra on the $M\times N$ torus requires a proper understanding of the representation theory of this algebra. This was achieved for the $Q$-Potts model by Jacobsen and Richard \cite{RJ07} and by Aufgebauer and Kl\"umper for quantum spin chains \cite{AK10}. For the loop model, we set the weight of a non-contractible loop to $\alpha$, independent of its winding numbers around the torus. For $d >0$, let us denote by $\Tr_{d,j}$ the coefficients of the trace of $\Tb(u)^M$ on $\stanp_N^d$ in an expansion in $\omega$:
\be
\label{eq:traceofT}
\Tr\,\Tb(u)^M\Big|_{\stanp_N^d} = \sum_{j \in \mathbb Z}  \Tr_{d,j} \omega^j.
\ee
For finite $M$, the trace of $\Tb(u)^M$ is a Laurent polynomial in $\omega$, so there are finitely many contributions in the sum on the right-hand side of \eqref{eq:traceofT}. The partition function of the loop model for arbitrary $\alpha$ and $\beta$ then reads \cite{MDPR13}
\be
 \ZtorMN = \delta_{N\equiv 0\, \textrm{mod}\, 2}\ \tr\,\Tb(u)^M\Big|_{\stanp_N^0} +2 \hspace{-0.2cm} \sum_{\substack{1 \le d \le N \\ d \equiv N \Mod 2}} \sum_{j\in \mathbb Z} \ T_{j \wedge d}(\tfrac \alpha 2)\, \Tr_{d,j}\ , 
\ee
where $T_k(x)$ is the $k$-th Chebyshev polynomial of the first kind and $j \wedge d$ is the greatest common divisor of $j$ and $d$. Even if our current derivation does not reveal this, we know that $\ZtorMN\big|_{\alpha=1}=1$ for all parities of $M$ and $N$. For $\alpha = 2$, a remarkable simplification occurs: $T_{j\wedge d}(1) = 1$ and 
\be
\ZtorMN\Big|_{\alpha = 2} = \hspace{-0.2cm} \sum_{\substack{0 \le d \le N \\ d \equiv N\, \textrm{mod}\, 2}} \hspace{-0.2cm}(2-\delta_{d,0})\, \Tr\,\Tb(u)^M\Big|_{\stanp_N^d, \,\omega = 1}\ .
\ee
The torus conformal partition function is then obtained from the scaling limit of the finite-size partition function:
\be
\label{eq:Zmodifiedtrace}
\eE^{MN f_{\text{b}}(u)} \ZtorMN\Big|_{\alpha = 2} \xrightarrow{M,N \rightarrow \infty} \Ztor(q,\bar q) =  \sum_{\substack{d\ge 0\\[0.05cm] d \equiv N\, \textrm{mod}\, 2}}\hspace{-0.2cm}(2-\delta_{d,0})\, \ZdNP
\ee
with $q$ given in terms of the lattice data in \eqref{eq:qper}.

From here, the computation splits between the odd and even parity of $N$. From \eqref{eq:Zdresults}, we extend the definition of $Z_d(q,\bar q)$ to $d<0$. Using the relation $Z_{d}(q,\bar q) = Z_{-d}(q,\bar q)$, we find
\be
\Ztoreven = \sum_{d \in 2 \mathbb Z} Z_d(q,\bar q), \qquad \Ztorodd = \sum_{d \in 2 \mathbb Z + 1} Z_d(q,\bar q).
\ee
These can in turn be written in terms of the $u(1)$ characters $\varkappa_j(q) = \varkappa^6_j(q)$ given in \eqref{eq:u1chars}. Defining $y = \frac1{(q)_\infty(\bar q)_\infty}$, we find the following identities for $N$ odd:
\begin{subequations}
\begin{alignat}{2}
&y\sum_{d \in 6 \mathbb Z+3} \sum_{\ell \in \mathbb Z} q^{\Delta_{0,3\ell -d }} \bar q^{\Delta_{0,3 \ell}}= \varkappa_0(q)\varkappa_6(\bar q) + \varkappa_6(q)\varkappa_0(\bar q),
\\
&y\sum_{d \in 6 \mathbb Z+3} \sum_{\ell \in \mathbb Z} q^{\Delta_{1,3\ell -d }} \bar q^{\Delta_{1,3 \ell}}= 2 |\varkappa_3(q)|^2, 
\\
&y\Big(\sum_{d \in 6 \mathbb Z+1} \sum_{\ell \in \mathbb Z} q^{\Delta_{0,3\ell -d+2}} \bar q^{\Delta_{0,3 \ell+2}} + \sum_{d \in 6 \mathbb Z+5} \sum_{\ell \in \mathbb Z} q^{\Delta_{0,3\ell -d+1 }} \bar q^{\Delta_{0,3 \ell+1}} \Big)= 
2\varkappa_2(q)\varkappa_4(\bar q) + 2\varkappa_4(q)\varkappa_2(\bar q),
\\
&y\Big(\sum_{d \in 6 \mathbb Z+1} \sum_{\ell \in \mathbb Z} q^{\Delta_{1,3\ell -d+2}} \bar q^{\Delta_{1,3 \ell+2}} + \sum_{d \in 6 \mathbb Z+5} \sum_{\ell \in \mathbb Z} q^{\Delta_{1,3\ell -d+1 }} \bar q^{\Delta_{1,3 \ell+1}} \Big)= 
2|\varkappa_1(q)|^2 + 2|\varkappa_5(q)|^2.\label{papK1K5}
\end{alignat}
\end{subequations}
For $N$ even, we find
\begin{subequations}
\begin{alignat}{2}
&y\sum_{d \in 6 \mathbb Z} \sum_{\ell \in \mathbb Z} q^{\Delta_{0,3\ell -d }} \bar q^{\Delta_{0,3 \ell}}= |\varkappa_0(q)|^2 + |\varkappa_6(q)|^2, \label{eq:forexample}\\
&y\sum_{d \in 6 \mathbb Z} \sum_{\ell \in \mathbb Z} q^{\Delta_{1,3\ell -d }} \bar q^{\Delta_{1,3 \ell}}= 2 |\varkappa_3(q)|^2, \\
&y\sum_{d \in 6 \mathbb Z+2} \sum_{\ell \in \mathbb Z} q^{\Delta_{0,3\ell -d+1 }} \bar q^{\Delta_{0,3 \ell+1}} + y \sum_{d \in 6 \mathbb Z+4} \sum_{\ell \in \mathbb Z} q^{\Delta_{0,3\ell -d+2}} \bar q^{\Delta_{0,3 \ell+2}} = 2|\varkappa_2(q)|^2 + 2|\varkappa_4(q)|^2, \\
&y\sum_{d \in 6 \mathbb Z+2} \sum_{\ell \in \mathbb Z} q^{\Delta_{1,3\ell -d+1 }} \bar q^{\Delta_{1,3 \ell+1}} + y\sum_{d \in 6 \mathbb Z+4} \sum_{\ell \in \mathbb Z} q^{\Delta_{1,3\ell -d+2}} \bar q^{\Delta_{1,3 \ell+2}} = 2\varkappa_1(q)\varkappa_5(\bar q) + 2\varkappa_5(q)\varkappa_1(\bar q).\label{eq:K1K5}
\end{alignat}
\end{subequations}
These relations are obtained using simple algebraic manipulations. To illustrate, \eqref{eq:forexample} is proved as follows:
\begin{alignat}{2}
y\sum_{d \in 6 \mathbb Z} \sum_{\ell \in \mathbb Z} q^{\Delta_{0,3\ell -d }} \bar q^{\Delta_{0,3 \ell}} 
&= y \sum_{k,\ell \in \mathbb Z} q^{\Delta_{0,3\ell -6k}} \bar q^{\Delta_{0,3 \ell}} 
= y \sum_{r=0,1}\sum_{j,k \in \mathbb Z} q^{\Delta_{0,6(j-k)+3r}} \bar q^{\Delta_{0,6 j + 3r}} \nonumber\\
& = y \sum_{r=0,1}\sum_{i,j\in \mathbb Z} q^{\Delta_{0,6i+3r}} \bar q^{\Delta_{0,6 j + 3r}} = y \Big| \sum_{i\in \mathbb Z} q^{\Delta_{0,6i}} \Big|^2 + y \Big| \sum_{i\in \mathbb Z} q^{\Delta_{0,6i+3}} \Big|^2 \nonumber\\
& = |\varkappa_0(q)|^2 + |\varkappa_6(q)|^2.\label{eq:K1K1K5K5}
\end{alignat}
For $N$ even, \eqref{eq:K1K5} is the only seemingly anti-diagonal contribution. It can be changed to a diagonal contribution using the relation
\be
\label{eq:K1K5relation}
\varkappa_1(q)\varkappa_5(\bar q) + \varkappa_5(q)\varkappa_1(\bar q) = |\varkappa_1(q)|^2 + |\varkappa_5(q)|^2-1,
\ee
which itself follows from $\varkappa_1(q)-\varkappa_5(q) = 1$. Using \eqref{eq:K1K5relation}, one can also write \eqref{papK1K5} as an anti-diagonal contribution.
Our final expressions for the conformal torus partition functions are
\begin{subequations}
\label{eq:subfinalZ}
\begin{alignat}{2}
\Ztorodd &=\varkappa_0(q)\varkappa_6(\bar q) + 2(-1)^M \varkappa_1(q)\varkappa_5(\bar q) + 2 \varkappa_2(q)\varkappa_4(\bar q) + 2(-1)^M |\varkappa_3(q)|^2 \nonumber\\[0.1cm]
&\hspace{1cm}+ 2 \varkappa_4(q)\varkappa_2(\bar q) + 2(-1)^M \varkappa_5(q)\varkappa_1(\bar q) + \varkappa_6(q)\varkappa_0(\bar q) + 2(-1)^M, \\[0.2cm]
\Ztoreven &= |\varkappa_0(q)|^2+ 2(-1)^M |\varkappa_1(q)|^2 + 2 |\varkappa_2(q)|^2+2(-1)^M |\varkappa_3(q)|^2 \nonumber\\[0.1cm]
&\hspace{1cm}+2 |\varkappa_4(q)|^2+ 2(-1)^M |\varkappa_5(q)|^2 + |\varkappa_6(q)|^2 -2(-1)^M.
\end{alignat}
\end{subequations}
These can be written in compact form:
\be
\label{eq:finalZ}
\Ztorodd = \sum_{j=0}^6 \Big((-1)^{M j} d_j^6 \varkappa_j(q)\varkappa_{6-j}(\bar q)\Big) + 2(-1)^M, \quad \Ztoreven = \sum_{j=0}^6 \Big((-1)^{M j} d_j^6 |\varkappa_j(q)|^2\Big) - 2(-1)^M,
\ee
with $d_j^n$ defined in \eqref{eq:dj}.
In terms of the $u(1)$ characters, the odd and even parities of $N$ thus correspond to different mixtures of diagonal and anti-diagonal sectors. For $M$ and $N$ even, the result agrees with the conjecture \eqref{Zpp} of Pearce and Rasmussen\cite{PRcosetGraphs11} with $n_{2,3} = -1$ so that $Z_{p,p'}(q)=Z^{\text{Circ}}_{p,p'}(q)$. 
It should also be noted that the partition functions \eqref{eq:finalZ} for $M$ odd are not genuine conformal partition functions, since some coefficients in the $q,\bar q$ expansions are negative.

\paragraph{Modular invariance and covariance.}
The $u(1)$ characters behave as follows under the $T$ and $S$ transformations of the modular group:
\begin{subequations}
\label{eq:transforms}
\begin{align}
\label{eq:Ttransform}
T:\qquad\varkappa^{n}_j(\eE^{2 \pi \ir(\tau+1)}) &= \exp\big(2 \pi \ir (\tfrac{j^2}{4n}-\tfrac1{24})\big) \varkappa^{n}_j(\eE^{2 \pi \ir\tau}),
\\[0.15cm]
\label{eq:Stransform}
S:\qquad\ \ \varkappa^{n}_j(\eE^{-2 \pi \ir/\tau}) &= \sum_{k=0}^{2n-1} S_{jk}\, \varkappa^{n}_k(\eE^{2 \pi \ir\tau}) = \frac{1}{\sqrt{2n}} \sum_{k=0}^{2n-1} \eE^{-\pi \ir j k/n} \varkappa^{n}_k(\eE^{2 \pi \ir\tau}).
\end{align}
\end{subequations}
It follows that diagonal and anti-diagonal terms in \eqref{eq:finalZ} transform trivially under $T$:
\begin{subequations}
\begin{alignat}{2}
|\varkappa^{n}_j(\eE^{2 \pi \ir(\tau+1)})|^2 &= |\varkappa^{n}_j(\eE^{2 \pi \ir(\tau)})|^2, \\[0.15cm]
\varkappa^{n}_j(\eE^{2 \pi \ir(\tau+1)})\varkappa^{n}_{n-j}(\eE^{-2 \pi \ir(\tau+1)}) &= (-1)^n \varkappa^{n}_j(\eE^{2 \pi \ir\tau})\varkappa^{n}_{n-j}(\eE^{-2 \pi \ir\tau}),
\end{alignat}
\end{subequations}
where we recall that $n = p p' = 6$ for percolation. The partition functions $\Ztorodd$ and $\Ztoreven$ are thus invariant under the action of $T$ for all parities of $M$ and $N$.

The $S$ matrix describing the transformations of the $u(1)$ characters is obtained from \eqref{eq:Stransform} and $\varkappa^{n}_j(q) =\varkappa^{n}_{2n-j}(q)$. For $n= pp' = 6$, in the basis $\{\varkappa_j(q),\, j = 0, \dots, 6\}$, we have
\be
S = \frac1{2 \sqrt 3}
\left(
\begin{smallmatrix}
 1 & 2 & 2 & 2 & 2 & 2 & 1 \\
 1 & \sqrt{3} & 1 & 0 & -1 & -\sqrt{3} & -1 \\
 1 & 1 & -1 & -2 & -1 & 1 & 1 \\
 1 & 0 & -2 & 0 & 2 & 0 & -1 \\
 1 & -1 & -1 & 2 & -1 & -1 & 1 \\
 1 & -\sqrt{3} & 1 & 0 & -1 & \sqrt{3} & -1 \\
 1 & -2 & 2 & -2 & 2 & -2 & 1
\end{smallmatrix}
\right).
\ee
Applying the $S$ transform to \eqref{eq:finalZ}, we find that the torus partition functions transform as follows:
\begin{subequations}
\begin{alignat}{2}\label{eq:covariance}
\Ztoroddodd (\eE^{-2 \pi \ir/\tau}) &= \Ztoroddodd (\eE^{2 \pi \ir\tau}),\\[0.15cm]
\Ztoroddeven (\eE^{-2 \pi \ir/\tau}) &= \Ztorevenodd (\eE^{2 \pi \ir\tau}),\\[0.15cm]
\Ztoreveneven (\eE^{-2 \pi \ir/\tau}) &= \Ztoreveneven (\eE^{2 \pi \ir\tau}).
\end{alignat}
\end{subequations}
Therefore the partition functions for $M,N$ both odd or both even are modular invariant, whereas the mixed cases are modular covariant, as they map to one another under the $S$ transformation.

%
\section{Conclusion}\label{sec:conclusion}
%
In this paper, we derive and analyse truncated $T$- and $Y$-systems of functional relations satisfied by the transfer matrix eigenvalues of critical bond percolation considered as the loop model ${\cal LM}(2,3)$. Using analyticity properties of the eigenvalues and empirically based selections rules, we solve non-linear integral equations in the form of $D_3$ TBA equations to obtain exact expressions for the finite-size corrections and conformal data. These calculations are carried out for both the geometry of the strip with the boundaries consisting of simple half-arcs,
and for periodic boundary conditions. Formulating selection rules encoding the patterns of zeros for all finite excitations, we give explicit expressions for finite-size spectrum generating functions in the various sectors. On the strip, our refined finitized characters reproduce the finitized Kac characters of \cite{PRZ06}. Additionally, fixing the weight of the non-contractible loops to $\alpha=2$, we obtain in the continuum scaling limit the conformal cylinder and torus partition functions.

Our expressions for the conformal cylinder partition functions are to be compared to those found using Coulomb gas arguments in \cite{SB89,C06}. In particular, the partition function given in \cite[equation 1]{C06} with $g = \frac23$ almost coincides with our result \eqref{eq:Zcylgen}, with the difference that \eqref{eq:Zcylgen} considers separately the partition functions for the $N$ odd and $N$ even cases, whereas the expression given in \cite{C06} is the sum of the two. The Coulomb gas argument is easily fixed by noting that for the dense loop model, the number of contractible loops has the same parity as $N$, and this produces the correct partition functions.

Our expressions for the conformal torus partition functions depend on the parities of the lattice dimensions $M$ and $N$. In each of the four cases, $\Ztor(q,\bar q)$ is a simple sesquilinear form \eqref{eq:subfinalZ} in the $u(1)$ characters $\varkappa_j^6(q)$. Somewhat surprisingly, for $M,N$ both even, the MIPF 
\be
Z_{2,3}(q)=Z^{\text{Circ}}_{2,3}(q)= |\varkappa_0(q)|^2+ 2 |\varkappa_2(q)|^2+2 |\varkappa_3(q)|^2 
+2 |\varkappa_4(q)|^2+ 2[\varkappa_1(q)\varkappa_5(\bar q) + \varkappa_5(q)\varkappa_1(\bar q)] + |\varkappa_6(q)|^2
\ee
is non-diagonal and not of diagonal $A$-type.  Consequently, it differs from the diagonal A-type MIPF of the triplet model~\cite{GRW10}. Nevertheless, it is of the form conjectured by Pearce and Rasmussen \cite{PRcosetGraphs11} as in \eqref{Zppa}:
\be
Z_{p,p'}(q)=Z^{\text{Proj}}_{p,p'}(q)+n_{p,p'} Z^{\text{Min}}_{p,p'}(q),\qquad n_{p,p'} \in {\mathbb Z}\label{Z=ZnZ}
\ee
since our lattice derivation shows that $n_{2,3} = -1$. Based on the assumption of a diagonal A-type MIPF, Pearce and Rasmussen instead conjectured $n_{p,p'}=2$ for all $p,p'$.
Knowing what we now know about the action of the modified trace, the result $n_{2,3} = -1$ can be understood by observing that, if $Z_{2,3}(q)$ has the form in \eqref{Z=ZnZ}, its $q,\bar q$ expansion starts out as
\be
\label{eq:Z23expansion}
Z_{2,3}(q)=(q\bar q)^{-1/24} + (1+n_{2,3}) + 2 (q\bar q)^{1/8} + 2 (q\bar q)^{1/3} + \cdots 
\ee
The coefficient $n_{2,3}$ only appears in the constant term, with $(1+n_{2,3})$ counting the number of states with conformal weights $\Delta = \bar \Delta = 0$, namely the Razumov-Stroganov eigenstates.

The MIPF found from the loop model with $\alpha = 2$ and $M, N$ both even is in fact equal to the partition function of the six-vertex model on a torus with no twist. Indeed, the spin-chain representation of $\eptl_n(\alpha,\beta)$ with magnetisation $m$ has the same spectra \cite{MDSA13no1}, including degeneracies, as the standard module $\stanp_n^d$ with $d=\frac m 2$. Because the representations with magnetisation $\pm m$ are isomorphic, taking the normal trace of $\Tb(u)$ in the full spin-chain representation precisely produces \eqref{eq:Zmodifiedtrace}: The sector with $m=0$ is singly degenerate, whereas all others are doubly degenerate. Since there are no Razumov-Stroganov eigenstates for $N$ even in the untwisted case \cite{RS01,RS01no2,S01}, the expansion \eqref{eq:Z23expansion} implies that $n_{2,3} = -1$. 
Our result \eqref{eq:Zdresults} for $Z_d(q,\bar q)$ should then be compared with the XXZ partition function \cite{FSZ87,ABGR88,AGR89}. An explicit comparison with \cite[equation 1.8]{AGR89} specialised to $h=\frac38$, $Q = \frac d2$, $l = 0$ and $z=q$ reveals that the partition functions coincide for $M$ even. For $M$ odd, however, \eqref{eq:Zdresults} incorporates factors of $(-1)^M$ to give the correct conformal partition functions for $M$ odd.

Ultimately, one of our goals is to derive the conformal torus partition functions for all the logarithmic minimal models $\mathcal{LM}(p,p')$, for all parities of $M$ and $N$. It is then natural to compare the results found here for percolation with those for critical dense polymers, corresponding to $\mathcal{LM}(1,2)$. For $M$ even, the conformal torus partition functions of this model were found in \cite{PRVcyl2010,MDPR13} and expressed in terms of $\mathcal W$-irreducible characters. In \cref{sec:CDPpartitionfunctions}, we present the conformal partition functions for all parities of $M$ and $N$ in terms of $u(1)$ characters. For $M$ and $N$ both even, we note that the torus partition function is just $Z^{\textrm{Circ}}_{1,2}(q)$, hinting at the possibility that for $\mathcal{LM}(p,p')$, the torus partition function for $M$ and $N$ even may equal $Z^{\textrm{Circ}}_{p,p'}(q)$. For the other parities, the results for the two models are quite different and it is not possible to identify a general pattern. Clearly, more data is required before the full picture can emerge describing the conformal partition function in this case, and the structure of the indecomposable representations in general.
From the lattice perspective, the study of the latter was indeed initiated in \cite{PRZ06} and \cite{RS07}.
This has become the subject of an extensive investigation in the following years, and yet much is still to be understood about the indecomposable structures \cite{Gurarie,GR08,GRW10,RGW14,GRSV15} of bulk logarithmic conformal field theories including percolation.

This work leaves open a plethora of avenues for future research on an analytic approach to the logarithmic minimal models ${\cal LM}(p,p')$. In particular, in this paper, we only consider boundary conditions on the strip corresponding to conformal Kac modules with highest weight $\Delta_{1,s}$. However, the methods of this paper should extend to the more general Kac modules with highest weight $\Delta_{r,s}$, which are realised on the lattice by including a seam on the boundary \cite{PRZ06,PTC2015,MDRD2015}. These methods should also extend to boundary conditions described by the one- and two-boundary Temperley-Lieb algebra~\cite{JS08,DJS09,PRT14,BPT2016}.
In all these cases, it is expected that the transfer matrix eigenvalues will satisfy the same universal $Y$-system, encoded by the Dynkin diagram $D_{p'}$. However, the analyticity properties will differ in the various cases, leading to the different Kac conformal weights describing the finite-size corrections. 
Ultimately, it would also be of interest to obtain the modular invariants of the dilute loop models. Needless to say, the analytic and combinatorial classification problems in all these challenges promise to be formidable.

\subsection*{Acknowledgments}

AMD was supported by the Belgian Interuniversity Attraction Poles Program P7/18 through the network DYGEST (Dynamics, Geometry and Statistical Physics) and by the FNRS fellowship CR28075116. 
AMD and AK acknowledge the support of the ERC grant {\it Loop models, integrability and combinatorics} and are grateful for the kind hospitality of Paul Zinn-Justin and the LPTHE where early stages of this work were done.
AMD and PAP acknowledge the hospitality of the University of Wuppertal and are grateful for the kind hospitality of Holger Frahm at the ITP in Hannover where later stages of this work were done.
All authors were supported by DFG through the program FOG 2316.
PAP thanks the APCTP, Pohang for hospitality during the writing of this paper.
The authors thank Paul Zinn-Justin, J\o rgen Rasmussen and Yacine Ikhlef for useful discussions. 

\bigskip

\bigskip

\appendix

%
\section{Integrals involving Rogers dilogarithms}\label{sec:Rogers}
%

In this section, we study the following integrals:
\begin{subequations}\label{defintegrals}
\begin{alignat}{2}
\mathcal I_1 & = \int_{0}^{\mathsf a^1(\infty)} \dd \mathsf a \bigg(\frac{\ln \left(1+ \mathsf a\right)} {\mathsf a} - \frac{\ln |\mathsf a|}{1+\mathsf a}\bigg),\\ 
\mathcal I_2 & = \int_{\mathsf a^2(-\infty)}^{\mathsf a^2(\infty)} \dd \mathsf a \bigg(\frac{\ln \left(1+ \eE^{3\ir\gamma}\mathsf a\right)} {\mathsf a} - \frac{\eE^{3\ir\gamma}\ln |\mathsf a|}{1+\eE^{3\ir\gamma}\mathsf a}\bigg),\\
\mathcal I_3 & = \int_{\mathsf a^2(-\infty)}^{\mathsf a^2(\infty)} \dd \mathsf a \bigg(\frac{\ln \left(1+ \eE^{-3\ir\gamma}\mathsf a\right)} {\mathsf a} - \frac{\eE^{-3\ir\gamma}\ln |\mathsf a|}{1+\eE^{-3\ir\gamma}\mathsf a}\bigg),
\end{alignat}
\end{subequations}
with 
\be
\mathsf a^1(\infty) = 4 \cos^2\gamma-1, \qquad \mathsf a^2(\infty) = 2 \cos\gamma,
\ee 
and consider two cases, namely $\mathsf a^2(-\infty) = \sigma=\pm 1$. All the integrals coming from the dilogarithm technique in
\cref{sec:fsc,sec:Pfsc} can be written in terms of $\mathcal K_\sigma(\gamma)$:
\begin{equation}
\mathcal K_\sigma(\gamma)=\mathcal I_1+\mathcal I_2+\mathcal I_3.
\end{equation}
Because $\mathcal K_\sigma(\gamma) = \mathcal K_\sigma(-\gamma) = \mathcal K_\sigma(\gamma+ 2\pi)$, we restrict $\gamma$ to the interval $[0,\pi]$.

We perform the calculations by taking derivatives of $\mathcal K_\sigma=\mathcal K_\sigma(\gamma)$ with
respect to $\gamma$. This removes all integrals and allows for calculations
with explicit rational functions and logarithms. In some instances, we need to distinguish
between the two cases $\sigma=\pm 1$.

First we perform integration by parts on the second integrand in each integral and obtain
\begin{subequations}
\begin{alignat}{2}
\mathcal I_1 & = 2\int_{0}^{\mathsf a^1(\infty)} \dd \mathsf a \frac{\ln
  \left(1+ \mathsf a\right)}{\mathsf a} -\ln(1+\mathsf a)\ln |\mathsf a|\Big|_0^{\mathsf a^1(\infty)}\ ,\\ 
\mathcal I_2 & = 2\int_{\mathsf a^2(-\infty)}^{\mathsf a^2(\infty)} \dd \mathsf a \frac{\ln \left(1+ \eE^{3\ir\gamma}\mathsf a\right)} {\mathsf a} - \ln\big(1+\eE^{3\ir\gamma}\mathsf a\big)\ln |\mathsf a|\Big|_{\mathsf a^2(-\infty)}^{\mathsf a^2(\infty)}\ ,\\
\mathcal I_3 & = 2\int_{\mathsf a^2(-\infty)}^{\mathsf a^2(\infty)} \dd \mathsf a \frac{\ln \left(1+ \eE^{-3\ir\gamma}\mathsf a\right)} {\mathsf a} - \ln\big(1+\eE^{-3\ir\gamma}\mathsf a\big)\ln |\mathsf a|\Big|_{\mathsf a^2(-\infty)}^{\mathsf a^2(\infty)}\ .
\end{alignat}
\end{subequations}
Next we use a substitution of the variable of integration $\mathsf a \mapsto
\eE^{\pm 3\ir\gamma}\mathsf a$ which moves the $\gamma$ dependence from the
integrand to the terminals, and use explicit expressions for the terminals in terms
of $\qo=\eE^{\ir\gamma}$:
\begin{subequations}
\begin{alignat}{2}
\mathcal I_1 & = 2\int_{0}^{\qo^2+1+\qo^{-2}} \dd \mathsf a \frac{\ln
  \left(1+ \mathsf a\right)}{\mathsf a} -\ln(1+\mathsf a)\ln |\mathsf a|\Big|_{0}^{\qo^2+1+\qo^{-2}}\ ,\\ 
\mathcal I_2 & = 2\int_{\sigma \qo^3}^{\qo^2+\qo^4} \dd \mathsf a \frac{\ln \left(1+\mathsf a\right)} {\mathsf a} - \ln\big(1+\qo^3\mathsf a\big)\ln |\mathsf a|\Big|_{\sigma}^{\qo+\qo^{-1}}\ ,\\
\mathcal I_3 & = 2\int_{\sigma \qo^{-3}}^{\qo^{-2}+\qo^{-4}} \dd \mathsf a \frac{\ln \left(1+\mathsf a\right)} {\mathsf a} - \ln\big(1+\qo^{-3}\mathsf a\big)\ln |\mathsf a|\Big|_{\sigma}^{\qo+\qo^{-1}}\ .
\end{alignat}
\end{subequations}
Note that the explicit log-log terms vanish at the lower terminals. The
derivatives with respect to $\qo$ read
\begin{alignat}{2}
\qo\frac{\dd}{\dd \qo}\mathcal K_\sigma=&
+2\frac{2 \qo^2-2 \qo^{-2}}{\qo^2+1+\qo^{-2}}\ln\big(\qo^2+2+\qo^{-2}\big)\nonumber\\
&+2\frac{2\qo^2+4\qo^{4}}{\qo^2+\qo^{4}}\ln\big(1+\qo^2+\qo^{4}\big)
-6\ln\big(1+\sigma \qo^3\big)\nonumber\\
&+2\frac{-2\qo^{-2}-4\qo^{-4}}{\qo^{-2}+\qo^{-4}}\ln\big(1+\qo^{-2}+\qo^{-4}\big)
+6\ln\big(1+\sigma \qo^{-3}\big)\nonumber\\
&-\frac{2 \qo^2-2 \qo^{-2}}{\qo^2+2+\qo^{-2}}\ln\big|\qo^2+1+\qo^{-2}\big|
-\frac{2 \qo^2-2 \qo^{-2}}{\qo^2+1+\qo^{-2}}\ln\big(\qo^2+2+\qo^{-2}\big)\nonumber\\
&-\frac{2 \qo^2+4 \qo^{4}}{1+\qo^2+\qo^{4}}\ln\big|\qo+\qo^{-1}\big|
-\frac{\qo- \qo^{-1}}{\qo+\qo^{-1}}\ln\big(1+\qo^2+\qo^{4}\big)\nonumber\\
&-\frac{-2 \qo^{-2}-4 \qo^{-4}}{1+\qo^{-2}+\qo^{-4}}\ln\big|\qo+\qo^{-1}\big|
-\frac{\qo-\qo^{-1}}{\qo+\qo^{-1}}\ln\big(1+\qo^{-2}+\qo^{-4}\big).
\label{ddoK}
\end{alignat}
Grouping terms and using
\begin{equation}
\ln\big(\qo^2+2+\qo^{-2}\big)=\ln\big(\qo+\qo^{-1}\big)^2=2\ln\big|\qo+\qo^{-1}\big|,
\end{equation}
we find
\begin{alignat}{2}
\qo\frac{\dd}{\dd \qo}\mathcal K_\sigma=&
\frac{7\qo+5\qo^{-1}}{\qo+\qo^{-1}}\ln\big(1+\qo^2+\qo^{4}\big)
-\frac{5\qo+7\qo^{-1}}{\qo+\qo^{-1}}\ln\big(1+\qo^{-2}+\qo^{-4}\big)\nonumber\\
&-2\frac{\qo-\qo^{-1}}{\qo+\qo^{-1}}\ln\big|\qo^2+1+\qo^{-2}\big|
+6\ln\big(1+\sigma \qo^{-3}\big)-6\ln\big(1+\sigma \qo^3\big).\label{ddoKv2}
\end{alignat}
Next we perform simplifications where care is taken when using the functional relations for the logarithm:
\begin{subequations}
\label{logids}
\begin{alignat}{2}
\ln\big(1+\qo^2+\qo^{4}\big)&=\ln\Big[\big(\qo^2+1+\qo^{-2}\big)\qo^2\Big]
=\ln\big|\qo^2+1+\qo^{-2}\big| + 2\ln \qo - n \pi\ir,\label{logid1}\\[0.1cm]
\ln\big(1+\qo^{-2}+\qo^{-4}\big)&=\ln\Big[\big(\qo^2+1+\qo^{-2}\big)\qo^{-2}\Big]
=\ln\big|\qo^2+1+\qo^{-2}\big| - 2\ln \qo + n \pi\ir,\label{logid2}\\[0.1cm]
\ln\big(1+\sigma \qo^{-3}\big)-\ln\big(1+\sigma \qo^3\big)&=-3\ln \qo+m \pi\ir,\label{logid3}
\end{alignat}
\end{subequations}
where the integers $m$ and $n$ are given by 
\be
n = \left\{\begin{array}{ll}
0 \hspace{0.1cm}& \gamma\in\ (0,\pi/3)\,,\\[0.15cm]
1 & \gamma\in\ (\pi/3,2\pi/3)\,,\\[0.15cm]
2 & \gamma\in\ (2\pi/3,\pi)\,,
\end{array}\right. \qquad
m = \left\{\begin{array}{lll}
0 \hspace{0.1cm}& \gamma\in\ (0,\pi/3), &\sigma = +1,\\[0.15cm]
2 & \gamma\in\ (\pi/3,\pi), &\sigma = +1,\\[0.15cm]
1 & \gamma\in\ (0,2\pi/3), &\sigma = -1,\\[0.15cm]
3 & \gamma\in\ (2\pi/3,\pi), &\sigma = -1.
\end{array}\right.
\ee

These integers are determined in the following way. The logarithms on the
left-sides of \eqref{logid1}-\eqref{logid3} are the same as in
\eqref{defintegrals} and have imaginary parts in $(-\pi,\pi]$ because of our choice of the branches. On the right-sides of \eqref{logid1}-\eqref{logid3}, a real logarithm of a positive
real number and $\ln\qo=\ir\gamma$ appear. 
The argument of the logarithms on the left-side of \eqref{logid1} and \eqref{logid2} is zero for $\gamma = \frac \pi 3, \frac {2\pi} 3$. The range $(0, \pi)$ therefore splits into the three subintervals on which $n$ takes constant integer values:
$(0,\frac \pi 3)$, $(\frac \pi 3,\frac {2\pi} 3)$ and $(\frac {2\pi} 3, \pi)$.
It is easy to check that the
integer $n$ has to be chosen in the above specified manner in order to
have the imaginary part of the right-sides of
\eqref{logid1} and \eqref{logid2} in $(-\pi,\pi]$.

Exponentiating both sides of \eqref{logid3}, we find that the integer $m$ is even for $\sigma = 1$ and odd for $\sigma = -1$. The arguments of the logarithms on the left-side of \eqref{logid3}
assumes the value $0$ for $\gamma=\pi/3\ (2\pi/3)$ for $\sigma=+1\ (-1)$. We treat $\sigma=+1$
first. The integer valued $m$ is constant in $[0,\pi/3)$ and in
    $(\pi/3,\pi)$. Inserting $\gamma=0$ and $\gamma=2\pi/3$ into
    \eqref{logid3} we find $m=0$ and $2$ respectively in these intervals. Next we treat $\sigma=-1$. The
    integer valued $m$ is constant in $(0,2\pi/3)$ and in
    $(2\pi/3,\pi]$. Inserting $\gamma=\pi/3$ and $\gamma=\pi$ into
    \eqref{logid3} we find $m=1$ and $3$ in these intervals.

Simplifying \eqref{ddoKv2} by use of \eqref{logids} yields
\begin{equation}
\qo\frac{\dd}{\dd \qo}\mathcal K_\sigma=6\ln \qo+(6m-12 n)\pi\ir.
\end{equation}
For the two cases, we find explicitly
\begin{subequations}
\begin{alignat}{3}
\qo\frac{\dd}{\dd \qo}\mathcal K_{+} &= 
\left\{\begin{array}{c}
 6\ln \qo \\[0.2cm]
 -12\pi\ir+6\ln \qo
\end{array}\right. \quad &&
\begin{array}{l}
0 \le \gamma \le \frac {2 \pi} 3, \\[0.2cm]
\frac {2 \pi} 3  \le \gamma \le \pi,
\end{array}
\\[0.3cm]
\qo\frac{\dd}{\dd \qo}\mathcal K_{-} &= 
\left\{\begin{array}{c}
6\pi\ir+ 6\ln \qo\\[0.2cm]
-6\pi\ir+ 6\ln \qo
\end{array}\right.
\quad &&
\begin{array}{l}
0 \le \gamma \le \frac {\pi} 3, \\[0.2cm]
\frac {\pi} 3  \le \gamma \le \pi.
\end{array}
\end{alignat}
\end{subequations}
Note that $\ln
\qo=\ir\gamma$ and $\frac{\dd}{\dd \gamma}=\ir \qo\frac{\dd}{\dd \qo}$. We next integrate with respect to $\gamma$:
\begin{subequations}
\begin{alignat}{3}
\mathcal K_{+} &= 
\left\{\begin{array}{c}
 C_+-3\gamma^2 \\[0.2cm]
 C_+-8\pi^2+12\pi\gamma-3\gamma^2
\end{array}\right. \quad &&
\begin{array}{l}
0 \le \gamma \le \frac {2 \pi} 3, \\[0.2cm]
\frac {2 \pi} 3  \le \gamma \le \pi,
\end{array}
\\[0.3cm]
\mathcal K_{-} &= 
\left\{\begin{array}{c}
C_--6\pi\gamma-3\gamma^2\\[0.2cm]
C_--4\pi^2+6\pi\gamma-3\gamma^2
\end{array}\right.
\quad &&
\begin{array}{l}
0 \le \gamma \le \frac {\pi} 3, \\[0.2cm]
\frac {\pi} 3  \le \gamma \le \pi,
\end{array}
\end{alignat}
\end{subequations}
where $C_\pm$ are integration constants and the required continuity at
$\gamma=\pi/3$ and $\gamma=2\pi/3$ has been imposed. The constants are
determined quite easily by noting that $\mathcal K_{1}(\frac \pi 3) = \mathcal K_{-1}(\frac {2\pi} 3) = 0$,
as for these values of $\gamma$ the upper and lower terminals of all integrals coincide. This yields
$C_+=\pi^2/3$ and $C_-=4\pi^2/3$.
We finally find
\begin{subequations}
\begin{alignat}{3}
\frac 1{8\pi^2}\mathcal K_+(\gamma) &= 
\left\{\begin{array}{c}
\frac1{24} - \frac38 \left(\frac \gamma \pi\right)^2 \\[0.2cm]
-\frac{23}{24} + \frac32 \left(\frac  \gamma \pi\right)- \frac38 \left(\frac \gamma \pi\right)^2
\end{array}\right. \quad &&
\begin{array}{l}
0 \le \gamma \le \frac {2 \pi} 3, \\[0.2cm]
\frac {2 \pi} 3  \le \gamma \le \pi,
\end{array}
\\[0.3cm]
\frac 1{8\pi^2}\mathcal K_-(\gamma) &= 
\left\{\begin{array}{c}
\frac1{6} - \frac34 \left(\frac  \gamma \pi\right) - \frac38 \left(\frac \gamma \pi\right)^2\\[0.2cm]
-\frac{1}{3} + \frac34 \left(\frac  \gamma \pi\right)- \frac38 \left(\frac \gamma \pi\right)^2
\end{array}\right.
\quad &&
\begin{array}{l}
0 \le \gamma \le \frac {\pi} 3, \\[0.2cm]
\frac {\pi} 3  \le \gamma \le \pi.
\end{array}
\end{alignat}
\end{subequations}

%
\section{Spectrum generating functions and characters}\label{sec:qstuff}
%

\subsection[Identities for $q$-binomials]{Identities for $\boldsymbol{q}$-binomials}\label{sec:qids}

The goal of this section is to derive the relation \eqref{eq:finaldoublebinomial} for the product of two Gaussian polynomials, which is 
useful in \cref{sec:charids} and \cref{sec:Pcpf}. The $q=1$ specialisations of the identities derived below are proven in \cite[Section 1.4]{R79}, and here we follow the same ideas. From the definition \eqref{eq:Gausspoly} of the $q$-binomial, $\smallqbinom{n}{m}$ is non zero for integers $m,n$ with $0 \le m \le n$. The definition of the $q$-binomial is extended to $n<0$ by using the definition \eqref{eq:Gausspoly}. This yields
\be
\qbinom{-a}{b}= (-1)^b q^{-\frac12b(b-1) -ab}\qbinom{a+b-1}{b},\qquad a\in \mathbb Z_{> 0}.\label{eq:minusqbin}
\ee
It follows that $\smallqbinom{n}{m}$ is non zero for $n<0$, $m\in \mathbb Z_{\ge 0}$. The $q$-binomials satisfy the relations
\begin{subequations}
\begin{alignat}{2}
\qbinom{a}{b} &= \qbinom{a}{a-b}\label{eq:id1}\\
\qbinom{a}{b}\qbinom{b}{c} &= \qbinom{a}{c}\qbinom{a-c}{b-c}\label{eq:id2}\\
\qbinom{a}{b} &= \qbinom{a-1}{b}+q^{a-b}\qbinom{a-1}{b-1}=q^b\qbinom{a-1}{b}+\qbinom{a-1}{b-1}
\label{eq:qbinnm}
\end{alignat}
\end{subequations}
as well as the $q$-Vandermonde identity:
\be
\qbinom{a+b}{c} = \sum_j \qbinom{a}{c-j} \qbinom{b}{j} q^{j(a-c+j)}\label{eq:qVan}.
\ee
Here and below, sums without bounds indicate that the variable is summed from $-\infty$ to $\infty$, with the summand being non-zero on a finite range only.

We first derive sum formulas for single $q$-binomials. A first relation is obtained as follows:
\be
\qbinom{n-p}{m} \overset{\textrm{\tiny\eqref{eq:qVan}}}{=} \sum_k \qbinom{n}{m-k}\qbinom{-p}{k}q^{k(n-m+k)} \overset{\textrm{\tiny\eqref{eq:minusqbin}}}{=} \sum_{k}(-1)^k \qbinom{n}{m-k}\qbinom{p+k-1}{k}q^{k(n-m+k)-\frac12k(k-1)-pk}.
\label{eq:id3}
\ee
A second relation is given by
\begin{alignat}{2}
\qbinom{n-p}{m} &\overset{\textrm{\tiny\eqref{eq:minusqbin}}}{=} (-1)^m q^{-\frac12m(m-1)-m(p-n)}\qbinom{p-n+m-1}{m} \nonumber\\
& \overset{\textrm{\tiny\eqref{eq:qVan}}}{=} (-1)^m q^{-\frac12m(m-1)-m(p-n)} \sum_k \qbinom{p}{m-k}\qbinom{-n+m-1}{k}q^{k(p-m+k)} \nonumber\\
&\overset{\textrm{\tiny\eqref{eq:minusqbin}}}{=}q^{-\frac12m(m-1)-m(p-n)} \sum_k (-1)^{m+k} \qbinom{p}{m-k}\qbinom{n-m+k}{k}q^{k(p-m+k)-\frac12k(k-1)-k(n-m+1)}\nonumber\\
& \ = \sum_{j}(-1)^j \qbinom{n-j}{m-j}\qbinom{p}{j} q^{\frac12j(j+1)+j(n-m-p)},
\end{alignat}
where we substituted $k = m-j$ at the last step.
By replacing $n-p$ with $n$, this identity can be rewritten as
\begin{alignat}{2}
\qbinom{n}{m}&=\sum_k(-1)^k \qbinom{n+p-k}{m-k}\qbinom{p}{k}q^{\frac12k(k+1)+k(n-m)} = \sum_k (-1)^k \qbinom{n+p-k}{n-m-k}\qbinom{p}{k}q^{\frac12k(k+1)+km} \nonumber\\
&\hspace{-0.2cm} \overset{\textrm{\tiny\eqref{eq:id1}}}{=} \sum_k  (-1)^k \qbinom{n+p-k}{m+p}\qbinom{p}{k}q^{\frac12k(k+1)+km} \label{eq:id4}
\end{alignat}
where we used $\smallqbinom{n}{m} = \smallqbinom{n}{n-m}$ at the second equality to replace $m$ by $n-m$.

Using these relations, we now derive an identity for the product of two Gaussian polynomials:
\begin{alignat}{2}
\qbinom{m}{p}\qbinom{n}{r} & = \qbinom{m-n+r+n-r}{p}\qbinom{n}{r} \overset{\textrm{\tiny\eqref{eq:qVan}}}{=} \sum_k \qbinom{m-n+r}{p-k}\qbinom{n-r}{k}\qbinom{n}{r} q^{k(m-n+r-p+k)}
\nonumber\\&
\hspace{-1.71cm}\overset{\textrm{\tiny\eqref{eq:id2}}}{=} \sum_k \qbinom{m-n+r}{p-k}\qbinom{r+k}{r}\qbinom{n}{r+k} q^{k(m-n+r-p+k)}
\nonumber\\&
\hspace{-1.65cm}\overset{\textrm{\tiny\eqref{eq:id4}}}{=} \sum_k \qbinom{m-n+r}{p-k}\qbinom{r+k}{r}q^{k(m-n+r-p+k)}\sum_j(-1)^{j}\qbinom{n+p-k-j}{p+r}\qbinom{p-k}{j}q^{\frac12j(j+1)+j(r+k)}
\nonumber\\&
\hspace{-1.5cm}= \sum_k \qbinom{m-n+r}{p-k}\qbinom{r+k}{r}q^{k(m-n+r-p+k)}\sum_i(-1)^{p-k-i}\qbinom{n+i}{p+r}\qbinom{p-k}{i}q^{\frac12(p-k-i)(p+k-i+1+2r)}
\nonumber\\&
\hspace{-1.71cm}\overset{\textrm{\tiny\eqref{eq:id1}}}{=} \sum_i \qbinom{n+i}{p+r}\sum_k(-1)^{p-k-i}\qbinom{r+k}{k}\qbinom{m-n+r}{p-k}\qbinom{p-k}{i}q^{k(m-n+r-p+k)+\frac12(p-k-i)(p+k-i+1+2r)}
\nonumber\\&
\hspace{-1.71cm}\overset{\textrm{\tiny\eqref{eq:id2}}}{=} \sum_i \qbinom{n+i}{p+r}\qbinom{m-n+r}{i}\sum_k(-1)^{p-k-i}\qbinom{r+k}{k}\qbinom{m-n+r-i}{p-i-k}q^{k(m-n+r-p+k)+\frac12(p-k-i)(p+k-i+1+2r)}
\nonumber\\&
\hspace{-1.65cm}\overset{\textrm{\tiny\eqref{eq:id3}}}{=} \sum_i \qbinom{n+i}{p+r}\qbinom{m-n+r}{i}(-1)^{p-i}\qbinom{m-n-i-1}{p-i}q^{\frac12i(i-1)+\frac12p(p+1)+pr-pi-ri}
\nonumber\\&
\hspace{-1.65cm}\overset{\textrm{\tiny\eqref{eq:minusqbin}}}{=}\sum_i \qbinom{n+i}{p+r}\qbinom{m-n+r}{i}\qbinom{n-m+p}{p-i}q^{\frac12i(i-1)+\frac12p(p+1)+pr-pi-ri-\frac12(p-i)(p-i-1)+(p-i)(m-n-i-1)}
\nonumber\\&
\hspace{-1.71cm}\overset{\textrm{\tiny\eqref{eq:id1}}}{=} \sum_i \qbinom{n+i}{p+r}\qbinom{m-n+r}{i}\qbinom{n-m+p}{n-m+i} q^{i^2+i(n-m-p-r)+mp-np+pr}.\label{eq:finaldoublebinomial}
\end{alignat}
The last line is an identity that is used multiple times in \cref{sec:charids,sec:Pcpf}.

\subsection{Character identities for the boundary case}\label{sec:charids}

In this section, we show that the finitized partition functions $\ZdN$ given in \eqref{eq:char.identities} are equal to the finitized Kac characters $\chit_{1,d+1}^{\textrm{\tiny$(N)$}}(q)$. Let $t \in \mathbb Z$ with $t \equiv N \Mod 2$. We define
\begin{subequations}
\label{eq:X1X2X3}
\begin{alignat}{2}
&X^1_{t} = \sum_{k\ge 0}\sum_{i=0}^{k} q^{i^2 + 2k (k-i+\frac12)+t(-t+i-k-1)}\qbinom{\frac{N+t}2+i}{2k+1}\qbinom{k}{i}\qbinom{k+1}{i+t+1},\label{eq:X1}\\
&X^2_{t} = \sum_{k\ge 1}\sum_{i=0}^{k-1} q^{i(i+1) + 2k (k-i-\frac12)+t(-t+i-k+1)}\qbinom{\frac{N+t}2+i}{2k}\qbinom{k-1}{i}\qbinom{k+1}{i+t+1},\\
&X^3_{t} = \sum_{k\ge 0}\sum_{i=0}^{k} q^{-1+i(i+1) + 2k (k-i-\frac12)+t(-t+i-k-2)}\qbinom{\frac{N+t}2+i}{2k}\qbinom{k}{i}\qbinom{k}{i+t+1}.
\end{alignat}
\end{subequations}
These quantities are useful because of the following identities:
\begin{subequations}
\label{eq:propsyo}
\begin{alignat}{2}
&d=3t: && \ZdN = q^{d(d-1)/6}(X^2_{t} - q^{d+1} X^3_{t} + \delta_{d,0}),\label{eq:prop0}\\[0.05cm]
&d=3t+1: && \ZdN = q^{d(d-1)/6}(X^3_{t-1} - q^{d+1} X^2_{-t-1}),\label{eq:prop1}\\[0.05cm]
&d = 3t+2: \qquad && \ZdN = q^{d(d-1)/6}(X^1_{t} - q^{d+1} X^1_{-t-2}).\label{eq:prop2}
\end{alignat}
\end{subequations}
In proving \eqref{eq:propsyo}, we start with the case $d \equiv 2 \Mod 3$ which is easiest. By using \eqref{eq:qnar} and substituting $j$ by $k=i+j+t$ in \eqref{eq:char2}, we find 
\begin{alignat}{2}
\ZdN = q^{d(d-1)/6} &\sum_{i,k} q^{i^2 + 2k (k-i+\frac12)+t(-t+i-k-1)}\qbinom{\frac{N+t}2+i}{2k+1}\bigg(\qbinom{k}{i}\qbinom{k+1}{i+t+1} - \qbinom{k+1}{i}\qbinom{k}{i+t+1}  \bigg).
\end{alignat}
The positive part is readily identified as $X^1_{t}$. After substituting $i$ for $j=i+t+1$ and setting $t =-s-2$ in the negative part, it is found to equal $-q^{-3(1+s)}X^1_{s} = -q^{d+1}X^1_{-t-2}$, ending the proof of \eqref{eq:prop0}.

For $d = 3t$ and $t>0$, applying \eqref{eq:qnar} to \eqref{eq:char2} yields four double-sums:
\begin{alignat}{2}
\ZdN &= q^{d(d-1)/6}\bigg( \sum_{i,k} q^{i(i+1)+2k(k-i-\frac12)+t(-t+i-k+1)}\qbinom{\frac{N+t}2}{2k}\Big(\qbinom{k-1}{i}\qbinom{k}{i+t}-\qbinom{k}{i}\qbinom{k-1}{i+t}\Big)\nonumber\\
&\hspace{-0.5cm} +  \sum_{i,\ell} q^{1+i(i+2)+2\ell(\ell-i-\frac12)+t(-t+i-\ell+2)}\qbinom{\frac{N+t}2}{2\ell}\Big(\qbinom{\ell-1}{i}\qbinom{\ell}{i+t+1}-\qbinom{\ell}{i}\qbinom{\ell-1}{i+t+1}\Big)\bigg)\label{eq:noname},
\end{alignat}
where the sum on the second line was obtained by substituting $\ell = k+1$. Using the identity \eqref{eq:qbinnm}, the first terms and second terms of each line of \eqref{eq:noname} respectively combine, resulting in
\begin{alignat}{2}
\ZdN &=q^{d(d-1)/6} \sum_{i,k} q^{i(i+1)+2k(k-i-\frac12)+t(-t+i-k+1)}\qbinom{\frac{N+t}2}{2k}\bigg(\qbinom{k-1}{i}\qbinom{k+1}{i+t+1}-\qbinom{k}{i}\qbinom{k}{i+t+1}\bigg) \nonumber\\
& = q^{d(d-1)/6}(X^2_{t}-q^{d+1} X^3_{t}),
\end{alignat}
ending the proof of \eqref{eq:prop0}.
The case $d=0$ is computed separately and the proof of \eqref{eq:prop0} in this case is a simple exercise in $q$-binomials, with the Razumov-Stroganov eigenvalue responsible for the extra factor $\delta_{d,0}$. The proof of \eqref{eq:prop1} uses the ideas of the previous two cases and is straightforward. 

The identity \eqref{eq:finaldoublebinomial} derived in \cref{sec:qids} allows us to simplify $X^1_t$, $X^2_t$ and $X^3_t$. The parameters $m$, $n$, $p$ and $q$ are specialised as follows:
\begin{subequations}
\begin{alignat}{5}
&\textrm{for } X^1_{t}:\qquad&&m=\tfrac{N-t-2}2,\quad &&n=\tfrac{N+t}2,\quad &&p=k-t,\quad&&r=k+t+1,\\
&\textrm{for } X^2_{t}:\qquad&&m=\tfrac{N-t-2}2,\quad &&n=\tfrac{N+t}2,\quad &&p=k-t,\quad&&r=k+t,\\
&\textrm{for } X^3_{t}:\qquad&&m=\tfrac{N-t-2}2,\quad&&n=\tfrac{N+t}2,\quad&&p=k-t-1,\quad&&r=k+t+1.
\end{alignat}
\end{subequations}
Starting from \eqref{eq:X1}, for $X^1_t$ we find
\begin{alignat}{2}
X^1_t &\overset{\textrm{\tiny\eqref{eq:finaldoublebinomial}}}{=} \sum_{k\ge 0} q^{k(k+1)-t(t+1)} \qbinom{\frac{N+t}2}{k+t+1}\qbinom{\frac{N-t-2}2}{k-t} = \sum_j q^{j(2t+1+j)}\qbinom{\frac{N+t}2}{j+2t+1}\qbinom{\frac{N-t-2}2}{j} \nonumber\\
& \hspace{-0.068cm}
\overset{\textrm{\tiny\eqref{eq:id1}}}{=}  \sum_j q^{j(2t+1+j)}\qbinom{\frac{N+t}2}{\frac{N-3t-2}2-j}\qbinom{\frac{N-t-2}2}{j} \overset{\textrm{\tiny\eqref{eq:qVan}}}{=} \qbinom{N-1}{\frac{N-3t-2}2}.
\label{eq:X1result}
\end{alignat}
Likewise for $X^2_t$: 
\begin{alignat}{2}
X^2_t &\overset{\textrm{\tiny\eqref{eq:finaldoublebinomial}}}{=} \sum_{k\ge 1} q^{k^2-t^2} \qbinom{\frac{N+t}2}{k+t}\qbinom{\frac{N-t-2}2}{k-t} = \bigg(\sum_{k\ge 0} q^{k^2-t^2} \qbinom{\frac{N+t}2}{k+t}\qbinom{\frac{N-t-2}2}{k-t}\bigg) - \delta_{t,0} \nonumber\\
& \overset{\phantom{\textrm{\tiny\eqref{eq:finaldoublebinomial}}}}{=} \bigg(\sum_j q^{j(2t+j)}\qbinom{\frac{N+t}2}{j+2t}\qbinom{\frac{N-t-2}2}{j}\bigg) - \delta_{t,0}
\nonumber\\
&\hspace{-0.068cm} \overset{\textrm{\tiny\eqref{eq:id1}}}{=} \bigg( \sum_j q^{j(2t+1+j)}\qbinom{\frac{N+t}2}{\frac{N-3t-2}2-j}\qbinom{\frac{N-t-2}2}{j}\bigg) - \delta_{t,0} \overset{\textrm{\tiny\eqref{eq:qVan}}}{=} \qbinom{N-1}{\frac{N-3t-2}2}- \delta_{t,0}
\label{eq:X2result}
\end{alignat}
where we used $q^{-t^2}\smallqbinom{\frac {N+t}2}{t}\smallqbinom{\frac {N+t}2}{-t}= \delta_{t,0}$ at the second equality.
The steps are identical for $X^3_t$:
\be
X^3_t \overset{\textrm{\tiny\eqref{eq:finaldoublebinomial}}}{=} \sum_{k} q^{k^2-(t+1)^2} \qbinom{\frac{N+t}2}{k+t+1}\qbinom{\frac{N-t-2}2}{k-t-1} 
= \sum_j q^{j(j+2t+2)}\qbinom{\frac{N+t}2}{j+2t+2}\qbinom{\frac{N-t-2}2}{j}
\hspace{-0.2cm}\overset{\textrm{\tiny\eqref{eq:id1}}}{=} \qbinom{N-1}{\frac{N-3t-4}2}.
\label{eq:X3result}
\ee
Combining \eqref{eq:X1result}, \eqref{eq:X2result} and \eqref{eq:X3result} with \eqref{eq:propsyo}, we find
\be
\ZdN = q^{d(d-1)/6}\bigg(\qbinom{N-1}{\frac{N-d}2}-q^{d+1}\qbinom{N-1}{\frac{N-d-4}2}\bigg)= q^{d(d-1)/6}\bigg(\qbinom{N}{\frac{N-d}2}-q^{d+1}\qbinom{N}{\frac{N-d-2}2}\bigg) = \XN_{1,d+1}(q), 
\ee
which holds for $d \equiv 0,1,2 \Mod 3$. The second equality is obtained (from right to left) using \eqref{eq:qbinnm}.

\subsection{Partition functions for the periodic case}\label{sec:Pcpf}

\paragraph{Explicit expressions.} The finitized partition functions $\ZdNP$ defined in \eqref{eq:ZdNP} are obtained by writing down the generating functions for each set in \eqref{eq:SRp} and summing over the corresponding values of $i$ and $j$. We note that the indices $i$ and $j$ in the selection rules always run over all possible values for which the corresponding sets $\BB{M}{L}{m}{n}{\ell}$ are well defined. This is also true for $k_1, k_2$ and $k_3$ in \eqref{eq:sep0}. For ease of notation, we omit to write the bounds of the sums over $i,j,k_2$ and $k_3$ and interpret the corresponding indices as running over $\mathbb Z$, understanding that only finitely many terms are non-zero. The same applies to $k_1$, except that the sum is split between the odd and even values, as they correspond to different values of $\sigma$, see \eqref{eq:sigmak1}. 

For $d \equiv 0 \Mod 3$, we use the prescription \eqref{eq:sep0} for the separation between upper and lower halves and write down the generating function \eqref{eq:Bgeneratingfunction} for each set in \eqref{eq:SRp0t}. Summing over these sets, we find
\begin{alignat}{2}
Z_{d=3t}^{\textrm{\tiny$(N)$}}(q,\bar q) &= \sum_{i,j,k_2,k_3} \!\Big(\sum_{k_1\,\textrm{even}} (q \bar q)^{-\frac1{24}}+  \sum_{k_1\,\textrm{odd}}(-1)^M (q \bar q)^{-\frac16}\Big) (q\bar q)^{E} q^{E_0}
\qqbinom{\lfloor \frac12(\frac{N+t}{2}+i)\rfloor - \epsilon}{k_1} \qqbinom{\lfloor{\frac {k_1}2}\rfloor}{k_2} \qqbinom{\lfloor{\frac {k_1+1}2}\rfloor}{k_3} 
\nonumber\\& \hspace{2cm}\times \qbinom{\lfloor \frac12(\frac{N+t}{2}+i+1)\rfloor + \epsilon}{2(i+j+t)-k_1}  \qbinom{i+j+t-\lfloor{\frac {k_1}2}\rfloor}{i-k_2}\qbinom{i+j+t-\lfloor{\frac {k_1+1}2}\rfloor}{i+t-k_3}
\label{eq:Z0}
\end{alignat}
where $E$ and $E_0$ are obtained from the proper specialisations of \eqref{eq:justE}:
\begin{subequations}
\begin{alignat}{2}
E &= \tfrac12(k_1^2+k_2^2+k_3^2-k_1k_2 - k_1 k_3),\label{eq:EE}\\[0.15cm]
E_0 &= i^2 - i k_1+ 2j(i+j-k_1+\tfrac12 k_2+\tfrac12k_3) + t(2i+3j - \tfrac32 k_1 + k_2 + \tfrac32 t).\label{E0}
\end{alignat}
\end{subequations}

For $d \equiv 1,2 \Mod 3$,  from \eqref{eq:SRp1t}, \eqref{eq:SRp2t} and \eqref{eq:SRp1}, $\ZdNP$ is split between contributions coming from the different subcases:
\begin{subequations}\label{eq:Z12separation}
\begin{alignat}{2}
Z_{d=3t+1}^{\textrm{\tiny$(N)$}}(q,\bar q) &= Z_1^{\bar \tB, \tB} + Z_1^{\bar \tA, \tB} + Z_1^{\bar \tB, \tA} + Z_1^{\bar \tA, \tA} + (-1)^M \delta_{d,1},\label{eq:Z1separation} \\[0.15cm]
Z_{d=3t+2}^{\textrm{\tiny$(N)$}}(q,\bar q) &= Z_2^{\bar\tA, \tA} + Z_2^{\bar\tA, \tB} + Z_2^{\bar\tB,  \tA} + Z_2^{\bar\tB, \tB}.
\label{eq:Z2separation}
\end{alignat}
\end{subequations}
Each $Z_k^{\tX, \tY}$, with $\tX \in \{\bar \tA, \bar \tB\}$, $\tY \in \{\tA, \tB\}$ and $k = 1,2$, is obtained by using the prescription \eqref{eq:sep12} for the separation between upper and lower halves, writing down the corresponding generating function \eqref{eq:Bgeneratingfunction}, and summing over the corresponding sets given by the selection rules: 
\begin{alignat}{2}
Z_k^{\tX, \tY} &= \sum_{i,j,k_2,k_3} \!\Big(\sum_{k_1\,\textrm{even}} (-1)^M+  \sum_{k_1\,\textrm{odd}} (q \bar q)^{\frac18}\Big) (q\bar q)^{E(\tX)} q^{E_k(\tX,\tY)} F_k(\tX, \tY)
\end{alignat}
with
\begin{subequations}
\begin{alignat}{2}
F_1(\bar \tB, \tB) &= 
\qqbinom{\lfloor \frac12(\frac{N+t-1}{2}+i)\rfloor - \epsilon}{k_1} \qqbinom{\lfloor{\frac {k_1-1}2}\rfloor}{k_2} \qqbinom{\lfloor{\frac {k_1}2}\rfloor}{k_3} \nonumber\\&  \hspace{0.8cm} \times
\qbinom{\lfloor \frac12(\frac{N+t-1}{2}+i+1)\rfloor + \epsilon}{2(i+j+t)-k_1}  \qbinom{i+j+t-1-\lfloor{\frac {k_1-1}2}\rfloor}{i-k_2}\qbinom{i+j+t-1-\lfloor{\frac {k_1}2}\rfloor}{i+t-1-k_3},\\[0.2cm]
F_1(\bar \tA, \tB) &= F_2(\bar \tB, \tA) = 
\qqbinom{\lfloor \frac12(\frac{N+t+1}{2}+i)\rfloor - \epsilon}{k_1} \qqbinom{\lfloor{\frac {k_1-1}2}\rfloor}{k_2} \qqbinom{\lfloor{\frac {k_1}2}\rfloor}{k_3} \nonumber\\& \hspace{2.25cm} \times
\qbinom{\lfloor \frac12(\frac{N+t+1}{2}+i+1)\rfloor + \epsilon}{2(i+j+t+1)-k_1}  \qbinom{i+j+t-\lfloor{\frac {k_1-1}2}\rfloor}{i-k_2}\qbinom{i+j+t-\lfloor{\frac {k_1}2}\rfloor}{i+t-k_3},\\[0.2cm]
F_1(\bar \tA, \tA) &=
\qqbinom{\lfloor \frac12(\frac{N+t+3}{2}+i)\rfloor - \epsilon}{k_1} \qqbinom{\lfloor{\frac {k_1-1}2}\rfloor}{k_2} \qqbinom{\lfloor{\frac {k_1}2}\rfloor}{k_3}  \nonumber\\&  \hspace{0.8cm} \times
\qbinom{\lfloor \frac12(\frac{N+t+3}{2}+i+1)\rfloor + \epsilon}{2(i+j+t+2)-k_1}  \qbinom{i+j+t+1-\lfloor{\frac {k_1-1}2}\rfloor}{i-k_2}\qbinom{i+j+t+1-\lfloor{\frac {k_1}2}\rfloor}{i+t+1-k_3},
\end{alignat}
\end{subequations}
and 
\begin{subequations}
\begin{alignat}{2}
F_2(\bar \tA, \tA) &= 
\qqbinom{\lfloor \frac12(\frac{N+t+2}{2}+i)\rfloor - \epsilon}{k_1} \qqbinom{\lfloor{\frac {k_1-1}2}\rfloor}{k_2} \qqbinom{\lfloor{\frac {k_1}2}\rfloor}{k_3}  \nonumber\\&  \hspace{0.8cm} \times
\qbinom{\lfloor \frac12(\frac{N+t+2}{2}+i+1)\rfloor + \epsilon}{2(i+j+t+1)-k_1}  \qbinom{i+j+t-\lfloor{\frac {k_1-1}2}\rfloor}{i-k_2}\qbinom{i+j+t-\lfloor{\frac {k_1}2}\rfloor}{i+t-k_3},\\[0.2cm]
F_2(\bar \tA, \tB) &= F_2(\bar \tB, \tA) = 
\qqbinom{\lfloor \frac12(\frac{N+t+2}{2}+i)\rfloor - \epsilon}{k_1} \qqbinom{\lfloor{\frac {k_1-1}2}\rfloor}{k_2} \qqbinom{\lfloor{\frac {k_1}2}\rfloor}{k_3}   \nonumber\\&  \hspace{0.8cm} \times
\qbinom{\lfloor \frac12(\frac{N+t+2}{2}+i+1)\rfloor + \epsilon}{2(i+j+t+2)-k_1}  \qbinom{i+j+t+1-\lfloor{\frac {k_1-1}2}\rfloor}{i-k_2}\qbinom{i+j+t+1-\lfloor{\frac {k_1}2}\rfloor}{i+t+1-k_3},
\\[0.2cm]
F_2(\bar \tB, \tB) &=
\qqbinom{\lfloor \frac12(\frac{N+t+2}{2}+i)\rfloor - \epsilon}{k_1} \qqbinom{\lfloor{\frac {k_1-1}2}\rfloor}{k_2} \qqbinom{\lfloor{\frac {k_1}2}\rfloor}{k_3}   \nonumber\\&  \hspace{0.8cm} \times
\qbinom{\lfloor \frac12(\frac{N+t+2}{2}+i+1)\rfloor + \epsilon}{2(i+j+t+3)-k_1}  \qbinom{i+j+t+2-\lfloor{\frac {k_1-1}2}\rfloor}{i-k_2}\qbinom{i+j+t+2-\lfloor{\frac {k_1}2}\rfloor}{i+t+2-k_3}.
\end{alignat}
\end{subequations}
The energies for the $\tA$ and $\bar \tA$ subcases are obtained from \eqref{eq:EA} under the proper specialisations of $m$, $n$ and $\ell$. The minimal energies for the $\tB$ and $\bar \tB$ subcases are likewise obtained from \eqref{eq:EB}. We find
\begin{subequations}
\begin{alignat}{2}
E(\bar \tA) &= \tfrac12 (k_1^2+k_2^2+k_3^2-k_1+k_2+k_3-k_1 k_2-k_1 k_3),\\[0.15cm]
E(\bar \tB) &= \tfrac12 (k_1^2+k_2^2+k_3^2+k_2+k_3-k_1 k_2-k_1 k_3),
\end{alignat}
\end{subequations}
and
\begin{subequations}
\begin{alignat}{2}
q^{E_1(\bar \tB, \tB)} & =q^{1 + i(i+1) - i k_1-\frac12k_1 - k_2 + 2j(i+j-k_1+\frac12 k_2+\frac12k_3+\frac 12) + t(2i+3j - \frac32 k_1 + k_2 + \frac32 t+\frac12)}, \\
q^{E_1(\bar \tA, \tB)} & =q^{2 + i(i+3) - i k_1- \frac32k_1 + 2j(i+j-k_1+\frac12 k_2+\frac12k_3+2) + t(2i+3j - \frac32 k_1 + k_2 + \frac32 t+\frac{7}2)}, \\
q^{E_1(\bar \tB, \tA)} & =q^{1 + i(i+2) - i k_1- \frac32k_1 + 2j(i+j-k_1+\frac12 k_2+\frac12k_3+\frac32) + t(2i+3j - \frac32 k_1 + k_2 + \frac32 t+\frac{5}2)}, \\
q^{E_1(\bar \tA, \tA)} & =q^{5 + i(i+4) - i k_1-\frac52k_1+k_2 + 2j(i+j-k_1+\frac12 k_2+\frac12k_3+3) + t(2i+3j - \frac32 k_1 + k_2 + \frac32 t+\frac{11}2)}, \\
q^{E_2(\bar \tA, \tA)} & =q^{1 + i(i+2) - i k_1-k_1 + 2j(i+j-k_1+\frac12 k_2+\frac12k_3+\frac 32) + t(2i+3j - \frac32 k_1 + k_2 + \frac32 t+\frac52)}, \\
q^{E_2(\bar \tA, \tB)} & =q^{7 + i(i+5) - i k_1- 3k_1 +k_2+ 2j(i+j-k_1+\frac12 k_2+\frac12k_3+\frac72) + t(2i+3j - \frac32 k_1 + k_2 + \frac32 t+\frac{13}2)}, \\
q^{E_2(\bar \tB, \tA)} & =q^{5 + i(i+4) - i k_1-3k_1 +k_2+ 2j(i+j-k_1+\frac12 k_2+\frac12k_3+3) + t(2i+3j - \frac32 k_1 + k_2 + \frac32 t+\frac{11}2)}, \\
q^{E_2(\bar \tB, \tB)} & =q^{15 + i(i+7) - i k_1-5k_1+2k_2 + 2j(i+j-k_1+\frac12 k_2+\frac12k_3+5) + t(2i+3j - \frac32 k_1 + k_2 + \frac32 t+\frac{19}2)}.
\end{alignat}
\end{subequations}

\paragraph{Scaling behavior for $\boldsymbol{d \equiv 0} \Mod \boldsymbol{3}$.}
We explicitly derive the formula \eqref{eq:Z0results} for the scaling limit of $Z_{d=3t}^{\textrm{\tiny$(N)$}}(q,\bar q)$.
We perform the calculation separately for the odd and even $k_1$ contributions in \eqref{eq:Z0}, namely we write
\be
Z_{d=3t}^{\textrm{\tiny$(N)$}}(q,\bar q) = Z_{0,\textrm{even}} + (-1)^M Z_{0,\textrm{odd}}.
\ee

We start with the even case. As discussed in \cref{sec:fsgf}, for $\epsilon \ll N$, the scaling behavior of \eqref{eq:Z0} is independent of $\epsilon$. In fact, $\epsilon$ can be chosen to depend on $i$, whose values are indeed much smaller than $N$ for the leading eigenvalues. For the same reason, one could also choose $\epsilon$ to depend on $j$, $k_1$, $k_2$ and $k_3$. Here we make a special choice of $\epsilon$ that allows the computation to go forward, namely we choose $\epsilon$ such that the first $\bar q$-binomial in \eqref{eq:Z0} does not depend on $i$. This allows us to perform the sums over $i$ and $j$ first, noting that their summands involve only powers of $q$ and not of $\bar q$. Defining $N_t = N+t$, we get
\be
Z_{0,\textrm{even}}=(q \bar q)^{-\frac1{24}}  \sum_{k_2,k_3} \sum_{k_1\,\textrm{even}} (q \bar q)^{E} \qqbinom{\lfloor \frac{N_t}4\rfloor}{k_1}\qqbinom{\frac{k_1}2}{k_2}\qqbinom{\frac{k_1}2}{k_3}  S_{0,\textrm{even}}
\ee
where 
\begin{alignat}{2}
S_{0,\textrm{even}} &= \sum_{i,j} q^{E_0} \qbinom{\lfloor \frac{N_t}4+\frac12\rfloor+i}{2(i+j+t)-k_1}\qbinom{i+j+t-\frac{k_1}2}{i-k_2}\qbinom{i+j+t-\frac{k_1}2}{i+t-k_3} \nonumber\\
& = \sum_{k,\ell} q^{E_0} \qbinom{\lfloor \frac{N_t}4+\frac12\rfloor+k_2+k}{2\ell-k_1}\qbinom{\ell-\frac{k_1}2}{k}\qbinom{\ell-\frac{k_1}2}{k+t+k_2-k_3}.
\end{alignat}
At the last equality, we substituted first $\ell = i+j+t$ and then $k = i-k_2$. In the last expression, the sums over $k$ and $\ell$ still run over $\mathbb Z$. Although the notation does not make it explicit, the expression \eqref{E0} for $E_0$ is understood as changing with every change of summation variables. It remains quadratic in the various parameters. The next step consists in using \eqref{eq:finaldoublebinomial} with
\be
m = \lfloor\tfrac{N_t}4 + \tfrac12\rfloor + k_3 - t, \quad n = \lfloor\tfrac{N_t}4 + \tfrac12\rfloor + k_2, \quad p = \ell - \frac{k_1}2 - k_2 + k_3  - t, \quad r = \ell - \frac{k_1}2 + k_2 - k_3  + t.
\ee
This yields
\be
S_{0,\textrm{even}} = \sum_\ell q^{E_0'} \qbinom{\lfloor\tfrac{N_t}4 + \tfrac12\rfloor + k_3 - t}{\ell - \frac{k_1}2 - k_2 + k_3  - t}\qbinom{ \lfloor\tfrac{N_t}4 + \tfrac12\rfloor + k_2}{\ell - \frac{k_1}2 + k_2 - k_3 + t}
\ee
where 
\be
E_0' = -\tfrac14 k_1^2 + \tfrac12 k_1 k_2 + \tfrac12 k_1 k_3 - k_2 k_3 + \ell^2 - \ell k_1 + \tfrac12 t^2 + t k_2 - t k_3.
\ee
This remaining sum can be evaluated using \eqref{eq:qVan}:
\be
S_{0,\textrm{even}} = \sum_j q^{E_0'}  \qbinom{\lfloor\tfrac{N_t}4 + \tfrac12\rfloor + k_3 - t}{j + 2k_3 - 2k_2 - 2t}\qbinom{ \lfloor\tfrac{N_t}4 + \tfrac12\rfloor + k_2}{j} = q^{E_0''} \specialqbinom{2\lfloor\tfrac{N_t}4 + \tfrac12\rfloor + k_2+k_3 - t}{\lfloor\tfrac{N_t}4 + \tfrac12\rfloor + 2k_2 - k_3 + t}.
\ee
Remarkably, $E_0''$ satisfies
\be
E_0'' + E = \tfrac32(k_3 -k_2-t)^2 = \Delta_{0,3(k_3-k_2-t)}+ \tfrac1{24}
\ee
with $E$ given in \eqref{eq:EE}. The right-side depends only on the difference between $k_3$ and $k_2$. Changing the summation variables to $k_3 = \ell + k_2$ and $k_1 = 2i$, we obtain
\be
Z_{0,\textrm{even}}=\bar q^{\,-\frac1{24}}  \sum_\ell q^{\Delta_{0,3(\ell-t)}} \sum_{i,k_2} \bar q^{E} \qqbinom{\lfloor \frac{N_t}4\rfloor}{2i}\qqbinom{i}{k_2}\qqbinom{i}{k_2 + \ell}  \specialqbinom{2\lfloor\tfrac{N_t}4 + \tfrac12\rfloor + 2k_2+\ell - t}{\lfloor\tfrac{N_t}4 + \tfrac12\rfloor + k_2 - \ell + t}.
\ee
The expression is thus reduced to sums of products of four binomials, with only one depending on $q$. While one may wish to reduce this expression further to a single sum with one binomial of each kind, this appears not to be feasible because the arguments of the remaining $q$-binomial involve both $\ell$ and $k_2$. However, both entries of this $q$-binomial scale linearly with $N$. We consider standard modules where the number $d$ of defects remains small as $N \rightarrow \infty$, namely values of $d$ such that $t \ll N$. Recalling that $k_2, \ell \ll N_t$ for large $N$, in the scaling limit we have
\be
\specialqbinom{2\lfloor\frac{N_t}4 + \frac12\rfloor + 2k_2+\ell - t}{\lfloor\tfrac{N_t}4 + \tfrac12\rfloor + k_2 - \ell + t} \xrightarrow{M,N \rightarrow \infty} \frac{1}{(q)_\infty}
\ee
and therefore
\be
Z_{0,\textrm{even}} \simeq \frac{\bar q^{\,-\frac1{24}}}{(q)_\infty} \sum_\ell q^{\Delta_{0,3(\ell-t)}} \sum_{i,k_2} \bar q^{E} \qqbinom{\lfloor \frac{N_t}4\rfloor}{2i}\qqbinom{i}{k_2}\qqbinom{i}{k_2 + \ell}.
\ee
Here, $X \simeq Y$ means that  $X$ and $Y$ are equal up to terms which go to zero in the scaling limit. The sums over $i$ and $k_2$ involve only powers of $\bar q$. The expression is not suitable for us to use \eqref{eq:finaldoublebinomial} to remove the sum over $i$, nor \eqref{eq:qVan} for the sum over $k_2$ because $E$ is not of the correct form. We however note that 
\be
\qqbinom{\lfloor \frac{N_t}4\rfloor}{2i} \simeq \qqbinom{\lfloor \frac{N_t}4\rfloor+k_2}{2i}
\ee
and
\be
\sum_{i,k_2} \bar q^{E} \qqbinom{\lfloor \frac{N_t}4\rfloor+k_2}{2i}\qqbinom{i}{k_2}\qqbinom{i}{k_2 + \ell} = \sum_i {\bar q}^{E_0'''} \qqbinom{\lfloor \frac{N_t}4\rfloor-\ell}{i-\ell} \qqbinom{\lfloor \frac{N_t}4\rfloor}{i+\ell} = \bar q^{\frac32 \ell^2} \specialqqbinom{2\lfloor \frac{N_t}4\rfloor - \ell}{\lfloor \frac{N_t}4\rfloor+\ell} \simeq \frac{\bar q^{\frac32 \ell^2}}{(\bar q)_\infty}
\ee
where we use \eqref{eq:finaldoublebinomial} and \eqref{eq:qVan} at the first and second equalities.
Using $\Delta_{0,3 \ell} = \frac32 \ell^2 - \frac1{24}$, we find
\be
Z_{0,\textrm{even}} \simeq \frac{1}{(q)_\infty (\bar q)_\infty} \sum_\ell q^{\Delta_{0,3 (\ell-t)}}\bar q^{\Delta_{0,3 \ell}}.
\ee
This is the first term in \eqref{eq:Z0results}.

The derivation of the scaling behavior of $Z_{0,\textrm{odd}}$ follows the same steps with only few modifications: 
\be
Z_{0,\textrm{odd}}=(q \bar q)^{-\frac1{6}}  \sum_{k_2,k_3} \sum_{k_1\,\textrm{odd}} (q \bar q)^{E} \qqbinom{\lfloor \frac{N_t}4\rfloor}{k_1}\qqbinom{\frac{k_1}2}{k_2}\qqbinom{\frac{k_1}2}{k_3}  S_{0,\textrm{odd}}
\ee
with
\begin{alignat}{2}
S_{0,\textrm{odd}} &= \sum_{i,j} q^{E_0} \qbinom{\lfloor \frac{N_t}4+\frac12\rfloor+i}{2(i+j+t)-k_1}\qbinom{i+j+t-\frac{k_1-1}2}{i-k_2}\qbinom{i+j+t-\frac{k_1+1}2}{i+t-k_3} \nonumber\\
& = \sum_{k,\ell} q^{E_0} \qbinom{\lfloor \frac{N_t}4+\frac12\rfloor+k_2+k}{2\ell-k_1}\qbinom{\ell-\frac{k_1-1}2}{k}\qbinom{\ell-\frac{k_1+1}2}{k+t+k_2-k_3}\nonumber\\
& = \sum_\ell q^{E_0'} \specialqbinom{\lfloor\tfrac{N_t}4 + \tfrac12\rfloor + k_3 - t}{\ell - \frac{k_1+1}2 - k_2 + k_3  - t}\specialqbinom{ \lfloor\tfrac{N_t}4 + \tfrac12\rfloor + k_2}{\ell - \frac{k_1-1}2 + k_2 - k_3  + t}\nonumber\\
& = q^{E_0''} \specialqbinom{2\lfloor\tfrac{N_t}4 + \tfrac12\rfloor + k_2+k_3 - t}{\lfloor\tfrac{N_t}4 + \tfrac12\rfloor + 2k_2 - k_3 + t + 1}  \simeq \frac{q^{E_0''}}{(q)_\infty}
\end{alignat}
with $E_0'$ and $E_0''$ adapted accordingly. In particular, $E_0''$ satisfies $E_0'' + E = \Delta_{1,3(k_3-k_2-t)}+ \tfrac1{6}$. Changing the summation variables to $k = 2i+1$ and $k_3 = k_2 +\ell$, we find
\begin{alignat}{2}
Z_{0,\textrm{odd}} &\simeq \frac{\bar q^{\,-\frac1{6}}}{(q)_\infty} \sum_\ell q^{\Delta_{1,3(\ell-t)}} \sum_{i,k_2} \bar q^{E} \qqbinom{\lfloor \frac{N_t}4\rfloor}{2i+1}\qqbinom{i}{k_2}\qqbinom{i+1}{k_2 + \ell}
\nonumber\\& 
\simeq \frac{\bar q^{\,-\frac1{6}}}{(q)_\infty} \sum_\ell q^{\Delta_{1,3(\ell-t)}} \sum_{i,k_2} \bar q^{E} \qqbinom{\lfloor \frac{N_t}4\rfloor+k}{2i+1}\qqbinom{i}{k_2}\qqbinom{i+1}{k_2 + \ell}\nonumber\\
& = \frac{\bar q^{\,-\frac1{6}}}{(q)_\infty} \sum_\ell q^{\Delta_{1,3(\ell-t)}} \sum_{i} \bar q^{E_0'''} \qqbinom{\lfloor \frac{N_t}4\rfloor-\ell}{i+1-\ell} \qqbinom{\lfloor \frac{N_t}4\rfloor}{i+\ell}\nonumber \\
& = \frac1{(q)_\infty} \sum_\ell q^{\Delta_{1,3(\ell-t)}}  \bar q^{\Delta_{1,3\ell}} \specialqqbinom{2\lfloor \frac{N_t}4\rfloor - \ell}{\lfloor \frac{N_t}4\rfloor+\ell-1} \simeq \frac{1}{(q)_\infty (\bar q)_\infty} \sum_\ell q^{\Delta_{1,3 (\ell-t)}}\bar q^{\Delta_{1,3 \ell}}.\label{eq:forlaststeps}
\end{alignat}
This ends the proof of \eqref{eq:Z0results}.

\paragraph{Scaling behavior for $\boldsymbol{d \equiv 1,2} \Mod \boldsymbol{3}$.} 
These cases are more complicated because of the various contributions to $\ZdNP$ in \eqref{eq:Z12separation}. For $d \equiv 1 \Mod 3$, we do the calculations only for $d>1$ for simplicity. The case $d=1$ uses the same arguments, with extra care given to the bounds of the sums and the contribution from the Razumov-Stroganov eigenvalue. Writing $d = 3t+1$ with $t>0$, each contribution to \eqref{eq:Z1separation} splits into an even and an odd part as
\be
\label{eq:noinspiration}
Z_{1}^{\tX,\tY} = (-1)^M Z_{1,\textrm{even}}^{\tX,\tY} + Z_{1,\textrm{odd}}^{\tX,\tY}\ , \quad \tX \in \{\bar\tA, \bar\tB\}, \quad \tY \in \{\tA, \tB\}.
\ee
Starting with the even contributions, as before, we choose $\epsilon$ such that the $\bar q$-binomials do not depend on $i$. Defining $N_t = N+t+1$, we obtain
\be
Z^{\tX,\tY}_{1,\textrm{even}}= \sum_{k_2,k_3} \sum_{k_1\,\textrm{even}} (q \bar q)^{E(\tX)} \qqbinom{\lfloor \frac{N_t}4\rfloor}{k_1}\qqbinom{\frac{k_1}2-1}{k_2}\qqbinom{\frac{k_1}2}{k_3}  S^{\tX,\tY}_{1,\textrm{even}}
\ee
with
\begin{subequations}
\begin{alignat}{2}
S^{\bar \tB, \tB}_{1,\textrm{even}} &= \sum_{i,j} q^{E_1(\bar \tB, \tB)} \qbinom{\lfloor \frac{N_t}4+\frac12\rfloor+i-1}{2(i+j+t)-k_1}\qbinom{i+j+t-\frac{k_1}2}{i-k_2}\qbinom{i+j+t-\frac{k_1}2-1}{i+t-k_3-1} \nonumber\\
& = \sum_{k,\ell} q^{E_1(\bar \tB, \tB)} \qbinom{\lfloor \frac{N_t}4+\frac12\rfloor+k_2+k-1}{2\ell-k_1}\qbinom{\ell-\frac{k_1}2}{k}\qbinom{\ell-\frac{k_1}2-1}{k+t+k_2-k_3-1},
\\[0.2cm]
S^{\bar \tB, \tA}_{1,\textrm{even}} &= \sum_{i,j} q^{E_1(\bar \tB, \tA)} \qbinom{\lfloor \frac{N_t}4+\frac12\rfloor+i}{2(i+j+t+1)-k_1}\qbinom{i+j+t+1-\frac{k_1}2}{i-k_2}\qbinom{i+j+t-\frac{k_1}2}{i+t-k_3} \nonumber\\
& = \sum_{k,\ell} q^{E_1(\bar \tB, \tA)} \qbinom{\lfloor \frac{N_t}4+\frac12\rfloor+k_2+k-1}{2\ell-k_1}\qbinom{\ell-\frac{k_1}2}{k-1}\qbinom{\ell-\frac{k_1}2-1}{k+t+k_2-k_3-1},
\\[0.2cm]
S^{\bar \tA, \tB}_{1,\textrm{even}} &= \sum_{i,j} q^{E_1(\bar \tA, \tB)} \qbinom{\lfloor \frac{N_t}4+\frac12\rfloor+i}{2(i+j+t+1)-k_1}\qbinom{i+j+t+1-\frac{k_1}2}{i-k_2}\qbinom{i+j+t-\frac{k_1}2}{i+t-k_3} \nonumber\\
& = \sum_{k,\ell} q^{E_1(\bar \tA, \tB)} \qbinom{\lfloor \frac{N_t}4+\frac12\rfloor+k_2+k}{2\ell-k_1}\qbinom{\ell-\frac{k_1}2}{k}\qbinom{\ell-\frac{k_1}2-1}{k+t+k_2-k_3},
\\[0.2cm]
S^{\bar \tA, \tA}_{1,\textrm{even}} &= \sum_{i,j} q^{E_1(\bar \tA, \tA)} \qbinom{\lfloor \frac{N_t}4+\frac12\rfloor+i+1}{2(i+j+t+2)-k_1}\qbinom{i+j+t+2-\frac{k_1}2}{i-k_2}\qbinom{i+j+t-\frac{k_1}2+1}{i+t-k_3+1} \nonumber\\
& = \sum_{k,\ell} q^{E_1(\bar \tA, \tA)} \qbinom{\lfloor \frac{N_t}4+\frac12\rfloor+k_2+k}{2\ell-k_1}\qbinom{\ell-\frac{k_1}2}{k-1}\qbinom{\ell-\frac{k_1}2-1}{k+t+k_2-k_3}.
\end{alignat}
\end{subequations}
It is not possible to simplify each $S^{\tX, \tY}_{1,\textrm{even}}$ individually using \eqref{eq:finaldoublebinomial}. One instead combines $S^{\bar \tB, \tB}$ and $S^{\bar \tB, \tA}$ using $E_1(\bar \tB, \tB) - E_1(\bar \tB, \tA) = j$ and \eqref{eq:qbinnm}:
\begin{alignat}{2}
S^{\bar \tB, \tB}_{1,\textrm{even}} + S^{\bar \tB, \tA}_{1,\textrm{even}} &= \sum_{k,\ell} q^{E_1(\bar \tB, \tA)} \qbinom{\lfloor \frac{N_t}4+\frac12\rfloor+k_2+k-1}{2\ell-k_1}\qbinom{\ell-\frac{k_1}2+1}{k}\qbinom{\ell-\frac{k_1}2-1}{k+t+k_2-k_3-1},\nonumber\\
& = \sum_\ell q^{E_1'(\bar \tB, \tA)}  \qbinom{\lfloor\tfrac{N_t}4 + \tfrac12\rfloor + k_3 - t}{\ell - \frac{k_1}2 - k_2 + k_3  - t}\qbinom{ \lfloor\tfrac{N_t}4 + \tfrac12\rfloor + k_2-1}{\ell - \frac{k_1}2 + k_2 - k_3  + t}\nonumber\\
& = q^{E_1''(\bar \tB, \tA)}\specialqbinom{2\lfloor\tfrac{N_t}4 + \tfrac12\rfloor + k_2+k_3 - t-1}{\lfloor\tfrac{N_t}4 + \tfrac12\rfloor + 2k_2 - k_3 + t} \simeq \frac{q^{E_1''(\bar \tB, \tA)}}{(q)_\infty}
\end{alignat}
with $E_1''(\bar \tB, \tA)$ satisfying $E_1''(\bar \tB, \tA)+E(\tB) = \Delta_{1,3(k_3-k_2-t)+1}$. The relations \eqref{eq:finaldoublebinomial} and \eqref{eq:qVan} were used for the second and third equality. Changing the summation variables in \eqref{eq:noinspiration} to $k_1 = 2i$ and $k_3 = k_2 + \ell$, we obtain
\be
Z_{1, \textrm{even}}^{\bar\tB,\tB} + Z_{1, \textrm{even}}^{\bar\tB,\tA} \simeq \frac1{(q)_\infty} \sum_\ell q^{\Delta_{1,3(\ell-t)+1}} \sum_{i,k_2} \bar q^{E(\tB)} \qqbinom{\lfloor \frac{N_t}4\rfloor}{2i}\qqbinom{i-1}{k_2}\qqbinom{i}{k_2+\ell}.
\ee
Likewise for $S^{\bar \tA, \tB}$ and $S^{\bar \tA, \tA}$, we use $E_1(\bar \tA, \tB) - E_1(\bar \tA, \tA) =j$ and find after simplification:
\be
S^{\bar \tA, \tB}_{1,\textrm{even}} + S^{\bar \tA, \tA}_{1,\textrm{even}} = q^{E_1''(\bar \tA, \tA)}\specialqbinom{2\lfloor\tfrac{N_t}4 + \tfrac12\rfloor + k_2+k_3 - t}{\lfloor\tfrac{N_t}4 + \tfrac12\rfloor + 2k_2 - k_3 + t + 2} \simeq \frac{q^{E_1''(\bar \tA, \tA)}}{(q)_\infty}
\ee
with $E_1''(\bar \tA, \tA)+E(\tB) = \Delta_{1,3(k_3-k_2-t-1)+1}$. With $k = 2i$, $k_2 = k_2'-1$ and $k_3 = k_2' + \ell$, we get
\be
Z_{1, \textrm{even}}^{\bar\tA,\tB} + Z_{1, \textrm{even}}^{\bar\tA,\tA} \simeq \frac1{(q)_\infty} \sum_\ell q^{\Delta_{1,3(\ell-t)+1}} \sum_{i,k'_2} \bar q^{E(\tB)} \qqbinom{\lfloor \frac{N_t}4\rfloor}{2i}\qqbinom{i-1}{k'_2-1}\qqbinom{i}{k'_2+\ell}.
\ee
Putting these results together, we find
\begin{alignat}{2}
Z_{1, \textrm{even}}^{\bar\tB,\tB}& + Z_{1, \textrm{even}}^{\bar\tB,\tA} + Z_{1, \textrm{even}}^{\bar\tA,\tB} + Z_{1, \textrm{even}}^{\bar\tA,\tA} 
\simeq \frac1{(q)_\infty} \sum_\ell q^{\Delta_{1,3(\ell-t)+1}} \sum_{i,k_2} \bar q^{E_1(\tA)} \qqbinom{\lfloor \frac{N_t}4\rfloor}{2i}\qqbinom{i}{k_2}\qqbinom{i}{k_2 + \ell} \nonumber \\
&\simeq \frac1{(q)_\infty} \sum_\ell q^{\Delta_{1,3(\ell-t)+1}} \sum_{i,k_2} \bar q^{E_1(\tA)} \qqbinom{\lfloor \frac{N_t}4\rfloor+k_2}{2i}\qqbinom{i}{k_2}\qqbinom{i}{k_2 + \ell} \nonumber\\
& = \dots = \frac1{(q)_\infty} \sum_\ell q^{\Delta_{1,3(\ell-t)+1}}  \bar q^{\Delta_{1,3\ell+2}} \specialqqbinom{2\lfloor \frac{N_t}4\rfloor - \ell}{\lfloor \frac{N_t}4\rfloor+\ell} \simeq \frac{1}{(q)_\infty (\bar q)_\infty} \sum_\ell q^{\Delta_{1,3 (\ell-t)+1}}\bar q^{\Delta_{1,3 \ell+2}}
\end{alignat}
where the last steps follow those in \eqref{eq:forlaststeps}. The result produces the second term in \eqref{eq:Z1results}.

The odd contributions are treated similarly. We thus only write down the intermediate results. Each contribution is written as
\be
Z^{\tX,\tY}_{1,\textrm{odd}}= (q \bar q)^{\frac18}\sum_{k_2,k_3} \sum_{k_1\,\textrm{odd}} (q \bar q)^{E(\tX)} \qqbinom{\lfloor \frac{N_t}4\rfloor}{k_1}\qqbinom{\frac{k_1-1}2}{k_2}\qqbinom{\frac{k_1-1}2}{k_3}  S^{\tX,\tY}_{1,\textrm{odd}}
\ee
with
\begin{subequations}
\begin{alignat}{2}
S^{\bar \tB, \tB}_{1,\textrm{odd}} & = \sum_{k,\ell} q^{E_1(\bar \tB, \tB)} \qbinom{\lfloor \frac{N_t}4+\frac12\rfloor+k_2+k-1}{2\ell-k_1}\qbinom{\ell-\frac{k_1+1}2}{k}\qbinom{\ell-\frac{k_1+1}2}{k+t+k_2-k_3-1},
\\[0.2cm]
S^{\bar \tB, \tA}_{1,\textrm{odd}} & = \sum_{k,\ell} q^{E_1(\bar \tB, \tA)} \qbinom{\lfloor \frac{N_t}4+\frac12\rfloor+k_2+k-1}{2\ell-k_1}\qbinom{\ell-\frac{k_1+1}2}{k-1}\qbinom{\ell-\frac{k_1+1}2}{k+t+k_2-k_3-1},
\\[0.2cm]
S^{\bar \tA, \tB}_{1,\textrm{odd}} & = \sum_{k,\ell} q^{E_1(\bar \tA, \tB)} \qbinom{\lfloor \frac{N_t}4+\frac12\rfloor+k_2+k}{2\ell-k_1}\qbinom{\ell-\frac{k_1+1}2}{k}\qbinom{\ell-\frac{k_1+1}2}{k+t+k_2-k_3},
\\[0.2cm]
S^{\bar \tA, \tA}_{1,\textrm{odd}} & = \sum_{k,\ell} q^{E_1(\bar \tA, \tA)} \qbinom{\lfloor \frac{N_t}4+\frac12\rfloor+k_2+k}{2\ell-k_1}\qbinom{\ell-\frac{k_1+1}2}{k-1}\qbinom{\ell-\frac{k_1+1}2}{k+t+k_2-k_3}.
\end{alignat}
These combine pairwise:
\end{subequations}
\be
S^{\bar \tB, \tB}_{1,\textrm{odd}} + S^{\bar \tB, \tA}_{1,\textrm{odd}} \simeq \frac{q^{ \Delta_{0,3(k_3-k_2-t)+1}-\frac18-E(\tB)}}{(q)_\infty},
\qquad
S^{\bar \tA, \tB}_{1,\textrm{odd}} + S^{\bar \tA, \tA}_{1,\textrm{odd}} \simeq \frac{q^{ \Delta_{0,3(k_3-k_2-t)-2}-\frac18-E(\tA)}}{(q)_\infty}.
\ee
The partial partition functions then satisfy
\begin{alignat}{2}
Z_{1, \textrm{odd}}^{\bar\tB,\tB} + Z_{1, \textrm{odd}}^{\bar\tB,\tA} &\simeq \frac{\bar q^{\frac18}}{(q)_\infty} \sum_\ell q^{\Delta_{0,3(\ell-t)+1}} \sum_{i,k_2} \bar q^{E(\tB)} \qqbinom{\lfloor \frac{N_t}4\rfloor}{2i+1}\qqbinom{i}{k_2}\qqbinom{i}{k_2+\ell},
\\
Z_{1, \textrm{odd}}^{\bar\tA,\tB} + Z_{1, \textrm{odd}}^{\bar\tA,\tA} &\simeq \frac{\bar q^{\frac18}}{(q)_\infty} \sum_\ell q^{\Delta_{0,3(\ell-t)+1}} \sum_{i,k_2} \bar q^{E(\tA)} \qqbinom{\lfloor \frac{N_t}4\rfloor}{2i+1}\qqbinom{i}{k_2-1}\qqbinom{i}{k_2+\ell},
\end{alignat}
and the final result is
\be
Z_{1, \textrm{odd}}^{\bar\tB,\tB} + Z_{1, \textrm{odd}}^{\bar\tB,\tA} + Z_{1, \textrm{odd}}^{\bar\tA,\tB} + Z_{1, \textrm{odd}}^{\bar\tA,\tA} 
\simeq \frac{1}{(q)_\infty (\bar q)_\infty} \sum_\ell q^{\Delta_{0,3 (\ell-t)+1}}\bar q^{\Delta_{0,3 \ell+2}},
\ee
as announced in \eqref{eq:Z1results}.

For $d \equiv 2 \Mod 3$, the derivation is done using the same ideas. One considers separately the odd and even $k_1$ contributions. For the even case (and likewise for the odd case), one chooses $\epsilon$ so only the $q$-dependent part depends on the sum label $i$. One writes down the $S^{\tX, \tY}_{2,\textrm{even}}$ corresponding to each $Z^{\tX,\tY}_{2,\textrm{even}}$, and then combines $S^{\bar \tA, \tA}_{2,\textrm{even}}$ with $S^{\bar \tA, \tB}_{2,\textrm{even}}$ and $S^{\bar \tB, \tA}_{2,\textrm{even}}$ with $S^{\bar \tB, \tB}_{2,\textrm{even}}$. Their scaling limits are evaluated using \eqref{eq:qVan} and \eqref{eq:finaldoublebinomial} as the ratio of a power of $q$ with $(q)_\infty$. Combining the four contributions and using \eqref{eq:qVan} and \eqref{eq:finaldoublebinomial} one last time, the final result is
\begin{subequations}
\begin{alignat}{2}
Z_{2, \textrm{even}}^{\bar\tB,\tB} + Z_{2, \textrm{even}}^{\bar\tB,\tA} + Z_{2, \textrm{even}}^{\bar\tA,\tB} + Z_{2, \textrm{even}}^{\bar\tA,\tA} 
&\simeq \frac{1}{(q)_\infty (\bar q)_\infty} \sum_\ell q^{\Delta_{1,3 (\ell-t)-1}}\bar q^{\Delta_{1,3 \ell+1}},
\\
Z_{2, \textrm{odd}}^{\bar\tB,\tB} + Z_{2, \textrm{odd}}^{\bar\tB,\tA} + Z_{2, \textrm{odd}}^{\bar\tA,\tB} + Z_{2, \textrm{odd}}^{\bar\tA,\tA} 
&\simeq \frac{1}{(q)_\infty (\bar q)_\infty} \sum_\ell q^{\Delta_{0,3 (\ell-t)-1}}\bar q^{\Delta_{0,3 \ell+1}},
\end{alignat}
\end{subequations}
consistent with \eqref{eq:Z2results}.

%
\section{Torus partition functions of critical dense polymers}\label{sec:CDPpartitionfunctions}
%

For critical dense polymers, namely $\mathcal{LM}(1,2)$, the torus conformal partition functions for even $M$ are written in terms of $\mathcal W$-irreducible characters in \cite{PRVcyl2010,MDPR13}. Here we express these results in terms of the $u(1)$ characters and present the results for $M$ odd. In this case, one must keep track of the overall sign $\varepsilon$ of each eigenvalue. In the scaling limit, the partition function in each standard module $\stanp_n^d$ then reads
\be
Z_{d}(q,\bar q) = \frac{(q \bar q)^{1/12}}{(q)_\infty(\bar q)_\infty} \sum_{\ell \in \mathbb Z} (-1)^{M \ell} q^{\Delta_{2\ell +d/2}} {\bar q}^{\Delta_{2\ell -d/2}}.
\ee
This holds for the odd and even parities of $N$ and for all values of $d$. The torus partition of the loop model with $\alpha = 2$ is computed from the modified trace \eqref{eq:Zmodifiedtrace}. After simplification, we find that it can be written in terms of the $u(1)$ characters $\varkappa^8_{j}(q)$:
\begin{subequations}
\begin{alignat}{2}
\Zcdpodd &= \sum_{j=1,3,5,7} d_j^8 \Big( |\varkappa^8_j(q)|^2 + (-1)^M \varkappa^8_j(q)\varkappa^8_{8-j}(\bar q)\Big),\\
\Zcdpeven &= \sum_{j=0,2,4,6,8} d_j^8 \Big( |\varkappa^8_j(q)|^2 + (-1)^M \varkappa^8_j(q)\varkappa^8_{8-j}(\bar q)\Big).
\end{alignat}
\end{subequations}
From the transformation laws \eqref{eq:transforms}, the cases with $M$ and $N$ of the same parity are found to be fully modular invariant, whereas the cases with opposite parities are invariant under the action of $T$ and covariant under the action of $S$, as is the case for percolation in \eqref{eq:covariance}. Using
\be
\varkappa_{2j}^{4n}(q) \pm \varkappa_{4n-2j}^{4n}(q) = \varkappa_j^n(q,\pm 1),
\ee
the torus partition function for critical dense polymers can also be written in terms of the $u(1)$-characters $\varkappa^2_j(q,\pm1)$, with $j$ integer and half-integer for $N$ even and odd respectively:
\begin{subequations}
\begin{alignat}{2}
\label{eq:Zcdptake 2}
\Zcdpodd &= 2 \varkappa^2_{1/2}(q,1)\varkappa^2_{1/2}(\bar q,(-1)^M) + 2 \varkappa^2_{3/2}(q,1)\varkappa^2_{3/2}(\bar q,(-1)^M),\\[0.15cm]
\Zcdpeven &= \varkappa^2_{0}(q,1)\varkappa^2_{0}\big(\bar q,(-1)^M\big) +2 \varkappa^2_{1}(q,1)\varkappa^2_{1}\big(\bar q,(-1)^M\big) +  \varkappa^2_{2}(q,1)\varkappa^2_{2}\big(\bar q,(-1)^M\big).
\end{alignat}
\end{subequations}
For $M$ and $N$ both even, the resulting conformal torus partition function is $Z^{\text{Circ}}_{1,2}(q)$.

\newpage
%
\section{Examples of patterns of zeros}
\subsection{Strip boundary conditions}
\label{sec:pats}
%

\begin{samepage}
\vspace{-0.5cm}
\begin{center}
\begin{equation*}

\end{equation*}
\captionof{figure}{The patterns of zeros for the ground states in $\stan_{14}^d$ and $\stan_{15}^d$.}
\label{fig:GSpatterns}
\end{center}
\vspace{-.6cm}

\begin{center}
\begin{alignat*}{2}
&\begin{array}{rcccccccc}
&
\psun
\begin{pspicture}[shift=-1.5](-0.4,-3.2)(1.4,0.5)
\psline{-}(0,-2.2)(0,0)(1,0)(1,-2.2)
\psdots[dotsize=0.09cm](0,-0.3)(0,-0.6)(0,-0.9)
\psdots[dotsize=0.09cm](0.5,-1.2)
\psdots[dotsize=0.09cm](1,-0.3)(1,-0.6)(1,-0.9)
\end{pspicture}
\begin{pspicture}[shift=-1.5](-0.4,-3.2)(1.4,0.5)
\psline{-}(0,-2.2)(0,0)(1,0)(1,-2.2)
\end{pspicture}
&
\psun
\begin{pspicture}[shift=-1.5](-0.4,-3.2)(1.4,0.5)
\psline{-}(0,-2.2)(0,0)(1,0)(1,-2.2)
\psdots[dotsize=0.09cm](0,-0.3)(0,-0.6)(0,-1.2)
\psdots[dotsize=0.09cm](0.5,-0.9)
\psdots[dotsize=0.09cm](1,-0.3)(1,-0.6)(1,-1.2)
\end{pspicture}
\begin{pspicture}[shift=-1.5](-0.4,-3.2)(1.4,0.5)
\psline{-}(0,-2.2)(0,0)(1,0)(1,-2.2)
\end{pspicture}
&
\psun
\begin{pspicture}[shift=-1.5](-0.4,-3.2)(1.4,0.5)
\psline{-}(0,-2.2)(0,0)(1,0)(1,-2.2)
\psdots[dotsize=0.09cm](0,-0.3)(0,-0.6)
\psdots[dotsize=0.09cm](0.5,-0.9)(0.5,-1.2)(0.5,-1.5)
\psdots[dotsize=0.09cm](1,-0.3)(1,-0.6)
\end{pspicture}
\begin{pspicture}[shift=-1.5](-0.4,-3.2)(1.4,0.5)
\psline{-}(0,-2.2)(0,0)(1,0)(1,-2.2)
\psdots[linecolor=black,fillcolor=lightgray,dotstyle=o,dotsize=0.09cm](0.5,-0.6)
\end{pspicture}
&
\psun
\begin{pspicture}[shift=-1.5](-0.4,-3.2)(1.4,0.5)
\psline{-}(0,-2.2)(0,0)(1,0)(1,-2.2)
\psdots[dotsize=0.09cm](0,-0.3)(0,-0.9)(0,-1.2)
\psdots[dotsize=0.09cm](0.5,-0.6)
\psdots[dotsize=0.09cm](1,-0.3)(1,-0.9)(1,-1.2)
\end{pspicture}
\begin{pspicture}[shift=-1.5](-0.4,-3.2)(1.4,0.5)
\psline{-}(0,-2.2)(0,0)(1,0)(1,-2.2)
\end{pspicture}
&
\psun
\begin{pspicture}[shift=-1.5](-0.4,-3.2)(1.4,0.5)
\psline{-}(0,-2.2)(0,0)(1,0)(1,-2.2)
\psdots[dotsize=0.09cm](0,-0.3)(0,-0.9)
\psdots[dotsize=0.09cm](0.5,-0.6)(0.5,-1.2)(0.5,-1.5)
\psdots[dotsize=0.09cm](1,-0.3)(1,-0.9)
\end{pspicture}
\begin{pspicture}[shift=-1.5](-0.4,-3.2)(1.4,0.5)
\psline{-}(0,-2.2)(0,0)(1,0)(1,-2.2)
\psdots[linecolor=black,fillcolor=lightgray,dotstyle=o,dotsize=0.09cm](0.5,-0.6)
\end{pspicture}
&
\psun
\begin{pspicture}[shift=-1.5](-0.4,-3.2)(1.4,0.5)
\psline{-}(0,-2.2)(0,0)(1,0)(1,-2.2)
\psdots[dotsize=0.09cm](0,-0.6)(0,-0.9)(0,-1.2)
\psdots[dotsize=0.09cm](0.5,-0.3)
\psdots[dotsize=0.09cm](1,-0.6)(1,-0.9)(1,-1.2)
\end{pspicture}
\begin{pspicture}[shift=-1.5](-0.4,-3.2)(1.4,0.5)
\psline{-}(0,-2.2)(0,0)(1,0)(1,-2.2)
\end{pspicture}
&
\psun
\begin{pspicture}[shift=-1.5](-0.4,-3.2)(1.4,0.5)
\psline{-}(0,-2.2)(0,0)(1,0)(1,-2.2)
\psdots[dotsize=0.09cm](0,-0.3)(0,-1.2)
\psdots[dotsize=0.09cm](0.5,-0.6)(0.5,-0.9)(0.5,-1.5)
\psdots[dotsize=0.09cm](1,-0.3)(1,-1.2)
\end{pspicture}
\begin{pspicture}[shift=-1.5](-0.4,-3.2)(1.4,0.5)
\psline{-}(0,-2.2)(0,0)(1,0)(1,-2.2)
\psdots[linecolor=black,fillcolor=lightgray,dotstyle=o,dotsize=0.09cm](0.5,-0.6)
\end{pspicture}
&
\psun
\begin{pspicture}[shift=-1.5](-0.4,-3.2)(1.4,0.5)
\psline{-}(0,-2.2)(0,0)(1,0)(1,-2.2)
\psdots[dotsize=0.09cm](0,-0.3)
\psdots[dotsize=0.09cm](0.5,-0.6)(0.5,-0.9)(0.5,-1.2)
\psdots[dotsize=0.09cm](1,-0.3)
\end{pspicture}
\begin{pspicture}[shift=-1.5](-0.4,-3.2)(1.4,0.5)
\psline{-}(0,-2.2)(0,0)(1,0)(1,-2.2)
\psdots[linecolor=black,fillcolor=lightgray,dotstyle=o,dotsize=0.09cm](0,-0.6)
\psdots[linecolor=black,fillcolor=lightgray,dotstyle=o,dotsize=0.09cm](1,-0.6)
\end{pspicture}
\\
\end{array}\\
\\[-0.5cm]&
\begin{array}{rcccccccc}
&
\psun
\begin{pspicture}[shift=-1.5](-0.4,-3.2)(1.4,0.5)
\psline{-}(0,-2.2)(0,0)(1,0)(1,-2.2)
\psdots[dotsize=0.09cm](0,-0.6)(0,-0.9)
\psdots[dotsize=0.09cm](0.5,-0.3)(0.5,-1.2)(0.5,-1.5)
\psdots[dotsize=0.09cm](1,-0.6)(1,-0.9)
\end{pspicture}
\begin{pspicture}[shift=-1.5](-0.4,-3.2)(1.4,0.5)
\psline{-}(0,-2.2)(0,0)(1,0)(1,-2.2)
\psdots[linecolor=black,fillcolor=lightgray,dotstyle=o,dotsize=0.09cm](0.5,-0.6)
\end{pspicture}
&
\psun
\begin{pspicture}[shift=-1.5](-0.4,-3.2)(1.4,0.5)
\psline{-}(0,-2.2)(0,0)(1,0)(1,-2.2)
\psdots[dotsize=0.09cm](0,-0.3)(0,-1.5)
\psdots[dotsize=0.09cm](0.5,-0.6)(0.5,-0.9)(0.5,-1.2)
\psdots[dotsize=0.09cm](1,-0.3)(1,-1.5)
\end{pspicture}
\begin{pspicture}[shift=-1.5](-0.4,-3.2)(1.4,0.5)
\psline{-}(0,-2.2)(0,0)(1,0)(1,-2.2)
\psdots[linecolor=black,fillcolor=lightgray,dotstyle=o,dotsize=0.09cm](0.5,-0.6)
\end{pspicture}
&
\psun
\begin{pspicture}[shift=-1.5](-0.4,-3.2)(1.4,0.5)
\psline{-}(0,-2.2)(0,0)(1,0)(1,-2.2)
\psdots[dotsize=0.09cm](0,-0.6)(0,-1.2)
\psdots[dotsize=0.09cm](0.5,-0.3)(0.5,-0.9)(0.5,-1.5)
\psdots[dotsize=0.09cm](1,-0.6)(1,-1.2)
\end{pspicture}
\begin{pspicture}[shift=-1.5](-0.4,-3.2)(1.4,0.5)
\psline{-}(0,-2.2)(0,0)(1,0)(1,-2.2)
\psdots[linecolor=black,fillcolor=lightgray,dotstyle=o,dotsize=0.09cm](0.5,-0.6)
\end{pspicture}
&
\psun
\begin{pspicture}[shift=-1.5](-0.4,-3.2)(1.4,0.5)
\psline{-}(0,-2.2)(0,0)(1,0)(1,-2.2)
\psdots[dotsize=0.09cm](0,-0.6)
\psdots[dotsize=0.09cm](0.5,-0.3)(0.5,-0.9)(0.5,-1.2)
\psdots[dotsize=0.09cm](1,-0.6)
\end{pspicture}
\begin{pspicture}[shift=-1.5](-0.4,-3.2)(1.4,0.5)
\psline{-}(0,-2.2)(0,0)(1,0)(1,-2.2)
\psdots[linecolor=black,fillcolor=lightgray,dotstyle=o,dotsize=0.09cm](0,-0.6)
\psdots[linecolor=black,fillcolor=lightgray,dotstyle=o,dotsize=0.09cm](1,-0.6)
\end{pspicture}
&
\psun
\begin{pspicture}[shift=-1.5](-0.4,-3.2)(1.4,0.5)
\psline{-}(0,-2.2)(0,0)(1,0)(1,-2.2)
\psdots[dotsize=0.09cm](0,-0.3)
\psdots[dotsize=0.09cm](0.5,-0.6)(0.5,-0.9)(0.5,-1.2)(0.5,-1.5)(0.5,-1.8)
\psdots[dotsize=0.09cm](1,-0.3)
\end{pspicture}
\begin{pspicture}[shift=-1.5](-0.4,-3.2)(1.4,0.5)
\psline{-}(0,-2.2)(0,0)(1,0)(1,-2.2)
\psdots[linecolor=black,fillcolor=lightgray,dotstyle=o,dotsize=0.09cm](0.5,-0.6)(0.5,-1.2)
\end{pspicture}
&
\psun
\begin{pspicture}[shift=-1.5](-0.4,-3.2)(1.4,0.5)
\psline{-}(0,-2.2)(0,0)(1,0)(1,-2.2)
\psdots[dotsize=0.09cm](0,-0.6)(0,-1.5)
\psdots[dotsize=0.09cm](0.5,-0.3)(0.5,-0.9)(0.5,-1.2)
\psdots[dotsize=0.09cm](1,-0.6)(1,-1.5)
\end{pspicture}
\begin{pspicture}[shift=-1.5](-0.4,-3.2)(1.4,0.5)
\psline{-}(0,-2.2)(0,0)(1,0)(1,-2.2)
\psdots[linecolor=black,fillcolor=lightgray,dotstyle=o,dotsize=0.09cm](0.5,-0.6)
\end{pspicture}
&
\psun
\begin{pspicture}[shift=-1.5](-0.4,-3.2)(1.4,0.5)
\psline{-}(0,-2.2)(0,0)(1,0)(1,-2.2)
\psdots[dotsize=0.09cm](0,-0.9)(0,-1.2)
\psdots[dotsize=0.09cm](0.5,-0.3)(0.5,-0.6)(0.5,-1.5)
\psdots[dotsize=0.09cm](1,-0.9)(1,-1.2)
\end{pspicture}
\begin{pspicture}[shift=-1.5](-0.4,-3.2)(1.4,0.5)
\psline{-}(0,-2.2)(0,0)(1,0)(1,-2.2)
\psdots[linecolor=black,fillcolor=lightgray,dotstyle=o,dotsize=0.09cm](0.5,-0.6)
\end{pspicture}
&
\psun
\begin{pspicture}[shift=-1.5](-0.4,-3.2)(1.4,0.5)
\psline{-}(0,-2.2)(0,0)(1,0)(1,-2.2)
\psdots[dotsize=0.09cm](0,-0.9)
\psdots[dotsize=0.09cm](0.5,-0.3)(0.5,-0.6)(0.5,-1.2)
\psdots[dotsize=0.09cm](1,-0.9)
\end{pspicture}
\begin{pspicture}[shift=-1.5](-0.4,-3.2)(1.4,0.5)
\psline{-}(0,-2.2)(0,0)(1,0)(1,-2.2)
\psdots[linecolor=black,fillcolor=lightgray,dotstyle=o,dotsize=0.09cm](0,-0.6)
\psdots[linecolor=black,fillcolor=lightgray,dotstyle=o,dotsize=0.09cm](1,-0.6)
\end{pspicture}
\\
\end{array}\\
\\[-0.5cm]&
\begin{array}{rcccccccc}
&
\psun
\begin{pspicture}[shift=-1.5](-0.4,-3.2)(1.4,0.5)
\psline{-}(0,-2.2)(0,0)(1,0)(1,-2.2)
\psdots[dotsize=0.09cm](0,-0.6)
\psdots[dotsize=0.09cm](0.5,-0.3)(0.5,-0.9)(0.5,-1.2)(0.5,-1.5)(0.5,-1.8)
\psdots[dotsize=0.09cm](1,-0.6)
\end{pspicture}
\begin{pspicture}[shift=-1.5](-0.4,-3.2)(1.4,0.5)
\psline{-}(0,-2.2)(0,0)(1,0)(1,-2.2)
\psdots[linecolor=black,fillcolor=lightgray,dotstyle=o,dotsize=0.09cm](0.5,-0.6)(0.5,-1.2)
\end{pspicture}
&
\psun
\begin{pspicture}[shift=-1.5](-0.4,-3.2)(1.4,0.5)
\psline{-}(0,-2.2)(0,0)(1,0)(1,-2.2)
\psdots[dotsize=0.09cm](0,-0.9)(0,-1.5)
\psdots[dotsize=0.09cm](0.5,-0.3)(0.5,-0.6)(0.5,-1.2)
\psdots[dotsize=0.09cm](1,-0.9)(1,-1.5)
\end{pspicture}
\begin{pspicture}[shift=-1.5](-0.4,-3.2)(1.4,0.5)
\psline{-}(0,-2.2)(0,0)(1,0)(1,-2.2)
\psdots[linecolor=black,fillcolor=lightgray,dotstyle=o,dotsize=0.09cm](0.5,-0.6)
\end{pspicture}
&
\psun
\begin{pspicture}[shift=-1.5](-0.4,-3.2)(1.4,0.5)
\psline{-}(0,-2.2)(0,0)(1,0)(1,-2.2)
\psdots[dotsize=0.09cm](0,-1.2)
\psdots[dotsize=0.09cm](0.5,-0.3)(0.5,-0.6)(0.5,-0.9)
\psdots[dotsize=0.09cm](1,-1.2)
\end{pspicture}
\begin{pspicture}[shift=-1.5](-0.4,-3.2)(1.4,0.5)
\psline{-}(0,-2.2)(0,0)(1,0)(1,-2.2)
\psdots[linecolor=black,fillcolor=lightgray,dotstyle=o,dotsize=0.09cm](0,-0.6)
\psdots[linecolor=black,fillcolor=lightgray,dotstyle=o,dotsize=0.09cm](1,-0.6)
\end{pspicture}
&
\psun
\begin{pspicture}[shift=-1.5](-0.4,-3.2)(1.4,0.5)
\psline{-}(0,-2.2)(0,0)(1,0)(1,-2.2)
\psdots[dotsize=0.09cm](0,-0.9)
\psdots[dotsize=0.09cm](0.5,-0.3)(0.5,-0.6)(0.5,-1.2)(0.5,-1.5)(0.5,-1.8)
\psdots[dotsize=0.09cm](1,-0.9)
\end{pspicture}
\begin{pspicture}[shift=-1.5](-0.4,-3.2)(1.4,0.5)
\psline{-}(0,-2.2)(0,0)(1,0)(1,-2.2)
\psdots[linecolor=black,fillcolor=lightgray,dotstyle=o,dotsize=0.09cm](0.5,-0.6)(0.5,-1.2)
\end{pspicture}
&
\psun
\begin{pspicture}[shift=-1.5](-0.4,-3.2)(1.4,0.5)
\psline{-}(0,-2.2)(0,0)(1,0)(1,-2.2)
\psdots[dotsize=0.09cm](0,-1.2)(0,-1.5)
\psdots[dotsize=0.09cm](0.5,-0.3)(0.5,-0.6)(0.5,-0.9)
\psdots[dotsize=0.09cm](1,-1.2)(1,-1.5)
\end{pspicture}
\begin{pspicture}[shift=-1.5](-0.4,-3.2)(1.4,0.5)
\psline{-}(0,-2.2)(0,0)(1,0)(1,-2.2)
\psdots[linecolor=black,fillcolor=lightgray,dotstyle=o,dotsize=0.09cm](0.5,-0.6)
\end{pspicture}
&
\psun
\begin{pspicture}[shift=-1.5](-0.4,-3.2)(1.4,0.5)
\psline{-}(0,-2.2)(0,0)(1,0)(1,-2.2)
\psdots[dotsize=0.09cm](0.5,-0.3)(0.5,-0.6)(0.5,-0.9)(0.5,-1.2)(0.5,-1.5)
\end{pspicture}
\begin{pspicture}[shift=-1.5](-0.4,-3.2)(1.4,0.5)
\psline{-}(0,-2.2)(0,0)(1,0)(1,-2.2)
\psdots[linecolor=black,fillcolor=lightgray,dotstyle=o,dotsize=0.09cm](0,-0.6)
\psdots[linecolor=black,fillcolor=lightgray,dotstyle=o,dotsize=0.09cm](0.5,-1.2)
\psdots[linecolor=black,fillcolor=lightgray,dotstyle=o,dotsize=0.09cm](1,-0.6)
\end{pspicture}
&
\psun
\begin{pspicture}[shift=-1.5](-0.4,-3.2)(1.4,0.5)
\psline{-}(0,-2.2)(0,0)(1,0)(1,-2.2)
\psdots[dotsize=0.09cm](0,-1.2)
\psdots[dotsize=0.09cm](0.5,-0.3)(0.5,-0.6)(0.5,-0.9)(0.5,-1.5)(0.5,-1.8)
\psdots[dotsize=0.09cm](1,-1.2)
\end{pspicture}
\begin{pspicture}[shift=-1.5](-0.4,-3.2)(1.4,0.5)
\psline{-}(0,-2.2)(0,0)(1,0)(1,-2.2)
\psdots[linecolor=black,fillcolor=lightgray,dotstyle=o,dotsize=0.09cm](0.5,-0.6)(0.5,-1.2)
\end{pspicture}
\\
\end{array}\\
\\[-0.5cm]&
\begin{array}{rccccccccc}
&
\psun
\begin{pspicture}[shift=-1.5](-0.4,-3.2)(1.4,0.5)
\psline{-}(0,-2.2)(0,0)(1,0)(1,-2.2)
\psdots[dotsize=0.09cm](0.5,-0.3)(0.5,-0.6)(0.5,-0.9)(0.5,-1.2)(0.5,-1.5)
\end{pspicture}
\begin{pspicture}[shift=-1.5](-0.4,-3.2)(1.4,0.5)
\psline{-}(0,-2.2)(0,0)(1,0)(1,-2.2)
\psdots[dotsize=0.09cm](0,-0.6)(0,-1.2)
\psdots[dotsize=0.09cm](0.5,-0.6)(0.5,-1.2)
\psdots[dotsize=0.09cm](1,-0.6)(1,-1.2)
\end{pspicture}
&
\psun
\begin{pspicture}[shift=-1.5](-0.4,-3.2)(1.4,0.5)
\psline{-}(0,-2.2)(0,0)(1,0)(1,-2.2)
\psdots[dotsize=0.09cm](0,-1.5)
\psdots[dotsize=0.09cm](0.5,-0.3)(0.5,-0.6)(0.5,-0.9)(0.5,-1.2)(0.5,-1.8)
\psdots[dotsize=0.09cm](1,-1.5)
\end{pspicture}
\begin{pspicture}[shift=-1.5](-0.4,-3.2)(1.4,0.5)
\psline{-}(0,-2.2)(0,0)(1,0)(1,-2.2)
\psdots[linecolor=black,fillcolor=lightgray,dotstyle=o,dotsize=0.09cm](0.5,-0.6)(0.5,-1.2)
\end{pspicture}
&
\psun
\begin{pspicture}[shift=-1.5](-0.4,-3.2)(1.4,0.5)
\psline{-}(0,-2.2)(0,0)(1,0)(1,-2.2)
\psdots[dotsize=0.09cm](0.5,-0.3)(0.5,-0.6)(0.5,-0.9)(0.5,-1.2)(0.5,-1.5)
\end{pspicture}
\begin{pspicture}[shift=-1.5](-0.4,-3.2)(1.4,0.5)
\psline{-}(0,-2.2)(0,0)(1,0)(1,-2.2)
\psdots[linecolor=black,fillcolor=lightgray,dotstyle=o,dotsize=0.09cm](0,-1.2)
\psdots[linecolor=black,fillcolor=lightgray,dotstyle=o,dotsize=0.09cm](0.5,-0.6)
\psdots[linecolor=black,fillcolor=lightgray,dotstyle=o,dotsize=0.09cm](1,-1.2)
\end{pspicture}
&
\psun
\begin{pspicture}[shift=-1.5](-0.4,-3.2)(1.4,0.5)
\psline{-}(0,-2.2)(0,0)(1,0)(1,-2.2)
\psdots[dotsize=0.09cm](0,-1.8)
\psdots[dotsize=0.09cm](0.5,-0.3)(0.5,-0.6)(0.5,-0.9)(0.5,-1.2)(0.5,-1.5)
\psdots[dotsize=0.09cm](1,-1.8)
\end{pspicture}
\begin{pspicture}[shift=-1.5](-0.4,-3.2)(1.4,0.5)
\psline{-}(0,-2.2)(0,0)(1,0)(1,-2.2)
\psdots[linecolor=black,fillcolor=lightgray,dotstyle=o,dotsize=0.09cm](0.5,-0.6)(0.5,-1.2)
\end{pspicture}
&
\psun
\begin{pspicture}[shift=-1.5](-0.4,-3.2)(1.4,0.5)
\psline{-}(0,-2.2)(0,0)(1,0)(1,-2.2)
\psdots[dotsize=0.09cm](0.5,-0.3)(0.5,-0.6)(0.5,-0.9)(0.5,-1.2)(0.5,-1.5)(0.5,-1.8)(0.5,-2.1)
\end{pspicture}
\begin{pspicture}[shift=-1.5](-0.4,-3.2)(1.4,0.5)
\psline{-}(0,-2.2)(0,0)(1,0)(1,-2.2)
\psdots[linecolor=black,fillcolor=lightgray,dotstyle=o,dotsize=0.09cm](0.5,-0.6)(0.5,-1.2)(0.5,-1.8)
\end{pspicture}
\\
\end{array}
\end{alignat*}
\captionof{figure}{\label{fig:pats82}The 28 patterns of zeros for $\stan_8^2$.}
\end{center}
\end{samepage}

\begin{center}
\begin{alignat*}{2}
&\begin{array}{rcccccccccc}
&
\psun
\begin{pspicture}[shift=-1.5](-0.4,-3.2)(1.4,0.5)
\psline{-}(0,-2.2)(0,0)(1,0)(1,-2.2)
\psdots[dotsize=0.09cm](0,-0.3)(0,-0.6)
\psdots[dotsize=0.09cm](0.5,-0.9)(0.5,-1.2)
\psdots[dotsize=0.09cm](1,-0.3)(1,-0.6)
\end{pspicture}
\begin{pspicture}[shift=-1.5](-0.4,-3.2)(1.4,0.5)
\psline{-}(0,-2.2)(0,0)(1,0)(1,-2.2)
\rput(-0.4,-2.5){\tiny{B}}
\end{pspicture}
&
\psun
\begin{pspicture}[shift=-1.5](-0.4,-3.2)(1.4,0.5)
\psline{-}(0,-2.2)(0,0)(1,0)(1,-2.2)
\psdots[dotsize=0.09cm](0,-0.3)(0,-0.9)
\psdots[dotsize=0.09cm](0.5,-0.6)(0.5,-1.2)
\psdots[dotsize=0.09cm](1,-0.3)(1,-0.9)
\end{pspicture}
\begin{pspicture}[shift=-1.5](-0.4,-3.2)(1.4,0.5)
\psline{-}(0,-2.2)(0,0)(1,0)(1,-2.2)
\rput(-0.4,-2.5){\tiny{B}}
\end{pspicture}
&
\psun
\begin{pspicture}[shift=-1.5](-0.4,-3.2)(1.4,0.5)
\psline{-}(0,-2.2)(0,0)(1,0)(1,-2.2)
\psdots[dotsize=0.09cm](0,-0.3)(0,-1.2)
\psdots[dotsize=0.09cm](0.5,-0.6)(0.5,-0.9)
\psdots[dotsize=0.09cm](1,-0.3)(1,-1.2)
\end{pspicture}
\begin{pspicture}[shift=-1.5](-0.4,-3.2)(1.4,0.5)
\psline{-}(0,-2.2)(0,0)(1,0)(1,-2.2)
\rput(-0.4,-2.5){\tiny{B}}
\end{pspicture}
&
\psun
\begin{pspicture}[shift=-1.5](-0.4,-3.2)(1.4,0.5)
\psline{-}(0,-2.2)(0,0)(1,0)(1,-2.2)
\psdots[dotsize=0.09cm](1,-0.6)(1,-0.9)
\psdots[dotsize=0.09cm](0.5,-0.3)(0.5,-1.2)
\psdots[dotsize=0.09cm](0,-0.6)(0,-0.9)
\end{pspicture}
\begin{pspicture}[shift=-1.5](-0.4,-3.2)(1.4,0.5)
\psline{-}(0,-2.2)(0,0)(1,0)(1,-2.2)
\rput(-0.4,-2.5){\tiny{B}}
\end{pspicture}
&
\psun
\begin{pspicture}[shift=-1.5](-0.4,-3.2)(1.4,0.5)
\psline{-}(0,-2.2)(0,0)(1,0)(1,-2.2)
\psdots[dotsize=0.09cm](1,-0.3)
\psdots[dotsize=0.09cm](0.5,-0.6)(0.5,-0.9)(0.5,-1.2)(0.5,-1.5)
\psdots[dotsize=0.09cm](0,-0.3)
\end{pspicture}
\begin{pspicture}[shift=-1.5](-0.4,-3.2)(1.4,0.5)
\psline{-}(0,-2.2)(0,0)(1,0)(1,-2.2)
\psdots[dotsize=0.09cm](0,-0.6)
\psdots[dotsize=0.09cm](0.5,-0.6)
\psdots[dotsize=0.09cm](1,-0.6)
\rput(-0.4,-2.5){\tiny{A}}
\end{pspicture}
&
\psun
\begin{pspicture}[shift=-1.5](-0.4,-3.2)(1.4,0.5)
\psline{-}(0,-2.2)(0,0)(1,0)(1,-2.2)
\psdots[dotsize=0.09cm](0,-0.6)(0,-1.2)
\psdots[dotsize=0.09cm](0.5,-0.3)(0.5,-0.9)
\psdots[dotsize=0.09cm](1,-0.6)(1,-1.2)
\end{pspicture}
\begin{pspicture}[shift=-1.5](-0.4,-3.2)(1.4,0.5)
\psline{-}(0,-2.2)(0,0)(1,0)(1,-2.2)
\rput(-0.4,-2.5){\tiny{B}}
\end{pspicture}
&
\psun
\begin{pspicture}[shift=-1.5](-0.4,-3.2)(1.4,0.5)
\psline{-}(0,-2.2)(0,0)(1,0)(1,-2.2)
\psdots[dotsize=0.09cm](0,-0.3)
\psdots[dotsize=0.09cm](0.5,-0.6)(0.5,-0.9)(0.5,-1.2)(0.5,-1.5)
\psdots[dotsize=0.09cm](1,-0.3)
\end{pspicture}
\begin{pspicture}[shift=-1.5](-0.4,-3.2)(1.4,0.5)
\psline{-}(0,-2.2)(0,0)(1,0)(1,-2.2)
\psdot[linecolor=black,fillcolor=lightgray,dotstyle=o,dotsize=0.09cm](0.5,-0.6)
\rput(-0.4,-2.5){\tiny{B}}
\end{pspicture}
&
\psun
\begin{pspicture}[shift=-1.5](-0.4,-3.2)(1.4,0.5)
\psline{-}(0,-2.2)(0,0)(1,0)(1,-2.2)
\psdots[dotsize=0.09cm](1,-0.6)
\psdots[dotsize=0.09cm](0.5,-0.3)(0.5,-0.9)(0.5,-1.2)(0.5,-1.5)
\psdots[dotsize=0.09cm](0,-0.6)
\end{pspicture}
\begin{pspicture}[shift=-1.5](-0.4,-3.2)(1.4,0.5)
\psline{-}(0,-2.2)(0,0)(1,0)(1,-2.2)
\psdots[dotsize=0.09cm](0,-0.6)
\psdots[dotsize=0.09cm](0.5,-0.6)
\psdots[dotsize=0.09cm](1,-0.6)
\rput(-0.4,-2.5){\tiny{A}}
\end{pspicture}
\\
\end{array}
\\
&
\begin{array}{rccccccccccccc}
&
\psun
\begin{pspicture}[shift=-1.5](-0.4,-3.2)(1.4,0.5)
\psline{-}(0,-2.2)(0,0)(1,0)(1,-2.2)
\psdots[dotsize=0.09cm](0,-0.9)(0,-1.2)
\psdots[dotsize=0.09cm](0.5,-0.3)(0.5,-0.6)
\psdots[dotsize=0.09cm](1,-0.9)(1,-1.2)
\end{pspicture}
\begin{pspicture}[shift=-1.5](-0.4,-3.2)(1.4,0.5)
\psline{-}(0,-2.2)(0,0)(1,0)(1,-2.2)
\rput(-0.4,-2.5){\tiny{B}}
\end{pspicture}
&
\psun
\begin{pspicture}[shift=-1.5](-0.4,-3.2)(1.4,0.5)
\psline{-}(0,-2.2)(0,0)(1,0)(1,-2.2)
\psdots[dotsize=0.09cm](0,-0.6)
\psdots[dotsize=0.09cm](0.5,-0.3)(0.5,-0.9)(0.5,-1.2)(0.5,-1.5)
\psdots[dotsize=0.09cm](1,-0.6)
\end{pspicture}
\begin{pspicture}[shift=-1.5](-0.4,-3.2)(1.4,0.5)
\psline{-}(0,-2.2)(0,0)(1,0)(1,-2.2)
\psdot[linecolor=black,fillcolor=lightgray,dotstyle=o,dotsize=0.09cm](0.5,-0.6)
\rput(-0.4,-2.5){\tiny{B}}
\end{pspicture}
&
\psun
\begin{pspicture}[shift=-1.5](-0.4,-3.2)(1.4,0.5)
\psline{-}(0,-2.2)(0,0)(1,0)(1,-2.2)
\psdots[dotsize=0.09cm](0,-0.9)
\psdots[dotsize=0.09cm](0.5,-0.3)(0.5,-0.6)(0.5,-1.2)(0.5,-1.5)
\psdots[dotsize=0.09cm](1,-0.9)
\end{pspicture}
\begin{pspicture}[shift=-1.5](-0.4,-3.2)(1.4,0.5)
\psline{-}(0,-2.2)(0,0)(1,0)(1,-2.2)
\psdots[dotsize=0.09cm](0,-0.6)
\psdots[dotsize=0.09cm](0.5,-0.6)
\psdots[dotsize=0.09cm](1,-0.6)
\rput(-0.4,-2.5){\tiny{A}}
\end{pspicture}
&
\psun
\begin{pspicture}[shift=-1.5](-0.4,-3.2)(1.4,0.5)
\psline{-}(0,-2.2)(0,0)(1,0)(1,-2.2)
\psdots[dotsize=0.09cm](0,-0.9)
\psdots[dotsize=0.09cm](0.5,-0.3)(0.5,-0.6)(0.5,-1.2)(0.5,-1.5)
\psdots[dotsize=0.09cm](1,-0.9)
\end{pspicture}
\begin{pspicture}[shift=-1.5](-0.4,-3.2)(1.4,0.5)
\psline{-}(0,-2.2)(0,0)(1,0)(1,-2.2)
\psdot[linecolor=black,fillcolor=lightgray,dotstyle=o,dotsize=0.09cm](0.5,-0.6)
\rput(-0.4,-2.5){\tiny{B}}
\end{pspicture}
&
\psun
\begin{pspicture}[shift=-1.5](-0.4,-3.2)(1.4,0.5)
\psline{-}(0,-2.2)(0,0)(1,0)(1,-2.2)
\psdots[dotsize=0.09cm](0,-1.2)
\psdots[dotsize=0.09cm](0.5,-0.3)(0.5,-0.6)(0.5,-0.9)(0.5,-1.5)
\psdots[dotsize=0.09cm](1,-1.2)
\end{pspicture}
\begin{pspicture}[shift=-1.5](-0.4,-3.2)(1.4,0.5)
\psline{-}(0,-2.2)(0,0)(1,0)(1,-2.2)
\psdots[dotsize=0.09cm](0,-0.6)
\psdots[dotsize=0.09cm](0.5,-0.6)
\psdots[dotsize=0.09cm](1,-0.6)
\rput(-0.4,-2.5){\tiny{A}}
\end{pspicture}
&
\psun
\begin{pspicture}[shift=-1.5](-0.4,-3.2)(1.4,0.5)
\psline{-}(0,-2.2)(0,0)(1,0)(1,-2.2)
\psdots[dotsize=0.09cm](0.5,-0.3)(0.5,-0.6)(0.5,-0.9)(0.5,-1.2)
\end{pspicture}
\begin{pspicture}[shift=-1.5](-0.4,-3.2)(1.4,0.5)
\psline{-}(0,-2.2)(0,0)(1,0)(1,-2.2)
\psdot[linecolor=black,fillcolor=lightgray,dotstyle=o,dotsize=0.09cm](0,-0.6)
\psdot[linecolor=black,fillcolor=lightgray,dotstyle=o,dotsize=0.09cm](1,-0.6)
\rput(-0.4,-2.5){\tiny{B}}
\end{pspicture}
&
\psun
\begin{pspicture}[shift=-1.5](-0.4,-3.2)(1.4,0.5)
\psline{-}(0,-2.2)(0,0)(1,0)(1,-2.2)
\psdots[dotsize=0.09cm](0,-1.2)
\psdots[dotsize=0.09cm](0.5,-0.3)(0.5,-0.6)(0.5,-0.9)(0.5,-1.5)
\psdots[dotsize=0.09cm](1,-1.2)
\end{pspicture}
\begin{pspicture}[shift=-1.5](-0.4,-3.2)(1.4,0.5)
\psline{-}(0,-2.2)(0,0)(1,0)(1,-2.2)
\psdot[linecolor=black,fillcolor=lightgray,dotstyle=o,dotsize=0.09cm](0.5,-0.6)
\rput(-0.4,-2.5){\tiny{B}}
\end{pspicture}
&
\psun
\begin{pspicture}[shift=-1.5](-0.4,-3.2)(1.4,0.5)
\psline{-}(0,-2.2)(0,0)(1,0)(1,-2.2)
\psdots[dotsize=0.09cm](0,-1.5)
\psdots[dotsize=0.09cm](0.5,-0.3)(0.5,-0.6)(0.5,-0.9)(0.5,-1.2)
\psdots[dotsize=0.09cm](1,-1.5)
\end{pspicture}
\begin{pspicture}[shift=-1.5](-0.4,-3.2)(1.4,0.5)
\psline{-}(0,-2.2)(0,0)(1,0)(1,-2.2)
\psdots[dotsize=0.09cm](0,-0.6)
\psdots[dotsize=0.09cm](0.5,-0.6)
\psdots[dotsize=0.09cm](1,-0.6)
\rput(-0.4,-2.5){\tiny{A}}
\end{pspicture}
\\
\end{array}\\
&
\begin{array}{rccccccccccccc}
&
\psun
\begin{pspicture}[shift=-1.5](-0.4,-3.2)(1.4,0.5)
\psline{-}(0,-2.2)(0,0)(1,0)(1,-2.2)
\psdots[dotsize=0.09cm](0,-1.5)
\psdots[dotsize=0.09cm](0.5,-0.3)(0.5,-0.6)(0.5,-0.9)(0.5,-1.2)
\psdots[dotsize=0.09cm](1,-1.5)
\end{pspicture}
\begin{pspicture}[shift=-1.5](-0.4,-3.2)(1.4,0.5)
\psline{-}(0,-2.2)(0,0)(1,0)(1,-2.2)
\psdot[linecolor=black,fillcolor=lightgray,dotstyle=o,dotsize=0.09cm](0.5,-0.6)
\rput(-0.4,-2.5){\tiny{B}}
\end{pspicture}
&
\psun
\begin{pspicture}[shift=-1.5](-0.4,-3.2)(1.4,0.5)
\psline{-}(0,-2.2)(0,0)(1,0)(1,-2.2)
\psdots[dotsize=0.09cm](0.5,-0.3)(0.5,-0.6)(0.5,-0.9)(0.5,-1.2)(0.5,-1.5)(0.5,-1.8)
\end{pspicture}
\begin{pspicture}[shift=-1.5](-0.4,-3.2)(1.4,0.5)
\psline{-}(0,-2.2)(0,0)(1,0)(1,-2.2)
\psdots[dotsize=0.09cm](0,-0.6)
\psdots[dotsize=0.09cm](0.5,-0.6)
\psdots[dotsize=0.09cm](1,-0.6)
\psdot[linecolor=black,fillcolor=lightgray,dotstyle=o,dotsize=0.09cm](0.5,-1.2)
\rput(-0.4,-2.5){\tiny{A}}
\end{pspicture}
&
\psun
\begin{pspicture}[shift=-1.5](-0.4,-3.2)(1.4,0.5)
\psline{-}(0,-2.2)(0,0)(1,0)(1,-2.2)
\psdots[dotsize=0.09cm](0.5,-0.3)(0.5,-0.6)(0.5,-0.9)(0.5,-1.2)(0.5,-1.5)(0.5,-1.8)
\end{pspicture}
\begin{pspicture}[shift=-1.5](-0.4,-3.2)(1.4,0.5)
\psline{-}(0,-2.2)(0,0)(1,0)(1,-2.2)
\psdots[dotsize=0.09cm](0,-1.2)
\psdots[dotsize=0.09cm](0.5,-1.2)
\psdots[dotsize=0.09cm](1,-1.2)
\psdot[linecolor=black,fillcolor=lightgray,dotstyle=o,dotsize=0.09cm](0.5,-0.6)
\rput(-0.4,-2.5){\tiny{A}}
\end{pspicture}
&
\psun
\begin{pspicture}[shift=-1.5](-0.4,-3.2)(1.4,0.5)
\psline{-}(0,-2.2)(0,0)(1,0)(1,-2.2)
\psdots[dotsize=0.09cm](0.5,-0.3)(0.5,-0.6)(0.5,-0.9)(0.5,-1.2)(0.5,-1.5)(0.5,-1.8)
\end{pspicture}
\begin{pspicture}[shift=-1.5](-0.4,-3.2)(1.4,0.5)
\psline{-}(0,-2.2)(0,0)(1,0)(1,-2.2)
\psdots[linecolor=black,fillcolor=lightgray,dotstyle=o,dotsize=0.09cm](0.5,-0.6)(0.5,-1.2)
\rput(-0.4,-2.5){\tiny{B}}
\end{pspicture}
\\
\end{array}
\end{alignat*}
\captionof{figure}{The 20 patterns of zeros for $\stan_8^4$}
\label{fig:pats84}
\end{center}

%
\subsection{Periodic boundary conditions}\label{sec:patsperiodic}
%

\vspace{-0.5cm}

\begin{center}
\begin{equation*}
%
\end{equation*}
\captionof{figure}{The patterns of zeros for the ground states in $\stanp_{12}^d$ and $\stanp_{13}^d$.}
\label{fig:GSpatternsp}
\end{center}

\begin{center}
\begin{alignat*}{2}
&\begin{array}{rcccccccc}
&
\psun
\begin{pspicture}[shift=-0.4](-0.4,-1.0)(1.4,1.0)
\psline{-}(0,-1)(0,1)\psline{-}(1,-1)(1,1)\psline[linestyle=dashed,dash=1pt 1pt]{-}(0,0)(1,0)
\psdots[dotsize=0.09cm](0,-0.3)(0,0)(0,0.3)
\psdots[dotsize=0.09cm](1,-0.3)(1,0)(1,0.3)
\end{pspicture}
\begin{pspicture}[shift=-0.4](-0.4,-1.0)(1.4,1.0)
\psline{-}(0,-1)(0,1)\psline{-}(1,-1)(1,1)\psline[linestyle=dashed,dash=1pt 1pt]{-}(0,0)(1,0)
\end{pspicture}
&
\psun
\begin{pspicture}[shift=-0.4](-0.4,-1.0)(1.4,1.0)
\psline{-}(0,-1)(0,1)\psline{-}(1,-1)(1,1)\psline[linestyle=dashed,dash=1pt 1pt]{-}(0,0)(1,0)
\psdots[dotsize=0.09cm](0,-0.3)(0,0.3)
\psdots[dotsize=0.09cm](0.5,-0.6)(0.5,0.6)
\psdots[dotsize=0.09cm](1,-0.3)(1,0.3)
\end{pspicture}
\begin{pspicture}[shift=-0.4](-0.4,-1.0)(1.4,1.0)
\psline{-}(0,-1)(0,1)\psline{-}(1,-1)(1,1)\psline[linestyle=dashed,dash=1pt 1pt]{-}(0,0)(1,0)
\psdots[linecolor=black,fillcolor=lightgray,dotstyle=o,dotsize=0.09cm](0.5,0)
\end{pspicture}
&
\psun
\begin{pspicture}[shift=-0.4](-0.4,-1.0)(1.4,1.0)
\psline{-}(0,-1)(0,1)\psline{-}(1,-1)(1,1)\psline[linestyle=dashed,dash=1pt 1pt]{-}(0,0)(1,0)
\psdots[dotsize=0.09cm](0,0)
\psdots[dotsize=0.09cm](0.5,-0.3)(0.5,0.3)
\psdots[dotsize=0.09cm](1,0)
\end{pspicture}
\begin{pspicture}[shift=-0.4](-0.4,-1.0)(1.4,1.0)
\psline{-}(0,-1)(0,1)\psline{-}(1,-1)(1,1)\psline[linestyle=dashed,dash=1pt 1pt]{-}(0,0)(1,0)
\psdots[linecolor=black,fillcolor=lightgray,dotstyle=o,dotsize=0.09cm](0,0)(1,0)
\end{pspicture}
&
\psun
\begin{pspicture}[shift=-0.4](-0.4,-1.0)(1.4,1.0)
\psline{-}(0,-1)(0,1)\psline{-}(1,-1)(1,1)\psline[linestyle=dashed,dash=1pt 1pt]{-}(0,0)(1,0)
\psdots[dotsize=0.09cm](0,-0.3)(0,-0.6)
\psdots[dotsize=0.09cm](0.5,0.3)(0.5,0.6)
\psdots[dotsize=0.09cm](1,-0.3)(1,-0.6)
\end{pspicture}
\begin{pspicture}[shift=-0.4](-0.4,-1.0)(1.4,1.0)
\psline{-}(0,-1)(0,1)\psline{-}(1,-1)(1,1)\psline[linestyle=dashed,dash=1pt 1pt]{-}(0,0)(1,0)
\psdots[linecolor=black,fillcolor=lightgray,dotstyle=o,dotsize=0.09cm](0.5,0.6)
\end{pspicture}
&
\psun
\begin{pspicture}[shift=-0.4](-0.4,-1.0)(1.4,1.0)
\psline{-}(0,-1)(0,1)\psline{-}(1,-1)(1,1)\psline[linestyle=dashed,dash=1pt 1pt]{-}(0,0)(1,0)
\psdots[dotsize=0.09cm](0,0.3)(0,0.6)
\psdots[dotsize=0.09cm](0.5,-0.6)(0.5,-0.3)
\psdots[dotsize=0.09cm](1,0.3)(1,0.6)
\end{pspicture}
\begin{pspicture}[shift=-0.4](-0.4,-1.0)(1.4,1.0)
\psline{-}(0,-1)(0,1)\psline{-}(1,-1)(1,1)\psline[linestyle=dashed,dash=1pt 1pt]{-}(0,0)(1,0)
\psdots[linecolor=black,fillcolor=lightgray,dotstyle=o,dotsize=0.09cm](0.5,-0.6)
\end{pspicture}
&
\psun
\begin{pspicture}[shift=-0.4](-0.4,-1.0)(1.4,1.0)
\psline{-}(0,-1)(0,1)\psline{-}(1,-1)(1,1)\psline[linestyle=dashed,dash=1pt 1pt]{-}(0,0)(1,0)
\psdots[dotsize=0.09cm](0,0)
\psdots[dotsize=0.09cm](0.5,-0.6)(0.5,-0.3)(0.5,0.3)(0.5,0.6)
\psdots[dotsize=0.09cm](1,0)
\end{pspicture}
\begin{pspicture}[shift=-0.4](-0.4,-1.0)(1.4,1.0)
\psline{-}(0,-1)(0,1)\psline{-}(1,-1)(1,1)\psline[linestyle=dashed,dash=1pt 1pt]{-}(0,0)(1,0)
\psdots[linecolor=black,fillcolor=lightgray,dotstyle=o,dotsize=0.09cm](0.5,-0.6)(0.5,0.6)
\end{pspicture}
&
\psun
\begin{pspicture}[shift=-0.4](-0.4,-1.0)(1.4,1.0)
\psline{-}(0,-1)(0,1)\psline{-}(1,-1)(1,1)\psline[linestyle=dashed,dash=1pt 1pt]{-}(0,0)(1,0)
\psdots[dotsize=0.09cm](0,-0.3)(0,0.6)
\psdots[dotsize=0.09cm](0.5,-0.6)(0.5,0.3)
\psdots[dotsize=0.09cm](1,-0.3)(1,0.6)
\end{pspicture}
\begin{pspicture}[shift=-0.4](-0.4,-1.0)(1.4,1.0)
\psline{-}(0,-1)(0,1)\psline{-}(1,-1)(1,1)\psline[linestyle=dashed,dash=1pt 1pt]{-}(0,0)(1,0)
\psdots[linecolor=black,fillcolor=lightgray,dotstyle=o,dotsize=0.09cm](0.5,0)
\end{pspicture}
&
\psun
\begin{pspicture}[shift=-0.4](-0.4,-1.0)(1.4,1.0)
\psline{-}(0,-1)(0,1)\psline{-}(1,-1)(1,1)\psline[linestyle=dashed,dash=1pt 1pt]{-}(0,0)(1,0)
\psdots[dotsize=0.09cm](0,-0.6)(0,0.3)
\psdots[dotsize=0.09cm](0.5,-0.3)(0.5,0.6)
\psdots[dotsize=0.09cm](1,-0.6)(1,0.3)
\end{pspicture}
\begin{pspicture}[shift=-0.4](-0.4,-1.0)(1.4,1.0)
\psline{-}(0,-1)(0,1)\psline{-}(1,-1)(1,1)\psline[linestyle=dashed,dash=1pt 1pt]{-}(0,0)(1,0)
\psdots[linecolor=black,fillcolor=lightgray,dotstyle=o,dotsize=0.09cm](0.5,0)
\end{pspicture}
\\
\end{array}
\\[0.4cm]
&
\begin{array}{rcccccccc}
&
\psun
\begin{pspicture}[shift=-0.4](-0.4,-1.0)(1.4,1.0)
\psline{-}(0,-1)(0,1)\psline{-}(1,-1)(1,1)\psline[linestyle=dashed,dash=1pt 1pt]{-}(0,0)(1,0)
\psdots[dotsize=0.09cm](0,-0.6)(0,0.6)
\psdots[dotsize=0.09cm](0.5,-0.3)(0.5,0.3)
\psdots[dotsize=0.09cm](1,-0.6)(1,0.6)
\end{pspicture}
\begin{pspicture}[shift=-0.4](-0.4,-1.0)(1.4,1.0)
\psline{-}(0,-1)(0,1)\psline{-}(1,-1)(1,1)\psline[linestyle=dashed,dash=1pt 1pt]{-}(0,0)(1,0)
\psdots[linecolor=black,fillcolor=lightgray,dotstyle=o,dotsize=0.09cm](0.5,0)
\end{pspicture}
&
\psun
\begin{pspicture}[shift=-0.4](-0.4,-1.0)(1.4,1.0)
\psline{-}(0,-1)(0,1)\psline{-}(1,-1)(1,1)\psline[linestyle=dashed,dash=1pt 1pt]{-}(0,0)(1,0)
\psdots[dotsize=0.09cm](0,0.3)
\psdots[dotsize=0.09cm](0.5,-0.3)(0.5,0)
\psdots[dotsize=0.09cm](1,0.3)
\end{pspicture}
\begin{pspicture}[shift=-0.4](-0.4,-1.0)(1.4,1.0)
\psline{-}(0,-1)(0,1)\psline{-}(1,-1)(1,1)\psline[linestyle=dashed,dash=1pt 1pt]{-}(0,0)(1,0)
\psdots[linecolor=black,fillcolor=lightgray,dotstyle=o,dotsize=0.09cm](0,0)(1,0)
\end{pspicture}
&
\psun
\begin{pspicture}[shift=-0.4](-0.4,-1.0)(1.4,1.0)
\psline{-}(0,-1)(0,1)\psline{-}(1,-1)(1,1)\psline[linestyle=dashed,dash=1pt 1pt]{-}(0,0)(1,0)
\psdots[dotsize=0.09cm](0,-0.3)
\psdots[dotsize=0.09cm](0.5,0.3)(0.5,-0)
\psdots[dotsize=0.09cm](1,-0.3)
\end{pspicture}
\begin{pspicture}[shift=-0.4](-0.4,-1.0)(1.4,1.0)
\psline{-}(0,-1)(0,1)\psline{-}(1,-1)(1,1)\psline[linestyle=dashed,dash=1pt 1pt]{-}(0,0)(1,0)
\psdots[linecolor=black,fillcolor=lightgray,dotstyle=o,dotsize=0.09cm](0,0.6)(1,0.6)
\end{pspicture}
&
\psun
\begin{pspicture}[shift=-0.4](-0.4,-1.0)(1.4,1.0)
\psline{-}(0,-1)(0,1)\psline{-}(1,-1)(1,1)\psline[linestyle=dashed,dash=1pt 1pt]{-}(0,0)(1,0)
\psdots[dotsize=0.09cm](0.5,-0.6)(0.5,-0.3)(0.5,0.3)(0.5,0.6)
\end{pspicture}
\begin{pspicture}[shift=-0.4](-0.4,-1.0)(1.4,1.0)
\psline{-}(0,-1)(0,1)\psline{-}(1,-1)(1,1)\psline[linestyle=dashed,dash=1pt 1pt]{-}(0,0)(1,0)
\psdots[dotsize=0.09cm](0,-0.6)(0.5,-0.6)(1,-0.6)(0,0.6)(0.5,0.6)(1,0.6)
\end{pspicture}
&
\psun
\begin{pspicture}[shift=-0.4](-0.4,-1.0)(1.4,1.0)
\psline{-}(0,-1)(0,1)\psline{-}(1,-1)(1,1)\psline[linestyle=dashed,dash=1pt 1pt]{-}(0,0)(1,0)
\psdots[dotsize=0.09cm](0.5,-0.6)(0.5,-0.3)(0.5,0.3)(0.5,0.6)
\end{pspicture}
\begin{pspicture}[shift=-0.4](-0.4,-1.0)(1.4,1.0)
\psline{-}(0,-1)(0,1)\psline{-}(1,-1)(1,1)\psline[linestyle=dashed,dash=1pt 1pt]{-}(0,0)(1,0)
\psdots[dotsize=0.09cm](0,-0.6)(0.5,-0.6)(1,-0.6)(0,0.6)(0.5,0.6)(1,0.6)
\end{pspicture}
&
\psun
\begin{pspicture}[shift=-0.4](-0.4,-1.0)(1.4,1.0)
\psline{-}(0,-1)(0,1)\psline{-}(1,-1)(1,1)\psline[linestyle=dashed,dash=1pt 1pt]{-}(0,0)(1,0)
\psdots[dotsize=0.09cm](0,0.3)
\psdots[dotsize=0.09cm](0.5,-0.6)(0.5,-0.3)(0.5,-0)(0.5,0.6)
\psdots[dotsize=0.09cm](1,0.3)
\end{pspicture}
\begin{pspicture}[shift=-0.4](-0.4,-1.0)(1.4,1.0)
\psline{-}(0,-1)(0,1)\psline{-}(1,-1)(1,1)\psline[linestyle=dashed,dash=1pt 1pt]{-}(0,0)(1,0)
\psdots[linecolor=black,fillcolor=lightgray,dotstyle=o,dotsize=0.09cm](0.5,-0.6)(0.5,0.6)
\end{pspicture}
&
\psun
\begin{pspicture}[shift=-0.4](-0.4,-1.0)(1.4,1.0)
\psline{-}(0,-1)(0,1)\psline{-}(1,-1)(1,1)\psline[linestyle=dashed,dash=1pt 1pt]{-}(0,0)(1,0)
\psdots[dotsize=0.09cm](0,-0.3)
\psdots[dotsize=0.09cm](0.5,-0.6)(0.5,0.0)(0.5,0.3)(0.5,0.6)
\psdots[dotsize=0.09cm](1,-0.3)
\end{pspicture}
\begin{pspicture}[shift=-0.4](-0.4,-1.0)(1.4,1.0)
\psline{-}(0,-1)(0,1)\psline{-}(1,-1)(1,1)\psline[linestyle=dashed,dash=1pt 1pt]{-}(0,0)(1,0)
\psdots[linecolor=black,fillcolor=lightgray,dotstyle=o,dotsize=0.09cm](0.5,0.6)(0.5,0)
\end{pspicture}
&
\psun
\begin{pspicture}[shift=-0.4](-0.4,-1.0)(1.4,1.0)
\psline{-}(0,-1)(0,1)\psline{-}(1,-1)(1,1)\psline[linestyle=dashed,dash=1pt 1pt]{-}(0,0)(1,0)
\psdots[dotsize=0.09cm](0.5,-0.6)(0.5,-0.3)(0.5,0.3)(0.5,0.6)
\end{pspicture}
\begin{pspicture}[shift=-0.4](-0.4,-1.0)(1.4,1.0)
\psline{-}(0,-1)(0,1)\psline{-}(1,-1)(1,1)\psline[linestyle=dashed,dash=1pt 1pt]{-}(0,0)(1,0)
\psdots[linecolor=black,fillcolor=lightgray,dotstyle=o,dotsize=0.09cm](0,-0.6)(0.5,0.6)(1,-0.6)
\end{pspicture}
\\
\end{array}
\\[0.4cm]
&
\begin{array}{rccccccccccccc}
&
\psun
\begin{pspicture}[shift=-0.4](-0.4,-1.0)(1.4,1.0)
\psline{-}(0,-1)(0,1)\psline{-}(1,-1)(1,1)\psline[linestyle=dashed,dash=1pt 1pt]{-}(0,0)(1,0)
\psdots[dotsize=0.09cm](0.5,-0.6)(0.5,-0.3)(0.5,0.3)(0.5,0.6)
\end{pspicture}
\begin{pspicture}[shift=-0.4](-0.4,-1.0)(1.4,1.0)
\psline{-}(0,-1)(0,1)\psline{-}(1,-1)(1,1)\psline[linestyle=dashed,dash=1pt 1pt]{-}(0,0)(1,0)
\psdots[linecolor=black,fillcolor=lightgray,dotstyle=o,dotsize=0.09cm](0,0.6)(0.5,-0.6)(1,0.6)
\end{pspicture}
&
\psun
\begin{pspicture}[shift=-0.4](-0.4,-1.0)(1.4,1.0)
\psline{-}(0,-1)(0,1)\psline{-}(1,-1)(1,1)\psline[linestyle=dashed,dash=1pt 1pt]{-}(0,0)(1,0)
\psdots[dotsize=0.09cm](0,0.6)
\psdots[dotsize=0.09cm](0.5,-0.6)(0.5,-0.3)(0.5,0)(0.5,0.3)
\psdots[dotsize=0.09cm](1,0.6)
\end{pspicture}
\begin{pspicture}[shift=-0.4](-0.4,-1.0)(1.4,1.0)
\psline{-}(0,-1)(0,1)\psline{-}(1,-1)(1,1)\psline[linestyle=dashed,dash=1pt 1pt]{-}(0,0)(1,0)
\psdots[linecolor=black,fillcolor=lightgray,dotstyle=o,dotsize=0.09cm](0.5,-0.6)(0.5,0.6)
\end{pspicture}
&
\psun
\begin{pspicture}[shift=-0.4](-0.4,-1.0)(1.4,1.0)
\psline{-}(0,-1)(0,1)\psline{-}(1,-1)(1,1)\psline[linestyle=dashed,dash=1pt 1pt]{-}(0,0)(1,0)
\psdots[dotsize=0.09cm](0,-0.6)
\psdots[dotsize=0.09cm](0.5,-0.3)(0.5,0.0)(0.5,0.3)(0.5,0.6)
\psdots[dotsize=0.09cm](1,-0.6)
\end{pspicture}
\begin{pspicture}[shift=-0.4](-0.4,-1.0)(1.4,1.0)
\psline{-}(0,-1)(0,1)\psline{-}(1,-1)(1,1)\psline[linestyle=dashed,dash=1pt 1pt]{-}(0,0)(1,0)
\psdots[linecolor=black,fillcolor=lightgray,dotstyle=o,dotsize=0.09cm](0.5,0)(0.5,0.6)
\end{pspicture}
&
\psun
\begin{pspicture}[shift=-0.4](-0.4,-1.0)(1.4,1.0)
\psline{-}(0,-1)(0,1)\psline{-}(1,-1)(1,1)\psline[linestyle=dashed,dash=1pt 1pt]{-}(0,0)(1,0)
\psdots[dotsize=0.09cm](0.5,-0.9)(0.5,-0.6)(0.5,-0.3)(0.5,0.3)(0.5,0.6)(0.5,0.9)
\end{pspicture}
\begin{pspicture}[shift=-0.4](-0.4,-1.0)(1.4,1.0)
\psline{-}(0,-1)(0,1)\psline{-}(1,-1)(1,1)\psline[linestyle=dashed,dash=1pt 1pt]{-}(0,0)(1,0)
\psdots[linecolor=black,fillcolor=lightgray,dotstyle=o,dotsize=0.09cm](0.5,-0.6)(0.5,0)(0.5,0.6)
\end{pspicture}
\\
\end{array}
\end{alignat*}

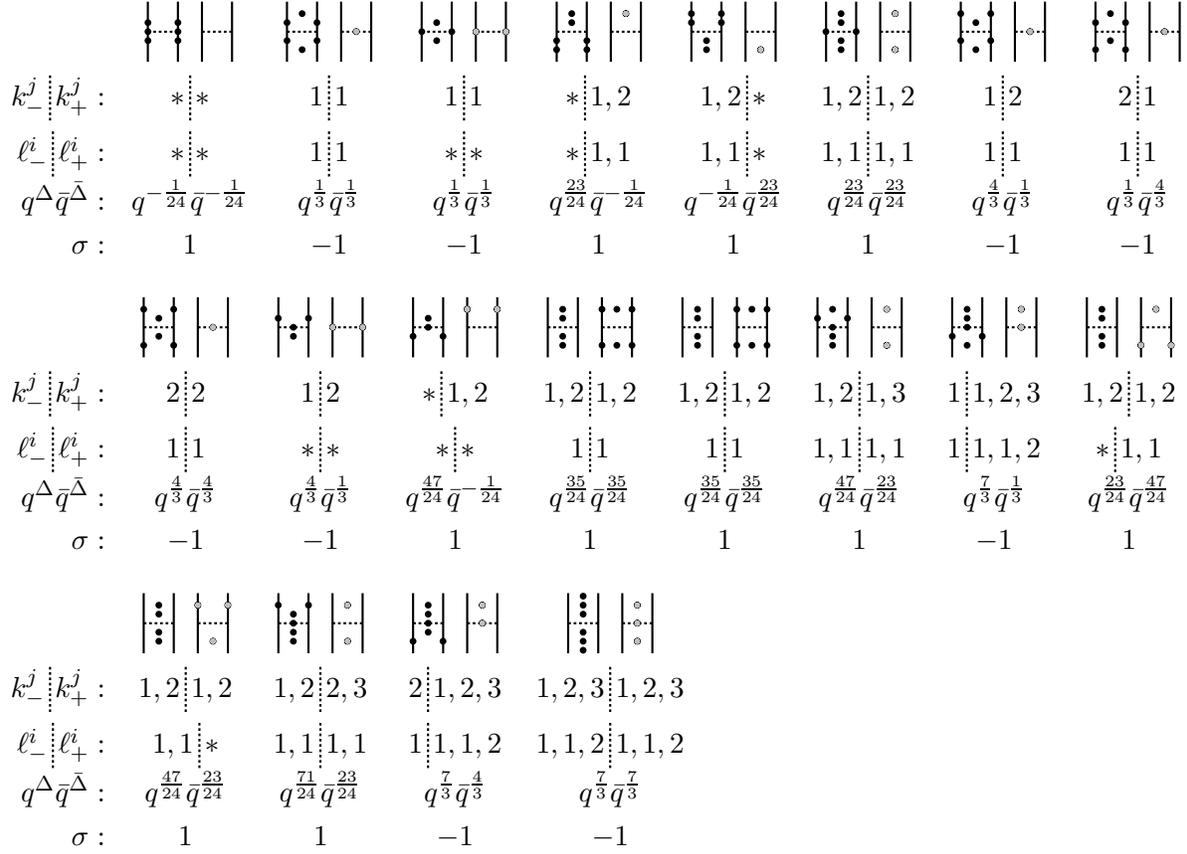
\captionof{figure}{The 20 patterns of zeros for $\stanp_6^0$ for $\alpha = 2$ and the corresponding data for $\epsilon = 0$}
\label{fig:patsP60}
\end{center}

\begin{center}
\begin{alignat*}{2}
&\begin{array}{rcccccccccc}
&
\psun
\begin{pspicture}[shift=-1.2](-0.4,-1.6)(1.4,1.6)
\psline{-}(0,-1)(0,1)\psline{-}(1,-1)(1,1)\psline[linestyle=dashed,dash=1pt 1pt]{-}(0,0)(1,0)
\psdots[dotsize=0.09cm](0,-0.3)(0,0.3)
\psdots[dotsize=0.09cm](0.5,-0.6)(0.5,0.6)
\psdots[dotsize=0.09cm](1,-0.3)(1,0.3)
\end{pspicture}
\begin{pspicture}[shift=-1.2](-0.4,-1.6)(1.4,1.6)
\psline{-}(0,-1)(0,1)\psline{-}(1,-1)(1,1)\psline[linestyle=dashed,dash=1pt 1pt]{-}(0,0)(1,0)
\rput(-0.4,1.3){\tiny{A}}
\rput(-0.4,-1.3){\tiny{$\bar \tA$}}
\end{pspicture}
&
\psun
\begin{pspicture}[shift=-1.2](-0.4,-1.6)(1.4,1.6)
\psline{-}(0,-1)(0,1)\psline{-}(1,-1)(1,1)\psline[linestyle=dashed,dash=1pt 1pt]{-}(0,0)(1,0)
\psdots[dotsize=0.09cm](0,-0.3)(0,-0.6)
\psdots[dotsize=0.09cm](0.5,0.3)(0.5,0.6)
\psdots[dotsize=0.09cm](1,-0.3)(1,-0.6)
\end{pspicture}
\begin{pspicture}[shift=-1.2](-0.4,-1.6)(1.4,1.6)
\psline{-}(0,-1)(0,1)\psline{-}(1,-1)(1,1)\psline[linestyle=dashed,dash=1pt 1pt]{-}(0,0)(1,0)
\rput(-0.4,1.3){\tiny{A}}
\rput(-0.4,-1.3){\tiny{$\bar \tA$}}
\end{pspicture}
&
\psun
\begin{pspicture}[shift=-1.2](-0.4,-1.6)(1.4,1.6)
\psline{-}(0,-1)(0,1)\psline{-}(1,-1)(1,1)\psline[linestyle=dashed,dash=1pt 1pt]{-}(0,0)(1,0)
\psdots[dotsize=0.09cm](0,0.3)(0,0.6)
\psdots[dotsize=0.09cm](0.5,-0.3)(0.5,-0.6)
\psdots[dotsize=0.09cm](1,0.3)(1,0.6)
\end{pspicture}
\begin{pspicture}[shift=-1.2](-0.4,-1.6)(1.4,1.6)
\psline{-}(0,-1)(0,1)\psline{-}(1,-1)(1,1)\psline[linestyle=dashed,dash=1pt 1pt]{-}(0,0)(1,0)
\rput(-0.4,1.3){\tiny{A}}
\rput(-0.4,-1.3){\tiny{$\bar \tA$}}
\end{pspicture}
&
\psun
\begin{pspicture}[shift=-1.2](-0.4,-1.6)(1.4,1.6)
\psline{-}(0,-1)(0,1)\psline{-}(1,-1)(1,1)\psline[linestyle=dashed,dash=1pt 1pt]{-}(0,0)(1,0)
\psdots[dotsize=0.09cm](0,-0.6)(0,0.3)
\psdots[dotsize=0.09cm](0.5,-0.3)(0.5,0.6)
\psdots[dotsize=0.09cm](1,-0.6)(1,0.3)
\end{pspicture}
\begin{pspicture}[shift=-1.2](-0.4,-1.6)(1.4,1.6)
\psline{-}(0,-1)(0,1)\psline{-}(1,-1)(1,1)\psline[linestyle=dashed,dash=1pt 1pt]{-}(0,0)(1,0)
\rput(-0.4,1.3){\tiny{A}}
\rput(-0.4,-1.3){\tiny{$\bar \tA$}}
\end{pspicture}
&
\psun
\begin{pspicture}[shift=-1.2](-0.4,-1.6)(1.4,1.6)
\psline{-}(0,-1)(0,1)\psline{-}(1,-1)(1,1)\psline[linestyle=dashed,dash=1pt 1pt]{-}(0,0)(1,0)
\psdots[dotsize=0.09cm](0,0.6)(0,-0.3)
\psdots[dotsize=0.09cm](0.5,0.3)(0.5,-0.6)
\psdots[dotsize=0.09cm](1,0.6)(1,-0.3)
\end{pspicture}
\begin{pspicture}[shift=-1.2](-0.4,-1.6)(1.4,1.6)
\psline{-}(0,-1)(0,1)\psline{-}(1,-1)(1,1)\psline[linestyle=dashed,dash=1pt 1pt]{-}(0,0)(1,0)
\rput(-0.4,1.3){\tiny{A}}
\rput(-0.4,-1.3){\tiny{$\bar \tA$}}
\end{pspicture}
&
\psun
\begin{pspicture}[shift=-1.2](-0.4,-1.6)(1.4,1.6)
\psline{-}(0,-1)(0,1)\psline{-}(1,-1)(1,1)\psline[linestyle=dashed,dash=1pt 1pt]{-}(0,0)(1,0)
\psdots[dotsize=0.09cm](0,0)
\psdots[dotsize=0.09cm](0.5,-0.6)(0.5,-0.3)(0.5,0.3)(0.5,0.6)
\psdots[dotsize=0.09cm](1,0)
\end{pspicture}
\begin{pspicture}[shift=-1.2](-0.4,-1.6)(1.4,1.6)
\psline{-}(0,-1)(0,1)\psline{-}(1,-1)(1,1)\psline[linestyle=dashed,dash=1pt 1pt]{-}(0,0)(1,0)
\psdots[linecolor=black,fillcolor=lightgray,dotstyle=o,dotsize=0.09cm](0.5,0)
\rput(-0.4,1.3){\tiny{A}}
\rput(-0.4,-1.3){\tiny{$\bar \tA$}}
\end{pspicture}
&
\psun
\begin{pspicture}[shift=-1.2](-0.4,-1.6)(1.4,1.6)
\psline{-}(0,-1)(0,1)\psline{-}(1,-1)(1,1)\psline[linestyle=dashed,dash=1pt 1pt]{-}(0,0)(1,0)
\psdots[dotsize=0.09cm](0.5,-0.6)(0.5,-0.3)(0.5,0.3)(0.5,0.6)
\end{pspicture}
\begin{pspicture}[shift=-1.2](-0.4,-1.6)(1.4,1.6)
\psline{-}(0,-1)(0,1)\psline{-}(1,-1)(1,1)\psline[linestyle=dashed,dash=1pt 1pt]{-}(0,0)(1,0)
\psdots[linecolor=black,fillcolor=lightgray,dotstyle=o,dotsize=0.09cm](0,0)(1,0)
\rput(-0.4,1.3){\tiny{A}}
\rput(-0.4,-1.3){\tiny{$\bar \tA$}}
\end{pspicture}
&
\psun
\begin{pspicture}[shift=-1.2](-0.4,-1.6)(1.4,1.6)
\psline{-}(0,-1)(0,1)\psline{-}(1,-1)(1,1)\psline[linestyle=dashed,dash=1pt 1pt]{-}(0,0)(1,0)
\psdots[dotsize=0.09cm](0,-0.6)(0,0.6)
\psdots[dotsize=0.09cm](0.5,-0.3)(0.5,0.3)
\psdots[dotsize=0.09cm](1,-0.6)(1,0.6)
\end{pspicture}
\begin{pspicture}[shift=-1.2](-0.4,-1.6)(1.4,1.6)
\psline{-}(0,-1)(0,1)\psline{-}(1,-1)(1,1)\psline[linestyle=dashed,dash=1pt 1pt]{-}(0,0)(1,0)
\rput(-0.4,1.3){\tiny{A}}
\rput(-0.4,-1.3){\tiny{$\bar \tA$}}
\end{pspicture}
\\
\end{array}
\\[0.3cm]
&
\begin{array}{rccccccccccccc}
&
\psun
\begin{pspicture}[shift=-1.2](-0.4,-1.6)(1.4,1.6)
\psline{-}(0,-1)(0,1)\psline{-}(1,-1)(1,1)\psline[linestyle=dashed,dash=1pt 1pt]{-}(0,0)(1,0)
\psdots[dotsize=0.09cm](0,0.3)
\psdots[dotsize=0.09cm](0.5,-0.6)(0.5,-0.3)(0.5,0)(0.5,0.6)
\psdots[dotsize=0.09cm](1,0.3)
\end{pspicture}
\begin{pspicture}[shift=-1.2](-0.4,-1.6)(1.4,1.6)
\psline{-}(0,-1)(0,1)\psline{-}(1,-1)(1,1)\psline[linestyle=dashed,dash=1pt 1pt]{-}(0,0)(1,0)
\psdots[linecolor=black,fillcolor=lightgray,dotstyle=o,dotsize=0.09cm](0.5,0)
\rput(-0.4,1.3){\tiny{A}}
\rput(-0.4,-1.3){\tiny{$\bar \tA$}}
\end{pspicture}
&
\psun
\begin{pspicture}[shift=-1.2](-0.4,-1.6)(1.4,1.6)
\psline{-}(0,-1)(0,1)\psline{-}(1,-1)(1,1)\psline[linestyle=dashed,dash=1pt 1pt]{-}(0,0)(1,0)
\psdots[dotsize=0.09cm](0,-0.3)
\psdots[dotsize=0.09cm](0.5,0.6)(0.5,0.3)(0.5,0)(0.5,-0.6)
\psdots[dotsize=0.09cm](1,-0.3)
\end{pspicture}
\begin{pspicture}[shift=-1.2](-0.4,-1.6)(1.4,1.6)
\psline{-}(0,-1)(0,1)\psline{-}(1,-1)(1,1)\psline[linestyle=dashed,dash=1pt 1pt]{-}(0,0)(1,0)
\psdots[linecolor=black,fillcolor=lightgray,dotstyle=o,dotsize=0.09cm](0.5,0.6)
\rput(-0.4,1.3){\tiny{A}}
\rput(-0.4,-1.3){\tiny{$\bar \tA$}}
\end{pspicture}
&
\psun
\begin{pspicture}[shift=-1.2](-0.4,-1.6)(1.4,1.6)
\psline{-}(0,-1)(0,1)\psline{-}(1,-1)(1,1)\psline[linestyle=dashed,dash=1pt 1pt]{-}(0,0)(1,0)
\psdots[dotsize=0.09cm](0.5,-0.6)(0.5,-0.3)(0.5,0.3)(0.5,0.6)
\end{pspicture}
\begin{pspicture}[shift=-1.2](-0.4,-1.6)(1.4,1.6)
\psline{-}(0,-1)(0,1)\psline{-}(1,-1)(1,1)\psline[linestyle=dashed,dash=1pt 1pt]{-}(0,0)(1,0)
\psdots[dotsize=0.09cm](0,0)(0.5,0)(1,0)
\rput(-0.4,1.3){\tiny{B}}
\rput(-0.4,-1.3){\tiny{$\bar \tA$}}
\end{pspicture}
&
\psun
\begin{pspicture}[shift=-1.2](-0.4,-1.6)(1.4,1.6)
\psline{-}(0,-1)(0,1)\psline{-}(1,-1)(1,1)\psline[linestyle=dashed,dash=1pt 1pt]{-}(0,0)(1,0)
\psdots[dotsize=0.09cm](0.5,-0.6)(0.5,-0.3)(0.5,0.3)(0.5,0.6)
\end{pspicture}
\begin{pspicture}[shift=-1.2](-0.4,-1.6)(1.4,1.6)
\psline{-}(0,-1)(0,1)\psline{-}(1,-1)(1,1)\psline[linestyle=dashed,dash=1pt 1pt]{-}(0,0)(1,0)
\psdots[dotsize=0.09cm](0,0)(0.5,0)(1,0)
\rput(-0.4,1.3){\tiny{A}}
\rput(-0.4,-1.3){\tiny{$\bar \tB$}}
\end{pspicture}
&
\psun
\begin{pspicture}[shift=-1.2](-0.4,-1.6)(1.4,1.6)
\psline{-}(0,-1)(0,1)\psline{-}(1,-1)(1,1)\psline[linestyle=dashed,dash=1pt 1pt]{-}(0,0)(1,0)
\psdots[dotsize=0.09cm](0,0.6)
\psdots[dotsize=0.09cm](0.5,-0.6)(0.5,-0.3)(0.5,0)(0.5,0.3)
\psdots[dotsize=0.09cm](1,0.6)
\end{pspicture}
\begin{pspicture}[shift=-1.2](-0.4,-1.6)(1.4,1.6)
\psline{-}(0,-1)(0,1)\psline{-}(1,-1)(1,1)\psline[linestyle=dashed,dash=1pt 1pt]{-}(0,0)(1,0)
\psdots[linecolor=black,fillcolor=lightgray,dotstyle=o,dotsize=0.09cm](0.5,0)
\rput(-0.4,1.3){\tiny{A}}
\rput(-0.4,-1.3){\tiny{$\bar \tA$}}
\end{pspicture}
&
\psun
\begin{pspicture}[shift=-1.2](-0.4,-1.6)(1.4,1.6)
\psline{-}(0,-1)(0,1)\psline{-}(1,-1)(1,1)\psline[linestyle=dashed,dash=1pt 1pt]{-}(0,0)(1,0)
\psdots[dotsize=0.09cm](0,-0.6)
\psdots[dotsize=0.09cm](0.5,-0.3)(0.5,0)(0.5,0.3)(0.5,0.6)
\psdots[dotsize=0.09cm](1,-0.6)
\end{pspicture}
\begin{pspicture}[shift=-1.2](-0.4,-1.6)(1.4,1.6)
\psline{-}(0,-1)(0,1)\psline{-}(1,-1)(1,1)\psline[linestyle=dashed,dash=1pt 1pt]{-}(0,0)(1,0)
\psdots[linecolor=black,fillcolor=lightgray,dotstyle=o,dotsize=0.09cm](0.5,0.6)
\rput(-0.4,1.3){\tiny{A}}
\rput(-0.4,-1.3){\tiny{$\bar \tA$}}
\end{pspicture}
&
\psun
\begin{pspicture}[shift=-1.2](-0.4,-1.6)(1.4,1.6)
\psline{-}(0,-1)(0,1)\psline{-}(1,-1)(1,1)\psline[linestyle=dashed,dash=1pt 1pt]{-}(0,0)(1,0)
\psdots[dotsize=0.09cm](0.5,-0.9)(0.5,-0.6)(0.5,-0.3)(0.5,0.3)(0.5,0.6)(0.5,0.9)
\end{pspicture}
\begin{pspicture}[shift=-1.2](-0.4,-1.6)(1.4,1.6)
\psline{-}(0,-1)(0,1)\psline{-}(1,-1)(1,1)\psline[linestyle=dashed,dash=1pt 1pt]{-}(0,0)(1,0)
\psdots[linecolor=black,fillcolor=lightgray,dotstyle=o,dotsize=0.09cm](0.5,-0.6)(0.5,0.6)
\rput(-0.4,1.3){\tiny{A}}
\rput(-0.4,-1.3){\tiny{$\bar \tA$}}
\end{pspicture}
\\
\end{array}
\end{alignat*}

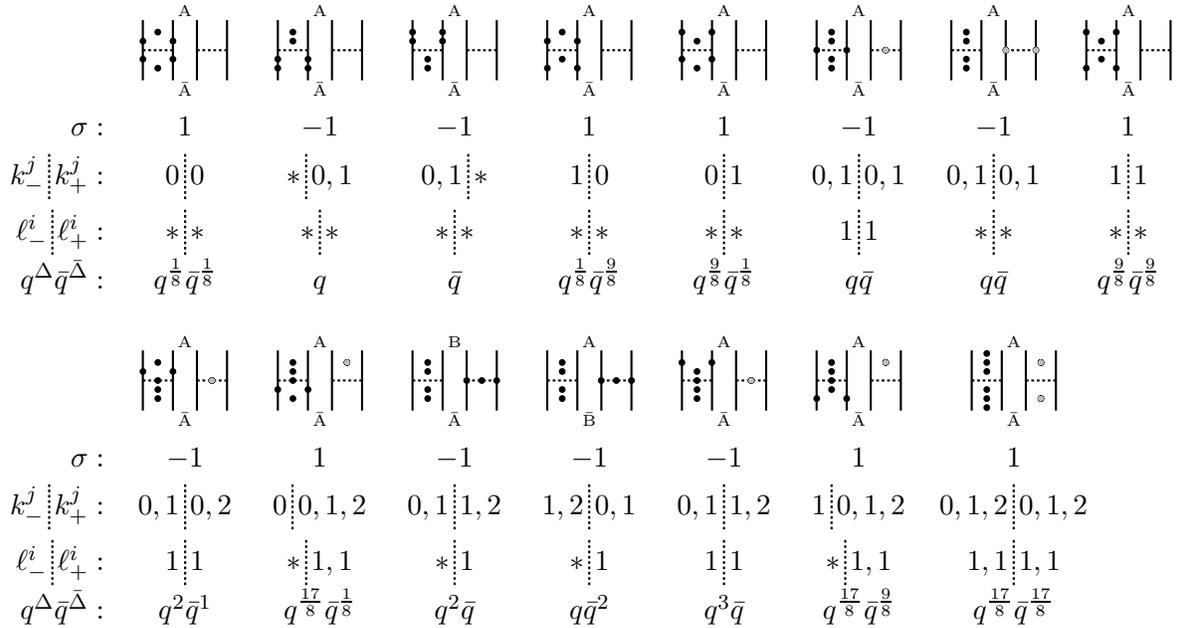
\captionof{figure}{The 15 patterns of zeros for $\stanp_6^2$ for $\omega = 1$ and the corresponding data for $\epsilon = 0$}
\label{fig:patsP62}
\end{center}

\newpage

%


\begin{thebibliography}{99} 
%

\bib{BroadHamm57} 
	S.~Broadbent, J.~Hammersley, 
	{\it Percolation processes I. Crystals and mazes}, 
	Proc. Camb. Phil. Soc. {\bf 53} (1957) 629.

\bib{KestonPerc82}
	H.~Kesten, 
	{\em Percolation theory for mathematicians}, 
	Springer Science (1982).

\bib{Stauffer92}
	D.~Stauffer, A.~Aharony, 
	{\it Introduction to Percolation Theory}, 
	Taylor and Francis (1992).

\bib{Grimmet97}
	G.~Grimmett, 
	{\em Percolation and disordered systems}, 
	Lectures on Probability Theory and Statistics, Springer (1997).

\bib{Saberi16}
	A.A.~Saberi,
	{\em Recent advances in percolation theory and its applications},
	Phys. Rep. {\bf 578} (2015) 1--32,
	\arxiv{1504.02898}{[cond-mat.stat-mech]}.

\bib{BaxBook}
	R.J.~Baxter, 
	{\it Exactly Solved Models in Statistical Mechanics},
	Academic Press (1982).

\bibitem{FMS}
	P.~Di Francesco, P.~Mathieu, D.~S\'en\'echal, 
	{\em Conformal Field Theory},
	Springer (1997).

\bib{Gurarie} 
	V.~Gurarie, 
	{\em Logarithmic operators in conformal field theory},
	Nucl. Phys. {\bf B410} (1993) 535--549,
	\arxiv{hep-th/9303160}{\!\!}.

\bib{SpecialIssue} 
	A.~Gainutdinov, D.~Ridout, I.~Runkel (Guest Editors),
	{\em Special issue on logarithmic conformal field theory}, 
	J. Phys. A: Math. Theor. {\bf 46(49)} (2013).

\bib{Harris}
	T.E.~Harris, 
	{\em A lower bound for the critical probability in a certain percolation process}, 
	Proc. Cambridge Philos. Soc. {\bf 56} (1960) 13--20.

\bib{Kesten80} 
	H.~Kesten,
	{\em The critical probability of bond percolation on the square lattice equals $1/2$}, 
	Commun. Math. Phys. {\bf 74} (1980) 41--59.

\bib{VJS12}
	R.~Vasseur, J.L.~Jacobsen, H.~Saleur,
	{\em Logarithmic observables in critical percolation},
	J. Stat. Mech. (2012) L07001, \arxiv{1206.2312}{[cond-mat.stat-mech]}.
	
\bib{DPSV13}
	G.~Delfino, M.~Picco, R.~Santachiara, J.~Viti,
	{\em Spin clusters and conformal field theory},
	J.~Stat.~Mech. (2013) P11011, \arxiv{1307.6123}{[cond-mat.stat-mech]}.

\bib{TL71}
	H.~Temperley, E.~Lieb,
 	{\em Relations between the ``percolation'' and ``colouring'' problem and other graph-theoretical problems associated with regular planar lattices: Some exact results for the ``percolation'' problem},
	Proc. Roy. Soc. London Ser.~{\bf A322} (1971) 251--280.
	
\bibitem{Jones}
	V.F.R.~Jones, 
	{\em Planar algebras I}, 
	\arxiv{math/9909027}{[math.QA]}.

\bib{HullPerc88}
	S.~Roux, E.~Guyon, D.~Sornette, 
	{\em Hull percolation}, 
	J. Phys. A {\bf 21} (1988) L475--L482. 

\bib{DuplantierHull}
	B.~Duplantier, 
	{\em Hull percolation and standard percolation}, 
	J. Phys. A {\bf 21} (1988) 3969--3973.

\bib{Cardy92} 
	J.~Cardy, 
	{\em Critical percolation in finite geometries}, 
	J. Phys. A {\bf 25} (1992) L201--L206,
	\arxiv{hep-th/9111026}{\!\!}; 
	{\em Conformal invariance and percolation}, 
	\arxiv{math-ph/0103018}{\!\!}.

\bib{LPSA94}
	R.~Langlands, P.~Pouliot, Y.~Saint-Aubin, 
	{\em Conformal invariance in two-dimensional percolation}, 
	Bull. Am. Math. Soc. {\bf 30} (1994) 1--61,
	\arxiv{math/9401222}{[math-ph]}.

\bib{CardyRGScaling}
	J.~Cardy,
	{\em Scaling and Renormalization in Statistical Physics}, 
	Cambridge Lecture Notes in Physics, 
	Cambridge University Press (1996).

\bib{denNijs79}
	M.P.M.~den Nijs, 
	{\em A relation between the temperature exponents of the eight-vertex model and $q$-state Potts model},
	J. Phys. A {\bf 12} (1979) 1857--68.

\bib{NienhuisRS80} 
	B.~Nienhuis, E.K.~Riedel, M.~Schick, 
	{\em Magnetic exponents of the two-dimensional $q$-state Potts model}, 
	J. Phys. A {\bf 13} (1980) L189--92.

\bib{CoulombGas1}
	B.~Nienhuis, 
	{\em Exact critical point and critical exponents of $O(n)$ models in two dimensions}, 
	Phys. Rev. Lett. {\bf 49} (1982) 1062--1065.

\bib{CoulombGas2}
	B.~Nienhuis, 
	{\em Critical behaviour of two-dimensional spin models and charge asymmetry in the Coulomb gas}, 
	J. Stat. Phys. {\bf 34} (1984) 731--761.

\bib{FSZ87}
	P.~di Francesco, H.~Saleur, J.-B.~Zuber, 
	{\em Relations between the Coulomb gas picture and conformal invariance of two-dimensional critical models}, 
	J. Stat. Phys. {\bf 49} (1987) 57--79;
	{\em Generalized Coulomb-gas formalism for two-dimensional critical models based on $SU(2)$ coset construction}, 
	Nucl. Phys. {\bf B300} (1988) 393--432.

\bib{Delfino1}
	G.~Delfino, J.~Cardy, 
	{\em Universal amplitude ratios in the two-dimensional $q$-state Potts model and percolation from quantum field theory},
	Nucl. Phys. {\bf B519} (1998) 551--578,
	\arxiv{hep-th/9712111}{\!\!}.

\bib{Delfino2}
	G.~Delfino, J.~Viti, J.~Cardy, 
	{\em Universal amplitude ratios of two-dimensional percolation from field theory}, 
	J. Phys. A {\bf 43} (2010) 152001,
	\arxiv{1001.5424}{[hep-th]}.

\bib{Saleur86} 
	H.~Saleur, 
	{\em New exact exponents for the two-dimensional self-avoiding walks}, 
	J. Phys. A: Math. Gen. {\bf 19} (1986) L807--L810.
 
\bib{Saleur87a}
	H.~Saleur, 
	{\em Conformal invariance for polymers and percolation}, 
	J. Phys. A: Math. Gen. {\bf 20} (1987) 455--470.
 
\bib{Saleur87b}
	H.~Saleur, 
	{\em Magnetic properties of the two-dimensional $n = 0$ vector model}, 
	Phys. Rev. {\bf B35} (1987) 3657--3660.

\bib{Duplantier86} 
	B.~Duplantier, 
	{\em Exact critical exponents for two-dimensional dense polymers},
	J. Phys. A: Math. Gen. {\bf 19} (1986) L1009--L1014.
 
\bib{DupSaleur86} 
	B.~Duplantier, H.~Saleur, 
	{\em Exact surface and wedge exponents for polymers in two dimensions}, 
	Phys. Rev. Lett. {\bf 57} (1986) 3179--3182.
 
\bib{SaleurDup87}
	H.~Saleur,  B.~Duplantier, 
	{\em Exact determination of the percolation hull exponent in two dimensions}, 
	Phys. Rev. Lett. {\bf 58} (1987) 2325--2328.

\bib{Saleur92} 
	H.~Saleur, 
	{\em Polymers and percolation in two dimensions and twisted $N=2$ supersymmetry},
	Nucl. Phys. {\bf B382} (1992) 486--531,
	\arxiv{hep-th/9111007}{\!\!}.
 
\bib{DeguchiFabMcCoy}
	T.~Deguchi, K.~Fabricius, B.M.~McCoy, 
	{\em The $sl_2$ loop algebra symmetry of the six-vertex model at roots of unity}, 
	J. Stat. Phys. {\bf 102} (2001) 701--736,
	\arxiv{cond-mat/9912141}{[cond-mat.stat-mech]}.

\bib{Deguchi}
	T.~Deguchi, 
	{\em The $sl_2$ loop algebra symmetry of the twisted transfer matrix of the six-vertex model at roots of unity}, 
	J. Phys. A {\bf 37} (2004) 347--358,
	\arxiv{cond-mat/0306498}{[cond-mat.stat-mech]}; 
	{\em Regular XXZ Bethe states as highest weight vectors of the $sl_2$ loop algebra at roots of unity}, 
	J. Phys. A {\bf 40} (2007) 7473--7508,
	\arxiv{cond-mat/0212217}{[cond-mat.stat-mech]}.
 
\bib{PRZ06} 
	P.A.~Pearce, J.~Rasmussen, J.-B.~Zuber, 
	{\em Logarithmic minimal models}, 
	J. Stat. Mech. (2006) P11017, 
	\arxiv{hep-th/0607232}{\!\!}.
	
\bib{PRT14}
	P.A.~Pearce, J.~Rasmussen, I.Y.~Tipunin, 
	{\em Critical dense polymers with Robin boundary conditions, half-integer Kac labels and ${\mathbb Z}_4$ fermions}, 
	Nucl. Phys. {\bf B889} (2014) 580--636,
	\arxiv{1405.0550}{[hep-th]}.

\bib{BPT2016}
	J.-E.~Bourgine, P.A.~Pearce, E.~Tartaglia, 
	{\em Logarithmic minimal models with Robin boundary conditions}, 
	J. Stat. Mech. (2016) 063104,
	\arxiv{1601.04760}{[hep-th]}.
	
\bib{CJV17}
	R.~Couvreur, J.L.~Jacobsen, R.~Vasseur,
	{Non-scalar operators for the Potts model in arbitrary dimension},
	\arxiv{1704.02186}{[cond-mat.stat-mech]}.	
	
\bib{PRcosetBranch13}
	P.A.~Pearce, J.~Rasmussen, 
	{\em Coset construction of logarithmic minimal models: branching rules and branching functions},
	J. Phys. A {\bf 46} (2013) 355402,
	\arxiv{1305.7304}{[hep-th]}.

\bib{MathieuRidout}
	P.~Mathieu, D.~Ridout, 
	{\em From percolation to logarithmic conformal field theory}, 
	Phys. Lett. {\bf B29} (2007) 120--129,
	\arxiv{0708.0802}{[hep-th]}.
 
\bib{MooreSeiberg}
	G.~Moore, N.~Seiberg, 
	{\em Classical and quantum conformal field theory}, 
	Commun. Math. Phys. {\bf 123} (1989) 177--254.

\bib{MDSA2011}
	A.~Morin-Duchesne, Y.~Saint-Aubin, 
	{\em The Jordan structure of two-dimensional loop models}, 
	J. Stat. Mech. (2011) P04007,
	\arxiv{1101.2885}{[math-ph]}.

\bib{MDSA13no2}
	A.~Morin-Duchesne, Y.~Saint-Aubin, 
	{\em Jordan cells of periodic loop models}, 
	J. Phys. A: Math. Theor. {\bf 46} (2013) 494013, 
	\arxiv{1302.5483}{[math-ph]}.

\bib{CardyModInv} 
	J.L.~Cardy, 
	{\em Operator content of two-dimensional conformally invariant theories}, 
	Nucl. Phys. {\bf B270} (1986) 186--204.

\bib{CIZ} 
	A.~Cappelli, C.~Itzykson, J.-B.~Zuber, 
	{\em Modular Invariant Partition Functions in Two Dimensions}, 
	Nucl. Phys. {\bf B280} (1987) 445--465.

\bib{Smirnov01} 
	S.~Smirnov, 
	{\em Critical percolation in the plane: conformal invariance, Cardy's formula, scaling limits}, 
	Comptes Rendus de l'Acad\'emie des Sciences, Series I - Mathematics, 
	{\bf 333} (2001) 239--244.

\bib{AAMRH}
	J.~Asikainen, A.~Aharony, B.B.~Mandelbrot, E.M.~Rauch, J.-P.~Hovi, 
	{\em Fractal geometry of critical Potts clusters}, 
	Euro. Phys. J. {\bf B34} (2003) 479--487, 
	\arxiv{cond-mat/0212216}{[cond-mat.dis-nn]}.

\bib{JS2005}
	W.~Janke, A.M.J.~Schakel, 
	{\em Fractal structure of spin clusters and domain walls in the two-dimensional Ising model}, 
	Phys. Rev. {\bf E71} (2005) 036703, 
	\arxiv{cond-mat/0410364}{[cond-mat.stat-mech]}.

\bib{JS2006}
	W.~Janke, A.M.J.~Schakel, 
	{\em Two-dimensional critical Potts and its tricritical shadow}, 
	Braz. J. Phys. {\bf 36} (2006) 708--716, 
	\arxiv{cond-mat/0612650}{[cond-mat.stat-mech]}.

\bib{SAPR2009}
	Y. Saint-Aubin, P.A.~Pearce, J.~Rasmussen, 
	{\em Geometric exponents, SLE and logarithmic minimal models}, 
	J. Stat. Mech. (2009) P02028,
	\arxiv{0809.4806}{[cond-mat.stat-mech]}.

\bib{Mandelbrot}
	B.B.~Mandelbrot, 
	{\em The Fractal Geometry of Nature}, 
	W.H.~Freeman and Co. (1982).

\bib{LSW01}
	G.F.~Lawler, O.~Schramm, W.~Werner, 
	{\em The dimension of the Brownian frontier is $4/3$}, 
	Math. Rev. Lett. {\bf 8} (2001) 13--24,
	\arxiv{math/0010165}{[math.PR]}.

\bib{Beffara}
	V.~Beffara,
	{\em Hausdorff dimensions for $\mbox{SLE}_6$}, 
	Ann. Prob. {\bf 32} (2004) 2606--2629,
	\arxiv{math/0204208}{[math.PR]}.

\bib{PR07} 
	P.A.~Pearce, J.~Rasmussen, 
	{\em Solvable critical dense polymers},
	J. Stat. Mech. (2007) P02015,
	\arxiv{hep-th/0610273}{\!\!}.
	
\bib{PRVcyl2010}
	P.A.~Pearce, J.~Rasmussen, S.P.~Villani, 
	{\em Solvable dense polymers on the cylinder}, 
	J. Stat. Mech. (2010) P02010,
	\arxiv{0910.4444}{[hep-th]}.

\bib{MD11} 
	A.~Morin-Duchesne, 
	{\em A proof of selection rules for critical dense polymers}, 
	J. Phys. A: Math. Theor. {\bf 44} (2011) 495003, 
	\arxiv{1109.6397}{[math-ph]}. 
	
\bib{PRV1210}
	P.A.~Pearce, J.~Rasmussen, S.P.~Villani,
	{\em Infinitely extended Kac table of solvable critical dense polymers},
	J. Phys. A: Math. Theor. {\bf 46} (2013) 175202,
	\arxiv{1210.8301}{[math-ph]}.
 
\bib{MDPR13}
 	A.~Morin-Duchesne, P.A.~Pearce, J.~Rasmussen, 
	{\em Modular invariant partition function of critical dense polymers}, 
	Nucl. Phys. {\bf B874} (2013) 312--357,
	\arxiv{1303.4895}{[hep-th]}.
	
\bib{PRcosetGraphs11}
	P.A.~Pearce, J.~Rasmussen, 
	{\em Coset graphs in bulk and boundary logarithmic minimal models}, 
	Nucl. Phys. {\bf B846} (2011) 616--649,
	\arxiv{1010.5328}{[hep-th]}.

\bib{MDPR14}
	A.~Morin-Duchesne, P.A.~Pearce, J.~Rasmussen, 
	{\em Fusion hierarchies, T-systems and Y-systems of logarithmic minimal models},
	J. Stat. Mech. (2014) P05012,
	\arxiv{1401.7750}{[math-ph]}.
	
\bib{PTC2015}
	P.A.~Pearce, E.~Tartaglia, R.~Couvreur, 
	{\em Kac boundary conditions of the logarithmic minimal models}, 
	J. Stat. Mech. (2015) P01018, 
	\arxiv{1410.0103}{[hep-th]}.

\bib{MDRD2015} 
	A.~Morin-Duchesne, J.~Rasmussen, D.~Ridout, 
	{\em Boundary algebras and Kac modules for logarithmic minimal models}, 
	Nucl. Phys. {\bf B899} (2015) 677--769,
	\arxiv{1503.07584}{[hep-th]}.

\bib{BR1989}
	V.V.~Bazhanov, N.~Reshetikhin,
	{\em Critical RSOS models and conformal field theory},
	Int. J. Mod. Phys. {\bf A4} (1989) 115--142.

\bib{Zam1991a}
	A.B.~Zamolodchikov,
	{\em On the thermodynamic Bethe ansatz equations for reflectionless ADE scattering theories},
	Phys. Lett. {\bf B253} (1991) 391--394.
 
\bib{Zam1991b}
 	A.B.~Zamolodchikov,
	{\em Thermodynamic Bethe ansatz for RSOS scattering theories}, 
	Nucl. Phys. {\bf B358} (1991) 497--523.
 
\bib{PK91}
	P.A.~Pearce, A.~Kl\"umper, 
	{\em Finite-size corrections and scaling dimensions of solvable lattice models: an analytic method}, 
	Phys. Rev. Lett. {\bf 66} (1991) 974--977.
 
 \bib{KP92}
	A.~Kl\"umper, P.A.~Pearce, 
	{\em Conformal weights of RSOS lattice models and their fusion hierarchies}, 
	Physica {\bf A183} (1992) 304--350.
 
\bib{KunibaNS9309}
 	A.~Kuniba, T.~Nakanishi, J.~Suzuki,
	{\em Functional relations in solvable lattice models I: functional relations and representation theory},
	Int. J. Mod. Phys. {\bf A9} (1994) 5215--5266,
	\arxiv{hep-th/9309137}{\!\!}.
 
\bib{KunibaNS9310}
	A.~Kuniba, T.~Nakanishi, J.~Suzuki,
	{\em Functional relations in solvable lattice models II},
	Int. J. Mod. Phys. {\bf A9} (1994) 5267--5312,
	\arxiv{hep-th/9310060}{\!\!}.
 
 \bib{KunibaNS1010}
	A.~Kuniba, T.~Nakanishi, J.~Suzuki,
	{\em $T$-systems and $Y$-systems in integrable systems},
	J. Phys. A: Math. Theor. {\bf 44} (2011) 103001,
	\arxiv{1010.1344}{[hep-th]}.
 
\bib{KLWZ1997} 
	I.~Krichever, O.~Lipan, P.~Wiegmann, A.~Zabrodin,
	{\em Quantum integrable models and discrete classical Hirota equations}, 
	Commun. Math. Phys. {\bf 188} (1997) 267--304, 
	\arxiv{hep-th/9604080}{\!\!}.
 
\bibitem{YangYang} 
	C.N.~Yang, C.P.~Yang, 
	{\em Thermodynamics of a one-dimensional system of bosons with repulsive delta-function interaction}, 
	J. Math. Phys. {\bf 10} (1969) 1115.

\bibitem{T71}
	M.~Takahashi,
	{\em One-dimensional Heisenberg model at finite temperature},
	Prog. Theor. Phys. {\bf 46} (1971) 401.
	
\bibitem{G71}
	M.~Gaudin,
	{\em Thermodynamics of the Heisenberg-Ising ring for $\Delta\ge 1$},
	Phys. Rev. Lett. {\bf 26} (1971) 1301.

\bibitem{Zam90} 
	A.B.~Zamolodchikov, 
	{\em Thermodynamic Bethe ansatz in relativistic models: Scaling $3$-state Potts and Lee-Yang models}, 
	Nucl. Phys. {\bf B342} (1990) 695--720.

\bibitem{Zam91} 
	A.B.~Zamolodchikov, 
	{\em On the thermodynamic Bethe ansatz equations for ADE reflectionless scattering theories}, 
	Phys. Lett. {\bf B253} 391--394; 
	{\em Thermodynamic Bethe ansatz for RSOS scattering theories}, 
	Nucl. Phys. \textbf{B358} (1991) 497--523; 
	{\em From tricritical Ising to critical Ising by thermodynamic Bethe ansatz}, 
	Nucl. Phys. \textbf{B358} (1991) 524--546; 
	{\em TBA equations for integrable perturbed $SU(2)_k\times SU(2)_1/SU(2)_{k+1}$ coset models}, 
	Nucl. Phys. \textbf{B366} (1991) 122--132.

\bib{CMP02} 
	C.~Chui, C.~Mercat, P.A.~Pearce, 
	{\em Integrable boundaries and universal TBA functional equations}, 
	Prog. Math. Phys. {\bf 23} (2002) 391--413,
 	\arxiv{hep-th/0108037}{\!\!}.
	
\bibitem{Sklyanin}  
	E.K.~Sklyanin, 
	{\em Boundary conditions for integrable quantum systems}, 
	J. Phys. {\bf A21} (1988) 2375--2389.

\bib{BPO96}
	R.E.~Behrend, P.A.~Pearce, D.L.~O'Brien, 
	{\em Interaction-round-a-face models with fixed boundary conditions: The ABF fusion hierarchy}, 
	J. Stat. Phys. {\bf 84} (1996) 1--48,
	\arxiv{hep-th/9507118}{\!\!}.

\bib{KP91}
	A.~Kl\"umper, P.A.~Pearce, 
	{\em Analytic calculation of scaling dimensions: Tricritical hard squares and critical hard hexagons}, 
	J. Stat. Phys. {\bf 64} (1991) 13--76.

\bib{OPW97}
	D.L.~O'Brien, P.A.~Pearce, S.O.~Warnaar, 
	{\em Analytic calculation of conformal partition functions: Tricritical hard squares with fixed boundaries}, 
	Nucl. Phys. {\bf B501} (1997) 773--799.

\bib{BDP15}
	Z.~Bajnok, O.~el Deeb, P.A.~Pearce,Ê
	{\em Finite-volume spectra of the Lee-Yang model},Ê
	JHEP {\bf 04} (2015) 73.

\bib{KSS98}
	A.~Kuniba, K.~Sakai, J.~Suzuki,
	{\em Continued fraction TBA and functional relations in XXZ model at root of unity},
	Nucl. Phys. {\bf B525} (1998) 597--626,
	\arxiv{math/9803056}{[math.QA]}.

\bib{OPW1996}
	D.L.~O'Brien, P.A.~Pearce, S.O.~Warnaar, 
	{\em Finitized conformal spectrum of the Ising model on the cylinder and torus}, 
	Physica A {\bf 228} (1996) 63--77.

\bib{BGN01}
	M.T.~Batchelor, J.~de Gier, B.~Nienhuis,
	{\em The quantum symmetric XXZ chain at $\Delta=-1/2$, alternating sign matrices and plane partitions},
	J. Phys. A: Math. Gen. {\bf 34} (2001) L265,
	\arxiv{cond-mat/0101385}{[cond-mat.stat-mech]}.
	
\bib{MBNM04}
	S.~Mitra, B.~Nienhuis, J.~de Gier, M.T.~Batchelor,
	{\em Exact expressions for correlations in the ground state of the dense $O(1)$ loop model},
	J. Stat. Mech. (2004) P09010,
	\arxiv{cond-mat/0401245}{[cond-mat.stat-mech]}.	

\bib{Jones83}
	V.~Jones,
	{\em Index for subfactors},
	Invent. Math. {\bf 72} (1983) 1--25.

\bib{M91}
	P.~Martin,
	{\em Potts models and related problems in statistical mechanics},
  	World Scientific, Singapore (1991).

\bib{GW93}
	F.~Goodman, H.~Wenzl,
	{\em The Temperley-Lieb algebra at roots of unity},
	Pacific J. Math. {\bf 161} (1993) 307--334.

\bib{W95}
	B.~Westbury,
	{\em The representation theory of the Temperley-Lieb algebras},
	Math. Zeit. {\bf 219} (1995) 539--565.

\bib{RSA14}
    	D.~Ridout, Y.~Saint-Aubin, 
    	{\em Standard modules, induction and the structure of the Temperley-Lieb algebra},
    	Adv.~Theo.~Math.~Phys. {\bf 18} (2014) 957--1041,
    	\arxiv{1204.4505}{[math-ph]}.
	
\bib{RJ06}	
	J.F.~Richard, J.L.~Jacobsen,
	{\em Character decomposition of Potts model partition functions. I. Cyclic geometry},
	Nucl. Phys. {\bf B750} (2006)
	250--264,
	\arxiv{math-ph/0605016}{[math-ph]}.

\bib{L91}
	D.~Levy,
	{\em Algebraic structure of translation-invariant spin-$\frac12$ XXZ and $q$-Potts quantum chains},
	Phys. Rev. Lett. {\bf 67} (1991) 1971--1974.

\bib{MS93}
	P.~Martin, H.~Saleur,
	{\em On an algebraic approach to higher dimensional statistical mechanics},
	Commun. Math. Phys. {\bf 158} (1993) 155--190,
	\arxiv{hep-th/9208061}{\!\!}.
 
\bib{GL98}
	J.J.~Graham, G.I.~Lehrer,
	{\em The representation theory of affine Temperley-Lieb algebras},
	Enseign. Math. {\bf 44} (1998) 173--218.

\bib{G98}
	R.M.~Green,
	{\em On representations of affine Temperley-Lieb algebras},
	CMS Conf. Proc. {\bf 24} (1998) 245--261.
 
\bib{EG99}
 	K.~Erdmann, R.M.~Green,
	{\em On representations of affine Temperley-Lieb algebras, II},
	Pac. J. Math. {\bf 191} (1999) 243--274,
	\arxiv{math/9811017}{[math.RT]}.
	
\bib{RS01}
	A.V.~Razumov, Y.G.~Stroganov,
	{\em Spin chains and combinatorics},
	J. Phys. A: Math. Theor. {\bf 34} (2001) 3185--3190,
	\arxiv{cond-mat/0012141}{[cond-mat.stat-mech]}.
	
\bib{RS01no2}
	A.V.~Razumov, Y.G.~Stroganov,
	{\em Spin chains and combinatorics: twisted boundary conditions},
	J. Phys. A: Math. Theor. {\bf 34} (2001) 5335--5340,
	\arxiv{cond-mat/0102247}{[cond-mat.stat-mech]}.	
	
\bib{S01}
	Y.G.~Stroganov, 
	{\em The importance of being odd}, 
	J. Phys. A {\bf 34} (2001) L179--L185,
	\arxiv{cond-mat/0012035}{[cond-mat.stat-mech]}.
		
\bibitem{RJ07} 
 	J.F.~Richard, J.L.~Jacobsen, 
	{\em Eigenvalue amplitudes of the Potts model on a torus},
 	Nucl. Phys.~{\bf B769} (2007) 
	256--274, 
	\arxiv{math-ph/0608055}{[math-ph]}.

\bib{AK10}
	B.~Aufgebauer, A.~Kl\"umper,
	{\em Quantum spin chains of Temperley-Lieb type: periodic boundary conditions, spectral multiplicities and finite temperature},
	J. Stat. Mech. (2010) P05018, 
	\arxiv{1003.1932}{[cond-mat.stat-mech]}.	
	
\bib{SB89}
	H.~Saleur, M.~Bauer,
	{\em On some relations between local height probabilities and conformal invariance},
	Nucl. Phys. {\bf B320} (1989) 591--624.

\bib{C06}
	J.~Cardy,
	{\em The $O(n)$ model on the annulus},
	J. Stat. Phys. {\bf 125} (2006)
	\arxiv{math-ph/0604043}.

\bib{GRW10} 
	M.R.~Gaberdiel. I.~Runkel, S.~Wood,
	{\em A modular invariant bulk theory for the $c=0$ triplet model},
	J. Phys. A {\bf 44} (2010)
	\arxiv{1008.0082}{[hep-th]}.
	
\bib{MDSA13no1}
	A.~Morin-Duchesne, Y.~Saint-Aubin, 
	{\em A homomorphism between link and XXZ modules over the periodic Temperley-Lieb algebra}, 
	J. Phys. A: Math. Theor. {\bf 46} (2013) 285207, 
	\arxiv{1203.4996}{[math-ph]}.	

\bib{ABGR88}
	F.C.~Alcaraz, M.~Baake, U.~Grimm, V.~Rittenberg,
	{\em Operator content of the XXZ chain},
	J. Phys. A: Math. Gen. {\bf 21} (1988) L117--120.

\bib{AGR89}
	F.C.~Alcaraz, U.~Grimm, V.~Rittenberg,Ê
	{\em The XXZ Heisenberg chain, conformal invariance and the operator content of $c<1$ systems},Ê
	Nucl. Phys. {\bf B316} (1989) 735--768.

\bib{RS07}
	N.~Read, H.~Saleur,
	{\em Associative-algebraic approach to logarithmic conformal field theories},
	Nucl.~Phys. {\bf B777} (2007) 316--351,
	\arxiv{hep-th/0701117}{\!\!}.
	
\bib{GR08}
	M.R.~Gaberdiel. I.~Runkel,
	{\em From boundary to bulk in logarithmic CFT},
	J. Phys. A: Math. Theor. {\bf 41} (2008) 075402,
	\arxiv{0707.0388}{[hep-th]}.

\bib{RGW14}
	I.~Runkel, M.R.~Gaberdiel, S. Wood,
	{\em Logarithmic bulk and boundary conformal field theory and the full centre construction},
	in Bai {\em et al.} (eds) Conformal Field Theories and Tensor Categories, Math. Lect. from Peking Univ., Springer (2014), 
	\arxiv{1201.6273}{[hep-th]}.	

\bib{GRSV15}
	A.M.~Gainutdinov, N.~Read, H.~Saleur, R.~Vasseur,
	{\em The periodic $s\ell(2|1)$ alternating spin chain and its continuum limit as a bulk logarithmic conformal field theory at $c=0$},
	J.~High~Energ.~Phys. {\bf 114} (2015),
	\arxiv{1409.0167}{[hep-th]}.		
		
\bib{JS08}	
	J.L.~Jacobsen, H.~Saleur,
	{\em Conformal boundary loop models},
	Nucl.~Phys. {B788} (2008) 137--166,
	\arxiv{math-ph/0611078}{\!\!}.		
	
\bib{DJS09}
	J.~Dubail, J.L.~Jacobsen, H.~Saleur,
	{\em Conformal two-boundary loop model on the annulus},
	Nucl.~Phys. {\bf B813} (2009) 430--459,
	\arxiv{0812.2746}{\!\!}.

\bib{R79}
	J.~Riordan,
	{\em Combinatorial identities},
	Robert E. Krieger Pub. Co (1979).		

\end{thebibliography}
\end{document}